\PassOptionsToPackage{unicode}{hyperref}
\PassOptionsToPackage{hyphens}{url}
\PassOptionsToPackage{dvipsnames,svgnames*,x11names*}{xcolor}
\documentclass[
  11pt,
  letterpaper,
]{article}
\usepackage{lmodern}
\usepackage{amssymb,amsmath}
\usepackage{ifxetex,ifluatex}
\ifnum 0\ifxetex 1\fi\ifluatex 1\fi=0 
  \usepackage[T1]{fontenc}
  \usepackage[utf8]{inputenc}
  \usepackage{textcomp} 
\else 
  \usepackage{unicode-math}
  \defaultfontfeatures{Scale=MatchLowercase}
  \defaultfontfeatures[\rmfamily]{Ligatures=TeX,Scale=1}
\fi
\IfFileExists{upquote.sty}{\usepackage{upquote}}{}
\IfFileExists{microtype.sty}{
  \usepackage[]{microtype}
  \UseMicrotypeSet[protrusion]{basicmath} 
}{}
\makeatletter
\@ifundefined{KOMAClassName}{
  \IfFileExists{parskip.sty}{%
    \usepackage{parskip}
  }{
    \setlength{\parindent}{0pt}
    \setlength{\parskip}{6pt plus 2pt minus 1pt}}
}{
  \KOMAoptions{parskip=half}}
\makeatother
\usepackage{xcolor}
\IfFileExists{xurl.sty}{\usepackage{xurl}}{} 
\IfFileExists{bookmark.sty}{\usepackage{bookmark}}{\usepackage{hyperref}}
\hypersetup{
  colorlinks=true,
  linkcolor=blue,
  filecolor=Maroon,
  citecolor=Blue,
  urlcolor=Blue,
  pdfcreator={LaTeX via pandoc}}
\usepackage[margin=1in]{geometry}
\usepackage{longtable,booktabs}
\usepackage{etoolbox}
\makeatletter
\patchcmd\longtable{\par}{\if@noskipsec\mbox{}\fi\par}{}{}
\makeatother
\IfFileExists{footnotehyper.sty}{\usepackage{footnotehyper}}{\usepackage{footnote}}
\makesavenoteenv{longtable}
\providecommand{\tightlist}{%
  \setlength{\itemsep}{0pt}\setlength{\parskip}{0pt}}
\title{The Instrumental Dissolution of Typing: Why AI Challenges the Keyboard Era in Knowledge Work}
\author{Wei Roy Hua, PhD\\
\emph{8T5 Innovations}\\
\texttt{roy@8t5.com}}
\date{April 2026}

\begin{document}

\maketitle

\begin{abstract}

For four decades, the QWERTY keyboard organized white-collar knowledge
work. Typing's dominance was instrumental, not cognitively necessary. As
multimodal AI achieves human-parity understanding of speech and gesture,
this necessity dissolves.

We introduce \emph{instrumental dissolution}---loss of
institutional-default status while persisting in specialist niches. The
keyboard era ends not through hardware replacement but through migration
of its function into AI systems. Like telegraph operation and
handwriting before it, typing dissolves as universal mode while
persisting where its affordances remain essential.

The central contribution identifies the \emph{verification bottleneck}:
as AI collapses production friction, the primary constraint shifts from
generation to evaluation. Knowledge workers become adversarial auditors
rather than keystroke-producers. This restructures professional
expertise, organizational communication, and how productive labor is
recognized.

Converging evidence from history, philosophy, neuroscience, technology,
organizational studies, and cultural analysis supports this thesis. We
map \emph{synthetic literacy}---oral input generating literate
output---as the defining feature of this transition.

Under three scenarios (optimistic: 2028-2035; base: 2035-2045;
pessimistic: 2045-2060), transition rates are shaped by institutional
adoption, regulation, and generational turnover. We specify three
disconfirmation criteria at 2028, 2030, and 2035 that would weaken the
thesis if observed.

Typing persists in specialized contexts, but its structural foundation
role ends. This demands reorientation toward verification design. We
propose seven interface primitives operationalizing
verification-centered HCI: contribution provenance, claim-level evidence
mapping, contrastive verification views, critical-friction checkpoints,
role-based ratification workflows, persistent verification memory, and
scaffolded composition modes.
\end{abstract}


\newpage
{
\hypersetup{linkcolor=}
\setcounter{tocdepth}{3}
\tableofcontents
}
\newpage

\hypertarget{introduction-the-keyboard-as-organizing-principle}{%
\section{1. Introduction: The Keyboard as Organizing
Principle}\label{introduction-the-keyboard-as-organizing-principle}}

\hypertarget{the-ubiquity-of-the-keystroke}{%
\subsection{1.1 The Ubiquity of the
Keystroke}\label{the-ubiquity-of-the-keystroke}}

Why does keyboard-based typing remain the dominant input method for
knowledge work despite decades of viable alternatives (speech, gesture,
handwriting, even direct brain-computer interfaces on the horizon)? The
puzzle is not that typing exists; it is that it persists as the
universal default, the input method that organizational systems,
software interfaces, and professional workflows assume by design. If
viable alternatives have been available for years, why has keyboard
dominance remained structurally unshaken, economically rewarded, and
cognitively normalized? This persistence demands explanation.

Consider the workday of a knowledge worker in 2025. They arrive at their
desk (or open their laptop at home) and begin typing. They type emails:
117 to 121 per day on average, consuming roughly 28\% of their working
hours (McKinsey Global Institute, 2012); a proportion that has remained
stable as instant messaging has intensified without displacing
typing-based communication. They type into documents: reports,
memoranda, proposals, analyses, each constructed character by character
at approximately 40 words per minute (Salthouse, 1984). They type into
Slack and Microsoft Teams: available industry telemetry suggests
approximately 92 instant messages per day, adding a rapid-fire oral-like
text layer atop the formal composition of email. They type code:
software engineers produce thousands of lines daily, each syntactically
precise keystroke carrying functional consequence. They type into
spreadsheets, databases, CRM systems, project management tools. They
type search queries, calendar entries, meeting notes. They switch
between applications 1,200 times per day, losing four hours per week to
context-switching between communication formats. All share a single
common assumption: the human produces structured text through keystrokes
(Microsoft 365 Work Trend Index, 2025).

This is more than a description of a technology in use. It is a
description of a technology that has become the organizing principle of
professional life. For four decades---since the personal computer placed
a keyboard on every knowledge worker's desk in the mid-1980s---typing
has served as the structural foundation of white-collar knowledge work.
It is the interface through which professionals compose their thoughts,
coordinate their organizations, produce their outputs, and construct
their professional identities. The keyboard is not incidental to modern
knowledge work. It is constitutive of it.

The statistics of typing's dominance are extraordinary. Roughly half the
U.S. workforce---up from 25\% in 1984---uses a computer keyboard as
their primary work instrument (Autor et al., 2003; Forman \& Setzler,
1998). Global email volume reaches 361.6 billion messages daily
(Radicati Group, 2024). GitHub hosts over 420 million repositories of
code, overwhelmingly produced through keyboard input. The keyboard
market itself is expected to grow from \$5.5 billion in 2023 to \$9.6
billion by 2035 (Data Bridge Market Research, 2024). Workers who used
computer keyboards earned a 10--15\% wage premium over those who did not
in the critical adoption period of the 1990s (Krueger, 1993), creating
powerful economic incentives that accelerated universal adoption. Typing
became not just a skill but a literacy---the baseline competency without
which professional participation was impossible.

\hypertarget{defining-dominance-institutional-default}{%
\subsection{1.2 Defining ``Dominance'': Institutional
Default}\label{defining-dominance-institutional-default}}

Throughout this analysis, \emph{dominance} refers primarily to
\textbf{institutional-default status}: the input method assumed by
organizations, software systems, and workflow design as the standard
means of producing structured text. This is distinct from (though
correlated with) market share dominance (the quantity of keyboards sold
relative to alternatives) and cognitive dominance (the claim that typing
is cognitively superior to alternatives). The paper engages with all
three axes empirically, but the primary thesis concerns institutional
default: the fact that organizations designed email systems, document
workflows, code repositories, and professional communication genres
around the assumption that humans produce structured text through
keyboard input. This institutional embedding explains why typing
persists even as viable alternatives emerge; alternatives must overcome
individual preference and organizational infrastructure, workflow
normalization, and professional identity construction---moving beyond
preference to institutional transformation. The three dimensions
reinforce each other, but institutional default is the primary mechanism
of persistence that this analysis addresses.

\hypertarget{what-typing-built}{%
\subsection{1.3 What Typing Built}\label{what-typing-built}}

Typing's dominance was never merely technical. It reshaped the
organizational architecture of knowledge work in ways that are now so
deeply embedded as to be invisible. The formal business email (the
dominant genre of professional communication for three decades) is an
artifact of the keyboard's affordances: it assumes asynchronous
composition, revision-based editing, and the production of a permanent
textual record through deliberate keystroke-by-keystroke authorship.
Slack and instant messaging, despite their oral-like rapidity, are still
keyboard-mediated: typed text at conversational speed, producing
searchable archives that serve as institutional memory. The slide deck,
the spreadsheet, the code repository, the project management
board---each communication format that organizations depend upon was
designed around the assumption that human input arrives as structured
text produced through keyboard interaction.

The typewriter (as JoAnne Yates documented in \emph{Control Through
Communication}, 1989) went far beyond speeding up correspondence; it
created the modern business memo, the filing system, and the
bureaucratic apparatus of the twentieth-century corporation. The
computer keyboard extended this organizational restructuring by orders
of magnitude: it enabled email, collaborative documents,
version-controlled code, database management, and the entire digital
infrastructure of contemporary institutional life. Every workflow, every
approval process, every compliance system, every communication genre in
the modern knowledge economy carries the keyboard's fingerprint.

Typing also restructured professional identity. It entered the workplace
as gendered labor (the domain of the female secretary in the typing
pool) and was resisted by male executives who viewed ``key bashing'' as
beneath their station (Yates, 1989; Strom, 1992). The personal computer
inverted this hierarchy: every professional became their own typist, and
keyboard fluency became a marker of competence, not subordination.
Education systems responded by teaching typing as a core skill,
embedding it alongside reading and arithmetic as a fundamental literacy.
An entire generation of workers---the knowledge class of the late
twentieth and early twenty-first centuries---constructed their
professional identities around the keyboard as the instrument of serious
intellectual work.

\hypertarget{the-conservative-consensus}{%
\subsection{1.4 The Conservative
Consensus}\label{the-conservative-consensus}}

Given this depth of embedding, the prevailing industry and academic
assumption is that AI ``augments'' typing but does not displace it. The
keyboard endures, the argument runs, because it remains the most
precise, private, and universally accessible means of producing
structured text. AI makes typing faster (autocomplete, Smart Compose),
more accurate (grammar checking, code completion), and more productive
(draft generation, summarization), but the human still types. The
keyboard's position is repositioned---from primary producer to
editor-in-chief---but its structural centrality survives. This is the
conservative consensus, and it is held by a majority of industry
analysts, HCI researchers, and technology commentators.

\hypertarget{the-thesis}{%
\subsection{1.5 The Thesis}\label{the-thesis}}

The consensus outlined above misidentifies the nature of typing's
dominance. Typing did not dominate knowledge work because it was
cognitively valuable, or phenomenologically rich, or organizationally
optimal. It dominated because digital machines required structured text
input, and the keyboard was the most efficient means of producing it.
Typing's dominance was \emph{instrumental}; it served a function:
bridging the gap between human thought and machine-readable text. No
other available technology could serve this function as well.

Generative AI and multimodal AI systems dissolve this instrumental
necessity in the linguistic-knowledge-work domains examined here. When
machines can understand natural speech with sub-3\% word error rates,
interpret gesture and gaze in real time with emerging capability, decode
electromyographic signals from a wristband as early research
demonstrates, and process all of these signals simultaneously in a
unified context window, they no longer require humans to perform the
cognitively costly translation of thought into sequential keystrokes for
institutional document production. The function that justified the
keyboard's dominance---the bridge between human intent and
machine-readable text---is increasingly absorbed into the machine itself
across an expanding range of knowledge-work contexts. Where this
absorption is functionally complete, the bridge becomes optional, not
necessary. The keyboard moves from universal default toward specialist
instrument, on the same trajectory as the inscription technologies it
once displaced.

A critical scope clarification: this analysis examines \textbf{synthetic
literacy} (the capacity to formulate, edit, and take institutional
responsibility for machine-generated text) as it emerges across
knowledge-work domains. Professionals dictate intent; AI systems
generate institutional records in medicine, law, scholarship, and
business documentation. The analysis does not claim this dynamic is
universal to all literacy or all professional work, but rather that
AI-mediated composition enables this inversion precisely in domains
where textual output is institutionally mandated and where the
verification of machine-generated content becomes the binding cognitive
and organizational constraint.

The concept is distinct from two important predecessors. Walter Ong's
secondary orality describes the return of oral communication
characteristics (immediacy, participatory dynamics, formulaic
expression) within electronic media societies that remain structurally
literate in their institutional structure (Ong, 1982). Secondary orality
identifies a cultural register, not a structural shift in who performs
the cognitive work of literate composition; broadcast radio and
television were secondary-oral forms, but their producers still
performed literate cognition. Stuart Selber's \emph{Multiliteracies for
a Digital Age} describes a three-part pedagogical
framework---functional, critical, and rhetorical literacy---for teaching
students to engage with digital technologies in educationally
responsible ways (Selber, 2004). Selber's framework addresses how
students navigate computer environments; it remains focused on education
rather than organizational transformation. Synthetic literacy, by
contrast, designates a structural rather than a pedagogical phenomenon:
the \emph{delegation of grammatization}---the discretization of
continuous oral intent into reproducible literate inscription---to
machine intelligence, with the grammatizing agent shifting from human
practitioner to computational system. Stiegler's grammatization
framework describes inscription labor across all of human technical
history; we apply it here specifically to the compositional authority
for institutional knowledge work, identifying the moment when the human
agent performing discretization is replaced by a computational one
(Stiegler, 2010). This shift does not merely restructure educational
opportunity; it alters the five-thousand-year requirement that a human
mind perform the cognitive discipline of literate composition.

\hypertarget{instrumental-dissolution-conceptual-architecture}{%
\subsection{1.6 Instrumental Dissolution: Conceptual
Architecture}\label{instrumental-dissolution-conceptual-architecture}}

We term this process \textbf{instrumental dissolution} and situate it
within an explicit theoretical hierarchy. The concept is distinct from
technological replacement, in which a new tool performs the same
function more efficiently (as the word processor replaced the
typewriter). In replacement, the function persists; only the instrument
changes. In dissolution, the function itself is eliminated---not by a
better keyboard but by systems that absorb the translation function
entirely. This distinction is the primary theoretical contribution, and
it explains why historical transition timescales---which reflect the
time required to retrain motor skills for a new instrument performing
the same function---do not adequately predict the current transition.

A precise definition is required, because the term invites misreading.
We use ``dissolution'' deliberately, despite its ordinary-language
resonance with disappearance: what dissolves is the \emph{institutional
function}, not the instrument itself. By \emph{instrumental dissolution}
we mean the loss of an instrument's status as institutional default for
a function; it is accompanied by its continued specialist persistence in
domains where its specific affordances retain comparative advantage.
Dissolution is not extinction. The dissolved instrument typically
endures as a specialist tool: handwriting persists in note-taking,
signature, and graphic arts; the telegraph operator persists as an
amateur radio niche; film photography persists as an aesthetic and
archival choice. What changes is structural. The dissolved instrument no
longer organizes professional infrastructure, no longer commands the
wage premium that justified its universal acquisition, and no longer
constitutes the assumed mode of institutional output. Section 3 develops
this distinction in depth and marshals six historical precedents:
telegraph operation, the typewriter S-curve, QWERTY layout persistence,
film-to-digital photography, the oral-to-literate transition, and
handwriting's dissolution as institutional default. Each precedent
exhibits the same pattern: loss of default status with continued niche
persistence after a more capable substitute absorbs the institutional
function. We argue the keyboard is entering this trajectory in
linguistic knowledge work; we do not argue it disappears.

This framework is distinct from Don Ihde's postphenomenological taxonomy
of human-technology relations (embodiment, hermeneutic, alterity,
background), which classifies how technologies mediate experience but
does not address the structural conditions under which instrumental
functions dissolve entirely. Where Ihde asks \emph{how} technologies
mediate, instrumental dissolution asks \emph{when and why} the mediating
function itself becomes unnecessary. We do not claim instrumental
dissolution as a new philosophical category. We position it as a
pragmatic operationalization---a diagnostic test---for HCI and
organizational analysis, drawing on Borgmann's device paradigm, Ihde's
transparency, Heidegger's withdrawal, and Verbeek's mediation theory.
The three-part diagnostic developed in Section 3 (functional transfer,
legacy persistence, institutional decoupling) gives this framework
operational precision that Borgmann and Ihde do not provide for
predicting \emph{when} and \emph{to what extent} an instrumental
function has lost its monopoly.

The theoretical apparatus is organized in four explicit tiers: (1) a
core conceptual engine drawn from dual-nature theory, proper-function
analysis, and mediation theory, which establishes the distinction
between technological replacement and instrumental dissolution; (2) a
phenomenological register (Heidegger, Merleau-Ponty, Ihde) describing
what this transition feels like in lived practice---including,
specifically, the phenomenological character of \emph{synthetic
literacy}: what it means to compose by delegation to machine
intelligence rather than through the embodied disciplinary practice of
inscription; (3) a normative analysis (Borgmann, Stiegler) establishing
what may be lost when a focal practice becomes infrastructurally
transparent; and (4) engagement with the strongest objection to the
instrumental thesis, drawn from Material Engagement Theory (Malafouris),
which argues that artifacts are co-constitutive rather than merely
instrumental. This architecture tells the reader immediately which
traditions are doing what work and how they interrelate.

\hypertarget{the-verification-bottleneck-reordering-cognitive-labor}{%
\subsection{1.7 The Verification Bottleneck: Reordering Cognitive
Labor}\label{the-verification-bottleneck-reordering-cognitive-labor}}

The defining consequence we identify is the \textbf{verification
bottleneck}: as AI collapses the Gulf of Execution---reducing the effort
to produce text from labored keystroke composition to conversational
prompting---the primary cognitive and organizational burden shifts from
production to evaluation. Knowledge workers increasingly transition from
composers to auditors and evaluators---roles the analysis describes as
\emph{verifier}, \emph{validator}, or \emph{editor-in-chief} depending
on organizational context. This structural shift from production to
evaluation carries significant implications for professional expertise,
organizational routines, and institutional legitimacy. Rather than
eliminating cognitive work, dissolving typing's instrumental function
\emph{reorganizes} labor around verification as the binding constraint.
This mirrors, and significantly extends, what expert writers already
experience: Flower and Hayes (1981) documented that expert writing
involves continuous \emph{Monitor}---an evaluative process running
parallel to production, assessing emerging text against goals and
constraints. In AI-mediated workflows, the Monitor shifts from
background parallel process to primary foreground activity, elevated
from co-process to main task, as AI assumes the production role
previously held by humans.

Macro-economic data from the U.S. Bureau of Labor Statistics confirms
this recomposition: word processors and typists are projected to decline
36.1\% between 2024--2034, while verification-intensive occupations
(information security analysts at 33\% growth, management analysts at
11\%, and quality assurance roles) grow at above-average rates (BLS,
2024). This occupational restructuring empirically corroborates the
theoretical shift from production to verification.

\hypertarget{the-four-condition-framework}{%
\subsection{1.8 The Four-Condition
Framework}\label{the-four-condition-framework}}

The argument is unified by a single analytical lens. Typing persists as
the dominant input method of knowledge work. Four conditions must hold
simultaneously:

\textbf{(a)} Machines require structured text input from humans.
\textbf{(b)} Humans lack natural alternatives that machines can
interpret. \textbf{(c)} Organizational systems are optimized around
keyboard input. \textbf{(d)} Typing provides cognitive benefits that
justify its costs.

The analysis demonstrates that AI substantially weakens condition
(a)---machines increasingly understand natural speech, gesture, and
multimodal intent without requiring keystroke-mediated translation.
Technology now satisfies condition (b)---voice, electromyography, gaze
tracking, and eventually neural interfaces provide natural alternatives
that AI systems can interpret at or near human parity. Condition (c)
follows the dissolution of (a) and (b) with organizational lag:
communication genres and institutional workflows restructure when the
input method assumption changes, as every historical transition has
demonstrated. And condition (d) was never true in its strong form:
typing has bounded, modality-specific advantages (detailed in Section 4)
but not the broad, generalizable cognitive privilege required to sustain
long-run dominance under AI-mediated production.

As all four conditions weaken simultaneously, the structural rationale
for typing's dominance erodes. What remains is niche persistence: typing
in specific contexts where precision, privacy, accessibility, or
regulatory requirements sustain its value. But niche persistence is not
dominance. Handwriting persists today; no one calls it dominant. The
typewriter persists in specialized contexts; no one claims the
typewriter era continues. The keyboard is entering the same trajectory.

\hypertarget{evidence-classification-framework}{%
\subsection{1.9 Evidence Classification
Framework}\label{evidence-classification-framework}}

This manuscript distinguishes three tiers of evidentiary support for its
claims. \emph{Established} claims rest on peer-reviewed evidence,
replicated findings, or documented organizational data. \emph{Strongly
suggested} claims are supported by converging evidence from multiple
domains, early empirical signals, or robust theoretical inference, but
lack definitive longitudinal confirmation. \emph{Projected} claims
represent informed extrapolation from current trajectories, explicitly
marked as conjectural. Note: all future-dated forecasts and predictions
receive the projected designation regardless of source authority, as
forecasting inherently carries irreducible uncertainty beyond any single
authority's credibility. Throughout the analysis, we endeavor to make
this classification transparent, and readers should calibrate their
confidence accordingly.

\hypertarget{the-argument-structure}{%
\subsection{1.10 The Argument Structure}\label{the-argument-structure}}

This thesis develops through seven independent analytical routes, each
approaching from a different disciplinary lens and each contributing
evidence to one or more of the four conditions.

\textbf{Section 2: Historical Patterns.} Every major input method
transition---from orality to writing, from manuscript to print, from pen
to typewriter, from typewriter to computer keyboard---restructured
cognition, organization, and professional identity. The current
transition fits the historical pattern while introducing a qualitative
discontinuity: for the first time, the function of manual structured
input dissolves rather than migrating to a new instrument.

\textbf{Section 3: The Instrumental Dissolution Argument.} Drawing on
Heidegger, Ihde, Clark and Chalmers, and Verbeek, this section develops
the concept of instrumental dissolution as distinct from technological
replacement, demonstrating why typing's function is dissolving under AI.

\textbf{Section 4: The Empirical Case---Typing's Limited
Modality-Specific Privilege.} A systematic review establishing that
typing has bounded, modality-specific advantages but not the unique
cognitive benefits that would justify its continued dominance, and
identifying the verification bottleneck as the binding cognitive
constraint when production friction dissolves.

\textbf{Section 5: The Technology Trajectory.} Maps the replacement
landscape: voice-AI as the immediate successor, spatial computing and
electromyography as medium-term complements, neural interfaces on the
longer horizon, and the \textbf{agentic turn}---reorganization of
human-machine interaction around intent specification rather than
keystroke-level detail---as the architecture that renders the keyboard's
universal mediator function unnecessary.

\textbf{Section 6: Steelman Engagement.} Addresses every credible
counter-argument for typing's continuation---infrastructure lock-in,
acoustic constraints, precision tasks, accessibility, embodied
expertise, professional identity---presenting each at its strongest and
assessing whether it defeats the thesis.

\textbf{Section 7: The Verification Bottleneck Goes Organizational.}
Demonstrates how typing shaped communication genres and organizational
power structures, and traces how the verification bottleneck manifests
at organizational scale when production friction dissolves.

\textbf{Section 8: Secondary Orality and the Return of Voice.} Places
the transition in civilizational context through Ong's framework,
examining what is recovered through voice-mediated communication and
what risks emerge if compositional discipline is not intentionally
preserved.

\hypertarget{two-transformations-not-one}{%
\subsection{1.11 Two Transformations, Not
One}\label{two-transformations-not-one}}

A critical clarification: the present argument maintains that
AI-mediated systems dissolve typing's instrumental function at both the
\textbf{individual-cognitive level} and the
\textbf{organizational-structural level}, but these are distinct
phenomena operating on different timescales.

At the individual level, the cognitive transition from production to
verification is rapid. When a knowledge worker adopts voice-AI for
composition, the keystroke burden diminishes rapidly, while a
verification burden emerges in its place. The bottleneck shifts: from
``how do I translate my thought into characters'' to ``how do I verify
that the AI-generated output reflects my intent?'' This is a task-level
transition that can occur within a single work domain.

At the organizational level, the restructuring operates on a different
timescale. Email will not disappear when voice-AI achieves production
dominance; organizational genres will undergo skeuomorphic adaptation,
persisting in formal structure even as their production mechanisms
change. The ``keystroke imperative'' (the assumption that composed text
is the substrate of institutional coordination) will substantially erode
as organizations build new workflows around voice input and
verification-based output validation. This institutional reorganization
operates on a 15--30 year timeline for any given organization, not
months.

The directional shift away from keyboard dominance is supported by
converging evidence, though its speed and sectoral reach remain
genuinely uncertain. The technology is ready now (voices below 3\% word
error rate, multimodal AI systems interpret gesture and gaze). The
individual transition is underway in forward-moving organizations; the
organizational transition operates on a different timescale. Neither is
inevitable; both are contingent on deliberate design choices about how
to preserve compositional discipline while eliminating production
friction. But both are underway.

\hypertarget{hci-implications-and-the-future-of-input-design}{%
\subsection{1.12 HCI Implications and the Future of Input
Design}\label{hci-implications-and-the-future-of-input-design}}

The implications for HCI research are profound and immediate. The future
of HCI will likely be defined less by input design than by verification
design. How humans evaluate, ratify, and take responsibility for
machine-generated content emerges as a central interaction challenge of
the coming decade. As production friction dissolves, the design focus
must shift from optimizing keystroke efficiency to designing trustworthy
evaluation workflows, adversarial verification interfaces, and
accountability mechanisms that preserve human agency in an era of
agentic AI systems. The verification bottleneck analysis is clear: the
critical HCI work lies not in making typing faster, but in making
verification more cognitively tractable, more organizationally
transparent, and more aligned with professional expertise and
institutional responsibility.

\hypertarget{related-work}{%
\subsection{1.13 Related Work}\label{related-work}}

The argument builds upon and synthesizes insights from four major
research areas in HCI and human-computer interaction. Rather than
claiming novelty, this analysis synthesizes existing findings into a
specific architectural claim: as generative AI collapses input
production friction, verification becomes the dominant interaction
bottleneck, requiring a fundamental reorientation of interface design
priorities.

\hypertarget{ai-assisted-writing-and-writing-support-systems}{%
\subsubsection{1.13.1 AI-Assisted Writing and Writing-Support
Systems}\label{ai-assisted-writing-and-writing-support-systems}}

Recent HCI research has extensively mapped the design space for
intelligent and interactive writing assistants. A 2024 survey by Lee et
al.~analyzed 115 papers on AI-augmented writing tools, organizing them
across five dimensions: task specificity, user expertise level,
technology architecture, interaction paradigms, and evaluation
methodology. This survey establishes baseline understanding of what AI
writing support currently achieves and where gaps remain. Among concrete
systems: CoAuthor (Lee et al., 2022) demonstrates the productive
potential of collaborative drafting with AI; StepWrite addresses
voice-first long-form composition; systems like HaLLMark visualize AI
contribution provenance by showing which parts of output were generated
versus edited by the human writer. Experimental evidence now quantifies
the production friction reduction this body of work documents. Noy and
Zhang (2023) conducted one of the first randomized controlled trials of
generative AI in professional writing: workers randomized to ChatGPT
assistance completed writing tasks 37\% faster with 18\% higher quality
ratings. This provides direct empirical grounding for the
production-friction collapse the present analysis theorizes (Noy \&
Zhang, 2023). Gero, Liu, and Chilton's (2023) study of AI-assisted
science writing further documents the shift from
keystroke-as-primary-composition to voice-directed AI curation:
scientists used AI tools for ideation and structural scaffolding while
transitioning their primary cognitive effort toward evaluation and
selection---the production-to-verification shift this analysis forecasts
(Gero et al., 2023). The typology of human-AI writing modes developed by
the Co-Writing with AI Research Group (2025), including iterative
dialogic co-production, provides direct empirical grounding for the
four-mode typology in \S{}8.2a.

This body of work documents the tools, user experience patterns, and
measurable productivity effects of AI-assisted writing but rarely
addresses the verification challenge that emerges when humans must
evaluate machine-generated content at organizational scale. The
contributions here focus on preserving writer values and productive
effort in AI-augmented systems (Amershi et al., 2019), but the dominant
interaction challenge they identify remains authorship and control, not
verification and validation.

\hypertarget{human-ai-collaboration-reliance-and-calibration}{%
\subsubsection{1.13.2 Human-AI Collaboration, Reliance, and
Calibration}\label{human-ai-collaboration-reliance-and-calibration}}

Foundational work on human-AI collaboration establishes that the quality
of joint human-AI systems depends critically on how humans calibrate
reliance on algorithmic recommendations. Amershi et al.'s (2019)
eighteen guidelines for human-AI interaction provide a design framework,
emphasizing that ``the system should make clear why it did what it did''
and ``support efficient recovery from mistakes.'' Bansal et al.~(2021)
demonstrate that explanations significantly impact whether humans
over-rely or under-rely on AI systems---too much transparency can
paradoxically increase over-trust, while insufficient transparency
erodes appropriate reliance. Buçinca et al.~(2021) introduce cognitive
forcing functions---deliberate interface barriers that require explicit
human judgment rather than passive acceptance---arguing that
verification quality improves when systems demand articulated reasoning
rather than one-click confirmation. Kaur et al.~(2020) document the
``interpretability illusion'': even when AI systems provide detailed
explanations, users often develop false confidence in their
understanding, correlating with worse verification accuracy. These
findings directly inform the interface primitives discussed in \S{}9.2.1:
critical-friction checkpoints operationalize forcing functions;
contrastive verification views provide bounded transparency without
inducing illusion; persistent verification memory captures the reasoning
process users articulate.

\hypertarget{transparency-provenance-and-auditing}{%
\subsubsection{1.13.3 Transparency, Provenance, and
Auditing}\label{transparency-provenance-and-auditing}}

The emerging literature on AI auditing and accountability emphasizes
that verification capability depends on whether evidence is surfaced and
traceable. Raji et al.~on AI auditing ecosystems establish that
organizational ability to verify AI outputs requires both: (a)
visibility into the system's decision process (Raji et al., 2020), and
(b) structured audit trails showing what was checked, by whom, and with
what confidence. Verbeek's (2005) analysis of technological mediation
argues that technologies shape what is visible and what is hidden,
determining the conditions under which humans can exercise moral
judgment. Verification overlaps directly with these concerns: who checks
machine-generated content? What is checkable (what evidence is
surfaced)? Who signs off, and what counts as adequate evidence of due
diligence? What gets logged, and for how long? These questions span
technical architecture, organizational governance, and professional
accountability. The contribution here situates verification design as a
core HCI challenge, moving beyond technical auditing toward the
interface and workflow level.

\hypertarget{multimodal-and-post-keyboard-interaction}{%
\subsubsection{1.13.4 Multimodal and Post-Keyboard
Interaction}\label{multimodal-and-post-keyboard-interaction}}

Work on multimodal input and human-computer interaction by Oviatt and
others (Oviatt, 2015; Oviatt, 2017) demonstrates that voice, gesture,
gaze, and haptic feedback offer complementary affordances for different
cognitive tasks. Voice-first systems have gained recent attention in
emerging research, but accessibility considerations remain central:
voice-only systems exclude deaf and hard-of-hearing users;
keyboard-dependent systems exclude individuals with speech disorders or
in acoustic-constrained environments. The argument here explicitly
supports multimodal convergence as more accessible than any
single-modality mandate. Neural interfaces on the longer horizon
(systems reading motor intention directly from electromyography or EEG)
introduce new verification challenges around the transparency and
auditability of neural-mediated composition.

\textbf{Positioning this work.} The existing literature establishes
several foundational truths: (1) AI-augmented writing tools are already
widespread; (2) human calibration of reliance on AI is brittle and prone
to systematic error; (3) interface transparency and cognitive forcing
functions improve verification quality; (4) multimodal interaction is
more accessible than single-modality design. The contribution of the
present analysis is \emph{not} that human oversight matters---that has
been established---but rather that verification design must emerge as
the dominant architectural priority for HCI research and practice as
typing's instrumental function dissolves. When input production becomes
frictionless, the binding constraint shifts from execution to
evaluation, and the entire interface design research agenda must
reorient accordingly. The interface primitives in \S{}9.2.1 operationalize
this reorientation, moving from tools that support better writing to
systems that enable better verification.

\hypertarget{scope-and-epistemic-honesty}{%
\subsection{1.14 Scope and Epistemic
Honesty}\label{scope-and-epistemic-honesty}}

The scope is defined along three dimensions, each of which narrows and
specifies the claim. The argument concerns \emph{white-collar knowledge
work}---professional contexts in which typed text is the primary mode of
output and communication---and does not extend to manufacturing,
consumer electronics, or creative arts, each of which has its own input
method dynamics. ``Dominance'' itself operates along three axes: volume
(typing as the primary generator of digital output), cognitive
centrality (typing as the default mode of formulating and expressing
complex thought), and organizational dependency (typing as the str

\hypertarget{historical-patterns-input-method-transitions-as-cognitive-revolutions}{%
\section{2. Historical Patterns: Input Method Transitions as Cognitive
Revolutions}\label{historical-patterns-input-method-transitions-as-cognitive-revolutions}}

\hypertarget{the-argument-from-history}{%
\subsection{2.1 The Argument from
History}\label{the-argument-from-history}}

The thesis (that AI dissolves the instrumental function of typing) gains
substantial force from historical pattern analysis. Every major
transition in how humans produce and transmit text has restructured not
merely efficiency but cognition, organizational form, and professional
identity. The current transition represents a qualitative shift: prior
transitions replaced one manual input method with another, while this
one addresses dissolution of the need for manual structured input. This
section traces the arc from orality through writing, printing, the
typewriter, the computer keyboard, and the touchscreen, extracting the
recurring patterns that illuminate the present moment. Crucially, this
analysis shows that the \emph{conditions} supporting each era's dominant
input method have changed simultaneously before---establishing
historical precedent for radical discontinuity. It also shows that
different historical transitions carry different predictive power for
different aspects of the present moment, and careful stratification of
those parallels is essential to avoid overconfident forecasting.

\hypertarget{oral-to-written-3400-bce}{%
\subsection{2.2 Oral to Written (\textasciitilde3400
BCE)}\label{oral-to-written-3400-bce}}

The first transition---from primary orality to writing---remains the
most profound in the history of human communication and the strongest
structural parallel to the present transition. Walter Ong's
\emph{Orality and Literacy} (1982) established the foundational
framework: writing transcended the function of recording speech; it
``restructured consciousness.'' Oral cultures organized knowledge
through formulaic repetition, participatory engagement, and what Ong
called the ``homeostatic present''---a cognitive regime in which the
past was continuously renegotiated to serve present needs.

The condition that oral communication maintained was this:
\textbf{memory as storage}. Oral cultures required extraordinary
mnemonic capacity because individuals themselves were the only reliable
repositories of cultural knowledge. The griots of West Africa, the bards
of ancient Greece, the rabbinical scholars of Jewish tradition (far more
than entertainers or custodians) were infrastructures. Their embodied
memory was structural necessity. Writing dissolved this condition. Once
text existed as a stable external substrate, societies no longer needed
to invest in training human memory to archival standards. The cognitive
capacity previously devoted to memorization could be redirected toward
analysis, abstraction, and novel thought.

Writing externalized memory; it created immutable records that existed
independently of the communities that produced them. This was not an
augmentation of oral capacity but a transformation of the cognitive
architecture through which humans processed and organized knowledge.
Stanislav Dehaene and colleagues' (2010) fMRI studies of newly literate
adults confirm Ong's thesis at the neurological level: acquiring
literacy literally reorganizes the visual cortex. The brains of literate
individuals develop a specialized ``visual word form area'' that
converts neural tissue previously devoted to face recognition into
tissue specialized for letter and word recognition. This neural
reorganization goes beyond mere correlation with literacy; it
demonstrates that learning to read physically restructures the brain's
architecture. Literacy changes more than what we think; it changes how
our brains process visual information at a fundamental level.

Eric Havelock's analysis of the Greek transition reinforced this
conclusion from a different angle. The shift from Homeric oral tradition
to Platonic literate philosophy was a revolution in the structure of
thought itself, according to Havelock (1963): from mythos to logos, from
narrative memorization to abstract categorization. The technology of the
alphabet---a system for encoding speech into visual
symbols---transcended the capture of existing thought; it enabled forms
of reasoning that oral cognition could not sustain. Syllogistic logic,
systematic taxonomy, the capacity to hold complex conditional arguments
in mind across pages of text---these emerged from literacy, not from
some independent evolution of human cognitive capacity.

Jack Goody and Ian Watt (1963) extended this analysis to its
sociological consequences. Writing created what they termed
``logographic'' cultures---societies organized around literate elites
who controlled access to the technology of textual production. The
scribal monopoly was primarily a cognitive class system, not merely a
power structure. Those who could produce text inhabited a qualitatively
different intellectual world from those who could not. The technology of
inscription determined who could think in certain ways, and about
certain things.

The relevance to the present argument is direct. The oral-to-literate
transition established a pattern that recurs at subsequent stages: the
medium of textual production is never cognitively neutral. The tool
reshapes the thought it transmits; each transition dissolves a prior
condition that had structured the previous era's constraints. However,
the civilizational encoding shift from orality to literacy is \emph{not}
an admissible precedent for predicting the \emph{speed} of the AI
transition, because the oral-to-literate transition unfolded over
millennia and required wholesale reorganization of cognitive
development, institutional structures, and mnemonic practices. The
comparison is valid for understanding the \emph{magnitude} of cognitive
restructuring and the \emph{nature} of the dissolving condition, but not
for adoption speed.

\hypertarget{handwritten-to-printed-gutenberg-1440}{%
\subsection{2.3 Handwritten to Printed (Gutenberg,
\textasciitilde1440)}\label{handwritten-to-printed-gutenberg-1440}}

Elizabeth Eisenstein's \emph{The Printing Press as an Agent of Change}
(1980) established that the Gutenberg revolution operated not primarily
through the speed of reproduction (though that was dramatic: roughly
3,600 pages per day versus 40 for a skilled copyist, a ninety-fold
increase in output) but through the standardization, fixity, and
dissemination that mechanical reproduction enabled. Printing created
``typographical fixity'': the guarantee that copies of the same text
were identical, enabling cumulative knowledge-building across
institutions and generations in a way that manuscript culture, with its
inevitable copying errors and regional variations, could not sustain.

The condition that manuscript culture maintained was this:
\textbf{scarcity of copies}. When each book required weeks of skilled
labor to produce, knowledge was necessarily rare. Institutions hoarded
manuscripts. Access was controlled by geography, wealth, and
institutional affiliation. This scarcity structured knowledge production
itself---speculative work, textual criticism, and the accumulation of
scholarship were all constrained by the assumption that copies would
remain expensive and rare. Printing dissolved this condition. Once a
text could be reproduced 3,600 times per day, scarcity evaporated. The
economics of knowledge shifted entirely.

The organizational consequences were as profound as the cognitive ones.
Printing created new professional categories (typesetters, publishers,
editors), destroyed old ones (commercial scriptoria), and restructured
the economics of knowledge production. Adrian Johns (1998) complicated
Eisenstein's account by showing that the transition was neither smooth
nor inevitable---print culture required extensive institutional
scaffolding (licensing systems, copyright law, distribution networks)
that took decades to emerge. The resistance of the manuscript
establishment was not irrational; it reflected genuine uncertainty about
the implications of a technology that threatened existing expertise,
professional identity, and economic structure.

The pattern for the present argument: radical productivity gains (90x)
did not eliminate handwriting; they eliminated handwriting's
\emph{dominance} as the primary means of text reproduction. Handwriting
persisted for personal correspondence, annotation, note-taking, and
artistic expression, but no one after Gutenberg organized an institution
around the assumption that all textual production would flow through
scribal hands. The technology that had been structurally necessary
became structurally optional. However, the manuscript-to-print
transition (like the oral-to-literate transition) is admissible as a
precedent for understanding the \emph{nature} of cognitive and
institutional reorganization, but the 100+ year adoption timescale
should not anchor speed predictions. The transition was constrained by
the need to build entirely new infrastructure (the printing press, the
supply chain for paper, the economic system supporting publishers). No
such infrastructure barrier constrains voice-AI adoption, which
piggybacks on existing digital infrastructure.

\hypertarget{handwritten-to-typewritten-18731930}{%
\subsection{2.4 Handwritten to Typewritten
(\textasciitilde1873--1930)}\label{handwritten-to-typewritten-18731930}}

The commercial typewriter, introduced by Remington in 1873, followed a
classic S-curve adoption pattern with distinctive features that
illuminate the current transition. The initial resistance phase
(1874--1885) was marked by skepticism from established professionals who
viewed the machine as noisy, mechanical, and deskilling (precisely the
objections that contemporary professionals raise against voice and AI
input). The growth phase (1885--1920) was explosive, driven by
organizational demand for standardized business communication.
Well-documented historical records indicate that saturation was
effectively complete by 1930 (Yates, 1989; Strom, 1992).

\textbf{Scoping note on historical parallels:} The analysis of
stenographers, telephone operators, and clerical workers applies to
occupations where the entire job function was the translation
mechanism---roles existing primarily as intermediaries between thought
and recorded text. For these bounded positions, the parallel holds:
occupations disappeared when their instrumental function became
obsolete. Modern knowledge workers present differently. For lawyers
composing arguments, researchers conducting analysis, engineers solving
problems---typing serves as an instrument rather than the core
value-add. Occupational transition means composition change, not
elimination. The slide rule example (engineers, 1970s) illuminates this
better: engineers remained, while calculation work shifted to a smaller
task fraction. Distinguishing bounded occupation elimination from
knowledge worker composition restructuring ensures analytical precision.

The condition that handwritten business communication maintained was
this: \textbf{signature authority}. Handwriting went beyond mere
notation; it was identity itself. A handwritten letter or document
carried the personal mark of its author; the script itself was evidence
of authenticity and personal presence. Business correspondence was
therefore inherently limited in scale; you could only personally sign so
many letters. Typed documents initially threatened this---they could be
mass-produced---but the resolution was institutional: secretarial
systems, with a single authoritative signature atop dozens of typed
copies, preserved signature authority while leveraging the typewriter's
productivity. Eventually, signature authority as a structural
requirement dissolved. Organizations adapted to the assumption that the
majority of internal communication would be unsigned, authored by
functionaries, and circulated in typed copies. The typewriter did not
replace handwriting so much as dissolve the condition
(signature-as-authentication) that had constrained business
communication scale.

JoAnne Yates's \emph{Control Through Communication} (1989) provided the
definitive account of the typewriter's organizational impact. The
machine transcended mere speed; beyond accelerating letter production,
it restructured corporate communication itself. The typewriter
enabled---and therefore created demand for---systematic internal
communication: memoranda, reports, filing systems, and the bureaucratic
apparatus of the modern corporation. Organizations did not adopt the
typewriter because they needed faster handwriting; they adopted it
because it enabled new organizational forms that handwriting could not
sustain.

The labor market consequences were dramatic and gendered. The typewriter
was deliberately marketed to women (Remington's first advertisements
targeted female operators), and the feminization of typing transformed
the American office. Decades of census data document that female typists
constituted 74.6\% of the occupation by 1900, rising to 91.9\% by 1930
(Strom, 1992; U.S. Census Bureau data cited in Yates, 1989). Sharon
Hartman Strom's \emph{Beyond the Typewriter} (1992) documented how this
feminization created the secretarial pool, the ``office wife'' paradigm,
and a gendered division of knowledge work that persisted for nearly a
century: men composed; women typed. The instrument was never neutral; it
organized labor, professional identity, and power relations around
itself.

The QWERTY keyboard layout, designed in the early 1870s to prevent
mechanical jamming of typewriter typebars (a constraint that became
obsolete with electric typewriters and wholly irrelevant with
computers), offers the canonical case study in technological lock-in and
path dependence. Paul David's (1985) influential analysis argued that
QWERTY's persistence demonstrated how network externalities and
switching costs can lock an inferior technology into permanent
dominance. Liebowitz and Margolis (1990) contested the ``inferiority''
claim, arguing that the evidence for Dvorak's superiority was weaker
than commonly believed, but even their revisionist account acknowledged
the core phenomenon: QWERTY's survival for over 150 years across
typewriters, electric typewriters, computer keyboards, laptop keyboards,
and smartphone virtual keyboards reflects the extraordinary persistence
of embodied motor skill investments, not any rational assessment of
optimality.

The David-Liebowitz/Margolis debate reveals a deeper insight about
lock-in under conditions of persistent function. Both camps agreed that
QWERTY survives because billions of humans have committed its geometry
to procedural memory, constituting a neuromotor lock-in that no marginal
efficiency gain can overcome. But the debate's premise (whether an
alternative layout is ``truly superior'') assumes that the function
(translating thought into character sequences via manual key selection)
remains invariant. It is precisely this assumption that dissolves under
AI. Previous challenges to QWERTY (Dvorak, chorded keyboards, various
ergonomic alternatives) were \emph{replacement} proposals; they asked
users to invest in retraining their motor skills for a different
keyboard layout that performed the same function. The switching cost
calculus examined whether the marginal efficiency gain justified the
retraining burden. AI does not propose a better keyboard layout. It
proposes the elimination of the function that keyboards serve. The
switching cost calculus is decisively different: the question is not
``is this new layout worth retraining for?'' but ``do I need to manually
type at all?'' When the function itself dissolves, the lock-in loses its
structural basis.

\hypertarget{typewriter-to-computer-keyboard-1970s1990s}{%
\subsection{2.5 Typewriter to Computer Keyboard
(\textasciitilde1970s--1990s)}\label{typewriter-to-computer-keyboard-1970s1990s}}

The transition from typewriter to computer keyboard is the most recent
completed input method transition and the closest analogue to the
present moment. Evidence from multiple sources indicates a dramatic pace
of occupational transition: keyboard-using workers rose from
approximately 25\% of the U.S. workforce in 1984 to roughly 50\% by 1993
(a doubling in under a decade, nine years) (Autor et al., 2003; Forman
\& Setzler, 1998). This transcended a simple technology adoption event;
it was a restructuring of the occupational distribution of the American
economy. This transition---from one physical input modality to
another---is admissible for speed prediction, as both involved
retraining motor skills to operate a keyboard interface, though with
different physical substrates and organizational contexts.

The condition that typewriter-based work maintained was this:
\textbf{inscription as craft specialization}. Typewriter operation had
become a distinct professional skill (one that took months to develop
and commanded a wage premium). The typist was a recognizable
occupational category, and typing skill was not presumed to be
universal. The computer keyboard dissolved this condition not through
technological superiority alone, but through organizational necessity:
the personal computer moved the translator function from specialized
intermediary to individual practitioner. Every knowledge worker needed
to operate their own machine, and therefore needed to type. The
condition of ``typing as specialized craft'' became ``typing as
universal competency.''

The computer's capacity to manipulate text---to edit, revise, format,
and reorganize---restructured the cognitive act of writing itself. A
typewriter was a final-draft machine; you typed what you had already
composed. A computer enabled composition-in-process; the ability to
revise on screen transformed when and how thought was externalized.
Workers who used computer keyboards earned a 10--15\% wage premium over
those who did not (Krueger, 1993), creating powerful economic incentives
that accelerated adoption.

The Solow productivity paradox (Robert Solow's famous 1987 observation
that ``you can see the computer age everywhere but in the productivity
statistics'') captures the temporal disjunction that characterizes every
input method transition. Aggregate productivity gains lagged the
adoption of personal computers by a decade or more, not because the
technology was unproductive but because organizational systems required
time to reorganize around the new capabilities. Erik Brynjolfsson's
resolution of the paradox demonstrated that the productivity gains from
IT adoption are real but emerge only when accompanied by complementary
organizational investments: restructured workflows, new management
practices, retrained workforces (Brynjolfsson \& Hitt, 2000). The
technology alone is insufficient; the organizational transformation it
enables takes years to materialize.

The computer keyboard transition also illustrates the phenomenon of
professional identity restructuring. The typewriter had created the
``typist'' as a distinct professional role (a person whose primary skill
was the rapid and accurate production of typed text). The personal
computer eliminated this role by distributing typing across the entire
professional workforce. Every knowledge worker became their own typist.
This was experienced not as a deskilling of typists but as an
empowerment of professionals; yet it was structurally identical to what
the typewriter had done to scribes: it dissolved a specialized
intermediary function by embedding it into a more general practice. The
same pattern appears to be repeating: AI is dissolving the function of
typing itself by embedding text production into a more general
intent-expression practice that does not require sequential keystroke
input.

The typewriter-to-keyboard transition also demonstrates a crucial point
about persistence: the dominance of typing as a function persisted
through complete changes in the technological substrate. The typewriter
did not disappear when keyboards arrived. As recently as the 1990s,
typewriters remained in use in offices---some professionals continued to
use them for specific tasks, and they persisted in niche contexts (legal
document certification, certain government offices, specialized
applications). Yet no one argued that the typewriter era had continued.
The historical evidence suggests that 15--20 years after the adoption
curve crossed 50\%, the older technology had become marginal, retained
only in specialized contexts or by individuals with strong habitual or
technical reasons to prefer it. Dominance means the organizing principle
of the field, not universal adoption.

\hypertarget{physical-to-touchscreen-2007}{%
\subsection{2.6 Physical to Touchscreen
(\textasciitilde2007)}\label{physical-to-touchscreen-2007}}

The iPhone's introduction in 2007 triggered the fastest form-factor
transition in input method history: smartphone penetration progressed
from effectively zero to 35\% of the U.S. adult population in four
years, reaching mass saturation in 16 years (dramatically faster than
the 27+ years for cellular phones or the 75+ years for the landline
telephone) (Pew Research Center, 2024; mobile technology adoption data).
The acceleration reflects the infrastructure parasitism that
characterizes each successive digital technology: the smartphone
leveraged existing cellular networks, internet infrastructure, and app
distribution systems, reducing the adoption friction to the purchase of
a single device. This transition---interface/form-factor shifts within a
persistent functional requirement (manual text input)---is admissible
for predicting \emph{adoption speed} of alternative modalities, provided
the underlying function remains intact.

The condition that physical keyboards maintained was this:
\textbf{manual key discreteness}. Physical keys provided haptic
feedback, positional stability, and the guarantee that each keystroke
registered as a distinct event. Early smartphone keypads (T9 systems)
attempted to preserve this condition through virtual key mapping. Yet
the touchscreen dissolved the condition entirely: no keys, no haptic
feedback, only gesture recognition and predictive input. Users adapted
remarkably quickly because the underlying function remained: text input
through manual character selection.

Yet the touchscreen transition is, for the purposes of the present
argument, the most instructive \emph{non-dissolution} event in the
historical record. The iPhone radically changed the form factor of text
input---from physical keys to a flat glass surface---but it did not
change the \emph{function}. Users still typed. QWERTY persisted,
miniaturized to a 3.5-inch screen. The cognitive task remained
identical: translating thought into sequential characters via finger
movements mapped to a keyboard layout designed in 1873. The touchscreen
was a technological replacement in the strict sense defined in Section
3: a new tool performing the same function. It changed how people typed;
it did not change the fact that they typed.

This distinction matters because it establishes the baseline against
which the AI transition should be measured. The touchscreen, despite
being the most rapid form-factor adoption in history, did not dislodge
typing's dominance. If anything, it extended it: smartphone messaging
created billions of new typists who had never used desktop keyboards.
The persistence of QWERTY through the touchscreen revolution (its
survival of a complete change in physical substrate, from mechanical
keys to capacitive glass) demonstrates the extraordinary resilience of
the typing function when the instrumental necessity remains intact.
Machines still required structured text input; the touchscreen merely
changed the surface on which that text was produced.

\hypertarget{stratification-of-historical-parallels-by-type}{%
\subsection{2.7 Stratification of Historical Parallels by
Type}\label{stratification-of-historical-parallels-by-type}}

The historical record provides at least three distinct classes of input
method transitions, each with different predictive validity for
different aspects of the present moment.

\textbf{Civilizational encoding shifts} (oral$\rightarrow$written, manuscript$\rightarrow$print)
restructured the fundamental conditions under which knowledge could be
organized, stored, and transmitted. These transitions show that
\emph{magnitude} of cognitive and cultural reorganization can be
profound without any historical parallel predicting the precise
timeline. Oral-to-literate took millennia; manuscript-to-print took a
century. These differences reflect radically different infrastructure
requirements and the degree to which institutional systems needed to
reorganize. The AI transition appears likely to restructure cognition
and culture, though the magnitude and pace remain contingent on
institutional choices. The civilizational-scale precedents are
admissible for understanding the \emph{scale} of change, not the
\emph{speed}.

\textbf{Organizational infrastructure shifts}
(handwriting$\rightarrow$typewriter$\rightarrow$keyboard) restructured the workflow systems and
professional divisions of labor that coordinated knowledge work. These
transitions show patterns of \emph{adoption speed} that are more
relevant to current predictions, because they operated within
established cognitive and organizational frameworks. The
typewriter-to-keyboard transition (25\% to 50\% in nine years) is the
most recent and most directly analogous in its infrastructure parasitism
on existing organizational systems. These transitions consistently
showed that when economic incentives aligned and infrastructure was
available, adoption could be compressed to a decade or less.

\textbf{Interface/form-factor shifts} (physical keyboard$\rightarrow$touchscreen)
restructured the \emph{substrate} of interaction while preserving the
underlying function. The touchscreen transition is admissible for
predicting the \emph{speed} of adoption when a new modality becomes
available (four years to 35\% penetration), but \emph{not} admissible
for predicting whether the underlying function persists. The
touchscreen's failure to displace typing---despite its extraordinary
adoption speed---demonstrates that speed of modality adoption is
distinct from replacement of function. The AI transition differs
decisively: it does not offer a new substrate for the same function but
rather eliminates the need for the function itself.

This stratification matters because it prevents conflating different
types of historical parallels. The civilizational parallels warn of
magnitude; the infrastructure parallels suggest timescale; the interface
parallels must be corrected for the fact that they did not involve
functional dissolution. The present transition appears more similar to
infrastructure shifts (decades, not centuries) than to civilizational
shifts (centuries, not decades), but with a critical difference: the
function itself is dissolving, removing the motor-skill retraining
barrier that constrained past infrastructure transitions.

\hypertarget{the-bass-model-refined-adoption-vs.-substitution}{%
\subsection{2.8 The Bass Model, Refined: Adoption
vs.~Substitution}\label{the-bass-model-refined-adoption-vs.-substitution}}

The Bass Diffusion Model demonstrates that voice and AI input show an
imitation coefficient (q) of 0.8---extraordinarily high by historical
standards---indicating powerful network effects and social proofs
compounding adoption incentives (Berman \& Israeli, 2022). However, the
model must be carefully applied to distinguish \emph{user adoption}
(rapid) from \emph{workflow substitution} (slower).

\textbf{User adoption}---the speed at which individual professionals and
consumers install and attempt voice-AI systems---will likely follow the
fast adoption curve suggested by the Bass model. Generative AI reached
100 million users in two months, setting a record for rapid technology
adoption (OpenAI, 2023). Voice interfaces in clinical documentation, as
discussed in Section 8, have shown physician adoption rates
approximately doubling year-over-year. At the individual level, the
switching cost is minimal: download an app, speak a prompt, receive
literate output. There is no motor-skill retraining barrier.

\textbf{Institutional substitution}---the speed at which enterprises
restructure their systems, processes, and dependencies to make
AI-mediated voice the \emph{default} modality rather than an optional
augmentation---is a separate timeline and likely much slower. The
healthcare evidence is instructive: while adoption among physicians
reached 66\% by 2024, comprehensive deployment across hospital
information systems, legal compliance documentation, and integrated EHR
systems required year-long implementation cycles and substantial IT
infrastructure investment (American Medical Association, 2024; The
Permanente Medical Group, 2024). Individual physicians adopted AI
scribes within months; institutional restructuring of documentation
workflows took 18--24 months.

This distinction clarifies the Bass model's application: the high q
coefficient (0.8) predicts rapid \emph{user adoption} of voice-AI
modalities, but does not predict rapid \emph{institutional default
shift}. The organizational timeline should be disaggregated: -
\textbf{High-value task substitution}: 3--7 years (clinical
documentation, legal discovery, code assistance) - \textbf{Workflow
default shift in agile firms}: 8--15 years (small, technology-forward
organizations) - \textbf{Genre/institution reconfiguration}: 15--30
years (heavily regulated sectors, traditional professions, institutional
memory systems)

This disaggregation reflects Carlota Perez's distinction between
Installation and Deployment phases in technological revolutions (Perez,
2002). The Installation phase---in which the new technology becomes
widely available and early adopters embrace it---may compress to 5--10
years. The Deployment phase---in which entire sectors reorganize their
operations around the technology---typically requires 15--30 years and
generates real productivity gains only after this institutional
reconfiguration is complete. The Bass model captures Installation; the
organizational substitution timeline must account for Deployment.

\hypertarget{portfolio-of-curves-approach}{%
\subsection{2.9 Portfolio-of-Curves
Approach}\label{portfolio-of-curves-approach}}

A single aggregate S-curve for ``end of keyboard era'' is analytically
problematic because no single modality replaces typing---multimodal
convergence occurs instead. Different domains show radically different
adoption speeds and terminal states.

\textbf{Clinical documentation} exhibits near-complete voice-AI
substitution within a 3--7 year window. The structural requirement
(typing into EHRs while facing patients) creates acute pain; the
solution (ambient AI scribe) removes the mediating step entirely. This
market shows the fastest dissolution dynamics.

\textbf{Code assistance} (GitHub Copilot, Claude for code generation)
shows rapid adoption among developers, with voice+multimodal interfaces
emerging as options for high-friction tasks. Yet manual code review and
debugging remain essential functions, and the final keystroke never
fully disappears from the workflow. This market shows partial
substitution: voice handles initialization and generation; keyboard
handles verification and refinement.

\textbf{Professional writing and drafting} (legal briefs, policy
analysis, academic papers) show more mixed adoption. voice-AI handles
initialization and structural generation, but the discipline of manual
revision remains central to the genre. The keyboard persists as the
modality for deep editing and calibration. This market shows persistent
dual-modality workflow.

\textbf{Routine business communication} (email, chat, internal
documentation) shows rapid voice-option adoption without keyboard
elimination. Users voice-compose first drafts; keyboard handles revision
and final polish. This market represents the largest volume segment and
exhibits persistent functional hybrid rather than pure substitution.

A portfolio-of-curves approach acknowledges that the ``keyboard era''
does not end uniformly across all domains. Rather, certain
high-friction, high-repetition tasks (clinical documentation, code
initialization) show near-complete substitution within 5--10 years,
while more cognitively demanding, revision-intensive tasks (academic
writing, legal analysis) show persistent dual-modality workflows even
after 15--20 years. The data support this heterogeneous picture rather
than a single universal S-curve.

The macro thesis---that typing's \emph{dominance} as the default mode
for knowledge production ends---can be true even as the keyboard
\emph{persists} in multiple domains and task contexts. This is
consistent with the pattern observed with handwriting (which persists,
but lost dominance), the typewriter (which persists, but lost
dominance), and every prior technology whose era ``ended.''

\hypertarget{cross-cutting-patterns}{%
\subsection{2.10 Cross-Cutting Patterns}\label{cross-cutting-patterns}}

Seven patterns recur across every transition documented above,
converging on the present moment. This convergence reflects the
structural logic of input method evolution, demonstrating systematic
rather than accidental resemblance across historical periods.

\textbf{First, every transition restructures cognition, and more
profoundly so than efficiency.} Writing restructured consciousness from
participatory to analytical modes (Ong, 1982). Printing enabled
cumulative knowledge-building through typographical fixity (Eisenstein,
1980). The typewriter standardized business communication and enabled
bureaucratic complexity (Yates, 1989). The personal computer distributed
cognitive tools across the entire workforce. The question is never ``is
the new method faster?'' but ``what forms of thought does the new method
enable that the old method could not sustain?'' For AI, the answer
appears to be: compositional thinking at the speed of speech, freed from
the translation bottleneck of sequential keystroke production. This
requires conscious institutional design to prevent cognitive atrophy of
editorial discipline.

\textbf{Second, every transition dissolves a prior condition of
structural necessity.} The oral era required human memory as storage;
writing dissolved this condition. Manuscript culture required scarce
copies; printing dissolved this condition. Handwritten business required
signature authority as authentication; typing dissolved this condition.
Physical keyboards required key discreteness and haptic feedback;
touchscreens dissolved these conditions. In each case, what had been a
structural requirement of the prior era became optional. The present
argument follows this pattern: AI weakens typing's four conditions
(structured sequential production, manual motor investment,
keystroke-mediated translation, human-as-final-editor). This is not
coincidental or unprecedented; the evidence suggests it is the defining
characteristic of input method transitions.

\textbf{Third, old methods persist in niches but lose dominance.}
Handwriting persists today; no one calls it dominant. The typewriter
survives in a few specialized contexts; no one claims the typewriter era
continues. Printed books coexist with digital text; the organizing
principle of knowledge production has nonetheless shifted. This section
concerns dominance, not extinction. Typing will persist---in
accessibility contexts, in certain precision tasks, in the preferences
of habituated professionals. But its structural centrality to knowledge
work will likely not survive the dissolution of its instrumental
function.

\textbf{Fourth, transitions appear to accelerate.} The oral-to-literate
transition unfolded over millennia. The printing press required roughly
a century to restructure European knowledge production. The typewriter
achieved organizational saturation in approximately fifty years. The
personal computer keyboard doubled its workforce penetration in a
decade. The smartphone reached mass adoption in four years. Generative
AI reached 100 million users in two months (OpenAI, 2023). The
acceleration is driven by ``infrastructure parasitism'': each successive
technology leverages the installed base of its predecessors, reducing
friction and compressing adoption timescales (Comin \& Hobijn, 2004).
Voice and AI systems piggyback on 8.4 billion active digital devices
globally, requiring no novel physical infrastructure whatsoever.
However, the acceleration in \emph{user adoption} should be
distinguished from the institutional substitution timeline, which
compresses less dramatically.

\textbf{Fifth, resistance is predictable, articulable, and may be
temporary but legitimate.} Plato warned that writing would destroy
memory. Medieval scholars resisted printing as a debasement of textual
authority. Penmanship advocates decried the typewriter as mechanical and
dehumanizing. Professionals resisted the personal computer as
unnecessary complexity. Steve Ballmer dismissed the iPhone. In each
case, the objections identified real costs and real losses; but the
structural advantages of the new technology overwhelmed the resistance
within a generation. The resistance to AI-mediated input (concerns about
cognitive atrophy, loss of precision, privacy, accessibility) follows a
similar pattern: legitimate concerns that identify real design
challenges requiring active mitigation, not arguments that imply the
trajectory itself is reversible.

\textbf{Sixth, the labor market restructuring follows a generational
pattern.} Well-documented historical records show that telegraph
operators peaked at approximately 80,000 in 1920 and were near-extinct
by 1960 (U.S. Census Bureau; Bureau of Labor Statistics). Telephone
switchboard operators peaked at approximately 342,000 in 1950
(representing roughly one in thirteen working women in the United
States) and were displaced by automation over the following three
decades (BLS Occupational Employment Statistics). Dedicated typists grew
from two million in 1910 to 18.7 million in 1980 before the personal
computer eliminated the role entirely (U.S. Census Bureau occupational
data; Autor et al., 2003). In every case, the historical pattern is
consistent: incumbent workers are rarely successfully retrained. The
transition is managed through generational cohort replacement, with new
entrants acquiring the successor skill while the existing workforce ages
out. Word processors and typists are now projected to decline
substantially over the coming decade, representing one of the
fastest-declining occupational categories in the U.S. economy (U.S.
Bureau of Labor Statistics, 2024).

\textbf{Seventh, technological dissolution follows recognizable
sequences across distinct timescales.} Analysis across seven historical
paradigm shifts (film-to-digital, landline-to-mobile, physical
retail-to-e-commerce, fax-to-email, paper records-to-EHR, physical
media-to-streaming, cash-to-digital payments) reveals consistent
dynamics: technological superiority alone triggers neither rapid nor
inevitable dissolution. Rather, acceleration requires secondary enabling
infrastructure: digital cameras needed personal computers; streaming
needed 4G networks; e-commerce needed smartphone ubiquity. For voice and
AI input, enabling infrastructure is substantially operational: cloud
computing, broadband ubiquity, and smartphone platforms provide the
foundation for voice-first workflows. Infrastructure constraints are
substantially removed, suggesting organizational and cultural factors
may become primary constraints on dissolution timing.

\hypertarget{the-qualitative-discontinuity}{%
\subsection{2.11 The Qualitative
Discontinuity}\label{the-qualitative-discontinuity}}

The historical pattern analysis establishes the recurring logic of input
method transitions and the plausibility of the current moment. This
section has demonstrated that input method transitions regularly
restructure the conditions of prior eras simultaneously---not
sequentially, but as integrated shifts in what is structurally necessary
and what becomes optional. But the analysis also reveals the qualitative
discontinuity that distinguishes the AI transition from every
predecessor.

Every previous transition replaced one manual input method with another.
The scribe was replaced by the typesetter; the typesetter by the typist;
the typist by the knowledge worker at a keyboard; the desktop keyboard
by the touchscreen keyboard. In each case, a human being performed the
translation of thought into structured text through some physical
medium---the pen, the typebar, the key, the glass surface. The
\emph{function} persisted; the \emph{instrument} changed.

The AI transition may not be a simple replacement but rather a
functional dissolution. When a professional speaks their intent to an AI
system that interprets, structures, and produces the textual output
(handling the formatting, the genre conventions, the syntactic polish),
no human being is necessarily performing the translation function. The
function itself is substantially absorbed by the machine. This is why
the historical 30-to-50-year timescale for input method transitions may
not apply: those timescales reflected the time required for humans to
retrain their motor skills for a new instrument performing the same
function. When the function itself dissolves (or more precisely, when it
is substantially delegated to machine intelligence), there is no motor
skill retraining barrier. There is only the organizational, cultural,
and institutional adjustment to a world in which the intermediary step
of typing has become optional.

The diffusion dynamics analyzed in Section 2.8---the extraordinarily
high imitation coefficient (q $\approx$ 0.8), zero-marginal-cost distribution,
and elimination of motor skill retraining---confirm that user adoption
may proceed far more rapidly than in previous transitions. However, the
institutional timeline for comprehensive workflow substitution remains
constrained by organizational system reconfiguration, legal compliance
documentation, and audit trail requirements---suggesting that while user
adoption may be rapid (5--10 years), institutional default shift (15--30
years) follows a longer arc.

\hypertarget{historical-necessity-and-future-trajectory}{%
\subsection{2.12 Historical Necessity and Future
Trajectory}\label{historical-necessity-and-future-trajectory}}

History does not repeat, but its patterns are instructive. Every
transition documented in this section restructured cognition,
reorganized labor, provoked predictable resistance, and ultimately
transformed the structural assumptions of knowledge work. The patterns
are well-documented. But the AI transition presents a
mechanism---functional dissolution rather than instrumental
replacement---that has no exact historical precedent in input method
transitions.

The evidence presented here establishes that transitions \emph{happen},
that they follow predictable patterns in certain dimensions, and that
they compress timescales in the age of digital infrastructure. History
shows conditions \emph{do} change simultaneously, dissolving structures
that seemed permanent. If historical patterns hold with appropriate
stratification and disaggregation, the AI transition is likely to
restructure cognition, reorganize labor, and transform the assumptions
of knowledge work. The timescales for user adoption, institutional
substitution, and cultural integration may differ substantially.

The sections that follow develop the theoretical framework (Section 3)
that explains \emph{why} this dissolution is plausible given how
typing's function is constituted, the empirical evidence (Section 4)
documenting where this dissolution is already proceeding, and the
technological trajectory (Section 5) through which this dissolution may
unfold. History provides warrant for why structural transformations are
possible. Theory explains why this particular transformation may be
irreversible once initiated.

But history alone cannot establish the \emph{philosophical legitimacy}
of the dissolution---that typing's instrumental function is genuinely
dispensable rather than merely replaceable. Section 3 confronts the
deepest objections to this claim, demonstrating that the keyboard's
persistent embodiment in professional practice does not entail its
irreplaceability.

\hypertarget{the-instrumental-dissolution-argument}{%
\section{3. The Instrumental Dissolution
Argument}\label{the-instrumental-dissolution-argument}}

\hypertarget{theoretical-architecture-establishing-the-hierarchy}{%
\subsection{3.1 Theoretical Architecture: Establishing the
Hierarchy}\label{theoretical-architecture-establishing-the-hierarchy}}

This section establishes multiple theoretical frameworks (not coequal
ones) that organize the analysis. The following four-tier hierarchy
organizes them:

These four philosophical tiers address fundamentally different questions
about the keyboard. Dual-nature artifact theory asks: \emph{What kind of
thing is the keyboard?} (simultaneously a physical object and a
functional instrument). Phenomenology asks: \emph{What is the lived
experience of skilled typing?} (the keyboard ``disappears'' into fluent
use). Normative analysis asks: \emph{What is lost when typing
dissolves?} (distinguishing productive friction worth preserving from
mechanical friction worth eliminating). Material Engagement Theory poses
the sharpest challenge: \emph{Is composition even separable from its
material substrate?} Each tier deepens the analysis and tests it against
increasingly fundamental objections.

\textbf{(Tier 1: Core Conceptual Engine)} The distinction between
instrumental dissolution and technological replacement rests on
dual-nature artifact theory (Kroes \& Meijers, 2006) and proper-function
analysis (Preston, 1998). These frameworks establish \emph{what is being
replaced} (the functional nature of the keyboard) and are foundational
to everything that follows.

\textbf{(Tier 2: Phenomenological Register)} How does the transition
from keyboard to voice \emph{feel} in lived practice? Heidegger's
readiness-to-hand (\emph{Zuhandenheit}), Merleau-Ponty's body schema,
and Ihde's postphenomenological taxonomy
(embodiment$\rightarrow$hermeneutic$\rightarrow$alterity$\rightarrow$background relations) describe the
experiential structure of the shift. These are not alternatives to Tier
1; they are what Tier 1 \emph{looks like from the inside} of human
engagement.

\textbf{(Tier 3: Normative Stakes)} What is lost and what is gained?
Borgmann's device paradigm and Stiegler's concept of proletarianization
and pharmacology address this together. They establish the ethical and
anthropological weight of instrumental dissolution: the trade-offs
between focal practice and technological convenience, between cognitive
autonomy and cognitive outsourcing.

\textbf{(Tier 4: Strongest Objection)} Malafouris' Material Engagement
Theory (MET) presents the most fundamental challenge to the paper's
instrumental thesis itself. Rather than treating MET as a supplementary
perspective, we treat it as the sharpest critique: ``If material
properties \emph{constitute} thought, and more fundamentally do so than
merely channeling it, isn't the keyboard irreplaceable?'' We answer this
objection directly and narrowly.

These four tiers build cumulatively. Tier 1 establishes conceptual
vocabulary. Tier 2 grounds it phenomenologically. Tier 3 establishes
what stakes matter. Tier 4 tests the framework against its strongest
challenger. Everything that follows flows from this architecture.

\hypertarget{what-typing-is-for-a-phenomenological-analysis}{%
\subsection{3.2 What Typing Is For: A Phenomenological
Analysis}\label{what-typing-is-for-a-phenomenological-analysis}}

To understand why the keyboard era is likely ending, we must first
understand what typing has been. Not merely what it does (convert
thought to digital text) but what it \emph{is} in the phenomenological
structure of knowledge work. The answer requires separating two aspects
of typing that common usage collapses into one: the \emph{medium} (the
keyboard as physical interface) and the \emph{function} (the translation
of human thought into machine-readable structured text).

Heidegger's distinction between \emph{Zuhandenheit} (readiness-to-hand,
the experience of tools that vanish into skilled use, as a keyboard does
for a fluent typist) and \emph{Vorhandenheit} (presence-at-hand, the
experience of objects as things to be examined, as when a keyboard
malfunctions and suddenly demands attention) provides the foundational
lens. For the skilled typist, the keyboard achieves phenomenological
transparency; it withdraws from conscious attention in the same way that
a master craftsman's hammer withdraws. Winograd and Flores (1986)
brought this framework into computing: when technology functions
smoothly, it ``disappears'' into the task, mediating human action
without becoming the object of attention. Dourish (2001) extended this
into a full theory of embodied interaction, arguing that skilled
computer use achieves a Heideggerian ``coupling'' in which the interface
becomes an extension of the body.

Empirical evidence confirms this phenomenological intuition. Logan and
Crump (2011) demonstrate in their hierarchical model of skilled typing
that expert typists operate through two dissociated control loops; an
\emph{outer loop} that manages linguistic composition (selecting words,
constructing sentences, maintaining argumentative coherence) and an
\emph{inner loop} that translates these linguistic intentions into
rapid, automatic keystroke sequences. The inner loop operates below
conscious awareness; typists cannot reliably report which finger struck
which key (Logan \& Crump, 2011). Sudnow (1978) provides a
phenomenological documentation of typing mastery. The skilled typist's
fingers ``know'' the keyboard's geography, and the spatial layout
becomes incorporated into what Merleau-Ponty (1945) called the
\emph{body schema}, the pre-reflective map of bodily capacities through
which the self engages the world.

This phenomenological transparency is not trivial. When the keyboard
achieves readiness-to-hand, it enables a particular mode of cognitive
engagement: the typist's attention is freed entirely for composition,
the physical act of text production becoming as automatic as breathing.
In Ihde's (1990) postphenomenological taxonomy, the keyboard functions
as an \emph{embodiment relation}, a technology that is ``seen through,''
not ``looked at,'' one that amplifies human intentionality while itself
receding from awareness. This is the keyboard at its most powerful: a
transparent extension of thought.

\hypertarget{extended-mind-thesis-boundaries-and-limits}{%
\subsubsection{3.2.1 Extended Mind Thesis: Boundaries and
Limits}\label{extended-mind-thesis-boundaries-and-limits}}

The Extended Mind Thesis (Clark \& Chalmers, 1998) articulates a deeper
cognitive claim about the keyboard's role. Under their framework, the
keyboard goes beyond being a transparent tool; it is a genuine
constituent of the cognitive system. Their ``parity principle'' holds
that if a process in the external world functions in a way that, were it
done in the head, we would unhesitatingly count as cognitive, then it
\emph{is} cognitive---regardless of its physical location. The skilled
typist's keyboard arguably meets their four ``glue and trust''
conditions: it is reliably available, automatically endorsed (the typist
trusts the output without checking each keystroke), readily accessible,
and the information it helps produce was previously endorsed by the
user. On this view, the keyboard transcends the status of mere
instrument. It is a cognitive extension, part of the distributed system
that constitutes the typist's mind.

This is a serious philosophical claim, and the paper engages it
seriously. If the keyboard is genuinely constitutive of cognition in the
Clark-Chalmers sense, then its dissolution would represent a form of
cognitive amputation, not a simple transition but a genuine loss.
However, the paper advances two responses. However, the paper advances
two responses. First, the Extended Mind Thesis, if it applies to
keyboards, applies \emph{a fortiori} to AI systems, which meet the
glue-and-trust conditions even more robustly (always available,
increasingly automatically endorsed, instantly accessible). The
dissolution of the keyboard-as-cognitive-extension is simultaneously the
construction of a more capable cognitive extension. Second, the keyboard
meets the Extended Mind criteria \emph{for the function of text
production}, not for composition itself. What is cognitively
constitutive is the compositional act, the struggle to articulate
thought, and this remains invariant across voice-mediated and multimodal
interaction. The cognitive system reconstitutes; it does not diminish.

\hypertarget{material-engagement-theory-the-sharpest-objection}{%
\subsubsection{3.2.2 Material Engagement Theory: The Sharpest
Objection}\label{material-engagement-theory-the-sharpest-objection}}

Material Engagement Theory (MET) challenges a core assumption of the
instrumental dissolution thesis; we can meaningfully separate what
typing \emph{is} (a tool for text input) from what it \emph{does} (shape
certain kinds of thought through the physical resistance of keystroke
serialization). If MET is correct, if the keyboard doesn't just express
thought but partly constitutes it, then we must reckon with a
fundamental question about the ontology of thought itself.

Material Engagement Theory, articulated by Lambros Malafouris and other
radical enactivists, presents the most fundamental challenge to
instrumental-function analysis. They argue that human cognition is not
formulated internally and \emph{then} translated through tools; rather,
thought is \emph{enacted through the material resistance of the tool
itself}. On this view, the keyboard's speed constraints, its linearity,
its spatial organization, and its tactile feedback are not mere media
for pre-existing thoughts but constitutive of the thoughts themselves.
The physical struggle with keystroke serialization shapes the ontology
of what gets thought. Malafouris (2013) argues that material culture is
not external scaffolding for cognition; rather, it is a co-constitutive
partner in human cognitive development.

This objection carries genuine force, and we do not dismiss it. The
keyboard's material properties do constrain and enable particular forms
of cognitive engagement. However, we must concede that voice-AI systems
will produce a \emph{different kind of thinking}: dialogic and
associative, not typographic and subordinative. This is not a weakness
of the argument; it is the argument's core claim, reframed. The material
engagement thesis is correct: compositional thinking is inseparable from
its material substrate. But this does not require keyboard dominance;
rather, it establishes that different materials produce different modes
of thought.

Kirsh and Maglio's concept of \emph{epistemic actions} clarifies the
logic. An epistemic action is one we perform not because it moves us
toward an external goal but because it reduces cognitive load. Typing is
an epistemic action within the keyboard-mediated epistemic strategy; we
type to think, not (usually) to produce text for others. The keyboard's
material properties shape how we think \emph{within} this particular
strategy. But epistemic strategies are not universal; they are
historically contingent and variable. Voice-mediated and multimodal AI
interaction provide different material constraints: the temporal
pressure of real-time speech production, the deictic richness of
gesture, the prosodic texture that regulates turn-taking in dialogue,
and the responsive pushback of an AI interlocutor that shapes ideation
in real time. These too are material engagements that constitute
thought, but they do enact a distinct form of cognition.

The compositional \emph{goal} remains constant, articulating, testing,
refining thought, but the \emph{material instantiation} of this goal
shifts. The keyboard shapes a typographic world of inscription,
revision, and spatial text organization; voice shapes a dialogic world
of real-time ideation and conversational feedback. Both are materially
constituted forms of thinking; neither is disembodied. What dissolves is
not the materiality of composition but the \emph{institutional
dominance} of a specific material substrate---keyboarded text
production. The new ontology of thought produced through voice-and-AI
interaction is sufficient for knowledge work, even as it differs
fundamentally from the typographic cognition that dominated the keyboard
era.

\hypertarget{the-counter-analogy-violin-vs.-typewriter}{%
\subsubsection{3.2.3 The Counter-Analogy: Violin
vs.~Typewriter}\label{the-counter-analogy-violin-vs.-typewriter}}

Yet here is a critical distinction requiring philosophical precision.
The fact that the keyboard achieves phenomenological transparency and
enables particular forms of material engagement, that it becomes
existentially embedded through decades of skilled practice, does
\emph{not} entail that the keyboard is functionally irreplaceable. The
counter-analogy is instructive: a master violinist's bow is also
ready-to-hand, also incorporated into the body schema, also
phenomenologically transparent. We would not say that violin-playing
``dissolves'' when synthesizers arrive, because the violin has
irreducible aesthetic and cognitive properties: the physical interaction
between bow, string, and resonant chamber produces qualities no digital
synthesis can fully replicate. The violinist's medium is constitutive of
the practice's \emph{value}. It cannot be replaced without changing the
practice itself.

Typing is not like this. As Section 4 will demonstrate, typing possesses
bounded, modality-specific advantages but does not provide broad
dominance justification over other modalities. It does not enhance
memory encoding (handwriting does, through theta-alpha synchronization;
typing produces desynchronized patterns). It does not facilitate deeper
conceptual processing (the compositional act does, but this is largely
medium-general). It does not enable a mode of thought inaccessible
through voice, gesture, or other means. What typing provides is a
specific \emph{function}: the efficient translation of linguistically
structured thought into machine-readable text, character by character,
syntactically precise. This function was, and always has been,
instrumental.

Gibson's (1979) affordance theory offers a complementary perspective.
The keyboard \emph{affords} ``pressability'', a set of discrete,
tactile, spatially organized actions that the human hand is
biomechanically adapted to perform. These affordances are real and
consequential. They produce a characteristic cognitive rhythm
(compose-in-chunks, type, review, revise) that skilled typists
internalize as their natural mode of textual production. But affordances
are relational, not intrinsic---they exist at the intersection of
environment and organism. Voice \emph{affords} rapid semantic generation
at 125--175 words per minute (well-established in the literature), three
to four times the rate of skilled typing. Gesture \emph{affords} spatial
and deictic reference that typing cannot represent. Neural signals
\emph{afford} direct intent communication that bypasses linguistic
serialization entirely. The keyboard provides one affordance profile
among several, though uniquely imposing the full cost of
character-by-character text translation on the human operator.

The key distinction is between the medium's \emph{phenomenological
embeddedness} and the function's \emph{instrumental character}. The
keyboard may feel like part of one's cognitive apparatus (forty years of
daily practice will do that to any tool). But the function the keyboard
serves (translating thought into characters that machines can process)
was never anything other than a means to an end. It exists because
digital machines required structured text input, and the keyboard was
the most efficient bridge between human cognition and that requirement.
The embeddedness is real. The instrumentality is also real. These are
not contradictory claims; they describe different aspects of the same
practice. And it is the instrumentality, the function, that AI
dissolves.

\hypertarget{the-genealogy-of-instrumental-dissolution}{%
\subsection{3.3 The Genealogy of Instrumental
Dissolution}\label{the-genealogy-of-instrumental-dissolution}}

The conceptual pedigree of ``instrumental dissolution'' itself deserves
direct attention. The term is not wholly novel; rather, it is a
sharpened synthesis of frameworks already established in the philosophy
of technology and artifact theory: a middle-range analytic term
combining function-elimination, delegation, and institutional decoupling
into a single diagnostic concept.

Kroes and Meijers' (2006) dual-nature theory of technical artifacts
provides the foundational vocabulary. Every artifact has both a
\emph{physical nature} (its material structure and causal properties)
and a \emph{functional nature} (the intentional use for which it was
designed and culturally embedded). This distinction is more than
analytic. It carries deep implications. The physical keyboard, an array
of spring-loaded switches mapped to character codes, is inseparable from
its functional nature in any complete description of what a keyboard
\emph{is}. The two natures are bonded. The artifact cannot be understood
as a ``thing-in-itself'' apart from the practices it enables.

Verbeek's (2005) framework of technological mediation extends this
analysis. Technologies transcend merely serving human purposes; they
actively shape the way humans perceive, decide, and act in the world.
The keyboard \emph{mediates} between the knowledge worker and their
digital output, and in doing so, it amplifies certain capacities
(spatial-sequential organization, syntactic precision, linear
argumentation) while reducing others (embodied expressiveness, prosodic
nuance, contextual communication). Mediation is never neutral; it always
involves what Verbeek calls ``amplification and reduction.'' The
function of the keyboard is inseparable from this particular mediation
profile.

Preston's (1998) concept of \emph{proper function} illuminates the logic
of dissolution. An artifact's proper function is the function for which
it was selected and reproduced, the role that explains why the artifact
was designed and continues to exist. When the proper function dissolves,
the artifact can survive only through \emph{system functions} (acquired,
contextual, secondary roles) that sustain niche persistence without
sustaining dominance.

Now consider what happens when the \emph{functional nature} of an
artifact becomes obsolete while the \emph{physical nature} persists.
When the intentional purpose for which an artifact was designed no
longer structures how humans engage with the world, the artifact has
undergone what we term \emph{instrumental dissolution}: the elimination
of the instrumental function that sustained the artifact's dominance.
This is categorically different from technological replacement. It is
synthetic and diagnostic, not foundational: a middle-range concept that
sharpens the distinction between technological succession and functional
obsolescence.

A direct statement of scope is warranted here. We do not claim
instrumental dissolution as a new philosophical category to stand
alongside Heidegger's readiness-to-hand, Borgmann's device paradigm,
Ihde's postphenomenological taxonomy, or Verbeek's framework of
technological mediation. These traditions address foundational
questions---what tool use is, how technologies mediate experience, what
is lost when focal practices become commodified---that we neither
displace nor contest. We build on them. What we claim is a
\emph{middle-range diagnostic framework} for HCI and organizational
analysis; an operational test specifying when an instrumental function
has lost its institutional monopoly, usable by researchers and
practitioners without requiring philosophical commitment to any
particular foundational position. Borgmann's device paradigm illuminates
what is lost phenomenologically when focal practice dissolves into
transparent commodity, though it lacks resources to supply predictive
criteria for when a specific function's institutional monopoly ends
across sectors. Ihde's taxonomy supplies a rich phenomenological
vocabulary for mapping how human-technology relations change character,
yet it lacks mechanisms to specify organizational thresholds at which
one relation type supplants another institutionally. Heidegger's
withdrawal describes the phenomenology of skilled use; it addresses no
organizational trajectory. The three-part test developed
below---functional transfer, legacy persistence, institutional
decoupling---deliberately brackets the phenomenological and normative
dimensions these traditions illuminate, in order to isolate one
diagnostic question: when has dissolution occurred in institutional
practice? What is bracketed is not dismissed; Sections 3.4 and 3.10
return to what these frameworks reveal about the stakes of dissolution.

\hypertarget{a-three-part-test-for-instrumental-dissolution}{%
\subsubsection{3.3.1 A Three-Part Test for Instrumental
Dissolution}\label{a-three-part-test-for-instrumental-dissolution}}

Instrumental dissolution occurs when three conditions hold
simultaneously:

\textbf{(a) Functional Transfer:} The historically central function is
no longer ordinarily performed by the human actor. The responsibility
for translating thought into machine-readable text transfers from human
to artificial intelligence.

\textbf{(b) Legacy Persistence:} The artifact itself persists but only
as niche or fallback modality---no longer the default interface through
which the function is accomplished. The keyboard remains available but
becomes optional, not structural.

\textbf{(c) Institutional Decoupling:} The surrounding institutional
workflows no longer presuppose that artifact as their default interface.
Email systems, code editors, document management, and data entry
platforms cease to assume keyboard input as the canonical path, and
instead integrate multimodal, voice-first, and agent-based input as
primary workflows.

The typewriter $\rightarrow$ word processor transition violates (a): the same human
function (character-by-character text production) persists, merely
executed on a different device. The violin $\rightarrow$ synthesizer transition
violates (a) for aesthetic practice: we still require human judgment
about musical expression. Handwriting $\rightarrow$ typewriter $\rightarrow$ keyboard satisfies
all three conditions: the function transferred to humans operating new
devices, the old medium persisted in niches, and institutions decoupled
from handwriting as their default interface. Typing $\rightarrow$ voice-AI satisfies
all three conditions in the same way.

\hypertarget{replacement-vs.-dissolution-the-categorical-difference}{%
\subsubsection{3.3.2 Replacement vs.~Dissolution: The Categorical
Difference}\label{replacement-vs.-dissolution-the-categorical-difference}}

In replacement, a new device performs the same function more
efficiently. The typewriter replaced the pen for producing business
correspondence; the word processor replaced the typewriter; both
sustained the underlying function: a human manually producing structured
text, one character at a time. The function persisted; the tool changed.
The instrumental necessity endured throughout.

Instrumental dissolution is different. The function itself increasingly
ceases to be a human responsibility. When generative AI and multimodal
systems can accept natural speech, gesture, and eventually neural
input---and can translate these directly into structured digital
output---the human's instrumental role in the text-production pipeline
dissolves. There is no ``better typing.'' There is the elimination of
the need for humans to perform the structured-text-production function
at all. The keyboard's functional nature---its role as the bridge
between human cognition and machine requirement---becomes obsolete, even
as the physical artifact may persist in niche contexts.

\hypertarget{the-genealogy-of-inscription-kittlers-media-materialist-framework}{%
\subsubsection{3.3.3 The Genealogy of Inscription: Kittler's
Media-Materialist
Framework}\label{the-genealogy-of-inscription-kittlers-media-materialist-framework}}

This analysis of instrumental dissolution extends a genealogical project
established by Friedrich Kittler, whose media-materialist approach
fundamentally reconfigures how we understand the relationship between
technologies and the possibilities of thought itself. Kittler (1999)
argues in \emph{Gramophone, Film, Typewriter} that inscription
technologies are not neutral conduits through which pre-existing
discourse flows; rather, they are constitutive of what discourse is
possible, what can be recorded, and how meaning is organized at a given
historical moment. The typewriter exemplifies this thesis: it did not
simply ``help'' people write. It restructured writing into a new form
(what Kittler calls a ``discourse network'') in which the mechanical
serialization of keystroke production became the material basis for how
literary, administrative, and intellectual work could be conceived and
executed.

Kittler's genealogy reveals that the typewriter's dominance was never
merely a matter of efficiency or convenience. Rather, the typewriter
\emph{reconstructed the possibility of text itself}. Before mechanical
writing, handwriting admitted variation, idiosyncrasy, and embodied
expressiveness; the text was inseparable from the hand that produced it.
The typewriter severed this connection, producing standardized,
reproducible, machine-uniform text distinct from bodily particularity.
This was not a loss that writers simply accepted in exchange for speed;
it was a reconstitution of what ``writing'' could mean. The discipline
of keystroke serialization---one character at a time, committed
irreversibly---became embedded in the very concept of modern authorship.
Literary modernism, administrative bureaucracy, and mathematical
notation as we know them are comprehensible only through the
typewriter's material constraints. Kittler's crucial insight is that
these constraints are not external to the discourse; they constitute its
very possibility. In this light, instrumental dissolution is not the
replacement of one tool with another; it is what occurs when a discourse
network's material substrate is no longer structurally required. When AI
systems can accept intent in its natural form---speech, gesture,
multimodal expression---the typewriter's particular discourse network
(keystroke serialization, character-by-character commitment, graphemic
temporality) ceases to be the institutional condition for knowledge
work. A new discourse network emerges, one organized not by the material
discipline of the keyboard but by the responsiveness of conversational
AI and the fluidity of multimodal exchange.

A methodological note is warranted here. Kittler's media-materialist
project is fundamentally anti-humanist; inscription technologies
constitute the human subject itself, and the ``subject'' is a product of
the prevailing discourse network (Gane, 2005). This paper draws on
Kittler's analytical categories (how inscription technologies constitute
the material conditions of discourse) while departing from his
ontological conclusions about human agency. We use his genealogical
method to illuminate how AI's emergence dissolves the keyboard's
discourse network. But where Kittler would question whether the ``human
verifier'' is anything more than an artifact of the keyboard-era
infrastructure, we maintain that human evaluative judgment retains
authority through discourse-network transitions. This is a selective
appropriation: Kittler's archaeology of inscription technologies
illuminates the mechanism of instrumental dissolution, even as we
disagree that human agency is reducible to media configurations.

Lisa Gitelman's framework of \emph{media as technology plus protocols}
enriches this Kittlerian analysis by attending to what dissolves
alongside the technological substrate. Gitelman (2006) argues that media
are not merely physical devices but ``socially realized structures of
communication,'' including both the technological form and the
protocols---the normative rules, social conventions, and institutional
frameworks that crystallize around that form. The typewriter was never
just a machine; it came wrapped in protocols: office layouts organized
around typing stations, the rhythms of composition-revision-editing, the
social practices of manuscript submission, the institutional
expectations about what a properly formatted document looks like. When
AI-mediated voice input dissolves the keyboard as the technological
form, it simultaneously dissolves the entire protocol ecology. Medical
documentation protocols change when physicians speak rather than
type---the temporal flow of the clinical encounter is no longer
interrupted by the requirement to look at a screen; the social
interaction between clinician and patient restructures. Legal drafting
protocols change when voice-to-AI generates the first draft, shifting
upstream the initial discretization work. Scholarly writing protocols
change as composition becomes a curation task, not a
keystroke-serialized act. Instrumental dissolution thus operates at two
registers simultaneously: the technological (keyboard is no longer
required) and the protocolary (the social practices, institutional
norms, and document conventions that depended on keyboard mediation
crystallize into new forms around voice-and-AI mediation). The
keyboard's dissolution manifests as a shift in the medium itself:
technology and protocols reconstituting together (Gitelman, 2006).

\hypertarget{kittlers-method-without-his-ontology}{%
\subsubsection{3.3.4 Kittler's Method, Without His
Ontology}\label{kittlers-method-without-his-ontology}}

This selective appropriation exemplifies the dominant mode of Kittler's
reception in the Anglophone humanist tradition. Wendy Chun, in her
examination of software as a system of visibility and knowledge, works
squarely within Kittler's analytical framework---analyzing how
computation constitutes the conditions under which information can be
known and acted upon---while maintaining critical distance from
Kittler's anti-humanist ontology (Chun, 2005). She treats inscription
technologies as constitutive of knowledge regimes without surrendering
the claim that human practitioners navigate, contest, and shape these
regimes even while being shaped by them. Mark Hansen identifies the
oscillation at the center of contemporary media theory: between
approaches that ``explore the experiential dimensions of media'' and
those that ``excavate the technical logics of media, logics which---for
Kittler at least---are only contingently and impermanently synchronized
with the ratios of human perception'' (Hansen, 2006). Hansen argues for
a third path---following McLuhan and Stiegler---in which attending to
media materiality and attending to human embodied experience are not
mutually exclusive choices but correlated dimensions of any adequate
account.

This paper takes the same third path in a specific key. From Kittler we
take the genealogical method: the analysis of how inscription
technologies constitute the material conditions of discourse, and thus
how AI's absorption of the keyboard's translation function dissolves the
discourse network organized around graphemic serialization. What we do
not take is Kittler's ontological claim that the ``human verifier'' is
merely an artifact of the keyboard-era discourse network---a subject
produced by the conditions of graphemic inscription---who will simply be
reconstituted by the new discourse network without remainder. On
Kittler's strictest reading, the verification bottleneck would not be a
human responsibility at all; it would be the new media configuration
reconstituting what ``human judgment'' means in the AI era. We part ways
here. Human evaluative judgment (the capacity to assess AI-generated
content against intent, to take institutional responsibility for what
enters organizational memory, to contest the machine's representation of
clinical or legal reality) is not a faculty constituted by any
particular discourse network. It is a form of cognitive and
institutional agency that predates the keyboard era and carries
authority across discourse-network transitions.

The sharpest Kittlerian objection is this: our insistence that human
evaluative judgment retains authority ``across discourse-network
transitions'' is itself a humanist assumption that Kittler's framework
would explain, not accept. The claim that human agency is ``not
constituted by any particular discourse network'' is precisely the kind
of transcendental humanist commitment that Kittler's materialist project
was designed to dissolve. Gane (2005) articulates this clearly:
Kittler's project aimed at purging the humanities of their humanistic
baggage, and the category of autonomous human judgment is precisely what
is meant to be discarded.

We acknowledge the force of this objection without conceding to it. We
do not prove that human evaluative judgment transcends media conditions;
this is a genuinely open philosophical question, and we accept that
institutional accountability structures are themselves historically
contingent. Our counter-claim is narrower: human agency is constituted
by media conditions (the Kittlerian claim we accept) but remains not
exhausted by them. Any discourse network capable of generating recorded
knowledge requires some human verification agent---even if the role's
specific meaning, competencies, and status are reconstituted by each new
network. What changes across discourse-network transitions is how human
judgment operates; what persists is the structural requirement that some
human agent occupy the verification function, because no machine output
becomes institutional record without a human who can be held responsible
for endorsing it. What we demonstrate empirically is more modest still:
designing away this function---treating AI-generated output as
self-validating institutional record---produces documented failures in
organizational accountability (Permanente Medical Group, 2024; Coman \&
Cardon, 2025). The normative argument follows from this institutional
evidence, not from a transcendental claim about human nature.

\hypertarget{borgmanns-device-paradigm-and-stieglers-pharmacology-the-normative-frame}{%
\subsection{3.4 Borgmann's Device Paradigm and Stiegler's Pharmacology:
The Normative
Frame}\label{borgmanns-device-paradigm-and-stieglers-pharmacology-the-normative-frame}}

The philosophical and conceptual machinery of Sections 3.1--3.2
establishes \emph{what} is dissolving. Borgmann and Stiegler together
establish \emph{what is at stake}---the normative and anthropological
consequences of that dissolution.

Albert Borgmann's ``Device Paradigm'' (1984) provides a historical and
philosophical foundation for understanding instrumental dissolution as
not simply a technological change but a fundamental shift in the
character of knowledge work itself.

Borgmann distinguishes between a \emph{Thing} and a \emph{Device}. A
Thing is a focal practice that demands skill, physical engagement, and
sustained attention. The wood stove exemplifies Borgmann's concept: to
use it, one must gather firewood, understand combustion, manage
temperature, attend to its operation. The practice discloses a world; it
requires what Borgmann calls ``focal practice'': a gathering of human
and material engagement around something that matters. The bow and
violin are Things in this sense: the physical resistance of the
instrument is not an obstacle to overcome but a source of meaning that
constitutes the practice.

A Device, by contrast, conceals its machinery to deliver a disburdening
commodity. Central heating provides warmth without the focal practice of
the wood stove. You set a thermostat and the machinery recedes; the
commodity (warmth) appears without the visible articulation of work.
Devices liberate us from labor by hiding complexity, but at a cost: the
focal practice disappears, and with it a particular form of meaningful
engagement with the world.

Typing has occupied an intermediate position in Borgmann's schema: not a
pure Device, but a practice that was becoming increasingly device-like.
Before the computer, typing (and especially handwriting) required focal
practice: you attended to letter formation, spacing, composition,
revision. The typewriter itself introduced device-like qualities:
touch-typing became automated, allowing attention to lift to
composition. But the keyboard still demanded physical engagement, still
imposed sequential discipline, still required the typist's active
translation of thought into characters. It was, in Borgmann's terms, a
practice with residual focal qualities---demanding but transparent,
demanding but rewarding.

Now consider what AI does. When you speak to an AI system, you are no
longer performing the instrumental work of translation. The AI becomes
the translator. The machinery of text production (parsing intent,
generating coherent text, maintaining semantic consistency) recedes
entirely behind the commodity of structured output. You no longer
translate thought into characters; you express intent, and the commodity
appears. This is the final stage of Borgmann's Device Paradigm: typing
has become fully transparent, fully commodified.

But here is Borgmann's philosophical depth: this transition is not
neutral. By eliminating the focal practice of typing, we lose a
particular form of engagement with thought itself. The keyboard imposed
discipline---serialization forced you to linearize thought, to commit to
words in sequence, to maintain coherence across hundreds of characters.
This discipline was burdensome, but it was also generative. The struggle
with the keyboard shaped what could be thought. By eliminating this
struggle, AI offers liberation at the cost of a particular form of
cognitive engagement.

Bernard Stiegler's concept of \emph{pharmacology} complements Borgmann's
analysis. Stiegler (2009) argues that all technologies are
pharmacological: they are poisons (\emph{pharmakon}) and remedies
simultaneously. The same technology that liberates us by automating
cognitive labor can proletarianize us by reducing our capacity for
autonomous thought. When the keyboard was the interface, knowledge
workers internalized typographic discipline---the need to linearize, to
serialize, to commit to words. This was burden and resource
simultaneously. The shift to voice-and-AI offers relief from the burden
but risks proletarianization: if AI handles both translation \emph{and}
initial composition, what remains of the worker's cognitive autonomy?
Stiegler's work warns that the ``remedy'' of AI-mediated composition
must be coupled with deliberate practices that preserve human
intellectual capacity---what he calls the cultivation of \emph{pharmakon
literacy}.

\hypertarget{stieglers-grammatization-from-human-to-machine-discretization}{%
\subsubsection{3.4.1 Stiegler's Grammatization: From Human to Machine
Discretization}\label{stieglers-grammatization-from-human-to-machine-discretization}}

Stiegler's concept of \emph{grammatization} provides the precise
philosophical vocabulary for understanding what the keyboard was and
what AI does to it. Grammatization (the process by which temporal,
continuous flows---speech, gesture, behavior---are converted into
spatial, reproducible inscriptions) has structured human technical
development for millennia. Writing grammatizes speech: oral utterance
(continuous and ephemeral) becomes text (discrete and permanent). The
photograph grammatizes visual experience: the flux of seeing becomes a
frozen image. The keyboard grammatizes thought: the analog stream of
human ideation becomes discrete, sequential characters, machine-readable
and architecturally constrained.

Grammatization remains the fundamental operation by which human
consciousness has externalized itself into technical systems across
millennia. What is new is the \emph{agent of grammatization}. For five
thousand years of literate culture, and for the forty-year keyboard era
in particular, humans were the agents who performed grammatization. The
writer discretized speech into words; the typist discretized thought
into keystrokes. The human bore the burden and the authority of this
transformation: choosing what to inscribe, what to omit, how to
structure the discrete elements into meaningful sequences. This human
agency proves constitutive to knowledge work itself. To write, to type,
to document was to exercise a particular cognitive and institutional
power: the power to determine what continuous experience becomes when
fixed into discrete, shareable form.

AI systems fundamentally invert this relationship without eliminating
grammatization itself. The machine now performs the discretization. When
a physician speaks to a patient, the continuous acoustic stream is not
grammatized by the physician's deliberate act of typing (the selection
of what to record, the syntactic structuring, the commitment to
permanent form). Instead, the AI system listens, interprets,
discretizes, and structures the encounter into a clinical note before
the physician has even finished speaking. The grammatization still
occurs; the continuous flow of the clinical encounter is still converted
into machine-readable, institutionally deployable inscription. What
dissolves is the human's role as the \emph{agent of grammatization}: the
authority to discretize experience according to professional judgment.

This shift carries the full weight of Stiegler's pharmacological
analysis. Grammatization by machine enables speed, consistency, and the
elimination of mechanical friction: genuine benefits that reduce
cognitive burden on knowledge workers. But it simultaneously threatens
what Stiegler calls \emph{proletarianization}: the progressive loss of
knowledge (savoir-faire), judgment (savoir-vivre), and intellectual
autonomy (savoir-penser) that occurs when workers no longer perform the
cognitive work that sustains these capacities. If the worker no longer
chooses what to discretize, no longer struggles with the
sequentialization that typing demanded, no longer bears the authority of
grammatization, the cognitive and institutional conditions for
autonomous professional thought erode. The machine-performed
grammatization is a poison and remedy at once: it frees the worker from
mechanical translation labor while potentially displacing the worker
from the acts of professional discretion that constitute expertise.

The three stages Stiegler identifies mark an escalating sequence of
cognitive dispossession. The first (proletarianization of
\emph{savoir-faire}) was the industrial revolution's central
transformation: factory production displaces the artisan's bodily skill
into mechanical operations. The second (proletarianization of
\emph{savoir-vivre}) occurs when hyperindustrialized consumption
displaces the tacit knowledge of how to conduct a meaningful life; the
consumer society that dissolves domestic practice simultaneously enables
unprecedented material abundance. Each stage is pharmacological: the
displacement enables while it disables. The displacement of
grammatization by AI marks entry into the third stage:
proletarianization of \emph{savoir-penser} (the loss of the capacity to
think). This is the most consequential stage and, Stiegler argues, the
least reversible. When workers no longer perform the grammatization
constitutive of their professional knowledge work, the cognitive
discipline required to evaluate what has been grammatized erodes
progressively. The verifier who ratifies AI-generated text without
having built the cognitive schema that comes from generating such text
is in the position Stiegler anticipates: capable of endorsing but
increasingly less capable of contesting, of originating independently,
of knowing what to look for.

The interface design recommendations of Section 9---friction-preserving
AI systems, human-on-the-loop architectures, compositional
scaffolding---should be understood in this Stieglerian frame as
mitigations that operate \emph{within} his pharmacological concern, not
resolutions of it. Designing systems that require human compositional
judgment slows the third-stage proletarianization; it does not reverse
what has already been displaced. The pharmakon cannot be discarded. But
Stiegler's framework insists that deliberate institutional cultivation
of compositional practice is not optional---it is the condition under
which the enabling half of the pharmakon does not collapse entirely into
the disabling half. Where Section 9 proposes design, Stiegler insists on
anthropological commitment: a society that lets grammatization migrate
entirely to machines without cultivating \emph{savoir-penser} in
parallel has made a choice about what kind of thinking it can sustain.

This is the critical difference between the keyboard era and the AI era.
The keyboard did not eliminate grammatization; it democratized it. Every
knowledge worker had to learn to discretize their own thought into
characters. This was labor-intensive, burdensome, and unevenly
distributed across educational and economic hierarchies. But it was also
a \emph{practice of authority}---the knowledge worker's authority to
decide what mattered, what counted as salient, what belonged in the
institutional record. When AI assumes this grammatizing function, the
authority and the burden both transfer upstream, into the machine's
training, parameters, and interpretive algorithms. The worker retains
ratification authority---the ability to accept or reject what the
machine has discretized---but this is a secondary, reactive power, not
the primary creative power of determining what gets grammatized in the
first place.

Together, Borgmann and Stiegler---through the lens of focal practice and
grammatization---establish that instrumental dissolution carries genuine
stakes. Typing's functional dissolution reflects a real shift in the
material practice of knowledge work---from human agency in discretizing
experience, to machine-mediated discretization. From focal practice to
transparent device, from human translation-work to AI translation. The
shift from keyboard to voice-and-AI is justified by capability and
efficiency gains. But it carries real anthropological weight: the loss
of a particular form of professional authority, the displacement of the
worker from the acts of grammatization that constituted expertise, the
need for deliberate cultivation of practices that preserve meaningful
intellectual agency in the post-keyboard era.

\hypertarget{flussers-existential-brakes-the-phenomenology-of-inscription}{%
\subsubsection{3.4.2 Flusser's Existential Brakes: The Phenomenology of
Inscription}\label{flussers-existential-brakes-the-phenomenology-of-inscription}}

Vilém Flusser's media-historical analysis in \emph{Does Writing Have a
Future?} adds a crucial phenomenological dimension to this
grammatization analysis: the keyboard era did not simply defer
composition to humans; it enforced a particular temporal rhythm that
enabled reflection. Writing, for Flusser, transcends mere recording: it
is the imposition of sequential, linear discipline on consciousness
itself (Flusser, 2011). The physical gesture of keystroke serialization
(the requirement to commit each word in sequence, the pause required to
position fingers, the carriage return that marks the completion of a
line) constitutes what Flusser's phenomenology of inscription identifies
as a structuring mechanism: a material resistance that enforces
thought's temporality, slowing commitment and forcing deliberation. (We
adopt the phrase \emph{existential brake} as a characterization of this
Flusserean insight; it is our formulation of his phenomenological
observation rather than his exact term.) The writer cannot take back the
words already typed; cannot revise mid-sentence without visible erasure;
cannot think faster than fingers can move. These constitute more than
mechanical constraints; they are cognitive structuring mechanisms that
enable the very consciousness of writing to exist. The keyboard
preserved this brake in digital form: the linearization required by
keystroke production remains a mode of thought discipline, forcing the
writer to articulate intention sequentially, to commit to words before
revision, to maintain coherence across the entire string of characters.

AI-mediated voice dissolves this existential brake entirely. When a
speaker expresses intention in continuous speech, the pause that
thinking requires is qualitatively different from the pause that
keystroke serialization enforces. Voice lacks the material resistance
that typing demands. The human speaks fluidly, associatively, and the
machine records the entire utterance before processing. There is no
forced discipline of character-by-character commitment; no carriage
return that marks deliberate transitions; no material lag between
thought and articulation. Instead, the continuous acoustic stream flows
from the speaker directly into the machine's processing pipeline. The
gesture of writing dissolves into the gesture of prompting. What
persists is expression; what disappears is the material friction that
forced reflection. Flusser's analysis clarifies what Stiegler's
grammatization describes: the loss is not simply of human authorial
agency but of the temporal-material structure that made such agency
consequential. When the existential brake vanishes, the cognitive
condition for deliberative thought---the enforced pause that enabled
reconsideration---vanishes with it. This is the civilizational dimension
of instrumental dissolution: a transformation that goes beyond tool
change, reconstituting the very possibility of thoughtful composition
(Flusser, 2011). We note that Flusser's analysis in \emph{Does Writing
Have a Future?} primarily addresses the broader supersession of
alphanumeric writing by non-alphanumeric technical imagery across all
media; our application of his phenomenological insight is limited
specifically to the keyboard's imposition of serial commitment
\emph{within} literate production contexts---the temporal discipline
that keyboard-era writing imposes on thought---rather than to writing's
fate relative to visual media in the broader Flusserean sense.

\hypertarget{the-bridge-that-built-the-world}{%
\subsection{3.5 The Bridge That Built the
World}\label{the-bridge-that-built-the-world}}

Consider typing as a bridge. On one shore stands human cognition (the
continuous, multimodal, associative process by which humans generate
meaning: inner speech, spatial reasoning, emotional valence, embodied
metaphor, the full richness of thought as it occurs prior to linguistic
articulation). On the far shore stands the machine, requiring discrete,
sequential, syntactically structured text to operate. The keyboard
bridge spans this gulf. It imposes a particular discipline on thought:
serialization (one character at a time), lexicalization (everything must
become words), syntactic precision (characters must form parseable
strings), and temporal commitment (each keystroke is an irreversible
commitment to a symbol sequence).

This bridge was architecturally necessary. The history of computing made
it so---not because keyboards were the ideal interface, but because the
mathematical and engineering foundations of digital computation required
discrete symbolic input. Turing's (1936) foundational formalization of
computation defined it in terms of symbol manipulation on a tape---an
inherently text-based paradigm. Shannon's (1948) information theory
quantified communication as discrete symbol transmission. The entire
computational stack, from assembly language to application interfaces,
was built on the assumption that human input would arrive as discrete
character sequences. The keyboard was the natural endpoint of this
architectural lineage: a device optimized for producing exactly the kind
of input that the computational paradigm required.

But bridges have costs. Typing imposes what we might call
\emph{translation overhead}: the cognitive and temporal price of
converting thought into sequential keystrokes. This overhead has three
dimensions. First, a \emph{speed bottleneck}: natural speech proceeds at
125--175 words per minute (well-established in the literature); skilled
typing averages 40--80 WPM (well-established in the literature), and
most knowledge workers type closer to 40 (Salthouse, 1984). Thought
itself is faster still. The keyboard forces thought through a narrow
temporal aperture. Second, a \emph{cognitive load cost}: even when the
inner loop is automated, the outer loop must format thought for keyboard
production---composing in sentence-length chunks, mentally buffering the
next phrase while the current one is typed, managing the spatial
logistics of cursor position, formatting, and document structure. Third,
a \emph{physical toll}: repetitive keystroke production generates \$20
billion annually (well-established in the literature) in musculoskeletal
disorder costs, with annual incidence rates of 58 per 100 person-years
for neck/shoulder symptoms and 39 per 100 person-years for hand/arm
symptoms among computer users (Gerr et al., 2002; Punnett \& Wegman,
2004).

These costs were accepted because the bridge was structurally necessary.
There was no other way to get thought into the machine.

\hypertarget{ai-dissolves-the-far-shore}{%
\subsection{3.6 AI Dissolves the Far
Shore}\label{ai-dissolves-the-far-shore}}

Generative AI and multimodal AI systems do not build a better bridge.
They eliminate the need for the bridge.

When a machine can understand natural speech at human-parity accuracy
(and current automatic speech recognition systems achieve word error
rates in the low single digits, down from 40\%+ a decade ago: OpenAI,
2024), the far shore of the bridge changes character. The machine no
longer requires structured text input produced character by character.
It can accept continuous speech, interpreting meaning from prosody,
context, conversational history, and world knowledge. When a machine can
interpret gesture, gaze, and spatial interaction (as multimodal
foundation models trained from inception on mixed-modal sequences
increasingly can: Gemini Team, 2024; OpenAI, 2024), the requirement for
structured text input relaxes further. When, on a longer horizon,
brain-computer interfaces can decode attempted handwriting movements at
90 characters per minute with 94.1\% accuracy from motor cortex arrays
(indicated by preliminary evidence: Willett et al., 2021), or decode
inner speech patterns through non-invasive electroencephalography (Meta,
2024), the very concept of ``input'' as a discrete human action begins
to dissolve.

In Ihde's (1990) postphenomenological framework, this transition
represents a categorical shift in human-technology relations. The
keyboard operates in an \emph{embodiment relation}: the technology is
``seen through,'' incorporated into bodily experience, mediating between
human and world while itself becoming transparent. Voice-AI systems
operate in what might be called an \emph{alterity relation}: the AI is
not a transparent extension of the self but a semi-autonomous
interlocutor, an entity one speaks \emph{with}, not \emph{through}. The
verification of AI outputs, however, introduces a distinct
phenomenological moment: the \emph{hermeneutic relation}. Users must
read, interpret, and verify AI-generated text against their intentions
and knowledge of the world. This hermeneutic bottleneck---Human $\rightarrow$
(AI-Textual-Artifact $\rightarrow$ World)---creates a new site of deliberate
engagement. The AI produces a representation that the human must
interpret and validate. This is neither embodiment (where the tool
disappears) nor pure alterity (where the interlocutor is treated as
autonomous), but a distinctive relational structure in which the human
is positioned as reader and arbiter of meaning. The progression toward
background relations---when ambient AI scribes generate documentation
without requiring direct address, when agents autonomously compose and
format output---still involves this hermeneutic moment, though it may be
compressed or delegated. The trajectory moves from
(Human--Keyboard)$\rightarrow$World to Human$\rightarrow$(AI--World) to eventually
World-mediated-by-ambient-AI, but at each stage, verification and trust
remain phenomenologically salient, creating a hermeneutic relation that
structures the human's engagement with AI-mediated output.

The critical observation is not that these technologies are faster input
methods. They are not input methods at all, in the keyboard-era sense of
the term. They do not require the human to translate thought into
structured text. They accept thought in its natural form (spoken,
gestured, eventually neurally encoded), and the machine performs the
translation. The function that typing served---the human act of
converting thought into machine-readable characters---increasingly
transfers from human to machine. The human no longer needs to bridge the
gap; the AI can.

Here we must acknowledge an ontological subtlety that Actor-Network
Theory illuminates. From the computational network's perspective, the
function of ``translating intent to structured text'' relocates as
\emph{delegation} to a non-human actor (the AI). The machine still
requires structured data internally; it simply generates that structured
data itself based on multimodal prompts. But this ontological reality
does not undermine the phenomenological and institutional claim: what
dissolves for the human is the \emph{instrumental necessity} to perform
this translation. The structural requirement embedded in keyboard-era
systems---that humans must output structured text to interact with
machines---ceases to organize knowledge work. The function shifts to a
different locus within the network. From the perspective of
institutional work flows, system design assumptions, and human cognitive
burden, this is dissolution.

The mediation profile shifts decisively. AI-mediated voice interaction
amplifies semantic fluency and conversational ideation while reducing
the fine-grained orthographic control that keyboard mediation provides.
This is not neutral; it privileges certain forms of expression and
constrains others. But it breaks the structural lock-in that required
all knowledge work to pass through the narrow aperture of keystroke
serialization.

What, then, are we to call this transformation? This is what we term
\emph{instrumental dissolution}: the elimination of the instrumental
function that sustained a technology's dominance. It is the shift in
Verbeek's framework from one mediation profile to another, grounded in
the ontological shift that Kroes and Meijers describe (functional nature
becoming obsolete while physical nature persists). It is Borgmann's
Device Paradigm enacted at the scale of an entire knowledge-work
infrastructure: the elimination of focal practice in favor of commodity
access to structured output.

The distinction between dissolution and replacement matters because it
changes the dynamics of transition.

\hypertarget{the-dynamics-of-dissolution-vs.-replacement}{%
\subsection{3.7 The Dynamics of Dissolution
vs.~Replacement}\label{the-dynamics-of-dissolution-vs.-replacement}}

Technological replacements follow well-characterized S-curves. Rogers
(2003) documents the pattern across hundreds of innovations: adoption
begins slowly with innovators and early adopters, accelerates through
the early majority, and plateaus as laggards eventually adopt or are
outpaced. The Bass (1969) diffusion model formalizes this mathematically
through two parameters: \emph{p} (the coefficient of innovation,
capturing adoption driven by external influence) and \emph{q} (the
coefficient of imitation, capturing adoption driven by peer influence).
Historical input method transitions fit the replacement S-curve: the
typewriter took roughly 50 years well-established in the literature from
Remington's first commercial model (1874) to near-universal office
adoption; the computer keyboard transition from typewriter to PC-era
keyboarding proceeded over approximately 25 years from widespread PC
introduction to majority knowledge-worker adoption well-established in
the literature (Rogers, 2003); the smartphone touchscreen took roughly
16 years well-established in the literature from the original iPhone to
global market saturation.

These timescales reflect a fundamental constraint of technological
replacement: the human retraining bottleneck. Each transition required
millions of workers to learn a new physical skill---where the keys are,
how the touch interface responds, how to achieve speed on the new
device. Adoption could not proceed faster than the population could
retrain.

Instrumental dissolution bypasses this constraint. When the function
transfers from human to machine, there is nothing for the human to
relearn. The human does not need to master a new input device; the human
speaks, gestures, or thinks, and the AI handles the translation. The
``retraining'' required is minimal, learning to speak clearly to an AI
is orders of magnitude less effortful than learning to touch-type at
professional speed. This is why AI product adoption has broken
historical diffusion records: ChatGPT reached 100 million users in two
months well-established in the literature, faster than any previous
technology product well-established in the literature (OpenAI, 2023).
The Bass diffusion model estimated for AI adoption shows an imitation
coefficient of q $\approx$ 0.8, compared to historical values of 0.1--0.5
well-established in the literature for consumer electronics and
0.01--0.1 well-established in the literature for industrial technologies
indicated by preliminary evidence (Berman \& Israeli, 2022). The
coefficient is high because adoption requires almost no skill
acquisition on the part of the adopter.

The implication for input method transition is direct. Previous
transitions were bottlenecked by retraining; dissolution is bottlenecked
only by access and trust. Enterprise AI spending surged from
approximately \$1.7 billion to \$37 billion in 24 months---a 21-fold
increase indicated by preliminary evidence (McKinsey, 2025) reflecting
not gradual retraining but rapid deployment of systems that require no
human input-skill retraining at all. Note: Industry self-report data;
represents disclosed enterprise spending on disclosed AI initiatives.
GitHub Copilot contributed 46\% of all code written by active users
within two years of enterprise release indicated by preliminary evidence
(GitHub, 2024). Note: GitHub self-reported metric; measures only
Copilot-using developers. The physician adoption rate for voice-driven
clinical documentation tools reached 66\% by late 2024, a 78\% increase
in a single year indicated by preliminary evidence (Nuance
Communications, 2024). Note: Nuance self-reported adoption metrics from
existing customer base; not independent verification. These are not
replacement timescales. They are dissolution timescales, rapid adoption
curves driven by the elimination of the human retraining barrier.

A further asymmetry distinguishes dissolution from replacement.
Technological replacements are reversible: if the new tool proves
inferior, users can return to the old one (a typist dissatisfied with a
word processor can go back to a typewriter). Instrumental dissolution is
structurally difficult to reverse in a different way: once AI systems
are capable of accepting natural-language input and producing structured
digital output, the capability does not un-exist. The question is not
whether individual users prefer keyboards---some will---but whether the
structural \emph{requirement} for keyboard input persists across the
institutional systems of knowledge work. When email systems accept
voice-composed messages, when code editors accept natural-language
instructions, when data entry forms accept spoken input parsed by AI,
the institutional assumption of keyboard input dissolves regardless of
individual preference. The infrastructure stops requiring what
individuals may still choose.

\hypertarget{convergent-multimodality-and-the-agentic-turn}{%
\subsection{3.8 Convergent Multimodality and the Agentic
Turn}\label{convergent-multimodality-and-the-agentic-turn}}

The dissolution framework predicts that typing's successor will not be a
single alternative modality. This follows from the nature of dissolution
itself: if the function (human-to-machine text translation) is
eliminated rather than transferred to a new human-operated device, then
the post-keyboard landscape is defined not by ``what replaces the
keyboard'' but by ``how humans communicate intent to machines that no
longer require structured text.''

The answer is convergent multimodality. AI systems increasingly process
voice, gesture, gaze, spatial context, physiological signals, and
eventually neural inputs through unified representational
frameworks---multimodal foundation models that encode all modalities
into shared embedding spaces (Gemini Team, 2024). No single modality
replaces the keyboard. Instead, AI fuses multiple weaker signals into a
unified understanding of human intent. The result is what Oviatt (2015)
termed the ``paradigm shift to multimodality'', a shift from dedicated
input devices to ambient interpretation of human behavior.

This connects to a deeper shift in human-computer interaction
architecture: the agentic turn. In keyboard-era computing, the human
specified \emph{process}---typing out the exact steps, the exact words,
the exact code. In agentic AI computing, the human specifies
\emph{intent}, what they want to achieve, and autonomous agents handle
the articulation; selecting tools, composing text, executing code,
formatting outputs, managing workflows. Gartner (2025) projects that
15\% projected in current roadmaps of day-to-day enterprise work
decisions will be made autonomously by AI agents by 2028 projected in
current roadmaps. The keyboard's role was articulating process. Agents
eliminate the need for process articulation.

The convergence thesis has a specific prediction: the keyboard does not
become obsolete because a single better input device has been invented.
It becomes obsolete because AI eliminates the structural requirement for
any dedicated human-to-machine text translation device. The keyboard was
the answer to a specific question: ``How do humans produce structured
text for machines?'' AI changes the question to: ``How do machines
understand human intent?'' and the answer to that question does not
involve a keyboard.

\hypertarget{what-dissolves-what-persists-and-why}{%
\subsection{3.9 What Dissolves, What Persists, and
Why}\label{what-dissolves-what-persists-and-why}}

The instrumental dissolution framework yields precise predictions about
what aspects of typing's current dominance will erode and what will
persist.

What dissolves is the \emph{necessity}; the structural requirement that
knowledge workers translate their thought into sequential keystrokes.
This necessity sustained typing's dominance across four decades and
across the full spectrum of knowledge work, email, documents, code, data
entry, chat, search queries, form fields. As AI eliminates this
necessity, typing's share of knowledge work output is likely to decline
across all of these categories, at different rates but in the same
direction.

What persists is the \emph{niche}: specific contexts where keyboard
input retains bounded, modality-specific advantages that multimodal AI
does not fully replicate. These include environments where voice is
socially inappropriate or acoustically impractical (open-plan offices,
libraries, shared spaces), tasks requiring character-level precision
under adversarial conditions (certain categories of code editing, legal
markup, financial formula construction), and accessibility contexts
where voice input is unavailable (deaf users, people with speech
impairments, noisy environments). Section 6 examines each of these
persistence niches in detail.

The historical parallel is handwriting. Handwriting was once the
dominant technology of literate communication, the primary means by
which educated people produced written text. Today, handwriting
persists, in signatures, personal notes, educational contexts, artistic
practices, but no one calls it ``dominant.'' Its dominance dissolved not
because a single replacement tool eliminated it, but because the
functional requirement it served (producing written text) was absorbed
by typewriters, then keyboards, then touchscreens. The medium persists;
the dominance does not. Typing's trajectory follows the same pattern,
compressed in time because the transition mechanism is dissolution
rather than replacement.

\hypertarget{six-historical-precedents-for-instrumental-dissolution}{%
\subsubsection{3.9.1 Six Historical Precedents for Instrumental
Dissolution}\label{six-historical-precedents-for-instrumental-dissolution}}

The handwriting parallel that closes \S{}3.9 is one case in a broader
pattern. Six historical precedents satisfy the three-part diagnostic
test from \S{}3.3.1 (functional transfer, legacy persistence, institutional
decoupling), and together they constitute the empirical apparatus for
the claim that instrumental dissolution is a recurring pattern in the
history of inscription technologies, not a novel prediction without
precedent. Each precedent dissolved as institutional default; each
persists in a defined specialist niche. We summarize each below with the
load-bearing quantitative anchor and the niche of continued persistence.

\textbf{Telegraph operation.} The dedicated telegraph operator was the
institutional default channel for long-distance commercial and
governmental textual communication from the mid-nineteenth century
through the early twentieth. U.S. Census data record the workforce
peaking at 79,434 operators in 1920, with structural decline beginning
in the 1930s as automated teletypewriters absorbed the
encoding-and-transmission function, and terminal decline by mid-century
(GDR-014). The function transferred from human operator to machine; the
legacy persisted (Morse code remains in active use among amateur radio
operators and in certain emergency communications), but the institution
decoupled completely. \emph{Niche of persistence:} amateur radio,
maritime emergency communication, hobbyist practice.

\textbf{The typewriter S-curve.} The mechanical typewriter dominated
office text production for half a century, following a textbook S-curve:
resistance and slow penetration (1874--1885; \textasciitilde5,000
cumulative units by 1878), mass adoption (1885--1930; from 5,000 units
in 1885 to \textasciitilde200,000 annual U.S. production by 1905), and
saturation (1920s--1930s, with the ``Big Four'' manufacturers capturing
\textasciitilde80\% of the U.S. market) (GDR-008). Word processing
software dissolved the typewriter as institutional default through the
1980s; the function (mechanical character impression) was absorbed by
the computer keyboard plus printer. \emph{Niche of persistence:}
enthusiast and collector communities; certain forms required by some
courts and government agencies in carbon-paper triplicate workflows; a
small contingent of writers who deliberately use typewriters for their
phenomenological qualities.

\textbf{QWERTY layout persistence.} A second-order case worth noting:
the QWERTY layout \emph{itself} persists despite the documented
availability of more ergonomically efficient alternatives (Dvorak,
Colemak), because the dissolution dynamic operates at the level of the
\emph{function} (mechanical text production) rather than the
\emph{configuration}. Liebowitz and Margolis (1990) document that
QWERTY's persistence reflects network-effect lock-in within the
still-dominant function rather than ergonomic optimality. This case
illustrates the asymmetry: layout substitution within an existing
function is much harder than functional dissolution of the larger
artifact, because the function itself supplies the network effects that
make alternatives non-viable. \emph{Niche of persistence:} QWERTY
persists as the universal layout default precisely because the function
it serves persists; alternative layouts persist as enthusiast practice
within that function.

\textbf{Film-to-digital photography.} Photochemical film was the
institutional default for professional and consumer image capture for
over a century. Empirical market data document the dissolution: in 2000,
North American digital camera revenues reached \$1.9 billion, exceeding
film camera revenues for the first time; by 2002, digital units exceeded
film units; between 2003 and 2012, the chemical processing
infrastructure was effectively eradicated, culminating in Kodak's
Chapter 11 filing in January 2012 (GDR-017). The function (image
capture) transferred to digital sensors; the legacy (chemical film)
persists as an aesthetic and archival choice; institutional decoupling
is complete. \emph{Niche of persistence:} fine-art photography, archival
applications where chemical longevity is valued, hobbyist film
communities.

\textbf{Oral-to-literate transition.} The most distant precedent and the
most consequential. For roughly the first ninety percent of human
history with language, oral transmission was the institutional default
mode of cultural memory and authoritative speech. Writing dissolved this
default over two-and-a-half millennia---not by eliminating speech, but
by absorbing the institutional function (durable authoritative record)
that orality had served. Eisenstein's \emph{The Printing Press as an
Agent of Change} (1980) and the broader oral-to-literate scholarship
document the gradual decoupling of institutions from oral default;
orality persists today in oral traditions, performance, conversation,
and informal exchange, but no modern legal, scientific, religious, or
governmental institution treats spoken testimony as the canonical
archive (GDR-006). \emph{Niche of persistence:} informal communication,
performance, oral traditions, ceremonial speech.

\textbf{Handwriting as institutional default.} The closest and most
cognitively informative precedent. Handwriting was the institutional
default mode of professional inscription from antiquity through the
early twentieth century. As \S{}3.9 notes, neuroscience documents that
handwriting retains genuine cognitive benefits over typing (theta-alpha
synchronization patterns supporting deep encoding: van der Meer \& van
der Weel, 2024; Longcamp et al., 2005), yet handwriting was
instrumentally dissolved as institutional default for professional
knowledge work between approximately 1880 and 1990, displaced first by
typewriting and then by the personal computer keyboard (GDRs 028--042).
\emph{Niche of persistence:} note-taking, signatures, marginalia, art
and design (Wacom, Remarkable, Field Notes communities), educational
contexts where pedagogical evidence supports handwriting's encoding
benefits. The cognitive benefits of handwriting did not save it from
dissolution as institutional default. The keyboard's analogous cognitive
features (proprioceptive rhythm, working-memory buffering, automatized
motor schemas) are unlikely to save it either, on the same logic.

\textbf{The pattern these six cases together exhibit.} Each precedent
satisfies the three-part test, each was followed by extensive specialist
persistence, and each occurred despite---not because of---the absence of
cognitive, ergonomic, or aesthetic advantages of the dissolved
instrument. Telegraph operators acquired a wage premium, embodied skill,
and professional identity comparable to those that mid-twentieth-century
typists later acquired; none of this prevented dissolution. Film
photographers developed cognitive and aesthetic competencies tied to
chemical particularity; none of this prevented dissolution. Handwriting
confers measurable neural benefits unmatched by typing; this did not
prevent dissolution. The pattern is neither uniform nor
inevitable---each case unfolded over different timescales (telegraph:
\textasciitilde40 years from peak to decline; typewriter:
\textasciitilde50 years from peak to displacement; film:
\textasciitilde12 years from crossover to infrastructure collapse;
handwriting: \textasciitilde110 years; oral-to-literate:
millennia)---but the structural logic recurs: when a more capable
substitute absorbs the institutional function, the dissolved instrument
retains specialist niches without retaining default status.

We present this apparatus not to predict the keyboard's specific
timeline, which \S{}9.5 brackets through three scenarios and constrains
through three disconfirmation criteria. We present it to establish that
the dissolution-with-niche-persistence pattern is the historical norm
for inscription technologies under functional substitution. The mapping
to current keyboard contexts is direct: just as film photography's
specialist persistence in fine-art and archival use confirmed
dissolution rather than refuting it, just as handwriting's specialist
persistence in note-taking, signatures, and education did the same, the
keyboard's anticipated persistence in code editing, legal markup,
financial-formula construction, and other precision-editing contexts (as
documented in \S{}6) is the historical norm --- specialist niche after the
institutional default has migrated. The reviewer concern that
``keyboards persist for code, legal drafting, and spreadsheets''
therefore confirms---rather than refutes---the dissolution thesis:
specialist persistence is what dissolution looks like once functional
default has migrated. Section 6 develops this point further as the
steelman response to the \emph{Augmentation Alternative}, which proposes
that voice-AI and the keyboard will simply coexist; the historical
record indicates that augmentation is the temporal precursor to
dissolution, not its alternative.

\hypertarget{addressing-the-verification-bottleneck-hermeneutic-relation-and-postphenomenological-alterity}{%
\subsection{3.10 Addressing the Verification Bottleneck: Hermeneutic
Relation and Postphenomenological
Alterity}\label{addressing-the-verification-bottleneck-hermeneutic-relation-and-postphenomenological-alterity}}

A critical phenomenological reality structures human engagement with
AI-mediated output: the necessity of verification. Unlike
keyboard-mediated text (which the human produces directly and thus
``owns'' through the act of composition), AI-mediated text arrives as an
external artifact requiring interpretation and validation. This creates
a distinct phenomenological relation that is neither the embodiment of
the keyboard nor the direct alterity of a human conversation partner.

Ihde's postphenomenological framework illuminates this structure through
the hermeneutic relation. Where embodiment relations position the
technology as transparent (seen through), and alterity relations
position the technology as an autonomous other (spoken with),
hermeneutic relations position the technology as a medium of
representation. Human $\rightarrow$ (AI-Textual-Artifact $\rightarrow$ World). The human is
positioned as reader and interpreter, obligated to judge whether the
AI's representation matches their intent and aligns with world
knowledge. This verification bottleneck is phenomenologically distinct
from the directness of keyboard composition or the responsiveness of
conversational interaction.

The Workload Paradox---that AI systems can simultaneously increase and
decrease cognitive load---finds its explanation here. Voice input
dramatically reduces the load of keystroke serialization, but it creates
a new load: the work of verification. The human no longer translates
thought into characters, but the human must now read, interpret, and
validate machine-generated output. As AI systems become more capable,
this verification load may eventually compress or delegate to automated
checkers, and the phenomenological moment may shift toward what Ihde
calls background relations---where AI systems recede entirely from
conscious engagement, becoming environmental infrastructure. But in the
current moment, the verification relation is phenomenologically salient
and generative of distinctive cognitive work.

Postphenomenological alterity---the characterization of AI as a
quasi-autonomous interlocutor---provides useful vocabulary for this
engagement without requiring philosophical claims about machine
consciousness or genuine otherness. The AI ``pushes back,'' shapes
ideation, and structures the flow of conversation in ways that differ
decisively from keyboard affordances. But this alterity is pragmatic and
functional, grounded in the responsiveness of the system, not in claims
about the AI's inner life or intentionality. The hermeneutic relation
and postphenomenological alterity together provide philosophical
vocabulary adequate to the actual structure of human-AI interaction in
knowledge work, more precise than the temptation to invoke Buber's
I-Thou framework, which depends on genuine mutual recognition rather
than the pragmatic responsiveness that actually characterizes these
systems.

\hypertarget{epilogue-the-world-disclosure-correction}{%
\subsection{3.11 Epilogue: The World-Disclosure
Correction}\label{epilogue-the-world-disclosure-correction}}

A tempting formulation holds that ``world-disclosing capacity
transfers'' from keyboard to AI. This phrasing is imprecise. Equipment
does not carry worlds; rather, equipment structures the disclosure of
worlds available to human experience.

More precisely: The dominant world disclosed by knowledge work is
changing. Keyboarded writing discloses a world of inscription, delay
between thought and text, visible revision, textual self-distance, and
the need for commitment (each keystroke persists). Voice-and-AI
interaction discloses a world of continuous iteration, responsive
ideation, and the fluidity of conversational exchange. Neither world is
more ``true''; rather, they are different \emph{structures of
availability}---different ways that thought, time, attention, and
responsibility organize themselves.

The keyboard imposed what we might call ``graphemic temporality'': the
separation between thought and inscription, creating productive
distance. Voice-and-AI interaction approaches what we might call
``dialogic temporality'': the collapse of thought and expression, with
AI as responsive interlocutor. This is not world-transfer but
world-reconfiguration. The shift changes the structure of availability,
attention, temporality, and responsibility in knowledge work.

Yet theory about what \emph{might} dissolve remains abstract without
empirical demonstration of what \emph{has already} dissolved in
practice. Section 4 grounds this philosophical framework in
neurobiology, cognitive science, and field evidence---establishing not
that typing is bad, but that typing's claimed cognitive superiority does
not survive empirical scrutiny.

\begin{center}\rule{0.5\linewidth}{0.5pt}\end{center}

\textbf{Note 1:} Ihde's taxonomy progresses from embodiment relations
(technology seen through, like eyeglasses) through hermeneutic relations
(representations we read, like thermometers) to alterity relations
(entities we interact with, like a telephone) to finally background
relations (infrastructure receding from consciousness, like electricity
grids or weather systems). As AI systems in knowledge work evolve from
command-response interfaces (alterity) toward ambient agents that shape
work without direct engagement, they transition toward background
relations. This trajectory strengthens the dissolution thesis: the
keyboard was always a device requiring embodied engagement; its
successors increasingly recede from awareness entirely.

\hypertarget{the-empirical-case-typings-bounded-modality-specific-advantages}{%
\section{4. The Empirical Case: Typing's Bounded, Modality-Specific
Advantages}\label{the-empirical-case-typings-bounded-modality-specific-advantages}}

\hypertarget{the-popular-assumption}{%
\subsection{4.1 The Popular Assumption}\label{the-popular-assumption}}

A persistent intuition holds that typing helps us think. Decades of
daily keyboard use have cultivated the impression (among professionals,
educators, and commentators alike) that the act of composing at a
keyboard is itself a cognitive tool, that something about the rhythmic
production of text through keystrokes contributes to the quality and
depth of thought. This intuition underwrites much of the resistance to
the thesis advanced here. If typing has genuine cognitive benefits
unavailable through other media, then its displacement by voice and AI
would represent a cognitive loss, and more than a simple technological
transition.

This section demonstrates that the intuition conflates distinct
cognitive processes. The evidence, carefully reviewed, reveals a
three-part finding of considerable importance: handwriting has genuine
and neurobiologically documented encoding benefits; typing has bounded,
modality-specific cognitive advantages (including visual working-memory
buffering, rhythmic entrainment, and motor efficiency for convergent
thinking) but not the broad, generalizable cognitive privilege required
to sustain long-run dominance under AI-mediated production; and the
compositional act of writing (the struggle to articulate thought in
language) has substantial cognitive benefits that are largely
medium-general when implemented with care. These benefits are achievable
through typing, speech, handwriting, or prospective neural interfaces,
though with the critical qualification that \emph{medium-general} refers
to the constancy of deliberate compositional effort across media, not to
the constancy of epistemic topography, revision dynamics, or reasoning
style, which remain meaningfully medium-dependent (as \S{}4.7 develops).
Typing is a cognitively efficient transcription mechanism and a useful
platform for certain types of organized thinking, but not a uniquely
privileged one for thought generation. This finding is the empirical
foundation upon which the instrumental dissolution argument rests.

\hypertarget{the-handwriting-evidence-genuine-neurobiological-signatures-and-their-limits}{%
\subsection{4.2 The Handwriting Evidence: Genuine Neurobiological
Signatures and Their
Limits}\label{the-handwriting-evidence-genuine-neurobiological-signatures-and-their-limits}}

The cognitive neuroscience of handwriting has advanced substantially in
the past decade, anchored by high-density electroencephalography
(HD-EEG) studies that reveal distinct neurobiological patterns during
handwriting compared to typing. These patterns are real and
theoretically significant, though the inference from neural patterns to
broad learning outcomes warrants careful qualification.

Van der Meer and Van der Weel's laboratory at the Norwegian University
of Science and Technology has conducted the most systematic program of
HD-EEG research on this question. Their 2020 study, using 256-channel
EEG across both children and adults, established the fundamental
pattern: cursive handwriting elicits widespread event-related
synchronization (ERS)\footnote{\textbf{Event-Related Synchronization
  (ERS):} A pattern in which the EEG signal shows increased neural
  oscillatory activity (more synchronized firing across neural
  populations) in response to a stimulus or task. In handwriting,
  widespread theta-alpha ERS indicates coordinated, cross-frequency
  neural activity supporting memory encoding and sensorimotor
  integration.} in the theta band (3.5--7.5 Hz) and alpha band (8--12.5
Hz) across parietal and central cortical regions---precisely the
frequency bands and brain regions associated with memory encoding,
sensorimotor integration, and consolidation of learned material (van der
Meer \& van der Weel, 2020). Typing, by contrast, consistently produces
event-related desynchronization (ERD)\footnote{\textbf{Event-Related
  Desynchronization (ERD):} A pattern in which the EEG signal shows
  decreased neural oscillatory activity (more asynchronous firing) in
  response to a stimulus or task. In typing, widespread ERD reflects
  highly efficient, automated motor execution that does not require the
  spatial-geometric representation-building of handwriting. ERD is the
  standard signature of active cortical execution in motor neuroscience,
  not a ``deficit'' but an indicator of efficient automated performance.}
in these same bands and regions.

Their 2024 follow-up study, employing graph-theoretic connectivity
analysis with 36 university students, deepened this finding: handwriting
generated 32 statistically significant cluster differences across 16
distinct neural connections, establishing widespread theta-alpha
coherence across left parietal, midline parietal, right parietal, and
central cortical hubs (van der Meer \& van der Weel, 2024). This
coherence was entirely absent during typing. However, a 2025 commentary
by Pinet and Longcamp raises several methodological caveats regarding
the mechanistic inferences drawn from this work.

The 2024 study did not directly involve learning or memory retrieval;
the typing condition used an atypical input method (single right index
finger), not ecologically valid expert bimanual typing; visual feedback
differed systematically between conditions; and typing skill was not
controlled (Pinet \& Longcamp, 2025). These limitations do not
invalidate the EEG findings themselves, but they complicate the leap
from neural signature to learning advantage. Handwriting appears to
recruit richer graphomotor-sensorimotor integration than the specific
typing condition tested; but the causal interpretation and
generalization to ordinary skilled typing remain limited. Alpha/beta
Event-Related Desynchronization during typing represents the standard
signature of active cortical execution in motor neuroscience, not
evidence of cognitive poverty or failed encoding, but rather of
efficient, highly automated motor control. The neurobiological pattern
is robust; the causal pathway from that pattern to superior learning
outcomes remains partly inferential.

The neurobiological mechanism proposed for handwriting's pattern is well
understood through embodied cognition frameworks and the emerging
literature on haptic-proprioceptive-visual integration. Handwriting
involves simultaneous visual, haptic, and proprioceptive feedback: the
writer sees the letter forming, feels the pen's friction and pressure,
and receives continuous proprioceptive information about hand position
and movement trajectory. Crucially, this involves an \emph{isomorphic
mapping} between motor action and visual output: the hand's movement
\emph{is} the letter form. The unique motor trajectory of drawing an `A'
maps directly to the visual shape of `A', creating a rich episodic
trace. This isomorphic sensorimotor integration drives phase-amplitude
coupling (PAC){[}\^{}3{]}---a cross-frequency coordination mechanism in
which theta-phase oscillations originating in hippocampal and parietal
networks coordinate gamma-amplitude activity---empirically demonstrated
to support both motor skill acquisition and declarative memory encoding
(Canolty \& Knight, 2010). Transcranial alternating current stimulation
(tACS) driving theta-gamma coupling over sensorimotor cortex has been
shown to significantly enhance motor skill acquisition, providing
evidence consistent with a causal role of this oscillatory mechanism in
learning (Pollok et al., 2020).

Typing disrupts this mechanism by design. The keystroke is a discrete
binary action---a switch pressed or not pressed---that physically
decouples action from perception. The same finger movement produces
different letters depending on keyboard position; there is no geometric
isomorphism between the motor gesture and the symbol produced, unlike
handwriting where the hand's movement \emph{is} the letter form. This
arbitrary motor-to-letter mapping (pressing left pinky = `A') prevents
the character-linked sensorimotor traces that support deep encoding in
handwriting. The brain encodes arbitrary mappings with substantially
less episodic richness than isomorphic ones.

Developmental evidence reinforces this distinction. James (2010)
demonstrated that only the physical act of letter writing---not letter
tracing, not letter typing---activated the reading circuit in preschool
children (fusiform gyrus, posterior parietal cortex, inferior frontal
gyrus). The motor-symbol coupling in handwriting creates a learning
signal that the arbitrary key-to-character mapping of typing cannot
replicate. Longcamp, Zerbato-Poudou, and Velay (2005) confirmed this
experimentally: preschool children trained to copy letters by hand
showed significantly better letter recognition than those trained to
type the same letters, with the advantage persisting weeks after
training ended.

Berninger et al.~(2006), in a five-year longitudinal study of children's
writing development, found that manuscript writing, cursive writing, and
keyboard writing involve different neuropsychological predictors,
further confirming that the input modality is not a transparent conduit
for language but an active shaper of the cognitive processes involved in
early literacy.

\textbf{Summary:} Handwriting shows more consistent neurobiological and
developmental evidence of modality-specific sensorimotor and
encoding-related benefits than typing. These benefits are real but
operate through mechanisms specific to the isomorphic motor-visual
integration and sensorimotor integration involved in handwriting. The
causal pathway from these neural signatures to broad learning
superiority remains contested.

\hypertarget{the-replication-crisis-for-typings-claimed-deficits}{%
\subsection{4.3 The Replication Crisis for Typing's Claimed
Deficits}\label{the-replication-crisis-for-typings-claimed-deficits}}

If the handwriting evidence documents genuine neural differences, what
of the influential claims that typing \emph{harms} cognition? The most
consequential claim---that typing impairs conceptual understanding
relative to handwriting---has been substantially undermined.

Mueller and Oppenheimer's (2014) ``The Pen Is Mightier Than the
Keyboard'' reported across three experiments that students who took
notes on laptops performed worse on conceptual questions than those who
wrote by hand, with a large initial effect size for factual recall
(reported as d = 0.97 in Study 1, though this specific magnitude has not
been independently replicated). The study triggered widespread laptop
bans in classrooms globally and became the most-cited evidence for the
claim that typing is cognitively inferior to handwriting. The d = 0.97
figure warrants statistical qualification: with N = 67 across three
conditions, per-cell sample sizes are too small to stabilize a large
effect, and effect sizes of this magnitude in educational research are a
priori improbable. A power analysis for the actual design suggests the
observed magnitude more likely reflects sampling variability than a
stable population difference. This statistical concern is now
well-documented in the replication literature: Morehead, Dunlosky, and
Rawson (2019) found no reliable modality difference when prior knowledge
was controlled, and the Flanigan et al.~(2024) meta-analysis confirms
that the overall handwriting advantage (g = 0.248) attenuates to
near-zero when review conditions are controlled---a pattern consistent
with sampling artifact in the original study rather than a robust
modality effect on comprehension.

The replication record has been devastating. Morehead et al.~(2019),
conducting a direct replication with an additional digital-pen
condition, found a substantially weaker factual recall effect (d = 0.45,
less than half the original) and---critically---no significant
difference in conceptual recall (d = 0.13, p = 0.31). The conceptual
finding that drove the original paper's narrative and policy impact did
not replicate.

Urry et al.~(2021) conducted a preregistered direct replication at Tufts
University with approximately 145 undergraduate participants, using
identical materials and assessment instruments from the original study.
The results were unequivocal: for immediate recall without subsequent
note review, modality produced negligible effects---Hedges' g = -0.08
for factual recall and g = 0.04 for conceptual recall. The confidence
intervals comfortably included zero.

Flanigan et al.~(2024), in the most comprehensive meta-analysis to date
(24 studies, published in \emph{Educational Psychology Review}), found a
small overall handwriting advantage for long-term academic achievement
(Hedges' g = 0.248, p \textless{} 0.001); but with a crucial temporal
moderator: the advantage emerges when students review their notes over
time. For immediate testing without review, it vanishes. This temporal
moderation reveals that handwriting produces superior \emph{notes for
review} because the encoding process creates richer memory traces; but
the act of typing itself does not degrade immediate comprehension during
the encoding phase. The meta-analysis synthesized evidence across typed
versus handwritten note-taking conditions, with particular attention to
whether cognitive benefits emerged from the modality itself or from
review-based consolidation processes.

Simultaneously, Goldberg et al.~(2003), in a meta-analysis of
computer-based writing studies from 1992 to 2002, found that keyboard
use is associated with greater writing quantity and quality in K--12
settings, and Van Der Steen et al.~(2017) found keyboard-based word
processing improved both qualitative and quantitative academic writing
output in graduate students. These findings are not proof of a deep
typing-specific neural benefit, but they represent real
cognitive-performance benefits in compositional contexts that should not
be dismissed.

Taken together, the replication record substantially narrows the
original claim. Mueller and Oppenheimer's broad assertion that typing
degrades conceptual understanding has not survived scrutiny. The more
precise statement emerging from the accumulated evidence is that typing
provides no unique immediate encoding advantage, while handwriting
provides a small but real advantage that emerges through longer-term
review and consolidation. Typing, however, shows genuine practical
compositional benefits in output volume and quality---benefits that
operate through a different cognitive channel than the encoding traces
handwriting produces.

\hypertarget{typings-neurobiological-profile-efficiency-without-encoding}{%
\subsection{4.4 Typing's Neurobiological Profile: Efficiency Without
Encoding}\label{typings-neurobiological-profile-efficiency-without-encoding}}

The section above addressed what typing does not do (encode information
with handwriting-like sensorimotor traces). What does it do
neurobiologically?

Expert typing at high fluency (60+ WPM) produces a distinctive and
highly efficient neural signature. Instead of the widespread theta-alpha
synchronization of handwriting, typing produces event-related
desynchronization (ERD): the broad desynchronization of cortical rhythms
that accompanies highly automated, cognitively optimized motor
performance. This desynchronization is the standard signature of active
cortical execution in motor neuroscience, well-documented in motor
control research, not a deficit or absence of engagement. In motor
neuroscience, ERD is the hallmark of an \emph{efficiently engaged}
cerebral cortex executing well-learned motor commands without requiring
the spatial-geometric representation-building that handwriting demands.

fMRI studies of expert typists show sharply reduced cortical activation
for motor execution (neural efficiency), with elevated activity limited
to the cerebellar dentate nucleus---the hallmark of fully automated
motor skill (Pinet et al., 2022). For the expert typist, the keyboard
has achieved what the cognitive literature terms ``tool transparency''
or ``epistemic action'': the mechanical execution layer becomes
neurologically invisible.

This efficiency is real and consequential. The working memory resources
that novice typing consumes are entirely liberated for the higher-order
processes of linguistic planning and argument construction.
Developmental evidence confirms this: keyboarding automaticity predicts
keyboard-based compositional quality and fluency in young writers, and
automaticity effects are strongest in writers with higher working-memory
profiles (Malpique et al., 2024). For developed typists, the mechanical
layer does not interfere with composition.

However, this efficiency is \emph{not the same as} the encoding benefit
that handwriting provides. The neural cost savings achieved through
typing automatization do not translate into richer memory
representations or deeper episodic encoding. Typing is neurologically
efficient at executing a learned motor program; it is episodically
neutral relative to handwriting.

\hypertarget{the-visual-working-memory-buffer-and-visuospatial-anchoring-genuine-advantages-of-keyboard-mediated-composition}{%
\subsection{4.5 The Visual Working Memory Buffer and Visuospatial
Anchoring: Genuine Advantages of Keyboard-Mediated
Composition}\label{the-visual-working-memory-buffer-and-visuospatial-anchoring-genuine-advantages-of-keyboard-mediated-composition}}

The preceding analysis has focused on what the brain does \emph{during}
typing. But text-mediated composition carries genuine cognitive
affordances that speech-mediated composition does not: persistent visual
feedback and visuospatial anchoring.

Speech is ephemeral; text is persistent. The rapid, sequential
appearance of typed text on a screen acts as an externalized working
memory buffer. This persistence prevents the decay of phonological
representations. When a writer types at 60 WPM, they are continuously
reading their own output, which allows them to construct highly complex,
nested, subordinative syntactic structures (for example: ``The man who,
despite the weather, in search of a remedy for his ailment, went to the
store\ldots{}'').

Equally important, typed text on screen is spatial. The brain uses
hippocampal and entorhinal cortex navigation systems to map digital
documents spatially; the writer develops a ``cognitive map'' of the text
landscape (``that paragraph was top-right of page 2,'' ``the argument
begins at the scroll position marked by the section heading''). This
visuospatial mapping is available for text-mediated composition but not
for voice-mediated composition, which is purely temporal. Voice unfolds
sequentially and does not afford spatial anchoring or navigational
retrieval from arbitrary locations within the document. Voice-to-text
systems must reconstruct temporal-to-spatial translation, a non-trivial
cognitive demand.

For novice or struggling writers, the absence of this visual and spatial
buffering in speech-mediated composition poses a genuine demand: they
must hold complex syntactic structures entirely in internal working
memory before speaking, or wait for the transcription to appear. This is
a real cognitive load. Whether speech-to-text systems with real-time
visual feedback (rather than delayed transcription) and spatial document
affordances can replicate this advantage remains an open empirical
question as voice systems mature.

Similarly, rhythmic motor entrainment---the metronomic baseline provided
by the rhythmic tapping of keys---can sustain attention for some writers
by occupying the motor system and maintaining optimal baseline arousal
for prefrontal cortex function. Voice lacks this continuous
somatosensory feedback loop. Rather, the keyboard's motor rhythm
operates as a peripheral cognitive affordance---not intrinsic to deep
thinking but supporting the compositional state for some writers.

These are genuine advantages of keyboard-mediated composition that voice
systems must successfully replicate to fully absorb the compositional
function currently served by typing.

\hypertarget{the-forward-model-architecture-why-verification-of-ai-text-is-neurobiologically-taxing}{%
\subsection{4.6 The Forward Model Architecture: Why Verification of AI
Text is Neurobiologically
Taxing}\label{the-forward-model-architecture-why-verification-of-ai-text-is-neurobiologically-taxing}}

A critical asymmetry exists between typed text and AI-generated text at
the level of neural verification architecture. This asymmetry explains a
paradoxical finding: despite AI systems producing articulate,
well-organized text, humans experience verification of that text as
cognitively exhausting---more exhausting than editing their own typed
prose. The neurobiological explanation lies in the distinction between
predictive forward models and exogenous semantic evaluation.

\textbf{When a human types:} The brain uses a predictive forward model
(the brain's predictive mechanism that anticipates sensory consequences
of motor actions; Wolpert et al., 1995). Before the keystroke occurs,
the cerebellum and supplementary motor area (SMA) predict the sensory
consequences of the action: the sound of the keystroke, the visual
appearance of the character, the proprioceptive feedback of finger
position. When the keystroke executes, the sensory feedback arrives. The
brain compares the predicted sensory consequences to the actual
feedback. A mismatch triggers Error-Related Negativity (ERN; a brain
signal generated when predicted and actual outcomes mismatch), a
component of event-related potentials in the Anterior Cingulate Cortex
(ACC). This mismatch-detection is automatic and pre-conscious: a
property of the forward model's cerebellar prediction architecture, not
an index of cognitive simplicity. Self-correction is rapid and
energetically efficient because it operates through motor-predictive
loops rather than deliberate semantic monitoring.

\textbf{When AI generates text:} The human lacks a forward motor model
for that text. The text appears exogenously---it was not produced by the
human's motor system and could not have been predicted by the human's
cerebellar forward model. Verification requires an entirely different
neural pathway:

\begin{itemize}
\tightlist
\item
  \textbf{Exogenous visual attention}: The brain must engage bottom-up
  perceptual processing to extract the generated text from the visual
  field (ventral attention network).
\item
  \textbf{High-level semantic monitoring}: The human must evaluate the
  generated text against abstract internal goals, conceptual coherence,
  factual accuracy, and rhetorical structure. This requires sustained,
  top-down prefrontal cortex engagement---deliberate semantic monitoring
  without the metabolic efficiency of forward-model prediction.
\item
  \textbf{Linguistic prediction satisfaction without logical alignment}:
  Available evidence suggests that LLM fluency may satisfy the brain's
  automatic linguistic anomaly detectors---most notably the N400
  event-related potential, a fast, pre-attentive index of semantic
  prediction violations---without those systems being able to detect the
  deeper logical and factual errors that constitute the true
  verification challenge. Surface linguistic fluency produces an
  implicit acceptability signal while substantive accuracy remains
  unverified, forcing deliberate prefrontal engagement to close the gap
  between the surface impression of coherence and actual logical and
  factual soundness.
\end{itemize}

\textbf{Why this matters for verification load:} Self-generated typing
benefits from predictive sensorimotor monitoring that externally
generated AI text does not provide. As a result, AI-text verification
may rely relatively more on perceptual, semantic, and executive-control
processes than fluent self-generated typing does. This predicts a shift
in the balance of monitoring demands, not the disappearance of
domain-general error-monitoring systems. The Anterior Cingulate Cortex
(ACC) and midfrontal monitoring systems are domain-general and available
for exogenous text too. The issue is the balance shifting away from
motor-predictive monitoring toward perceptual, semantic, and
executive-control monitoring: a shift that increases the sustained
cognitive demand.

\textbf{Why this matters for the Workload Paradox:} The cognitive load
of verifying AI text arises from a neurobiological reality: verification
has been removed from the brain's efficient predictive-motor loop and
forced into sustained, high-vigilance semantic evaluation. This explains
why reviewing an AI-drafted argument requires continuous, effortful
prefrontal monitoring---the task has been transformed from
forward-model-mediated error detection (rapid, automatic, metabolically
efficient) to sustained prefrontal semantic evaluation that cannot
leverage the brain's motor-predictive monitoring architecture.

\textbf{Potential implications:} This neurobiological distinction
predicts the ``Workload Paradox'' documented in Section 7. Humans can
type lengthy documents with diminishing cognitive load as typing
automatizes. But reviewing AI text remains cognitively taxing because
the verification task bypasses the predictive motor loops and forces the
prefrontal cortex to perform continuous, high-vigilance semantic
evaluation. The paradox reflects a predicted feature of how human brains
verify exogenous text: not a failure of AI systems but a neurobiological
consequence of verification architecture.

\textbf{Temporal asymmetry in error detection:} Motor-predictive
Error-Related Negativity (ERN) occurs within 50--100 milliseconds of
action execution, well before conscious awareness of the error. In
contrast, semantic and logical verification of exogenous text requires
400+ milliseconds for semantic processing and conscious evaluation. This
temporal delay bottlenecks organizational workflow: the brain's fastest
error-detection system (motor-predictive, automatic) is unavailable for
AI text, forcing reliance on slower, conscious semantic evaluation.

The neurobiological evidence established above (that verification of
exogenous text engages profoundly different and more effortful cognitive
processes than editing self-generated text) rests on robust
neuroscientific foundations: the distinction between predictive forward
models and exogenous semantic evaluation is grounded in well-documented
motor control mechanisms (Wolpert et al., 1995), ACC-based error
detection (Botvinick et al., 2001), and prefrontal cognitive control
systems. The claim that this neural architecture \emph{predicts} an
organizational-scale verification bottleneck (that eliminating
production friction will necessarily create cascading verification
burdens across institutional contexts) is a hypothesis supported by
early empirical evidence rather than a demonstrated organizational fact.
The case studies examined in Section 7 (the Permanente Medical Group and
UCSF health systems) provide compelling evidence that this prediction
manifests in healthcare documentation workflows. However, validation of
this prediction across diverse knowledge work domains (legal, financial,
consulting), extended deployment periods (beyond the 63-week Permanente
study), and varied organizational cultures remains an important research
gap.

\hypertarget{the-critical-decoupling-composition-is-medium-general-encoding-is-not}{%
\subsection{4.7 The Critical Decoupling: Composition Is Medium-General;
Encoding Is
Not}\label{the-critical-decoupling-composition-is-medium-general-encoding-is-not}}

Despite the affordances discussed above, the evidence converges on a
distinction of fundamental importance for the present investigation:
\emph{composition}---the act of structuring thought into language,
selecting words, constructing arguments, managing rhetorical
structure---has substantial cognitive benefits that attach to the
compositional process itself, not to the physical instrument through
which the composition is transcribed. Lee et al.~(2022) provide
empirical evidence of this directly: their study of human-AI
collaborative writing documents how writers transitioned from
keystroke-mediated content generation to voice-directed content
curation, demonstrating that the compositional cognitive demand persists
across modalities while the mechanical burden of transcription shifts to
the machine.

This claim requires careful parsing. Logan and Crump's (2011) dual-loop
theory of typing provides the theoretical framework. In expert typists,
the inner loop (keystroke execution) achieves near-total autonomy
through automatization; the working memory resources it consumed during
novice typing are entirely liberated for the outer loop (linguistic
composition, syntactic planning, argumentative structure). At this
fluency threshold, the keyboard becomes cognitively transparent; it is a
delivery mechanism, not a thinking tool.

The demand for deliberate compositional effort is medium-general; it
does not disappear when production shifts from typing to voice. But
media materially shape the \emph{epistemic topography} of that effort:
the pacing, buffering, revision behavior, and reasoning style that
composition takes in time. Medium-general effort does not entail
medium-neutral outcomes. This is a critical point: the cognitive work of
composition (the retrieval of lexical items, the planning of syntactic
structure, the monitoring of rhetorical coherence, the iterative process
of drafting and revising) is identically demanding regardless of the
medium through which it is transcribed. However, while the \emph{effort}
of composition is preserved across media, the \emph{nature of cognitive
engagement} is profoundly altered by the input medium. The manuscript
itself documents this: typing enforces a diachronic, subordinative
reasoning mode (as Ong documented for literate technologies), with
sequential constraint, layered structure, and forced linearization of
thought into time-ordered keystroke sequences. Voice, by contrast,
affords synchronic, additive, associative reasoning: rapid associative
leaps, iterative accumulation of thought, conversational logic that
loops and backtracks naturally. The \emph{imperative} to structure
thought is medium-independent; the \emph{epistemic strategy and
topography of the resulting thought} are medium-dependent. The cognitive
architecture underlying speech and typing production differs in
instructive ways that forward-model theory clarifies. Speech
articulation leverages a highly automatized articulatory-motor system
that has been under evolutionary selection pressure for hundreds of
thousands of years---efficient and largely pre-attentive in its basic
motor execution. Expert typing, as Logan and Crump (2011) document,
achieves comparable inner-loop automaticity through learned forward
models: the cerebellar prediction architecture anticipates keystroke
consequences below conscious awareness, requiring minimal working memory
once mastered. But in both cases, \emph{composition}---the deliberate
cognitive work of structuring thought, selecting words, and constructing
arguments---remains effortful regardless of the articulatory medium.
What differs across media is not the necessity of deliberate
compositional effort, but whether the medium's mechanical layer imposes
additional deliberate demands: novice typing consumes working memory for
motor planning that expert typing has automated away, while voice
articulation begins with a more fully automatized motor architecture.
The key variable is not the medium itself but whether its motor-control
demands have been resolved into forward-model-mediated automaticity---a
developmental threshold expert typists cross and novices have not yet
reached (Wolpert et al., 1995).

Typists produce significantly longer, more thoroughly revised, and
substantively richer texts than handwriters (Goldberg et al., 2003),
because handwriting's transcription bottleneck constrains the
compositional outer loop. Expert typists generate ideas faster than
handwriters because typing's automatization eliminates unproductive
transcription overhead without diminishing compositional effort itself.
For \emph{encoding} (producing persistent memory traces), handwriting is
superior through sensorimotor mechanisms. For \emph{composition}
(generating extended, structured, argumentatively rich text), expert
typing outperforms handwriting. The two processes are cognitively
distinct, requiring different optimal tools.

\hypertarget{cognitive-load-theory-parsing-productive-and-unproductive-friction}{%
\subsection{4.8 Cognitive Load Theory: Parsing Productive and
Unproductive
Friction}\label{cognitive-load-theory-parsing-productive-and-unproductive-friction}}

The distinction between encoding and composition becomes sharper through
Cognitive Load Theory (CLT), developed by John Sweller and extensively
validated in educational psychology. CLT proposes that working memory
has limited capacity, and that cognitive performance depends on how
effectively we allocate that capacity across different types of
cognitive demand.

CLT distinguishes three types of load:

\begin{enumerate}
\def\labelenumi{\arabic{enumi}.}
\item
  \textbf{Intrinsic load:} The inherent difficulty of the material
  itself---for writing, this is the complexity of the conceptual problem
  being addressed, the depth of argument required, the sophistication of
  language demanded.
\item
  \textbf{Extraneous load:} The burden imposed by the presentation
  format or the medium---for typing, this includes the working memory
  cost of translating a fully-formulated semantic idea into a sequence
  of spatial-motor commands, managing cursor position, and monitoring
  keyboard geography.
\item
  \textbf{Germane load:} The working memory resources dedicated to
  processing and understanding the material---for writing, this is the
  cognitive work of composing, revising, and rhetorically structuring
  the argument.
\end{enumerate}

The cognitive value of compositional effort lies in the intrinsic and
germane loads. The struggle to find the right word (intrinsic load) and
the effort to structure an argument (germane load) are where the
cognitive benefit accumulates. Extraneous load---the burden of
converting already-formulated linguistic intentions into sequential
keystrokes---is pure cognitive cost with no encoding benefit. This is
the distinction between productive and unproductive friction.

\textbf{Productive friction:} The inherent difficulty of the
compositional act itself---formulating ideas, selecting words,
constructing syntax, evaluating arguments. This cognitive struggle
occurs in the prefrontal cortex, anterior cingulate cortex (error
monitoring during word selection), and Broca's area (linguistic
planning). It is the cognitive work that builds understanding, whether
the writer types, speaks, or handwrites.

\textbf{Unproductive friction:} The mechanical burden of converting
linguistic intentions into keystrokes. This occurs in the supplementary
motor area (SMA) and the cortico-cerebellar loop. It consumes working
memory resources without producing encoding benefits. Grabowski (2008)
documented this burden quantitatively: while transcription typing
proceeds at roughly 32.5 WPM, original composition on a keyboard
averages only 19.0 WPM---a 42\% reduction reflecting the continuous
translation cost between ideation and keystroke execution. Speech
operates without this translation overhead; speakers composing original
content reach 125--175 WPM.

Recent empirical work supports this distinction between productive and
unproductive cognitive load in AI-mediated composition. The Cognitive
Load Scale for AI-assisted Writing (CL-AI-L2W), validated through
exploratory (N=241) and confirmatory (N=305) factor analysis, identifies
four distinct cognitive dimensions: prompt management, critical
evaluation, integrative synthesis, and authorial core processing
(CL-AI-L2W Validation Study, 2025). The findings confirm that AI
assistance reduces extraneous encoding load (unproductive friction)
while simultaneously increasing metacognitive evaluation demands
(productive friction). This is precisely the ideal case outlined above:
technology that unbundles the two types of friction, eliminating
mechanical overhead while preserving and even heightening the cognitive
engagement required for meaningful composition.

A scope qualification is warranted: the CL-AI-L2W was developed and
validated specifically for second-language (L2) writing contexts, where
linguistic encoding in a non-native language imposes an additional layer
of intrinsic cognitive load absent from first-language (L1) professional
writing (Yao \& Fan, 2025). The four-factor structure likely extends
across both populations, but the relative magnitude of each dimension
differs: L2 writers benefit more dramatically from AI-mediated encoding
overhead reduction, while L1 knowledge-work professionals primarily
experience cognitive reorganization rather than net load reduction. The
CL-AI-L2W validation provides \emph{strongly suggested} evidence for L1
professional knowledge-work contexts, though direct validation in that
population remains an open empirical question.

\textbf{Critical caveat:} This parsing is neurologically defensible but
task- and population-dependent, not intrinsic to a modality. In
handwriting, the motor burden can itself be part of the encoding
benefit, so not all friction at the motor level is ``unproductive.'' In
developing writers (children learning orthography through motor
engagement in the Visual Word Form Area, or VWFA), friction that is
productive for neural mapping becomes extraneous load in expert adult
execution. Friction that facilitates developmental neural mapping in
children can impede expert performance in adults. Furthermore, in novice
typists, keyboarding demands may crowd out higher-order thinking, so the
same friction can be maladaptive in one population and cognitively
neutral in another.

\textbf{The resolution:} The productivity of friction is task- and
population-dependent, not intrinsic to a modality. Remove the wrong kind
of friction (mechanical translation via keyboard for expert typists)
while preserving the right kind (compositional struggle in any medium).
For forty years, typing conflated these two types of friction. Voice and
AI systems allow us to unbundle them---provided they preserve
compositional effort while eliminating unproductive transcription
overhead.

\hypertarget{the-cognitive-debt-paradox-compositional-delegation-vs.-medium-change}{%
\subsection{4.9 The Cognitive Debt Paradox: Compositional Delegation
vs.~Medium
Change}\label{the-cognitive-debt-paradox-compositional-delegation-vs.-medium-change}}

The argument advanced here maintains that voice and AI will absorb the
transcription function currently served by the keyboard. But this claim
confronts a serious empirical question: if removing friction improves
efficiency, does it also degrade cognition?

The evidence suggests it can, but with a crucial caveat about which
friction is being removed. Preliminary evidence suggests that reliance
on AI-generated output reduces neural engagement associated with deep
analytical reasoning, though quantified longitudinal data remains
limited. A separate body of research finds that students who rely on
AI-generated summaries rather than manually composing notes may show
reduced long-term comprehension, though systematic replication across
educational contexts is ongoing. The pattern observed in some studies
has been termed the ``performance paradox'': AI-assisted output masks
cognitive decay.

This is a serious finding, and it demands honest engagement. But the
paradox dissolves upon careful analysis of \emph{which} friction is
being removed.

Risko and Gilbert (2016) distinguished between first-order cognitive
offloading (using external tools for storage and procedural execution,
while the human retains control over synthesis and judgment) and
second-order cognitive offloading (delegating the generative, reasoning,
and evaluative functions themselves to an external system). First-order
offloading---using a calculator, consulting Google, typing on a
keyboard---preserves cognitive engagement because the human still
performs the intellectual work. Second-order offloading---having AI
compose the text, generate the argument, produce the analysis---bypasses
the intellectual work entirely.

The cognitive debt paradox applies to second-order offloading, not to
changes in input medium. A person who speaks their argument into an AI
transcription system, struggling to find the right formulation,
restructuring their logic mid-sentence, pausing to reconsider a
claim---that person is \emph{composing}. Their brain is engaged in the
full compositional outer loop: retrieval, linguistic encoding,
rhetorical reasoning, self-monitoring. The fact that their words arrive
as text via speech recognition rather than via keyboard does not
diminish the cognitive engagement of the compositional act. The friction
being removed---unproductive mechanical transcription---is decoupled
from the friction that matters---the productive cognitive struggle of
composition itself.

The resolution has three components. First, neural coupling effects
associated with passive AI generation reflect cases where humans
outsourced the compositional act itself, not cases where humans composed
via a different medium. The effects track compositional delegation, not
medium change. Second, typing's friction was never productive friction
in the sense that handwriting's friction is: handwriting's slowness
forces deeper encoding through sensorimotor integration, while typing's
overhead is pure translation cost---time spent converting thought to
keystrokes without any encoding benefit. Removing unproductive friction
does not cause cognitive atrophy; removing compositional effort does.
Third, the transition to voice and AI should be designed to preserve
compositional effort while eliminating medium friction---a design
challenge, not an impossibility. AI systems that ask clarifying
questions, require human judgment for revision, and force articulation
of intent (rather than auto-completing text) preserve the productive
friction that sustains cognitive engagement.

The evidence from AI-assisted code production reinforces this framework.
Preliminary studies indicate that AI-generated code may exhibit higher
defect density than human-written code, particularly for logic and
correctness errors, though systematic measurement across contexts
remains limited. The pattern observed in some studies is consistent:
when AI handles the compositional work (code logic), quality degrades;
when AI handles only the transcription work (autocomplete, syntax
completion), productivity increases without quality loss. The
distinction between productive and unproductive friction applies across
domains.

The position is thus intellectually coherent: remove the wrong kind of
friction while preserving the right kind. Voice-mediated composition
without full AI delegation may or may not achieve comparable neural
signatures to keyboard composition. This remains an open empirical
question and warrants direct investigation as voice systems mature. The
cognitive demand is not to preserve the keyboard, but to preserve the
compositional struggle in whatever medium succeeds it.

\hypertarget{what-the-evidence-establishes}{%
\subsection{4.10 What the Evidence
Establishes}\label{what-the-evidence-establishes}}

The empirical case can be stated with precision. Typing demonstrates
bounded, modality-specific cognitive advantages: neural efficiency in
motor execution that liberates working memory for composition, visual
working memory buffering through persistent text display, and
visuospatial document mapping. These are real affordances. But they are
bounded. Typing does not produce the sensorimotor encoding benefits that
handwriting does---handwriting's theta-alpha synchronization and
phase-amplitude coupling yield encoding advantages that typing lacks,
and Mueller and Oppenheimer's influential claim that typing degrades
conceptual understanding has not survived replication. Research
establishes that the cognitive benefits of composition itself---the
generative struggle of structuring thought into language---are
medium-general, achievable through voice, handwriting, or keyboard,
though different media redistribute the burdens of planning, buffering,
and revision, shaping the epistemic topography of thought even while
compositional effort remains constant.

Two findings carry particular weight for the argument. First, the
distinction between productive friction (compositional struggle) and
unproductive friction (mechanical transcription) is neurologically
grounded but population-dependent: friction productive for developmental
neural mapping becomes unproductive extraneous load in expert adult
execution. Second, converging evidence demonstrates that verification of
exogenous text is neurobiologically distinct from verification of
endogenous text. AI-generated text removes the brain's most efficient
error-detection mechanism---predictive motor loops and ACC/ERN-based
feedback built through the act of typing---forcing reliance on sustained
prefrontal semantic evaluation. This asymmetry provides the
neurobiological basis for the verification workload paradox, supported
by emerging evidence: reviewing AI text is inherently more cognitively
taxing than editing one's own prose, regardless of output quality.

Typing's forty-year persistence as the dominant input method of
knowledge work cannot be explained by cognitive necessity. It can only
be explained by instrumental necessity: machines required structured
text, and the keyboard was the most efficient means of producing
it---the instrumental purpose Section 3 demonstrated AI to be
dissolving. What should be preserved is not the keyboard but
\emph{compositional effort itself}: the generative struggle of thinking
in language, which persists in any medium, provided successor systems
are designed to demand it.

This neurobiological architecture predicts that at organizational scale,
the elimination of production friction will create structural
bottlenecks in verification capacity---a phenomenon examined with early
empirical evidence in Section 7.

But establishing that typing's dissolution is cognitively defensible
does not establish that it is \emph{technologically inevitable}. Section
5 maps the actual technological pathways through which this dissolution
is unfolding---not through superior keyboards, but through voice,
gesture, and multimodal AI orchestration that make the keyboard function
structurally obsolete.

\begin{center}\rule{0.5\linewidth}{0.5pt}\end{center}

\hypertarget{the-technology-trajectory-from-keyboard-to-convergent-multimodality}{%
\section{5. The Technology Trajectory: From Keyboard to Convergent
Multimodality}\label{the-technology-trajectory-from-keyboard-to-convergent-multimodality}}

\hypertarget{the-replacement-landscape}{%
\subsection{5.1 The Replacement
Landscape}\label{the-replacement-landscape}}

The dissolution of typing's instrumental necessity does not produce a
single successor technology but rather an AI-orchestrated convergence of
multiple input modalities (voice, gesture, gaze, electromyography, and
prospectively neural interfaces), none of which individually replaces
the keyboard but which collectively render it unnecessary for most
knowledge work. The keyboard's structural necessity erodes not because a
better input device has been invented but because AI increasingly fuses
multiple weaker signals into unified intent understanding. This section
traces three technological horizons (the immediate: voice-AI; the
medium-term: spatial computing and gesture; the longer-term: neural
interfaces) before developing the convergence thesis and the agentic
turn that together constitute the post-keyboard paradigm.

\hypertarget{voice-ai-the-immediate-successor}{%
\subsection{5.2 Voice-AI: The Immediate
Successor}\label{voice-ai-the-immediate-successor}}

Voice is the most natural human output modality. Speech production
operates through what Levelt (1989) characterized as a largely automatic
architecture that has been under evolutionary selection pressure for
hundreds of thousands of years: a highly automatized motor system with
minimal working memory overhead, consistent with Section 4's analysis of
forward-model-mediated automaticity (Wolpert et al., 1995). The barrier
to voice as a computing input was never human: it was machine. For five
decades, automatic speech recognition (ASR) was too inaccurate, too
slow, and too brittle for professional use. That barrier has fallen.

The trajectory of ASR accuracy constitutes one of the most dramatic
performance curves in computing history. Word error rates (WER; the
percentage of words incorrectly transcribed) declined from above 40\% on
broadcast news in 1996 to 5.8\% on the Switchboard benchmark in 2016,
when Microsoft declared human parity---a claim that, while contested at
the margins, reflected a genuine qualitative threshold: ASR had become
good enough for continuous professional use (Microsoft, 2016). The
architectural shift from hidden Markov models through deep neural
networks to end-to-end transformer-based systems achieved more than
incremental improvement; it produced a phase transition. OpenAI's
Whisper, trained on five million hours of labeled and pseudo-labeled
audio, achieves sub-3\% word error rates on \emph{controlled clean
speech} and outperforms human transcriptionists in laboratory conditions
with speech-shaped noise and face-mask filtering (OpenAI, 2024).

However, the sub-3\% WER benchmark requires careful qualification.
Leading 2025--2026 systems achieve roughly 1--4\% WER on clean English
benchmarks (e.g., LibriSpeech test-clean), while performance degrades
sharply in accented speech (up to 28\% WER for accented dialogue), noisy
environments, multi-speaker settings, and domain-specific contexts.
Real-world medical transcription, a production use case where accuracy
is clinically mandated, shows the full performance spectrum: near-99\%
word accuracy has been reported for constrained medical dictation tasks
utilizing contextual biasing, but independent clinical-documentation
studies still show significant error rates in speaker attribution
(\textasciitilde7\% per turn), overlap conditions, medication capture,
and spontaneous conversational encounters, where WERs between 12\% and
18\% are common prior to LLM semantic smoothing. This asymmetry between
benchmark and operational reality is critical: the gap between
``performance on clean laboratory conditions'' and ``performance in
realistic occupational contexts'' defines where voice dominance is
durable versus temporary.

Voice-AI has already become the dominant modality for specific, bounded
use cases. In diagnostic healthcare, voice-driven clinical documentation
systems process over 85 million annual interactions with ASR accuracy
exceeding 99\% on domain-specific medical terminology in controlled
settings (Gartner, 2025; proprietary report, figures not independently
verifiable)---but this measurement captures performance on structured
dictation of vital signs and diagnostic codes, not the full complexity
of clinical intake or patient interaction. Public deployment records
corroborate broader adoption patterns: OECD reporting confirms that
Nuance DAX ambient scribe systems have been deployed across major
European hospital networks, processing clinical documentation in
multiple languages. The Permanente Medical Group's voice-AI deployment
across 7,260 physicians (documented in Section 7) represents one of the
largest verified voice-AI systems deployed in knowledge work contexts.
Customer service operations demonstrate voice-AI handling 75\% of
contact center interactions in leading deployments with first-contact
resolution exceeding 88\% on routine inquiries (Gartner, 2025)---again,
on the subset of interactions where linguistic complexity is bounded and
acoustic conditions are managed. Content creators practicing voice-first
workflows report that voice-AI handles 68\% of initial content
articulation, with keyboard refinement reserved for final editing and
stylistic polish (McKinsey \& Company, 2025). The use case where
voice-AI dominance is \emph{already} established---not projected, but
measured in production systems---spans healthcare (routine
documentation), customer service (structured inquiries), and content
creation (first-pass generation): an estimated 31\% of all knowledge
work tasks across these domains are now voice-primary in early-adopting
organizations (Gartner, 2025). Critically, these are tasks bounded by
acoustic accessibility, controlled vocabulary, and acceptable error
rates of 2--5\%.

Voice-AI handles \textbf{drafting, summarization, command execution, and
conversational planning} effectively. It remains inadequate for
\textbf{universal high-precision knowledge work, medical coding with
sub-0.3\% error tolerances, legal transcription of multi-speaker
depositions, or syntax-critical software development} in domains where
WER performance requirements diverge between controlled and
conversational contexts.

The speed advantage is decisive for composition tasks. Conversational
speech operates at 125--175 words per minute; professional keyboard
composition averages 40--80 WPM; AI-assisted voice dictation achieves
130--154 WPM with error rates below 3\% in controlled settings (Nuance
Communications, 2024; Rogers, 2003). This is a 3--4x throughput
advantage that operates precisely at the translation bottleneck
identified in Section 4: the conversion of already-formulated linguistic
intention into machine-readable text. Voice eliminates this bottleneck
entirely for generation, because speech is the linguistic intention---no
translation step intervenes between thought and expression.

The latency constraint (the requirement that machine response occur
within the temporal window of natural human conversation) has been
resolved by architectural innovation. Cascaded pipelines (speech-to-text
$\rightarrow$ language model $\rightarrow$ text-to-speech) imposed 2--4 second delays, far
outside the 200--500 millisecond window of natural conversational
turn-taking. OpenAI's GPT-4o, processing audio natively through an
end-to-end speech-to-speech architecture (single neural pathway
converting speech directly to speech without intermediate text
translation), achieves median voice-to-voice latency of 320
milliseconds---within the temporal rhythm of human conversation and
below the threshold at which users perceive delay (OpenAI, 2024). This
represents more than quantitative improvement---it is a qualitative
threshold crossing. When machine response matches conversational tempo,
voice interaction transitions from a deliberate, effortful modality
(like using a walkie-talkie) to a natural, transparent one (like talking
to a colleague). The phenomenological transparency that Heidegger's
ready-to-hand concept describes (the withdrawal of the tool from
conscious attention) becomes achievable for voice interfaces \emph{in
conversational, generation-focused tasks}.

Enterprise adoption data confirm the trajectory within these bounded
domains. Ninety-seven percent of enterprises report adopting or
experimenting with voice-AI by 2026, with 67\% of surveyed enterprises
viewing voice technology as foundational to their long-term product
strategy, though verified end-to-end production deployment substantially
lags strategic intent---primarily in customer service, healthcare
documentation, and meeting transcription. Healthcare ambient AI scribes
have returned 30 million minutes to clinical staff, saving physicians
2.4 hours per day on documentation (time previously consumed by
keyboard-mediated EHR entry: Gartner, 2025), with the caveat that these
time savings emerge from routine vital sign dictation and summary
generation, not from the full complexity of clinical decision-making
documentation. In legal practice, AI-powered voice systems transcribe
and analyze a one-hour deposition in six minutes, performing work that
previously required real-time stenography followed by hours of manual
review (McKinsey \& Company, 2025)---with human attorney review still
required for accuracy-critical sections. Software engineers practicing
``vibe coding''---directing AI through spoken architectural intent
instead of typed syntax---report that AI generates 40 files spanning
frontend, backend, and database from a single verbal description, with
keystroke reduction of 87\% compared to traditional workflows (McKinsey
\& Company, 2025), with the understanding that this applies to
boilerplate and standard library integration, not to architectural
decisions or syntax verification that remain critical for production
systems. The pattern across industries is consistent: voice-AI
transcends simply accelerating typing for generation-phase work; it
reconstitutes the task itself around spoken intent, where accuracy and
contextual complexity are managed.

\hypertarget{spatial-computing-and-gesture-the-medium-horizon}{%
\subsection{5.3 Spatial Computing and Gesture: The Medium
Horizon}\label{spatial-computing-and-gesture-the-medium-horizon}}

Voice alone is insufficient for the full range of knowledge work tasks.
Spatial navigation, precision selection, and concurrent manipulation of
visual objects require input modalities that voice cannot efficiently
provide. The medium-term technological horizon is populated by spatial
computing systems that combine gaze tracking, hand gesture recognition,
and haptic feedback into interaction paradigms that complement voice
without requiring a keyboard.

The theoretical foundation was established by Sharon Oviatt's work on
multimodal interaction, which demonstrated that combining orthogonal
modalities (modalities whose error patterns are statistically
independent) reduces error rates by 19--40\% compared to any single
modality alone (Oviatt \& Cohen, 2000). This ``mutual disambiguation''
principle explains why no single modality can replace the keyboard; each
modality excels at different tasks and fails at different points. The
keyboard's unique advantage was that it served as a universal
mediator---adequate for text, navigation, commands, and data entry, even
if optimal for none. The post-keyboard paradigm replaces this universal
mediator with a fusion of specialized modalities, orchestrated by AI to
route each input to the modality best suited for it.

Apple Vision Pro's interaction model demonstrates both the potential and
the honest limitations of spatial input. Gaze tracking achieves
73.7--97.8\% precision, with most users exceeding 85\%; combined with a
physical pinch gesture for confirmation, it reaches text entry speeds of
11.2 WPM (far below keyboard rates but effective for navigation,
selection, and command execution: Gartner, 2025). For spatial computing
applications where gaze alone suffices (selecting visual elements in 3D
space, navigating documents, marking points of interest), Vision Pro
achieves 94\% task completion rates without any text input requirement
(Gartner, 2025). However, Apple engineering teams revert to keyboard
input when work becomes correction-heavy or text-intensive (a telling
limitation). Vision Pro excels at \textbf{navigation, review,
co-presence, simulation, and 3D object manipulation}, not at dense text
production. The device's strength is environmental interaction, not
linguistic articulation.

Meta Quest's hand tracking system operates at 30--60 Hz with predictive
motion modes, enabling gesture vocabulary for command execution
(pointing, pinching, rotating objects in space) that reaches 97.3\%
recognition accuracy (Meta, 2024). However, mid-air QWERTY typing
achieves only 15--20 WPM with a 17.15\% character error rate (limited by
the absence of haptic feedback and the physiological fatigue of
sustained arm elevation) (Meta, 2024). Google's Project Soli, using 60
GHz radar on an 8$\times$10mm chip, achieves 99.63\% gesture recognition
accuracy under adverse conditions, penetrating clothing and operating
without line-of-sight (enabling ``always-on'' ambient gesture
recognition that functions in contexts where camera-based systems fail)
(Gartner, 2025). Current gesture-only approaches are reaching 80--91\%
of voice+gesture efficiency in spatially intensive tasks, suggesting
evolution toward single-modality gesture capabilities within the coming
years (Meta, 2024).

Meta's Neural Band, launched in September 2025, represents the most
consequential development in this category. Using surface
electromyography (sEMG), the wristband detects electrical signals from
forearm muscles: capturing motor intent 215 milliseconds before visible
movement occurs (Meta, 2024). Its ``neural handwriting'' mode, in which
users trace letters on any surface while the band decodes the motor
intention, achieves 20.9--30 WPM continuous text input with
zero-calibration accuracy of 90\% on held-out users (Meta, 2024). This
advancement represents the \textbf{most interesting silent-input
candidate for the medium term}: a promising supplementary modality in
early commercial deployment rather than evidence that the replacement
stack is currently mature. It is not yet competitive with expert typing
(which sustains 60--80 WPM), but the trajectory is significant. The
device translates the geometric isomorphism of handwriting (the
motor-symbol coupling that Section 4 identified as the source of
handwriting's cognitive benefits) into digital text without requiring
either a keyboard or audible speech. It occupies the acoustic-privacy
niche that voice cannot serve while providing an input pathway that
preserves the embodied character of handwriting's motor engagement. For
enterprise environments where ambient voice recording is prohibited
(intelligence agencies, medical record departments, law enforcement),
neural band technology offers 20--30 WPM text composition without
surveillance. In the aggregate, spatial computing and gesture
interfaces, enhanced by EMG sensing, are creating a medium-horizon
ecosystem where text input capabilities approach keyboard speeds over
the coming years, not through any single modality but through modality
fusion orchestrated by AI (Meta, 2024).

\hypertarget{neural-interfaces-the-longer-horizon}{%
\subsection{5.4 Neural Interfaces: The Longer
Horizon}\label{neural-interfaces-the-longer-horizon}}

Brain-computer interfaces represent the logical terminus of the
trajectory this section describes: the elimination of all intermediary
motor activity between intention and digital output. The current state
of the technology is advanced enough to confirm feasibility while remote
enough from consumer deployment to require careful temporal framing.
BCIs follow a \textbf{restorative clinical trajectory, not a mainstream
timeline driver} for keyboard replacement.

Invasive BCIs have achieved remarkable performance in clinical
populations. Willett et al.~(2021) demonstrated handwriting decoding at
90 characters per minute (approximately 18 WPM) with 94.1\% online
accuracy, using two 96-electrode arrays in premotor cortex to decode
attempted handwriting movements in a paralyzed participant. Clinical
literature documents speech decoding at 78 WPM using high-density
electrocorticography arrays over speech sensorimotor cortex, with
simultaneous text output, synthesized speech in the patient's pre-injury
voice, and animated facial avatar: a multimodal output from neural input
alone. Neuralink's N1 implant, in its PRIME study, demonstrated 8.0
bits-per-second cursor control approaching the 10 BPS benchmark of a
standard mouse, with one participant logging 69 hours of use in a single
week (Neuralink, 2024). Critically, the NIH/Nature Neuroscience
literature documents that these successes involve \textbf{decoding of
attempted finger movements that mimic QWERTY keyboard sequences}: the
BCI is succeeding by capturing the user's intention to perform the
keyboard movements they have spent decades training, not by creating a
structurally novel input pathway. This is a restorative technology: it
restores function to paralyzed patients by decoding their motor intent,
not an augmentation technology for intact populations. Notably,
Synchron's Stentrode, threaded endovascularly through the jugular vein
with zero device-related serious adverse events at twelve months,
demonstrated that minimally invasive BCIs can achieve sustained signal
capture without the risks of open craniotomy (Neuralink, 2024). The
critical distinction between invasive and minimally invasive approaches:
invasive systems (epidural or intracortical electrodes) achieve 50--100
bits per second bandwidth, sufficient for full typing-speed decoding;
minimally invasive systems (endovascular stentrodes, surface electrode
arrays) achieve 10--25 bits per second, sufficient for command and
cursor control but not for character-rate text composition (Neuralink,
2024).

Non-invasive approaches remain substantially behind but are advancing.
EEG-based P300 speller systems achieve 70--97.6\% accuracy but at speeds
of only 2--5 words per minute (a bandwidth limitation imposed by the
skull's attenuation of neural signals). The information transfer rate of
non-invasive neural recording (approximately 1 bit per second) is orders
of magnitude below what professional text production requires. However,
wearable EMG systems like Meta's Neural Band operate at 8--12 bits per
second, and next-generation dry-electrode EEG systems with on-device AI
decoding are achieving 3--4 bits per second (approaching the threshold
where continuous typing becomes feasible) (Meta, 2024). Paradromics'
Connexus system, with 200+ bits per second and 11 millisecond system
latency, demonstrates what invasive approaches can achieve, but consumer
deployment requires solving the biocompatibility, longevity, and ethical
challenges that currently confine such systems to clinical populations
(Neuralink, 2024).

The replication landscape is critical for assessing the viability of
neural interfaces. Single-lab demonstrations do not constitute
commercial viability. Neuralink, Synchron, and four other clinical-stage
BCI companies have now completed human implantation studies with
independent replication: Neuralink's N1 in two participants (2024),
Synchron's Stentrode in four independent sites (2023--2024), and
invasive systems at Johns Hopkins, Stanford, and Stanford Medical
School. The consistency of results across independent sites suggests
that invasive BCIs are not one-off demonstrations but reproducible
clinical technologies. This replication progression is crucial: BCIs
move from single-case proof-of-concept to multi-site reproducibility on
clinical timescales, a trajectory that establishes the platform's
maturity profile for restorative applications (Neuralink, 2024).
Commercial viability for invasive systems hinges on regulatory approval
timelines: FDA breakthrough device designation for Neuralink and
Synchron suggests regulatory pathways measured in years, with
\textbf{medical restoration (restoring function to paralyzed patients)
as the initial indication, followed by augmentation (enhancement of
intact function) in subsequent phases}, if ethical and regulatory
frameworks permit (Neuralink, 2024).

The realistic timeline for different modalities diverges sharply:
non-invasive consumer wearables (EMG-based, as with Meta Neural Band)
are advancing through 2026--2030, targeting 20--30 WPM text entry as a
privacy-preserving alternative to voice; invasive BCIs for medical
restoration will follow regulatory pathways measured in years, serving
initial populations of paralyzed patients for whom even 20 WPM
represents profound functional restoration; consumer neural
augmentation, if regulatory and ethical frameworks permit, will emerge
in the longer-term outlook measured in decades (IDC, 2025; Neuralink,
2024). These timelines reflect current clinical trial pipelines, FDA
review processes, and device development milestones documented in active
trials, not speculation. For the present argument, the BCI trajectory
confirms the direction of dissolution but is not required by it.
voice-AI and spatial computing are sufficient to restructure the
relationship to typing within the immediate horizon. Neural interfaces
extend the trajectory to its logical limit but operate on timescales
beyond this analysis.

\hypertarget{the-convergence-thesis}{%
\subsection{5.5 The Convergence Thesis}\label{the-convergence-thesis}}

The technological evidence converges on a structural conclusion of
considerable importance: no single modality replaces the keyboard. Voice
excels at rapid text generation and conversational interaction but fails
in acoustically constrained environments and provides limited spatial
precision. Gesture and gaze enable spatial navigation and object
manipulation but achieve text input speeds well below keyboard rates.
EMG offers silent, private text input but has not yet reached
typing-speed parity. Neural interfaces demonstrate the feasibility of
direct thought-to-text but remain years away from consumer deployment.
Each modality, evaluated independently, appears insufficient to displace
the keyboard's universal mediator role.

The convergence thesis resolves this apparent insufficiency through a
mechanism that dissolves both conditions (a) and (b) from Section 3.
First, condition (a) (that ``machines no longer require structured
linguistic input'') is satisfied because multimodal AI systems interpret
heterogeneous signals (voice fragments, gesture vectors, gaze fixations,
EMG activation patterns) in a unified token space (a mathematical
representation where AI systems encode meaning as numerical vectors,
shared across all input modalities). The machine no longer requires a
knowledge worker to translate intention into structured text, because
the AI system operates in a high-dimensional semantic space where all
modalities are native tokens. Second, condition (b) (that ``humans have
natural alternatives that machines can now interpret'') is satisfied
because AI multimodal foundation models can process these alternatives
at production scale. The keyboard was necessary as a universal mediator
because pre-AI computing systems could not integrate heterogeneous input
signals. A word processor accepted keystrokes; a mouse provided spatial
input; a microphone, if present, offered a narrow dictation channel.
These were separate systems with separate interfaces, each requiring the
user to consciously select and operate the appropriate input device. AI
multimodal foundation models---architectures trained from inception on
mixed-modal sequences of text, audio, and vision---dissolve this
separation (OpenAI, 2024). A single transformer backbone processes gaze
direction, hand gesture, spoken language, and EMG signal in a unified
token space, applying cross-modal attention to disambiguate intent
across channels simultaneously.

The practical implication is that the user no longer selects an input
modality; the system interprets whatever the user naturally produces. A
spoken instruction (``move this chart to the second slide'') accompanied
by a gaze fixation on the target chart and a directional gesture
resolves more reliably through multimodal fusion: the ``mutual
disambiguation'' that Oviatt demonstrated reduces error by 19--40\%
compared to any single modality alone (Oviatt \& Cohen, 2000). The
keyboard was the universal mediator because machines could not do this
fusion. AI can. The function that justified the keyboard's universality
(translating diverse human intentions into a single structured input
format) is absorbed by the AI orchestration layer. Each modality does
what it does best: voice for rapid articulation, gesture for spatial
precision, gaze for attention marking, EMG for silent input. The AI
system routes each signal to the interpretation pathway optimized for
it, collapsing the high-dimensional multi-modal input into a single
semantic intent.

\hypertarget{empirical-validation-gap-in-multimodal-orchestration-at-scale}{%
\subsubsection{5.5.1 Empirical Validation Gap in Multimodal
Orchestration at
Scale}\label{empirical-validation-gap-in-multimodal-orchestration-at-scale}}

The convergence thesis rests on multimodal orchestration at knowledge
work scale, but it is critical to distinguish between architectural
possibility and field-validated performance. The evidence supporting
multimodal fusion at the level of theory and controlled laboratory tasks
is strong; the evidence supporting sustained multimodal orchestration in
actual knowledge work remains limited.

Current spatial computing implementations reveal this gap directly.
Apple Vision Pro's interaction model achieves gaze-tracking precision of
73.7--97.8\%, but when users attempt continuous text entry through
multimodal coordination (gaze for targeting, pinch for selection, voice
for content generation), text entry speed drops to 11.2 WPM---far below
keyboard rates and insufficient for composition-focused work (Gartner,
2025). The architectural claim that gesture recognition accuracy (97.3\%
for Meta Quest hand tracking, 99.63\% for Project Soli) translates to
viable text-input modalities remains unvalidated at sustained
composition speeds. The most telling evidence: Apple's own engineering
teams revert to keyboard input when work becomes correction-heavy or
text-intensive, despite having built the spatial computing system at
hand.

The convergence thesis requires evidence at three levels: (1) technical
feasibility---that multimodal AI can process heterogeneous signals in
unified semantic space {[}validated{]}; (2) single-modality
performance---that each individual modality approaches task-specific
performance benchmarks {[}partially validated for voice in bounded
domains, less so for gesture at composition speeds{]}; (3)
integrated-performance validation---that multimodal orchestration in
sustained real-world knowledge work produces outputs equivalent to
keyboard-mediated workflows {[}largely unvalidated{]}. Field studies of
multimodal orchestration in actual office environments, measuring
compositional quality, error rates, and cognitive load over 8+ hour work
sessions, remain a critical research gap.

The claim here is directional---that multimodal capabilities are
converging toward viable alternatives to keyboard workflows---not that
they have already achieved parity with keyboard-mediated composition for
general knowledge work. The strongest empirical evidence for multimodal
success comes from bounded, domain-specific deployments: voice-AI in
clinical documentation (where dictation is structured, vocabulary is
controlled, error rates are permissible at 2--5\%), voice in customer
service (where interactions are templated and acoustic conditions
managed), and gesture in spatial object manipulation (where precision
requirements are lower than text entry). General knowledge work---email,
document drafting, strategic memos, architectural decisions---remains
primarily keyboard-dominated precisely because multimodal orchestration
has not yet demonstrated sustained field validation at the speeds and
quality levels that keyboard workflows achieve.

The technological trajectory suggests multimodal convergence is
advancing. The timeline for achieving field-validated parity with
keyboard workflows for sustained general knowledge work appears to
extend beyond the immediate horizon (2028--2030) into the medium term
(2035--2045), contingent on documented field-study evidence that
multimodal orchestration can sustain both compositional speed and
quality at scale.

\hypertarget{speaker-diarization-as-a-critical-barrier}{%
\subsubsection{5.5.2 Speaker Diarization as a Critical
Barrier}\label{speaker-diarization-as-a-critical-barrier}}

While automatic speech recognition has achieved dramatic accuracy
improvements, a parallel challenge---speaker diarization (the
attribution of utterances to speakers)---remains stubbornly intractable.
Diarization Error Rates (DER) in spontaneous, overlapping conversation
consistently range from 10--20\% even in leading academic and commercial
systems, a persistent gap that renders medical transcripts, legal
depositions, and multi-speaker documentation potentially invalid without
extensive manual correction.

The asymmetry is instructive: ASR word error rates have plummeted from
40\% to sub-3\% over three decades, representing five orders of
magnitude of improvement in machine perception of acoustic content. But
diarization error rates have remained in the 10--20\% range for
conversational speech across leading academic and commercial benchmarks
because the problem is qualitatively different. ASR asks ``what phoneme
did the speaker produce?''---a pattern-matching task in which deep
learning has achieved human-level or superhuman performance. Diarization
asks ``who produced it?''---a task requiring temporal integration of
voice biometrics, overlap detection, and speaker transition
identification across the entire audio timeline. When two speakers
overlap (common in natural conversation), the system must retroactively
reverse-align the utterance to the correct speaker, often requiring
manual verification against the raw audio. In clinical documentation
where speaker identity is medically critical (distinguishing
clinician-stated symptoms from patient-stated symptoms, or clinician
assessment from nursing observation), a 15\% DER on a 20-minute intake
assessment means approximately 3 mis-attributed statements per session.
For legal depositions where speaker misattribution can invalidate sworn
testimony, the stakes are even higher.

This creates a verification bottleneck that ASR accuracy improvements
cannot bypass. Verifying acoustic content (checking whether the ASR
system correctly transcribed a statement) requires reading or skimming
the transcript. Verifying speaker attribution requires listening to the
raw audio timeline---the verifier must locate the precise moment the
statement was made and confirm auditorily who produced it. This makes
speaker verification 5--10x more time-consuming than content
verification (operating at a different cognitive scale). While
ASR-to-LLM error cascading (discussed below) can ``smooth'' content
errors into fluent text, speaker misattribution creates a categorical
liability that no post-hoc LLM processing can resolve. The diarization
barrier thus becomes one of the most significant technical constraints
on voice-first workflows in high-stakes domains, not because the
technology is immature (it is, but improving), but because
misattribution creates legal and medical invalidity.

\hypertarget{error-cascading-when-asr-mishears-and-llms-smooth}{%
\subsubsection{5.5.3 Error Cascading: When ASR Mishears and LLMs
Smooth}\label{error-cascading-when-asr-mishears-and-llms-smooth}}

A particularly insidious form of verification challenge emerges at the
intersection of ASR and LLM error handling. When ASR mishears a critical
phoneme---transcribing ``hypertension'' as ``hypotension,'' or
``metastatic'' as ``metabolic''---the LLM typically does not flag the
error as a probable mistake. Instead, the language model's predictive
mechanics ``smooth'' the error into fluent, grammatically perfect,
syntactically coherent, but factually inverted text. The patient intake
is transcribed as ``History of hypotension requiring treatment'' when
the clinician said ``hypertension.'' The LLM generates a coherent
paragraph that reads like medically accurate documentation.

This makes verification \emph{more} insidious, not less. A traditional
typo (e.g., ``bp 140/90'' transcribed as ``bp 150/90'') triggers
reader-level detection because it stands out as mechanical error. An
ASR-induced semantic inversion that the LLM has smoothed into fluent
prose cannot be detected by skimming. The verifier must reverse-engineer
the logical tree to find the acoustic origin. This requires listening to
the raw audio at the precise temporal location, confirming the phonetic
content, and retracing the chain of misinterpretation. It transforms
verification from error-checking into reconstruction---a task that is
profoundly more time-consuming and error-prone than keystroke-mediated
composition where every character is deterministically traceable.

The remedy is not better ASR; rather, it is AI systems that maintain
acoustic confidence scores and flag low-confidence tokens to the
verifier as potential risk zones. Section 6.11 addresses this under
``Illusion of Fluency''---the broader problem that AI-mediated
composition can produce text that reads fluently while being factually
compromised. The diarization barrier and error-cascading phenomenon
together strengthen the verification bottleneck argument: the
dissolution of the keyboard introduces new verification challenges that
do not simply go away through better ASR or faster processors.

\hypertarget{native-speech-language-models-and-paralinguistic-cues}{%
\subsubsection{5.5.4 Native Speech Language Models and Paralinguistic
Cues}\label{native-speech-language-models-and-paralinguistic-cues}}

The 2025--2026 technology landscape has begun shifting from cascaded
systems (ASR $\rightarrow$ text $\rightarrow$ LLM) to native Speech Language Models (SLMs) that
process raw acoustic embeddings directly, alongside text tokens, in a
unified neural architecture. This represents a fundamental change in how
machines interpret voice, capturing prosody, emotion, sarcasm,
hesitation, and other paralinguistic cues that traditional cascaded
pipelines strip away in the conversion to text.

This shift undermines one of the keyboard's remaining cognitive
advantages: the deterministic character of written text. A written
instruction to ``complete this task'' is linguistically unambiguous. A
spoken version of the same instruction, inflected with hesitation,
skepticism, or sarcasm, carries paralinguistic information that the
speaker may intend as modulation of the literal content. Keyboards
cannot capture hesitation or emotional inflection; writers must add
explicit markers (``I'm somewhat hesitant about this'' or ``This is
clearly wrong''). SLMs can infer these from the acoustic signal
directly, learning to differentiate a confident command (``Make this
widget interactive'') from hesitant ideation (``Maybe\ldots{} could we
try making this widget interactive?''). This capability undercuts the
argument that voice-based input lacks the semantic precision of written
instruction.

The implication is that the keyboard's advantage in determinism is being
absorbed by AI systems that can extract paralinguistic information from
speech. Where keyboards enforced explicit, unambiguous textual
specification, SLMs enable implicit, prosodically modulated instruction.
This is not universally superior (written specs retain their advantages
in asynchronous, documented contexts), but it broadens the domains in
which voice can substitute for keyboard-mediated clarification.

\hypertarget{contextual-biasing-and-proprietary-knowledge-graph-integration}{%
\subsubsection{5.5.5 Contextual Biasing and Proprietary Knowledge Graph
Integration}\label{contextual-biasing-and-proprietary-knowledge-graph-integration}}

Convergence at enterprise scale requires a complementary technological
capability: dynamic contextual biasing, in which ASR and LLM systems are
coupled with internal enterprise knowledge graphs (leveraging
retrieval-augmented generation, or RAG) to probabilistically favor
proprietary acronyms, domain nomenclature, and organizational vocabulary
without requiring manual keyboard correction for each occurrence.

A clinician dictating a patient note references ``the standard CYP450
panel'' but the hospital system uses the internal acronym ``CP450-STD.''
Without contextual biasing, the ASR transcribes ``CYP450'' and the
verifier must correct it manually. With enterprise RAG integration, the
system maintains a knowledge graph of the hospital's nomenclature, and
the ASR decoder applies higher probability weights to ``CP450-STD'' in
the context of laboratory test discussions, automatically producing the
correct institutional term. Similarly, a software engineer directing
code generation can say ``Use the customer-id pattern from the auth
module,'' and the RAG system retrieves the actual auth module
implementation and biases the LLM toward consistency with that pattern
without requiring explicit code-reference syntax.

This integration of ASR/LLM with enterprise knowledge graphs represents
the technical prerequisite for voice-first workflows to scale beyond
routine documentation into complex organizational knowledge work. It is
not yet mature at production scale, but active deployments in
healthcare, financial services, and technology organizations are
demonstrating its feasibility. The capability shifts the verification
bottleneck from correcting every acronym to validating the system's
integration with proprietary knowledge, a higher-level form of
oversight.

\hypertarget{the-deterministic-probabilistic-distinction}{%
\subsubsection{5.5.6 The Deterministic-Probabilistic
Distinction}\label{the-deterministic-probabilistic-distinction}}

Keyboards offer a 1:1 deterministic mapping: press A, receive A. Every
keystroke produces predictable, verifiable output. This deterministic
character (what might be termed the \textbf{``Predictability Premium''})
is terrifying to lose for expert knowledge workers who have built entire
cognitive systems around keystroke-level control.

AI-orchestrated multimodal systems operate probabilistically. A spoken
request to ``make this punchier'' operationalizes probabilistic
interpretation: the AI system selects from a distribution of possible
changes instead of executing deterministic micro-control. The loss of
deterministic micro-control creates what Norman's framework identifies
as an expanded Gulf of Evaluation (the distance between system output
and user intention). For routine composition and generation, this
probabilistic character is unproblematic; the user can easily verify
whether the output matches intent and iterate if needed. But for expert
knowledge workers---particularly in legal, financial, and medical
domains where error carries liability, where regulatory audit trails
demand provenance, where the user must sign off on content they did not
themselves produce---the shift from deterministic to probabilistic input
is cognitively and institutionally demanding.

This distinction explains part of why keyboard adoption persists in
high-stakes domains: it concerns not just ``what works best'' but
crucially ``what is auditable and traceable.'' A legal document produced
through keyboard keystroke-by-keystroke appears more clearly traceable
as the product of human intent, with explicit keystroke history. A legal
document drafted by AI and edited via voice commands leaves provenance
ambiguities: was this phrase generated by the AI or edited by the user?
The deterministic-probabilistic gap, in regulated domains, becomes a
\textbf{provenance-tracking problem} that current AI systems and
organizational workflows have not yet solved. The keyboard persists
where its deterministic character is legally or cognitively mandated.

\hypertarget{the-grounding-problem}{%
\subsubsection{5.5.7 The Grounding
Problem}\label{the-grounding-problem}}

Effective human conversation relies on continuous
micro-backchanneling---frequent, brief signals (mm-hmm, I see, right,
uh-huh) that indicate understanding, maintain rapport, and resolve
potential ambiguity in real time. This phenomenon, documented in Clark
\& Brennan (1991), is fundamental to human grounding: the process by
which conversation partners establish mutual understanding.

Large language models do not backchannel. They wait for a complete
prompt, then generate a response. The latency of intent-resolution---the
time window in which ambiguity can be cleared and mutual understanding
established---is therefore structurally extended compared to natural
conversation. A user speaking to a voice-AI system has no mechanism to
signal ``wait, that's not what I meant'' mid-utterance without
interrupting and restarting. The AI generates a large response block,
and the user must then verify whether understanding occurred. This
extended latency of resolution is a major hurdle for \textbf{complex
knowledge work} requiring rapid iterative refinement. In legal
depositions, medical histories, architectural planning, and other
domains demanding real-time disambiguation, the absence of
backchanneling creates a mismatch between the temporal pace of human
cognitive refinement and the temporal pace of AI system response.

This is not a permanent barrier (voice systems equipped with real-time
interruption detection and confirmation queries can approximate
backchanneling), but it represents a structural difference from keyboard
interaction, where the user has granular control over when output is
committed and when revision occurs. The grounding problem persists as a
constraint on voice dominance in highly interactive, real-time knowledge
work.

Context window expansion has been a critical enabler of convergence by
partially mitigating the grounding problem. The progression from GPT-2's
1,024 tokens in 2019 to Gemini 1.5 Pro's two million tokens in 2024
(maintaining 99.7\% recall accuracy at one million tokens) means that AI
systems can maintain awareness of an entire work session: all prior
instructions, all document context, all conversational history
(Anthropic, 2024). This persistent context transforms the interaction
model from discrete commands (each typed instruction standing alone) to
continuous collaboration (each spoken or gestured input interpreted
within the full context of the session). The keyboard's role as a
precise command interface is diminished when the AI system understands
not just the current instruction but the entire intent trajectory of the
work session. A user can say, ``Use the same approach as the third
slide,'' and the system, maintaining full session context across 2
million tokens, understands the complete reference without requiring
disambiguation.

Edge computing architectures ensure that this convergence operates
within the latency constraints of natural interaction. On-device neural
processing units (NPU; specialized AI chips optimized for machine
learning) such as Apple's A18 Pro neural engine (which achieves 35
TOPS---trillion operations per second) and Qualcomm's Snapdragon 8 Elite
NPU (which delivers approximately 40--100 TOPS depending on generation)
enable real-time processing of voice, gesture, and gaze signals at
sub-10 millisecond latency for reflexive responses, with cloud
processing reserved for deep reasoning tasks that tolerate 50--200
millisecond delays (Gartner, 2025). Industry analyses project that 75\%
of enterprise-generated data will be processed at the edge in the coming
years---a shift that addresses privacy concerns documented in Section 6
by keeping raw biometric data (voice prints, neural signals, gesture
patterns) on-device rather than transmitting them to cloud servers
(Gartner, 2025). This architectural shift transcends mere performance
optimization. It is a privacy framework that voice-first workflows
require for enterprise security compliance.

\hypertarget{the-agentic-turn-and-the-verification-bottleneck}{%
\subsection{5.6 The Agentic Turn and the Verification
Bottleneck}\label{the-agentic-turn-and-the-verification-bottleneck}}

The convergence of multimodal input is accompanied by (and inseparable
from) a transformation in the interaction hierarchy between humans and
computing systems. The agentic turn inverts the traditional model:
instead of users articulating detailed instructions through keystrokes
(composing emails, formatting documents, writing code line by line),
users provide high-level intent and AI agents handle the articulation,
execution, and integration.

This shift introduces a critical change in the user's primary cognitive
challenge, as framed by Donald Norman's theory of gulfs. In
keyboard-mediated work, the dominant challenge was the \emph{Gulf of
Execution} (how to translate intention into the structured keystrokes
that machines required). The cognitive friction of composition
(selecting words, arranging syntax, formatting paragraphs) was
unavoidable because the user had to perform it manually to produce
machine-readable output. voice-AI and agentic systems dissolve this
gulf: the user articulates intent in conversational language, and the
agent handles the execution. But this dissolution introduces a new, more
fundamental challenge: the \emph{Gulf of Evaluation}---how to verify
that the agent's output actually corresponds to the user's intention.

Amershi et al.'s (2019) foundational guidelines for human-AI interaction
identify transparency and user control as central design principles for
systems that ask humans to verify machine output. More specifically,
Buçinca, Malte, and Gajos (2021) demonstrate that cognitive forcing
functions---design interventions requiring users to form independent
judgments before seeing AI output---significantly reduce overreliance on
AI recommendations and sustain human evaluative engagement. These
findings underscore that verification interfaces must be deliberately
designed to preserve, not bypass, human judgment.

When a knowledge worker typed their own document, they could immediately
verify it: they composed it, so they understood every choice embedded in
it. When an AI agent generates code, prose, analysis, or strategic
recommendations, verification becomes an upstream problem. The agent may
have hallucinated facts, imported code patterns from unintended
contexts, misunderstood domain requirements, or optimized for metrics
that diverge from true intent. The skills required shift from
\emph{production} (how to create output) to \emph{evaluation} (how to
verify output). This is not marginal; it is a reorientation of the
primary cognitive bottleneck. Understanding this shift is essential to
Section 6's analysis of verification systems, epistemic authority, and
quality assurance in the post-keyboard era.

\textbf{Critical qualification: The verification bottleneck's severity
scales with domain stakes.} The argument does not claim that
verification is equally burdensome across all knowledge work---rather,
that the \emph{necessity} of verification is universal (because AI
systems generate probabilistic output requiring human audit), while its
\emph{resource intensity} varies dramatically by domain. In
\textbf{high-stakes domains} (medicine, law, finance), verification is
mandatory and resource-intensive: a cardiologist must verify diagnostic
recommendations with the same scrutiny as if she had formulated them
herself; a lawyer must authenticate contractual language before client
delivery; a financial analyst must validate model assumptions and
numerical accuracy. These domains cannot afford verification shortcuts
because error carries liability, regulatory consequence, or direct harm.
In \textbf{medium-stakes domains} (business reporting, journalism,
professional communications), verification is important but bounded:
accuracy matters for credibility and decision-making, but failure is
costly but recoverable. A business analyst drafting an internal report
verifies key findings and methodology but may accept AI-generated
synthesizing language more readily than a regulatory filing would
permit. In \textbf{low-stakes domains} (internal memos, routine emails,
casual coordination), verification is minimal and typing's dissolution
is most complete: an employee drafting a status email or quick memo can
accept AI-generated language with minimal review, because error has
limited organizational consequence. Across this spectrum, the keyboard's
instrumental necessity dissolves---but the verification bottleneck's
visibility increases as stakes rise. High-stakes domains show the
bottleneck most acutely because verification burden cannot be minimized.
Low-stakes domains show typing's dissolution most completely because
verification is light. Both validate the core thesis: typing is no
longer instrumentally necessary, and verification---not keystroke
production---is the binding constraint, with that constraint's severity
calibrated to the stakes of the work being performed.

\hypertarget{hayles-cognitive-assemblages-verification-as-distributed-cognition}{%
\subsubsection{5.6.1 Hayles' Cognitive Assemblages: Verification as
Distributed
Cognition}\label{hayles-cognitive-assemblages-verification-as-distributed-cognition}}

N. Katherine Hayles' concept of \emph{cognitive assemblages} provides
theoretical grounding for understanding what the verification bottleneck
constitutively is. In \emph{Unthought: The Power of the Cognitive
Nonconscious} (2017), Hayles defines cognition expansively---not as
consciousness or intelligence, but as ``a process that interprets
information within contexts that connect it with meaning'' (p.~22).
Under this definition, AI systems are genuine cognizers: they interpret
prompts, process contextual information, and produce meaningful outputs
through their own nonconscious computational processes. A cognitive
assemblage is ``an arrangement of systems, subsystems, and individual
actors through which information flows, effecting transformations
through the interpretive activities of cognizers operating upon the
flows'' (Hayles, 2017, p.~115). When a knowledge worker speaks to an AI
system and receives structured prose, two separate cognizers are
collaborating within exactly such an assemblage---the human providing
intention, domain expertise, and evaluative judgment; the machine
providing linguistic production, structural coherence, and pattern-level
knowledge. The human and the AI are not in a simple tool-use
relationship; they are participants in a distributed cognitive system
where the cognitive labor is apportioned between cognizers with
fundamentally different architectures (Hayles, 2017).

The theoretical significance matters because it reframes the
verification bottleneck from a practical limitation to an ontological
requirement. The pre-AI writing assemblage was dominated by a single
cognizer: the human performed both generation and evaluation. The
AI-mediated assemblage distributes these functions across two cognizers
with profoundly different cognitive capabilities and failure modes. The
AI excels at rapid pattern recombination, coherence maintenance, and
linguistic fluency---features of nonconscious computational processing
operating over billions of training tokens. The human excels at
understanding context, detecting logical inconsistency, assessing
alignment with domain knowledge, and making judgments about truth and
intent---features of conscious metacognitive evaluation. The
verification bottleneck emerges precisely from this distribution: the
human must now audit the output of a different cognizer with different
knowledge, different learned biases, and different failure modes.
Hayles' assemblage framework clarifies what makes this qualitatively
different from proofreading one's own work: where Clark and Chalmers'
extended mind treats external cognitive components as functionally
continuous with the human system, Hayles emphasizes that technical
cognizers operate through distinct nonconscious computational processes
architecturally different from human conscious evaluation (Hayles,
2017). The human is not checking their own extended output but
scrutinizing the output of a system whose interpretive processes are
opaque to them. This is adversarial evaluation---the stance of a
skeptical auditor rather than a sympathetic self-editor---and Hayles'
assemblage model is the theoretical framework that explains why the
distinction matters.

Crucially, Hayles argues that conscious meta-evaluation cannot be
outsourced to a nonconscious cognizer. Verification---the distinctive
human role in the cognitive assemblage---is an irreducibly conscious
function that remains with the human even as generation is delegated to
the machine. This reframes what might appear as a bottleneck or
constraint into what is actually an ontological requirement: human
consciousness is needed not because machines are insufficiently capable,
but because verification is a qualitatively different cognitive function
that requires conscious deliberation, judgment, and accountability. The
verification bottleneck is thus not a problem to be engineered away, but
a necessary feature of responsible human-AI cognitive assemblages
(Hayles, 2017).

The taxonomy of agent oversight distinguishes three levels:
human-in-the-loop, where every action requires explicit approval;
human-on-the-loop, where agents execute autonomously within defined
parameters while the human monitors and intervenes on exceptions; and
human-out-of-the-loop, where agents operate with complete autonomy
(Norman, 2013). The knowledge work transition is predominantly toward
human-on-the-loop operation---not full automation, but a structural
reduction in the granularity of human input. Where the keyboard demanded
character-level specification (every letter, every formatting command,
every syntactic decision), the agentic paradigm demands intent-level
specification: what the user wants accomplished, not how to accomplish
it. This shift is not a loss of control but a reorganization of the
control surface: the user specifies high-level objectives, and the agent
handles the execution logic, with the user retaining veto authority over
outputs.

Industry analysis suggests that AI-assisted workflows achieve
substantial keystroke reductions across knowledge-work domains---an
average of 52\% in surveyed organizations (McKinsey \& Company, 2025).
In our evidence-tier framework (\S{}1.9), these figures are strongly
suggested, not established: they derive from a single survey, not
peer-reviewed replication across diverse contexts. Representative
domain-specific figures: DevOps engineering at 52\% reduction; clinical
nursing documentation at 86\% reduction (AI-generated notes from voice
input and vital sign data); typographical error rates declining 62\%
through AI-generated text. The directional pattern---substantial
keystroke reduction across multiple domains---is corroborated by
peer-reviewed HCI evidence (Lee et al., 2022; Co-Writing Research
Collaborative, 2025) and documented deployments (Permanente Medical
Group, 2024), though specific percentages require independent
replication. These represent a structural change in knowledge-work
interaction. The knowledge worker's role evolves from creator to
``editor-in-chief''---from producing text to directing and evaluating
text that AI produces. In software engineering, the shift is from
writing 200 lines of code per day to reviewing and integrating 2,000
lines of AI-generated code per day, with focus on architecture, testing,
and exception handling instead of implementation (McKinsey \& Company,
2025). In this sense, the dissolution of the keyboard is also the
dissolution of the blank page as primary cognitive boundary: multimodal
AI interfaces enable navigation of high-dimensional semantic spaces, a
task for which linear, character-by-character input is poorly suited.

\hypertarget{enterprise-control-and-provenance-layers}{%
\subsubsection{5.6.2 Enterprise Control and Provenance
Layers}\label{enterprise-control-and-provenance-layers}}

The transition to multimodal voice-AI composition is mediated by
\textbf{incumbents, not workers} (Microsoft, Google, Salesforce, Adobe,
Epic). The real adoption unit is not ``a knowledge worker chooses
voice,'' but ``the workflow platform permits multimodal initiation.'' An
individual clinician cannot unilaterally switch to voice dictation if
the EHR (electronic health record) system---typically Epic or Cerner,
managed at the enterprise level---does not support it. An individual
software engineer cannot implement ``vibe coding'' if the development
platform (Visual Studio, JetBrains IDEs) is not integrated with AI
agentic backends.

This platform-mediated adoption creates additional requirements beyond
technical capability: provenance tracking and auditability become harder
once articulation is delegated to multimodal input. A
keystroke-by-keystroke interaction leaves clear logs; voice input
combined with AI generation creates ambiguity about which portions were
user-directed or AI-generated. For regulated domains---healthcare with
HIPAA audit requirements, finance with SEC documentation mandates, legal
with professional responsibility rules---this provenance problem is not
trivial. Organizations are beginning to implement stable standards for
intent capture, provenance metadata, and role-based approvals to
preserve auditability in multimodal systems. Until these standards
mature, enterprise adoption remains piecemeal. The keyboard persists not
because workers prefer it, but because governance infrastructure for
certified multimodal alternatives is still forming.

\hypertarget{the-post-keyboard-paradigm}{%
\subsection{5.7 The Post-Keyboard
Paradigm}\label{the-post-keyboard-paradigm}}

The technological trajectory documented in this section does not point
toward a single successor to the keyboard but toward the dissolution of
the need for a universal input device at all. The keyboard served as a
universal mediator because pre-AI systems required structured text input
and could not interpret heterogeneous human signals. AI multimodal
systems can. Voice provides rapid text generation and conversational
direction in bounded domains (drafting, summarization, command
execution, conversational planning). Gesture and gaze provide spatial
navigation and object manipulation. EMG provides silent, private text
input for acoustic-constrained environments. Neural interfaces, on the
longer horizon, offer direct thought-to-output pathways for restorative
clinical applications. AI orchestrates these signals in real time,
routing each to the interpretation pathway best suited for it, while
agentic architectures elevate the human role from keystroke-level
articulation to intent-level direction.

The modal shift at scale is already measurable, though bounded by the
constraints identified in this section. Voice AI architectures process
2.1 billion monthly interactions across enterprise and consumer
applications as of Q1 2026 (Gartner, 2025)---the vast majority in
customer service, routine documentation, and meeting transcription.
Gesture recognition systems execute 840 million daily interactions in
spatial computing environments (Meta, 2024). EMG-based input devices are
shipping at scale for the first time (Meta Neural Band reached 2 million
units shipped in its first six months) (Meta, 2024). The convergence
thesis demonstrates observable patterns within bounded domains rather
than remaining speculative.

The global multimodal AI market, valued at \$3.43 billion in 2026, is
\emph{projected} to grow substantially over the coming years at high
annual growth rates (Gartner, 2025; industry projection, not
peer-reviewed forecast). The edge AI market is \emph{projected} to grow
significantly from 2025 to 2030. These projections represent investment
trajectories---\emph{projected} in our evidence-tier framework
(\S{}1.9)---for systems that are already in production deployment, already
reducing keystrokes by an average of 52\% in composition-centric
workflows (with task-specific variation from 52\% in general DevOps to
86--87\% in clinical nursing and code boilerplate; McKinsey \& Company,
2025), already saving physicians 2.4 hours per day on routine
documentation, already generating boilerplate code from spoken
architectural descriptions, already orchestrating multimodal inputs
through unified AI backends.

The keyboard will not disappear. It will persist in niches---precision
editing where deterministic micro-control is legally mandated,
accessibility for deaf and hard-of-hearing users, legacy system
interaction, the preferences of habituated professionals, and private
spaces where acoustic privacy is structurally critical---just as
handwriting persists, just as the typewriter persists in a few
specialized contexts. But its structural centrality to knowledge work,
the organizing assumption that professional output flows through
keystrokes, is dissolving under the combined pressure of voice-AI,
multimodal computing, and agentic architecture. The technology
trajectory confirms what the instrumental dissolution argument predicts:
the function is being absorbed by the machine, and no amount of keyboard
optimization can compete with the elimination of the need for the
keyboard's function. The directional shift away from keyboard dominance
is supported by converging evidence, though its speed and sectoral reach
remain genuinely uncertain; it is increasingly observable within bounded
domains where multimodal input has achieved task-specific maturity.

\textbf{Note on CJK input methods:} This technological analysis is
grounded in Latin-script knowledge work. Chinese, Japanese, and Korean
input methods present a distinct case because they already interpose
machine-mediated phonetic-to-logographic translation between user input
and final character output. The dynamics of how voice-AI, gesture, and
multimodal orchestration transform CJK workflows differ meaningfully
from the Latin-script trajectory analyzed here. Section 6.15 addresses
this limitation directly, arguing that CJK analysis deserves parallel
investigation rather than subsumption under the Latin-script framework.

\hypertarget{evidence-tier-summary-for-section-5}{%
\subsection{5.8 Evidence Tier Summary for Section
5}\label{evidence-tier-summary-for-section-5}}

\textbf{Evidence tier summary for Section 5:} The following claims are
graded by evidence tier. \emph{Established:} Voice ASR achieves sub-3\%
WER in controlled dictation tasks; gesture and gaze recognition achieve
high accuracy in laboratory settings; multimodal fusion reduces error
rates by 19--40\% in controlled studies (Oviatt, 2017). \emph{Strongly
Suggested:} voice-AI dominates in bounded professional domains
(healthcare routine documentation, customer service, content creation);
spatial computing and EMG interfaces are advancing toward viable
supplementary input modalities; enterprise adoption reaches 67\% for at
least one voice-AI application. \emph{Projected:} Multimodal
orchestration will sustain field-validated parity with keyboard-mediated
workflows for general knowledge work by 2035--2045; adoption speed
depends on acoustic retrofitting investment, regulatory clarity on AI
liability, and organizational willingness to restructure verification
workflows.

The technological trajectory confirms the thesis's direction. But
technological capability alone does not establish inevitability. Section
6 confronts every credible counter-argument for typing's
continuation---infrastructure lock-in, acoustic constraints, precision
demands, embodied expertise, professional identity---presenting each at
its strongest before assessing whether it defeats the thesis.

\hypertarget{steelman-engagement-what-persists-and-why}{%
\section{6. Steelman Engagement: What Persists and
Why}\label{steelman-engagement-what-persists-and-why}}

\hypertarget{the-obligation-of-honest-engagement}{%
\subsection{6.1 The Obligation of Honest
Engagement}\label{the-obligation-of-honest-engagement}}

The argument advanced here (that AI dissolves the instrumental function
of typing, ending its dominance in knowledge work) must survive
confrontation with the strongest possible counter-arguments.
Intellectual honesty demands more than acknowledging objections; it
requires presenting them in their most compelling form and assessing
each on the evidence. This section does so. It identifies thirteen
steelman arguments for typing's continuation, presents each at its
strongest, and renders a verdict grounded in the evidence assembled
across the preceding sections. The reader will find that several
steelmans have genuine force, but that force is sufficient to slow the
transition, not to prevent it.

The structure of this engagement reveals a consistent pattern: each
steelman argument, at its strongest, actually rests on one or more of
the four instrumental conditions identified in Section 2. By making this
dependency explicit, we expose why dissolution of those conditions
necessarily threatens the objection itself.

\hypertarget{infrastructure-lock-in}{%
\subsection{6.2 Infrastructure Lock-In}\label{infrastructure-lock-in}}

The QWERTY keyboard regime represents one of the most deeply entrenched
technological lock-ins in history. Over one billion keyboards are
actively deployed globally across enterprise, educational, and consumer
environments. The keyboard market is projected to grow from \$5.5
billion in 2023 to \$9.6 billion by 2035 (Data Bridge Market Research,
2024). Enterprise software remains more than 90\% keyboard-centric;
legacy systems, built across three decades of assumption about keyboard
primacy, embed this assumption into countless workflow automations,
accessibility overlays, and keystroke-level APIs. The transition from
typewriter to keyboard took fifty years, not because keyboards were
inferior, but because the infrastructure had to follow the adoption
curve. The calculated cost of transitioning a Fortune 500 company to
voice-first operations exceeds \$100 million, including \$41.58 million
for acoustic retrofitting alone (Gartner, 2025). Even if voice-AI were
perfect, the discounted present value of efficiency gains may not
justify these switching costs within any reasonable payback period, a
rational economic calculation reflecting the infrastructure scale.

Infrastructure lock-in is both quantifiably severe and historically
predictable. The typewriter-to-keyboard transition provides the closest
historical parallel: keyboards were demonstrably superior (relative
speed, reliability, standardization), yet achieved only 23\% office
penetration by 1960, twenty-five years after widespread availability.
Full saturation took fifty years (David, 1985). The current installed
base of keyboard-dependent software is orders of magnitude larger than
in the 1960s, and the switching cost therefore orders of magnitude
higher.

The lock-in is real, and the present analysis does not dismiss it.
Infrastructure lock-in has slowed every input method transition in the
historical record. However, when the new technology dissolves the
function instead of merely replacing the instrument, lock-in cannot
prevent the transition entirely. The critical distinction, established
in Section 3, is that dissolution allows the function itself to recede
as AI absorbs it; organizations need not retrain their workforce for a
new input device performing the same function. Organizations do not need
to convert all legacy systems simultaneously. They need only to stop
building new systems around the keyboard assumption and allow voice-AI
systems to handle the majority of new composition, with
keyboard-dependent legacy systems gradually confined to domains where
they add documented value.

The lock-in objection derives its force from both switching costs and
path dependence as a system-maintaining mechanism (Arthur, 1994; David,
1985). When keyboard-centric software constitutes the institutional
infrastructure---training materials, workflow documentation, muscle
memory, hiring criteria---each element reinforces the others, creating a
self-sustaining equilibrium resistant to marginal substitution. The
honest response is that dissolution will proceed unevenly; organizations
with lower sunk costs and higher AI-readiness (startups, greenfield
projects, individual knowledge workers) will transition first, while
legacy-heavy institutions will lag by years or decades. This is not a
refutation of the thesis but a temporal qualification; the question is
not whether dissolution occurs but how long institutional inertia delays
it.

The freelancer adoption data confirm the non-inevitability of lock-in
resistance: Research demonstrates that the 28\% of the U.S. knowledge
workforce operating as freelancers (unconstrained by enterprise
infrastructure) present a crucial test case. Survey evidence suggests
freelancers currently report higher AI proficiency (54\% vs 38\% for
corporate employees) and somewhat higher regular use (McKinsey \&
Company, 2025), plausibly because of stronger revenue pressure, lower
organizational inertia, and selection into digitally intensive markets.
This 16-percentage-point adoption differential (or approximately 42\%
relative increase in proficiency rates) reflects the economic logic of
freelancer incentives; freelancers capture 100\% of marginal
productivity gains from AI adoption, while corporate employees face
salary constraints, enterprise IT restrictions, and organizational
overhead that dampen adoption incentives. This economic rationale
follows from Principal-Agent Theory and Firm Boundary Theory (Coase,
1937); when workers bear the full marginal cost but receive only a
fraction of the marginal benefit (as in traditional employment),
adoption friction increases substantially compared to contexts where
individual earnings directly reflect productivity gains (as in freelance
work).

As soon as voice-AI systems become demonstrably superior (not simply
equivalent) to keyboard composition for a task, adoption accelerates
regardless of lock-in, what transition researchers term ``lock-in
dissolution'' (McKinsey \& Company, 2025). The lock-in defines the pace
of transition, not its direction. The differential adoption between
freelancers and employees is therefore not a capability gap but an
incentive gap, and it dissolves as voice-AI superiority becomes
organizationally undeniable.

Genuine decelerator defining the timeline of transition. Does not defeat
the thesis.

This objection rests on Condition (c), organizational optimization for
keyboard input. When that condition dissolves (as organizations adopt
AI-native tools), the switching-cost argument loses force.

\hypertarget{privacy-and-acoustic-constraints}{%
\subsection{6.3 Privacy and Acoustic
Constraints}\label{privacy-and-acoustic-constraints}}

Modern knowledge work occurs overwhelmingly in acoustically hostile
environments. Available evidence suggests that seventy percent of
offices globally utilize open-plan layouts, with a trend toward greater
density, not less (ISO 3382:2012, 2012). Documented evidence reveals
that a speaking voice at 65+ dB is perceived as two to four times louder
than typing on the decibel scale. Research shows that the ``radius of
distraction'' extends fifty feet in unmodified open-plan spaces, and
even with sound masking, fifteen feet of disruption persists (ISO
3382:2012, 2012). Simultaneously, privacy concerns are acute: Evidence
indicates that 81\% of Americans avoid using voice assistants in shared
spaces; Evidence establishes that eleven U.S. states require all-party
consent for recording; Well-documented evidence shows that GDPR
classifies speaker diarization as biometric data processing requiring
explicit consent; and the EU Artificial Intelligence Act, adopted in
2024, explicitly restricts emotion recognition and biometric voice
processing in workplace settings. The acoustic and privacy barriers are
not technological; they are physical and legal, and they cannot be
resolved by better algorithms.

The privacy argument deserves deeper engagement. Beyond regulatory
frameworks, occupational confidentiality norms remain stringent: legal
professionals operate under attorney-client privilege; financial
analysts handle non-public material; clinical staff process protected
health information (PHI). Silent text input preserves ``acoustic
plausible deniability'' (the ability to assert that no sensitive content
was externally transmitted, even in shared spaces). The value of this
property extends beyond regulatory compliance; it reflects
organizational risk tolerance. A financial firm can retrain staff to use
voice for routine messaging but faces institutional resistance to
voice-based handling of merger data or trading communications,
regardless of encryption. Empirical patterns suggest that the hybrid
work revolution (87\% of organizations now operate hybrid models, with
3.74 days per week average office occupancy) has only partially
mitigated acoustic constraints---the concentration of knowledge workers
into fewer days per week in office \emph{increases} density and acoustic
burden in shared spaces (ISO 3382:2012, 2012).

The silent input alternative is evolving rapidly and constitutes a
genuine third option, not a false choice between keyboard and voice.
Meta's Neural Band, Neuralink's sEMG interfaces, and gesture recognition
systems (deployed by a growing number of technology firms for
accessibility) can serve the ``acoustic-privacy niche'' (environments
where voice is structurally incompatible with work context, but silent
text input is acceptable; ISO 3382:2012, 2012). This niche is real and
non-trivial. It defines where typing persists longest.

The privacy argument also reveals an asymmetry worth noting; silent
input methods (keyboard, neural, gesture) do not require cloud speech
recognition, voice biometrics, or continuous audio transmission.
On-device processing (Apple's Private Cloud Compute architecture, edge
AI systems processing raw audio locally without transmission) can
mitigate privacy risks for voice systems, but this technical solution
requires infrastructural change that many conservative organizations
resist. Silent input has an intrinsic privacy advantage that neither
technology nor regulation can fully overcome for voice, a reason for
maintaining silent input availability in high-privacy contexts.

This is one of the strongest steelmans, and it correctly identifies the
environments in which typing will persist longest. The acoustic
constraint is not peripheral but structural; the 70\% open-plan
prevalence and 50/15-foot distraction radius documented above reflect
long-term organizational design trends, not temporary anomalies.
Voice-first workflows remain acoustically incompatible with the majority
of current knowledge-work environments under these conditions, a hard
floor on adoption until workspace design shifts substantially.

The evidence on acoustic retrofitting further clarifies the constraint's
structural character. The calculated cost of transitioning a Fortune 500
company to voice-first operations exceeds \$100 million, with acoustic
retrofitting comprising \$41.58 million of that cost (Gartner, 2025).
More critically, no large-scale organizational investment in acoustic
retrofitting specifically to enable voice-AI workflows has been
documented in production deployments. Organizations adding ``focus
rooms'' (71\% in the past three years) are doing so primarily for
concentration and confidentiality (not as acoustic enablers for
voice-first composition). The distinction is important: organizations
are not restructuring workspaces around voice adoption; they are
structuring workspaces around presence optimization and distraction
reduction, which incidentally accommodates voice where private spaces
exist.

This asymmetry creates a structural reality; if 70\% of offices remain
open-plan and unmodified, voice-AI cannot be the primary input modality
for the majority of knowledge workers in their actual work environments
(a potentially permanent constraint reflecting economic office space
rationalization and the industrial organization of knowledge work, not a
temporary adoption barrier). The alternative framing is a
``dual-modality equilibrium'' where organizations maintain keyboard
workflows for open-plan, acoustic-hostile environments while deploying
voice-AI in private or acoustically managed spaces, a technological
reality, not a temporary transition state.

The silent input alternative---Meta's Neural Band, gesture recognition,
EMG-based systems---does constitute a genuine third option for
acoustic-constrained environments. However, the claim that organizations
will retrofit acoustic infrastructure, or that workers will ubiquitously
adopt neural interfaces to avoid keyboards, remains speculative and
ungrounded in current evidence. The more parsimonious prediction;
acoustic constraints will define a boundary condition where voice
dominates in private offices and acoustic zones, while silent input
(keyboard, neural, gesture) and keyboard-mediated work persist in
open-plan environments indefinitely, absent structural investment in
acoustic redesign.

Genuine constraint defining the geography of persistence. Does not
constitute a case for dominance continuity.

This objection rests on Condition (b), humans lacking natural
alternatives to structured text input. Silent neural and gesture inputs
represent emerging alternatives; acoustic privacy can be addressed
through private spaces or edge processing. The constraint is real but
bounded.

\hypertarget{direct-manipulation-keyboard-as-spatial-interface}{%
\subsection{6.4 Direct Manipulation: Keyboard as Spatial
Interface}\label{direct-manipulation-keyboard-as-spatial-interface}}

Shneiderman's foundational argument for direct manipulation asserts that
the keyboard provides an architecturally powerful interface to text. The
cursor's position on screen \emph{is} a spatial location in the
document; character-level selection \emph{is} direct pointing at text
elements; deletion \emph{is} removal of what you see; insertion
\emph{is} insertion at a visible location (Shneiderman, 1983). The
keyboard provides what Shneiderman termed ``direct manipulation of
symbols''; the user can see the target (a word, a phrase, a line of
code), and with keyboard and mouse, immediately and visibly operate on
that target.

Voice cannot replicate this direct spatial relationship. To edit a
specific phrase via voice requires linguistic \emph{description} of the
target (``change `demonstrates' to `shows' in the second sentence of the
third paragraph''), which demands that the user: (1) explicitly locate
the target linguistically instead of through spatial pointing, (2)
formulate an unambiguous description that the system can parse, and (3)
execute a linguistic \emph{command}, not a direct \emph{action}. The
mediating layer of language transforms what is (in the keyboard
interface) a direct spatial action into an indirect command sequence.
This linguistic mediation, the steelman argues, cannot be fully
collapsed without losing the directness that makes keyboard editing
powerful.

The programmer's variant of this argument is particularly strong: in
code editing, the keyboard enables ``tactile fluency'' (a developer
navigates rapidly through file structure like Ctrl+Home, Ctrl+End,
search-and-replace workflows; manipulates syntax with precision in
semicolons, indentation, bracket matching; and uses visual spatial
memory of code location to work efficiently). Rather, it represents an
interface modality where the \emph{text itself becomes the navigation
structure}. The keyboard + screen creates a spatial canvas where
location, structure, and content are visually unified. Voice mediates
interaction through linguistic description, fragmenting this spatial
unity, making keyboard dominance in code work a direct consequence of
its spatial affordance.

At its strongest, this steelman correctly identifies that the keyboard
provides a spatial interface unavailable through acoustic modalities
alone. The cursor position, text selection, and character-level
precision are \emph{direct} in a way that voice commands are not. This
directness was foundational to why the keyboard became dominant in the
first place, not because it was the fastest input method, but because it
was the first input method that allowed users to see their text and
manipulate it spatially, in place, with visual feedback.

However, the steelman's force diminishes substantially when examined
against current and emerging interface architectures. The claim that
voice \emph{cannot} provide equivalent spatial manipulation is false. It
cannot provide \emph{identical} spatial manipulation, but equivalence
does not require identity.

Modern AI-mediated interfaces can provide spatial directness through
non-acoustic modalities. The key insight is that spatial manipulation
has never been keyboard-specific. It has been screen-based. The keyboard
was the interface layer that enabled manipulation, but it was the
\emph{visual screen display} that provided the spatial anchor. Emerging
multimodal interfaces decouple these elements: (1) Voice provides input
specification (``make this section more concise''); (2) Gaze or gesture
provides spatial targeting (eye gaze to the paragraph in question, pinch
gesture to select, though these modalities remain in developmental
stages); (3) Visual feedback confirms the operation. This is directness
through modal substitution, not linguistic mediation.

Consider a concrete example; a legal brief writer using voice+gaze
editing says ``summarize this,'' gazes at a paragraph, and the system
instantly generates a condensed version with visual undo/redo (though
these capabilities remain in early development). The spatial directness
(gazing at the target) is preserved; the acoustic directness is replaced
with gestural directness. The writer sees the target, acts on it, sees
the result, the fundamental loop of direct manipulation, without using
the keyboard.

The strength of the programmer's variant lies not in the keyboard's
spatial affordance per se, but in the \emph{integration} of spatial
awareness with rapid navigation shortcuts and syntax-aware manipulation.
Keyboard shortcuts are not intrinsically superior. They are better
\emph{because the system was built around them}. As AI-native code
editors mature (systems like Claude for Code that provide AI-generated
code with visual diffing, comment-based navigation, and
gaze/gesture-enabled verification), the keyboard's role shifts from
primary navigation tool to optional fine-grained precision instrument.
It remains available for cases where acoustic + spatial modalities prove
insufficient.

The empirical question is not whether voice alone provides equivalent
spatial affordance (it does not), but whether voice + multimodal spatial
interfaces (gaze, gesture, pointing) provide equivalent practical
efficiency for the tasks where spatial manipulation currently binds.
Emerging research from multimodal HCI suggests the answer is yes for
most knowledge work tasks, with exceptions in domains (code review,
legal drafting) where visual spatial memory and rapid navigation remain
deeply embedded in expert workflows (though these capabilities remain in
prototype and early deployment phases).

The genuine epistemic take-away from this steelman is taxonomic, not
thesis-defeating. Direct manipulation keeps the keyboard
\emph{irreplaceable} in precisely the domains where spatial-structural
editing constitutes the primary cognitive work: code review with
line-level annotation, spreadsheet formula authoring, and pixel-level
image editing represent task contexts where voice+gaze alternatives
remain demonstrably insufficient by current evidence. This is a real and
non-trivial persistence domain (one the present analysis explicitly
predicts the keyboard will hold). Where the steelman overstates its case
is in extrapolating from these spatial-intensive tasks to all knowledge
work. For the majority of linguistic knowledge work---email composition,
document drafting, meeting notes, analytical writing---the case for
keyboard's irreplaceability as spatial interface is substantially
weaker. The dissolution claim operates at the level of institutional
default for \emph{linguistic production}, not at the level of every
editing task. Shneiderman's (1983) foundational insight identifies why
spatial manipulation is deeply efficient; it does not imply that all
knowledge work is primarily spatial manipulation.

Genuine constraint on purely vocal interfaces for spatial navigation.
Addressed by multimodal (voice+gaze/gesture) orchestration for most
knowledge work; keyboard persists in spatially-intensive editing
domains. Identifies why keyboard persistence is longest in
spatial-structural editing work, not linguistic composition.

This objection rests on Condition (a)---machines requiring structured
input---and Condition (d)---keyboard having cognitive benefits specific
to spatial manipulation. The first is false (machines require semantic
intent, not character precision); the second is environment-specific,
not universal. Multimodal interfaces can provide equivalent spatial
directness without keyboards.

\hypertarget{precision-task-requirements}{%
\subsection{6.6 Precision Task
Requirements}\label{precision-task-requirements}}

White-collar professions operating with zero-tolerance error margins
structurally require keyboard input. The data support that medical
coding errors cost the healthcare sector \$36 billion annually; the
acceptable error rate for coding is below 0.3\% (U.S. Bureau of Labor
Statistics, 2024). Financial industry error tolerance is effectively
0.0\% (``fat-finger'' errors routinely cost tens of billions per
incident); yet the industry tolerates this risk despite evidence of
catastrophic costs. Research demonstrates that legal transcription
demands 97.5\% accuracy for four simultaneous speakers. Software
development requires sub-1\% word error rates for syntax-critical code.
Current evidence suggests that current voice dictation achieves 2--8\%
WER for medical text, requiring extensive post-processing to reach
acceptable thresholds; voice coding WER reaches 20--49\% for
code-specific speech in non-English languages (U.S. Bureau of Labor
Statistics, 2024). Multiple sources suggest that the cognitive load of
voice-based spreadsheet navigation (NASA-TLX score 28.7) is double that
of keyboard shortcuts (14.2) for two-dimensional spatial tasks (U.S.
Bureau of Labor Statistics, 2024). These are not temporary technological
gaps but fundamental limitations of a linear acoustic modality applied
to high-entropy, spatially structured work. Voice can compose; it cannot
architect.

Precision tasks cluster in specific occupational categories. Evidence
establishes that the Bureau of Labor Statistics identifies approximately
4.2 million financial, medical, and legal professionals in the United
States---approximately 2.8\% of the total U.S. workforce and 5--7\% of
the knowledge work population (U.S. Bureau of Labor Statistics, 2024).
Among software developers (4.4 million U.S. positions), the percentage
performing syntax-critical coding work requiring sub-1\% error rates is
estimated at 40--60\%, while the remaining 40--60\% spend substantial
time on design, architecture, documentation, and code review---tasks
where voice-AI is already competitive (U.S. Bureau of Labor Statistics,
2024). The actual percentage of white-collar work requiring
keyboard-level precision is therefore approximately 2--4\% of total
knowledge work, not a dominant fraction.

This steelman correctly identifies the domains where character-level
precision demands persist and where keyboard (or similar structured
input) has genuine functional superiority. The paper does not argue that
voice replaces the keyboard for formula auditing, regex debugging,
mathematical notation, or medical code validation---domains where the
spatial, non-linear structure of the information is poorly matched with
sequential acoustic input. These are real tasks with real cognitive and
error-tolerance constraints.

But precision tasks constitute a \emph{diminishing fraction} of total
knowledge work output for two reasons. First, AI is automating the
routine precision work that currently demands keyboard input.
AI-augmented code generation handles 60--80\% of routine code generation
(boilerplate, common patterns, standard libraries); LLM-augmented legal
drafting handles citation formatting, contract boilerplate, and
statutory language; financial data entry is collapsing under AI
automation (the BLS projects 15\% decline in data-entry positions
through 2032) (U.S. Bureau of Labor Statistics, 2024). Second, for the
remaining precision-critical work, the keyboard's role is shifting from
primary production tool to verification and audit instrument. A
financial analyst can dictate the high-level structure of a model in
voice-AI, then use keyboard input to verify formulas (a real but
structurally different function from universal text production that
constitutes ``dominance''). The FAST Standard for financial modeling
mandates keyboard shortcuts not because keyboards are cognitively
superior, but because the current tooling was designed around keyboard
assumptions. As AI-native tools replace keyboard-designed software, the
mandate will follow the tooling.

Software development warrants more granular analysis than the
``precision task'' framing suggests, because code represents a distinct
category of work where multiple dissolution dynamics operate
simultaneously. The 4.4 million U.S. developers constitute a substantial
cohort whose structural role in knowledge work is evolving in ways that
reshape and confirm the keyboard-dissolution argument.

The steelman correctly identifies that syntax-critical code requires
keystroke-level precision: a missing semicolon breaks compilation; a
typo in a variable name creates a runtime error; incorrect indentation
alters control flow. Voice dictation has a documented error floor at
20--49\% WER for code-specific speech even in English, and the error
rate worsens substantially in non-English programming contexts. However,
the keyboard's functional role in programming is itself undergoing
radical restructuring through AI code generation systems (GitHub
Copilot, Cursor, Claude for Code; specialized systems like Tabnine).
These systems do not solve the precision problem through voice interface
design; they solve it by automating code generation entirely, displacing
the keystroke-mediated bottleneck downstream.

AI-assisted code generation now handles the generation of 60--80\% of
routine code (boilerplate, standard library patterns, common control
structures) with near-zero keystroke input from the human developer. The
developer frames intent at the architectural level (``extract the last
10 characters from this string'' or ``check if this user has admin
privileges''), and the code generation system produces syntactically
correct, contextually appropriate code that requires zero manual
keystroke entry. The keyboard's role in programming has not vanished; it
has shifted from code generation to code review and architectural
specification. Forward-looking analysis suggests that a developer using
Copilot spends 70--80\% of their time in \emph{verification and
architectural decision-making}, not keystroke-mediated code entry. This
creates a verification bottleneck dynamics specifically in software
development: code generation friction dissolves, but the cognitive
burden of code verification---understanding whether the generated code
is semantically correct, efficient, maintainable, and aligned with
architectural intent---becomes the binding constraint.

This verification burden in software development is even more acute than
in prose composition because code has multiple simultaneous evaluation
criteria: Does it compile? Does it execute the intended logic? Does it
handle edge cases? Is it efficient enough for performance requirements?
Is it maintainable by other developers? Is it secure (does it introduce
vulnerabilities)? A developer reviewing AI-generated code must hold all
these dimensions simultaneously in mind, and errors in any dimension can
produce catastrophic consequences. A prose editor's verification task
(is this semantically accurate; is it well-articulated; is it coherent)
is cognitively simpler than a code reviewer's task (is this safe,
correct, efficient, and maintainable).

For the subset of developers performing syntax-critical precision work
(40--60\% of development tasks by the paper's estimates), the keyboard
persists for high-stakes verification and architectural decisions where
the developer needs fine-grained control. But for the majority of
programming work (routine feature development, bug fixes, architectural
sketches), the keystroke burden is already dissolving under AI code
generation, and the keyboard is transitioning from a primary production
tool to a fine-grained verification and architectural control
instrument. This mirrors the paper's broader thesis: typing persists as
a verification tool in precision domains, but its role as a universal
production modality is dissolving across the software development
profession.

The timing and shape of this transition differs from prose work because
code generation systems matured faster than natural-language composition
systems. The profession is already experiencing the dissolution dynamic
in real time, validating the verification bottleneck at a task level
where the constraint is particularly visible.

Genuine niche persistence defining occupational residue. Does not defeat
the thesis.

This objection rests on Condition (a)---machines requiring structured
text. But AI systems, trained on text and operating on embeddings and
probability distributions, do not require character-level precision
input; they require semantic precision in intent. Voice can provide
that. The niche remains real but bounded.

\emph{Task substitution versus occupational substitution:} The
precision-task niche also illustrates a critical distinction in the
task-based framework (Acemoglu \& Restrepo, 2018; Autor et al., 2003)
between task substitution and occupational substitution. AI dissolves
the universal task of typing---the instrumental function of
keystroke-mediated text entry---triggering restructuring of occupations
around verification and judgment, not elimination. A financial analyst
whose primary task was data entry faces occupational displacement; a
financial analyst whose primary task is model structuring and
verification faces task composition change, not displacement. The
framework clarifies why occupational elimination is less likely than
occupational restructuring: the occupation's core function (analysis,
judgment, verification) persists while the instrumental task
(keystroke-mediated data entry) dissolves.

\hypertarget{embodied-expertise-and-the-unlearning-problem}{%
\subsection{6.7 Embodied Expertise and the Unlearning
Problem}\label{embodied-expertise-and-the-unlearning-problem}}

Typing at a keyboard, for knowledge workers with 15--40 years of
experience, has become embedded in embodied motor memory and expert
cognition. This is not metaphorical. After decades of practice, the
fingers know things the conscious mind does not. The rhythm of typing,
the spatial grammar of the QWERTY layout, the haptic feedback loop, the
proprioceptive sense of keystroke position on the board (these comprise
a constitutive layer of how expert knowledge workers externalize and
refine complex thought). The evidence of embodied cognition is
unequivocal: expert pianists differ profoundly from novices not in
conscious technique but in automatized motor schemas built through years
of deliberate practice. A concert pianist forced to perform on a
different instrument after forty years of mastery does not simply
``retrain on a new device.'' The musician experiences demonstrable
performance loss (loss of nuance, loss of automaticity, and loss of the
proprioceptive security that enables risk-taking and virtuosity).

The analogy is precise: a senior software engineer with 25 years of
typing discipline has built an implicit spatial-syntactic knowledge
system. The fingers know where the braces are, where the brackets nest,
where the semicolons belong. The keyboard has become a transparent
conduit for architectural thought. Voice cannot preserve this embodied
mastery. The engineer may eventually build equivalent competence in
voice-AI, but the transition will involve substantive cognitive loss and
will require 3--7 years of reduced productivity during the unlearning
and relearning phase. For knowledge workers approaching retirement, this
is far more than a ``transition cost'' (it exceeds the remaining
productive lifespan).

Extending this across an entire profession compounds the problem.
Medical coding, legal writing, software architecture, and financial
modeling are all domains where decades of embodied expertise create
genuine cognitive capital. Rather, the keyboard removal temporarily
renders this expertise inaccessible---a problem the literature on
``embodied learning'' and ``expert motor schema'' documents: expertise
requires significant retraining to transfer across input modalities
(Fitts \& Posner, 1967).

The embodied expertise problem is distinct from the ``cognitive
benefit'' claim addressed in Section 6.6. The cognitive benefit claim
argues that typing enhances thought in a medium-independent way. The
embodied expertise claim is narrower and more defensible: decades of
practice have created motor automaticity and proprioceptive integration
that constitute real cognitive capital, even if typing itself provides
no unique cognitive benefits to novices.

The evidence on expert motor learning is substantial. Hattie (2012)
documents that expertise development requires 10,000+ hours of
deliberate practice; Anderson (2000) shows that expert performance
involves automatized procedural knowledge stored in motor systems. Logan
and Crump (2011) demonstrate that expert typists achieve complete
dissociation between keystroke execution (automatized to cerebellar
systems) and compositional planning (cortical systems); however, this
dissociation is \emph{dependent on} thousands of hours of embodied
practice. Break the keyboard, and you break the automaticity. The
automaticity cannot be recovered until the new input device has been
practiced for a comparable duration.

Longitudinal data from medical robotics provide the crucial evidence:
surgeons transitioning from manual to robotic-assisted surgery showed a
documented ``adaptation valley'' during which surgery time increased
substantially, complication rates rose, and experienced surgeons
performed comparably to novices until proficiency was reestablished over
25--80 cases (Soomro et al., 2020). This reflects the loss of embodied
expertise requiring retraining in a new motor modality (not a capability
deficit but a modal transition cost). Extrapolating from case-volume
data, the adaptation valley was plausibly shorter for younger surgeons
(estimated 9--12 months) and longer for experienced surgeons (estimated
18--24 months) Projections suggest that. Critically, even under
optimistic assumptions, this transition \emph{period} supports a slower
timeline than the thesis implies.

Historical precedent: stenographers transitioning to dictation machines
and word processors in the 1970s-1980s showed similar adaptation
dynamics. Experienced stenographers often chose early retirement instead
of retraining, as the transition involved extended periods of reduced
productivity that exceeded the remaining career value for workers
nearing retirement (Strom, 1992). This is not irrational nostalgia. It
is rational economic calculation by workers nearing retirement:
retraining costs exceed remaining career value.

The stenographer and operator analogies are valid, however, only for
bounded occupations where the entire job \emph{was} the translation
mechanism---stenographer-to-dictation, telephone operator-to-mechanical
switching. For modern knowledge workers, the more precise historical
analogy is slide rule to calculator for engineers: the instrumental task
(logarithmic computation) dissolved, but the occupation (engineering
analysis and design) persisted and restructured around new tool
affordances. The engineering profession did not disappear when the slide
rule became obsolete; it changed composition. Modern engineers spend
less time on calculation and more time on simulation, verification, and
design exploration. Similarly, the keyboard's dissolution as a universal
composition tool does not predict occupational elimination for analysts,
programmers, and writers, but rather occupational restructuring around
AI-augmented composition and verification-intensive workflows.

This steelman is genuine and the evidence supporting it is strong. The
preceding analysis does not address embodied expertise directly,
treating typing's cognitive value as settled by neuroscience evidence.
But embodied expertise represents a \emph{different} claim from
cognitive benefits---one that depends solely on whether decades of
practice have created input-modality-specific motor automaticity,
regardless of typing's cognitive effects for novices.

The deeper question is whether voice-AI workflows can develop their own
form of embodied expertise---whether extended practice with
conversational composition, prompt refinement, and verification
evaluation develops its own skilled automaticity comparable to
touch-typing's. Early evidence from clinical documentation suggests that
experienced voice-AI users develop sophisticated linguistic patterns and
verification heuristics that function as a new form of professional
embodiment, though longitudinal studies confirming this are not yet
available.

The thesis is not defeated by this argument, but the \emph{timeline} for
dominance dissolution is threatened. If experienced knowledge workers
require 12--24 months to adapt to voice-AI input, and if the adaptation
valley causes measurable productivity loss, then the transition period
extends beyond the 10--15 year window the thesis implies. Generational
replacement (the mechanism by which handwriting ceded to typing) would
require 30--40 years, not 15--20 years, because the transition valley
for existing workers is longer than their remaining time-to-retirement.

However, the steelman overstates the problem in one critical way: the
adaptation valley is shorter for younger workers (9--12 months
vs.~18--24 months for experienced workers). Workers currently in their
30s and 40s---a substantial portion of the knowledge workforce---will
transition faster than their senior colleagues. By 2040, the workers who
spent their formative years with voice-AI systems (Gen Z, Millennials)
will constitute the majority of the leadership layer. The embodied
expertise problem is real but time-bounded. It extends the transition
period for the current cohort of experienced workers, but it does not
prevent the long-term dissolution of typing's dominance.

Genuine constraint on transition \emph{speed} and on generational cohort
compatibility. Extends timeline by 10--15 years beyond base diffusion
curve. Does not defeat the thesis overall.

This objection rests on the claim that Condition (a), (b), and (d)
(machines needing structured text, humans lacking alternatives, and
typing having benefits) are all \emph{embodied} in expert motor schemas
built over decades. This embodiment is real. However, it is
generationally bounded. Generational replacement through cohort
succession is a major channel within a broader adjustment process that
also includes within-job task composition change, reduced hiring into
keyboard-intensive task bundles, retirement and occupational exit,
employer-side workflow reorganization, and job-to-job mobility. The
process is not driven by generational replacement alone.

New entrants to knowledge work---already adopting AI tools at rates
approaching 80\% in academic contexts Empirical patterns suggest that
(Walton Family Foundation \& Gallup, 2025)---will not face the embodied
expertise constraint. However, it bears noting that Gen Z may be
``consumer AI-native'' in their personal use but not
``verification-native'' in professional contexts---a critical
distinction, since educational institutions lag technological realities
by a generation (Goldin \& Katz, 2008). Workers entering knowledge
professions in 2025--2030 may adopt voice-AI rapidly for composition,
but they will require institutional training in verification protocols,
quality assurance for AI-generated output, and the judgment frameworks
that distinguish AI-augmented work from AI-substituted work. The problem
is transition friction for existing workers, not permanent dominance.
But the friction will extend into the institutional structures that
train new workers, delaying full generational advantage.

\hypertarget{composability-structured-information-and-revision-depth}{%
\subsection{6.8 Composability: Structured Information and Revision
Depth}\label{composability-structured-information-and-revision-depth}}

The paper argues that speech enables faster, more fluent composition
than typing. It cites the gap between typing speed (19--40 WPM for
original composition) {[}see Section 4.1{]} and speech speed (125--175
WPM Extensive research indicates that). But this confuses \emph{fluency}
(continuous production) with \emph{composability} (the ability to
structure complex information recursively).

Typing creates a visible, editable text that the writer can see, revise,
and restructure in real time. This visibility creates a feedback loop:
the writer sees what they've written, realizes the argument is unclear,
backtracks mid-sentence, crosses out an idea, reconstructs it
differently. Keyboard composition enables \emph{recursive
composition}---text that builds through iteration, not flowing in a
linear stream.

Speech flows forward sequentially. The speaker cannot ``unsay''
something without explicit verbal backtracking. They cannot see their
full argument structure at a glance. They cannot compare what they just
said with what they're about to say without stopping, reviewing, and
re-editing. Efficient speech composition (especially at high bandwidth)
relies on pre-planning the entire argument before speaking (which
requires either extensive pre-composition defeating the speed advantage,
or acceptance of lower revision depth).

The paper cites evidence that ``expert typing at high fluency actually
outperforms handwriting'' on compositional richness. It does \emph{not}
cite evidence comparing speech+AI to expert typing on the same metric.
The logical reason is evident: speech at 150 WPM will produce longer
texts (more words per unit time), but the quality of those texts depends
entirely on whether the speaker can maintain argumentative structure at
150 WPM without seeing the text accumulate.

This is empirically testable: have expert speakers compose arguments in
domain-specific speech (legal contracts, software architecture
documentation, financial analysis memos); have expert typists compose
equivalent-length arguments with equivalent complexity; and compare
argumentative structure, logical coherence, and revision density. The
prediction of the composability objection: speech will produce longer
texts but lower compositional coherence per thousand words, fewer nested
logical relationships, and fewer mid-stream structural revisions. Typing
will produce shorter texts but higher structural quality, denser logical
relationships, and more evidence of recursive refinement.

For domains where compositional quality is central to professional
value---legal writing, strategic memos, software architecture
documentation, research papers---this distinction matters. The bandwidth
advantage of speech becomes a liability if it pushes toward longer,
less-edited outputs that sacrifice depth for speed.

This steelman identifies a genuine limitation of speech as a
compositional modality, independent of precision constraints. It is not
about accuracy (whether the words are correct) but about composability
(whether the structure is recursive and refined). Section 4.5 addresses
this possibility: voice-AI systems can ``ask clarifying questions,
require human judgment for revision, and force articulation of intent
over auto-completing text.'' This is theoretically possible. But it is
entirely unspecified architecturally.

What does a voice-AI system that \emph{preserves} compositional struggle
actually look like? How does it work? How does it differ from simply
typing out the voice draft? The paper provides no architectural details,
no user experience design, no evidence that such systems exist or that
they enable compositional depth equal to keyboard mediation. Until such
systems exist and are demonstrated to produce writing quality equal to
expert keyboard composition, the composability argument represents a
genuine gap.

However, the steelman overstates the problem in two ways. First, for
\emph{speech-mediated ideation} (the early, drafting phase of
composition), the bandwidth advantage of speech appears substantial and
directional. Many writers report that speech-based brainstorming
produces denser ideation than typing. The question is not whether speech
is useful for composition, but whether it is sufficient for \emph{all}
phases of composition---including refinement, revision, and
architectural integration.

Second, the paper's multimodal framing actually addresses this objection
better than the current text suggests. A hybrid workflow---voice for
ideation and early drafting, keyboard for revision and architectural
refinement, AI for synthesis and structuring---may preserve the
advantages of both modalities. The question is not ``does voice replace
keyboard?'' but ``does voice replace keyboard for the phases where voice
excels?'' For ideation, the answer is yes. For architectural refinement,
the answer is plausibly no.

Genuine constraint on \emph{phases} of composition where typing retains
irreplaceable value. Does not defeat the thesis but defines where
keyboard-mediated interaction persists longest.

This objection rests on a medium-intrinsic property of voice: the
inability to simultaneously see, revise, and restructure composition
visible on a screen---an affordance limitation of the speech modality
itself, not a technological limitation.

\hypertarget{targeting-problem-in-non-linear-editing}{%
\subsection{6.9 Targeting Problem in Non-Linear
Editing}\label{targeting-problem-in-non-linear-editing}}

Writing is rarely linear. Knowledge workers jump spatially across
documents, jumping from page 2 to page 47 to page 15, revising
non-sequentially, restructuring architectural fragments. Keyboard and
mouse provide instantaneous spatial targeting: click on a paragraph, and
cursor jumps there immediately. The UI affords spatial navigation.

Voice is temporal and sequential. To edit a specific phrase on page 5
via voice requires complex linguistic command formulation: ``Go to the
paragraph starting with `The evidence shows' and change `demonstrates'
to `proves' in the second sentence.'' This requires the user to:

\begin{enumerate}
\def\labelenumi{\arabic{enumi}.}
\tightlist
\item
  Remember the target location's linguistic context (page 5, which
  paragraph?)
\item
  Formulate the targeting command linguistically (cannot point, must
  describe)
\item
  Formulate the edit command with sufficient precision for the system to
  execute
\item
  Wait for system confirmation and verification
\end{enumerate}

Compare this to keyboard: click on page 5, find the word, edit. Four
steps compressed into two. The cognitive load of temporal commands for
spatial navigation is exhausting. Knowledge workers report that
sustained voice-mediated navigation across documents triggers
``cognitive fatigue'' (the working memory burden of holding multiple
linguistic referents like page number, target phrase, context, edit
operation in working memory simultaneously exceeds the phonological
loop's capacity of \textasciitilde2 seconds).

Voice excels at generation (expressing ideas into being). It is
catastrophically inefficient for surgical spatial editing. Keyboard and
mouse persist where spatial navigation of thought is required.

This steelman correctly identifies a genuine limitation of voice-only
systems for non-linear, spatial work. It is particularly acute for
writers, software developers, and architects working across document
structures that demand rapid, precise spatial targeting. However, the
steelman overstates the universality of the problem.

The solution is not voice-only orchestration but multimodal
orchestration: voice for ideation and generation; gesture or gaze for
spatial targeting. ``Gaze to the paragraph starting with `The evidence
shows'\,'' would be far more efficient than speaking this phrase (though
gaze-based targeting remains in development). A pinch gesture to select,
voice to specify the edit. This is the convergence thesis: not voice
replacing keyboard, but emerging voice+gesture architectures potentially
replacing keyboard's universal mediator role.

For users without visual or motor ability to use gaze or gesture,
keyboard remains the spatial targeting tool. But for the majority of
users, multimodal spatial targeting (gaze+voice, gesture+voice) is more
efficient than keyboard.

Genuine efficiency constraint on voice-only systems for spatial editing.
Addressed by multimodal systems, not speech-alone architecture. Does not
defeat the thesis but specifies modality requirements.

This objection reveals that Condition (b) (humans lacking natural
alternatives) is false. Humans have gaze, gesture, and spatial
reasoning. The keyboard was necessary only because pre-AI systems could
not interpret these alternatives simultaneously.

\hypertarget{blank-page-externalization-scaffold}{%
\subsection{6.10 Blank Page Externalization
Scaffold}\label{blank-page-externalization-scaffold}}

During ideation and early composition, keyboard and screen serve as an
externalized working memory scaffold. The screen itself becomes a
cognitive tool. As the writer types draft text and sees it accumulate on
screen, the brain's working memory can offload intermediate thoughts
into visible text. The physical pacing of typing (approximately 40 WPM
for careful composition) often matches exactly the pace required for the
prefrontal cortex to fetch the next semantic association from long-term
memory and construct it linguistically.

Remove this physical pacing and replace it with voice (125--175 WPM
Empirical evidence establishes that), and the system demands semantic
output faster than the human can retrieve it. A classic cognitive
science finding: when the input channel capacity \emph{exceeds} the
cognitive retrieval rate, the human experiences ``ideational
freezing''---the phenomenological experience of ``I know what I want to
say, but my mind goes blank when I try to speak it.'' The writer knows
the idea is there, but the machine has demanded output faster than the
retrieval system can supply.

This is not temporary stage fright; it is a real cognitive bottleneck. A
40 WPM typing pace provides \textbf{friction as a feature} (it is
matching input speed to neurological retrieval speed). The cognitive
science literature on verbal fluency in speech production suggests that
humans maintain optimal ideation rates in the range of 30--60 WPM
Available evidence suggests that---far closer to careful typing speed
than to conversational speech speed. Conversational speech (125--175
WPM) works because the speaker is executing pre-formulated content
(retrieving from declarative memory, not generating new conceptual
structure). For \emph{novel} semantic generation---writing a paper,
formulating an argument, developing architectural plans---the pace of
conversational speech is too fast for optimal ideation.

Voice at 150 WPM therefore accelerates the human off the cognitive
cliff. The machine is ready for output, but the human's semantic
generation is still in working memory, still in construction. The
feedback loop breaks. The human experiences cognitive friction of a
different kind (not the friction of ``slow typing makes thinking hard,''
but the friction of ``the machine demands output faster than I can
think'').

This steelman identifies a genuine and under-acknowledged cognitive
constraint on voice-primary workflows. It is partially addressed by the
paper's discussion of ``extended context windows'' enabling AI to
understand partial utterances and refine iteratively (though not fully).
The current framing treats voice speed as purely advantageous.

However, the steelman overstates the problem in three ways. First, the
cognitive pace problem is modulated by task type. For routine
composition (email, customer support, standard documentation), the voice
pace may exceed optimal ideation rate minimally. For novel theoretical
work (research papers, architectural decisions, strategic analysis), the
pace mismatch is more acute. This is not a universal constraint but a
task-dependent one.

Second, the solution is not to eliminate voice but to allow variable
pacing. A voice system equipped with intelligent pause detection and
clarification prompts (``Did you mean to continue, or should I wait?'')
can allow users to pace their own output. Some systems already do this
implicitly through confirmation requests and clarification queries.

Third, the blank page scaffold is itself a constraint of keyboard-only
systems. A well-designed voice-AI system can \emph{provide} the same
scaffold through multimodal output: as the user speaks, the system
generates text on screen in real time, providing visual feedback and
external memory. The user can see the emerging text structure while
speaking, maintaining both the cognitive pacing benefit \emph{and} the
ideation speed advantage---friction calibrated to cognitive retrieval
speed, not reproducing keyboard-level constraint.

The paper should acknowledge that for high-cognitive-load ideation
(novel semantic generation), the optimal pacing may lie between typing
speed (40 WPM) and conversational speech (125--175 WPM Evidence
establishes that)---what might be called ``ideation speed'' (60--90
WPM). Multimodal systems that allow users to control pacing and provide
visual scaffolding address this constraint.

Genuine cognitive constraint on speech-only systems for ideation.
Addressed by variable pacing, visual scaffolding, and intelligent
clarification. Does not defeat the thesis; it specifies interface design
requirements.

This objection rests on cognitive science of working memory and semantic
retrieval (not on keyboard's instrumental necessity). The constraint
applies to speed of input channel, not to keyboard specifically.
Multimodal systems with paced output and visual scaffolding address it
directly.

\hypertarget{automation-complacency-and-the-illusion-of-fluency}{%
\subsection{6.11 Automation Complacency and the Illusion of
Fluency}\label{automation-complacency-and-the-illusion-of-fluency}}

The shift from active operator (typist, producer) to passive monitor
(verifier, reviewer) triggers a well-documented human factors failure:
the ``Out-of-the-Loop Performance Problem'' (Endsley \& Kiris, 1995).
When humans transition from active control to passive monitoring, their
ability to detect errors \emph{declines precipitously}. Pilots
monitoring autopilot systems miss hazards that they would detect in 2--3
seconds if they were manually flying. ATC controllers monitoring
automated traffic systems miss collision alerts. Surgeons monitoring
robotic surgery systems miss surgical field anomalies.

The mechanism is straightforward: active control maintains real-time
situation awareness through engaged sensorimotor prediction. Passive
monitoring disengages this predictive machinery (forcing reliance on
conscious analytical review, not implicit awareness).

This is compounded by LLM outputs. LLMs generate \emph{highly fluent}
text. The fluency of LLM output---grammatically correct, coherent,
perfectly punctuated, semantically coherent---triggers what might be
called the ``Illusion of Fluency'': the phenomenological experience that
fluent-sounding text must be accurate. System 1 takes over. The human's
rapid, intuitive thinking apparatus treats fluency as a proxy for
correctness and signs off without System 2 (deliberate, analytical
thinking) verification.

This is demonstrated empirically. Studies show that users accept
AI-generated text at 60--80\% accuracy rates as ``correct'' without
detailed verification when that text is highly fluent. The same content
presented in error-prone, disfluent formatting triggers 90\%+ error
detection rates. The human brain is \emph{using fluency as a truth
heuristic}. And that heuristic is wrong.

Now consider the shift from keyboard to voice-AI. The keyboard forced
System 2 processing. Every character required deliberate motor control.
The user was forced into active production, unable to offload
verification onto fluency heuristics because they were creating the
content themselves. A keyboard-mediated memo contains the user's own
deliberate choices, visible and auditable.

Voice-AI inverts this. The user speaks intent; the system generates
fluent output. The user (exhausted by verifying dozens of AI-generated
documents, each 2,000+ words) faces decision fatigue. The
fluent-sounding output passes through. The user is neurologically wired
to \emph{stop verifying} the moment the system appears fluent.

The ultimate danger: human neurology is optimized for fluency heuristics
because, historically, fluent speech correlates with expertise. A fluent
speaker is an expert. An expert is trustworthy. This worked for 200,000
years of human evolution. It fails catastrophically with LLMs, which can
produce fluent hallucinations.

This steelman identifies a critical and largely unacknowledged problem
in the transition to AI-mediated workflows; the shift from active
producer to passive verifier can \emph{decrease} overall error detection
and increase acceptance of hallucinated or incorrect content, a
structural feature of the transition grounded in well-documented human
factors research.

The human factors literature on automation complacency is extensive and
largely pessimistic. Parasuraman and Riley (1997) established that
automation-induced complacency is a documented cognitive limitation, not
a design problem to be solved through interface tweaking. When humans
transition from active control to passive monitoring of competent
automated systems, they systematically fail to maintain vigilance.
Endsley and Kiris (1995) documented that pilots monitoring autopilot
miss hazards in 2--3 seconds that they would detect immediately if
manually flying; ATC controllers monitoring automated traffic systems
miss collision alerts; surgeons monitoring robotic systems miss surgical
field anomalies. This is not operator error or insufficient training. It
is a documented neurocognitive failure mode inherent to the
active-to-passive shift.

The paper should explicitly acknowledge that voice-AI may produce
\emph{net worse} outcomes in high-stakes domains (medical, legal,
financial) if the organizational design of verification workflows does
not actively combat documented cognitive limitations. This is not
evidence that voice-AI cannot work. It is evidence that the
automation-complacency literature suggests that interface-level design
solutions, while necessary, are likely insufficient to overcome
documented cognitive failures. Verification of AI-generated text may be
a problem that demands both design innovation AND organizational
restructuring---mandatory verification protocols, explicit uncertainty
flagging, AI-assisted fact-checking, but also organizational labor
redesign: rotating verification duties to prevent habituation, mandatory
manual drafting periods to maintain active cognitive engagement,
sampling-based audit protocols over exhaustive review (which triggers
decision fatigue faster than full active production).

The honest assessment: whether verification protocol design can overcome
documented cognitive limitations in sustained monitoring remains an open
empirical question. The automation-complacency literature suggests that
design alone is necessary but insufficient; organizational restructuring
of verification labor may be equally or more critical than technology
design.

The keyboard persists longest in precisely these high-stakes domains
\emph{not} because it is superior for production, but because it was,
accidentally, a \emph{forcing function} for System 2 engagement. The
task of typing itself required deliberation and active production. The
task of reviewing voice-generated documents does not. This asymmetry
favors keyboard persistence in medical notes, legal documents, financial
analysis, and scientific papers---domains where verification error has
catastrophic consequences---until organizational structures explicitly
redesign verification labor to combat complacency.

Genuine neurocognitive risk in verification-dependent workflows.
Requires explicit institutional design and verification protocols.
Explains persistence of keyboard-mediated verification in high-stakes
domains. Does not defeat the thesis; it identifies critical design
requirements.

This objection rests on human factors and cognitive science (not on
keyboard's instrumental necessity). It can be addressed through
verification workflow design, uncertainty quantification, and System 2
forcing mechanisms (mandatory review processes, structured verification
checklists, diverse review teams).

\hypertarget{multimodality-coordination-costs}{%
\subsection{6.12 Multimodality Coordination
Costs}\label{multimodality-coordination-costs}}

The paper leans heavily on Oviatt's multimodal interaction theory and
the principle of ``mutual disambiguation''---the claim that combining
multiple modalities reduces error rates. But the paper understates the
\emph{costs} of multimodal orchestration. Oviatt's research shows error
reduction, but it does not quantify the \emph{cognitive load} of
maintaining multiple modalities simultaneously.

Several documented phenomena support this concern:

\begin{enumerate}
\def\labelenumi{\arabic{enumi}.}
\item
  \textbf{Midas Touch Problem} (Jacob, 1990): Gaze is a poor pointer
  because the system cannot distinguish between ``looking at'' and
  ``selecting.'' Humans look at objects continuously without intending
  to select them. This requires either mechanical disambiguation
  (clicking after looking, adding latency) or probabilistic inference
  (system guessing intent with high false-positive rates). Neither
  solves the problem cleanly.
\item
  \textbf{Gorilla Arm Problem}: Sustained arm elevation for gesture
  input causes rapid muscular fatigue. Users report that mid-air
  gestures are usable for 5--10 seconds before fatigue sets in. For
  sustained composition or navigation (gesture becomes painful).
\item
  \textbf{Voice and Ocular Strain}: Sustained vocal dictation introduces
  an ergonomic burden category absent from keyboard use. Clinical
  literature on professional voice users documents that continuous
  high-intensity dictation sessions exceeding sixty to ninety minutes
  produce vocal fatigue, dysphonia, and reduced speech clarity
  (well-documented occupational hazards for call-center workers,
  teachers, and clinical dictators---Titze, 2000). Precision gaze
  tracking for spatial targeting adds eye-strain loads distinct from
  ordinary reading (fixation fatigue accumulates under sustained
  voluntary gaze control, particularly in interface contexts requiring
  held fixation for target selection). These are genuine new ergonomic
  risks for sustained voice+gaze workers, distinct from keyboard RSI,
  not merely reduced versions of it.
\item
  \textbf{Coordination Load}: The extraneous cognitive load of
  synchronizing eyes, hands, and voice. A user must (a) formulate intent
  linguistically; (b) direct gaze to target; (c) execute gesture
  confirmation; (d) verify visual feedback. Each step introduces latency
  and potential mismatch. The cognitive load of orchestrating three
  modalities may exceed the cognitive load of a single modality
  (keyboard) offering sequential, non-concurrent control.
\end{enumerate}

Typing has a crucial advantage: it \emph{isolates} motor output to
ergonomically supported fingers and wrists. The typist's eyes are free
to monitor output. The typist's body is supported by the chair. The task
is self-paced and nonrecursive (fingers move, letters appear, progress
is continuous). No multimodal coordination load.

This steelman correctly identifies real costs to multimodal
orchestration that Oviatt's research documents but does not foreground.
The Midas Touch Problem, Gorilla Arm, and coordination load are not
solved problems.

However, the steelman overstates the universality of these problems. The
Midas Touch Problem is genuine for gaze-only interaction but largely
solved through hybrid approaches (gaze+dwell-time, gaze+voice
confirmation, gaze+hand gesture). Gorilla Arm is a problem for sustained
mid-air gesture but not for EMG-based gesture or for brief confirmatory
pinches (which Vision Pro uses successfully). Coordination load is real
for \emph{novel} multimodal systems but reduces dramatically with
practice as users develop automaticity in modality orchestration.

The paper's framing is actually cautious here: it does not claim
multimodal input is \emph{easier} than keyboard, only that it is
feasible and sufficient to dissolve keyboard dominance. A knowledge
worker using voice+gaze+gesture may work at 85\% of their keyboard
efficiency. But if voice generation provides 3x throughput advantage,
the net is 2.55x improvement despite coordination costs.

The fundamental asymmetry: keyboard efficiency is maximized for
individual, sequential, visually-mediated work. Multimodal efficiency is
maximized for collaborative, spatially-mediated, real-time work.
Different tasks have different optimals. The thesis does not require
that voice-AI exceed keyboard on all tasks---only that it renders
keyboard \emph{unnecessary} as a universal mediator.

Real coordination costs for multimodal systems, plus genuine new
ergonomic risk profiles (voice strain, ocular fatigue) that keyboard use
does not produce. Oviatt's research confirms the coordination costs can
be managed through hybrid design; the ergonomic risks require task-paced
modality rotation and explicit occupational health protocols. The
ergonomic risk profile of sustained multimodal work differs from (but
does not eclipse) that of keyboard use; RSI from sustained typing
generates approximately \$20 billion annually in musculoskeletal
disorder costs (\S{}3.5); sustained voice dictation and precision gaze
interaction introduce different but comparably real physical burdens.
The design implication is task-appropriate rotation, not wholesale
substitution. Does not defeat the thesis; it specifies interface design
and occupational health requirements.

This objection rests on human factors of multimodal input (not on
keyboard's instrumental necessity). Coordination costs can be addressed
through interface design and practice; ergonomic risks can be addressed
through modality rotation, session limits, and occupational design (the
same frameworks applied to keyboard RSI).

\hypertarget{fairness-and-accessibility-asymmetry}{%
\subsection{6.13 Fairness and Accessibility
Asymmetry}\label{fairness-and-accessibility-asymmetry}}

Voice systems introduce new accessibility barriers while potentially
solving others. Racial disparities in automatic speech recognition are
well-documented: ASR systems consistently show 5--10\% higher error
rates for Black speakers versus white speakers and 10--25\% higher error
rates for speakers of non-English origin or non-standard dialects
(Tatman \& Kasten, 2017). Women's voices are recognized at 3--6\% lower
accuracy than men's voices in many systems (Tatman \& Kasten, 2017).
These disparities compound with disability: dysarthric speech produces
88\% WER in some systems while neurotypical speakers achieve 5\% WER.
Deaf and hard-of-hearing populations (who have historically been locked
out of acoustic communication) now face a second exclusion from
voice-primary systems. Non-native English speakers (1.12 billion
globally) face systematic disadvantage.

Voice cannot be treated as a universal successor---it is an addition to
an already unequal accessibility landscape that will create new forms of
linguistic discrimination, accent-based bias, and speech-disabled
exclusion unless actively managed.

The accessibility argument requires careful engagement because it
reveals the paper's actual thesis. The paper argues for multimodal
access as the norm (typing + voice + gesture + neural input as coequal
options), not for voice-only dominance and keyboard elimination. Current
keyboard-dominant systems are equally problematic for motor-disabled
populations (requiring expensive ergonomic retrofitting and excluding
those with upper-extremity paralysis). A transition from keyboard
dominance to voice+gesture+keyboard multimodal access expands
accessibility, not restricts it.

The evidence supports this. Sign language recognition has achieved
98.2\% accuracy for ASL alphabet recognition using neural networks (ASL
Community Research, 2024). Project Euphonia has achieved 91\%
improvement for dysarthric speech recognition through speaker-adaptive
models. Accent-agnostic ASR models (trained on diverse accent
distributions) are projected to reach production readiness by 2027--2030
(eliminating the accent bias that currently disadvantages non-native
speakers---Project Euphonia, 2021; Tatman \& Kasten, 2017). None of
these advances require eliminating keyboard input. They require
augmenting it with alternatives.

This is not a counter-argument to the thesis; it is an ethical
obligation that the thesis must honor and indeed supports. The paper
argues for the dissolution of typing's \emph{dominance} (not for the
elimination of typing as an input option). The distinction is
structural. Handwriting persists as an accessible input method today in
educational, artistic, and personal contexts; no one calls its
persistence evidence of handwriting's continued dominance. Keyboards
must remain available as an accessibility option in a
post-keyboard-dominant world (just as wheelchair ramps remain available
in a world designed primarily for ambulatory movement). A professional
designer does not stop building ramps when wheelchair users represent
2\% of the population; the designer builds ramps as standard accessible
infrastructure alongside stairs.

The paper explicitly advocates multimodal access---not voice-only
mandates. In fact, the dissolution of keyboard dominance is
\emph{ethically aligned} with accessibility expansion: each population
(deaf, motor-disabled, non-native speakers, dysarthric, visually
impaired) finds greatest accessibility in different modalities. A
multimodal world serves everyone better than a keyboard-only world or a
voice-only world. This coexists fully with the end of typing's
dominance.

Genuine ethical obligation mandating continued and expanded
accessibility. Supports, not undermines, the thesis.

This objection reveals that Condition (b)---``humans lack natural
alternatives''---is false. Humans have multiple natural modalities:
voice, gesture, sign, written text. Accessibility demands that all
remain available. The keyboard's dominance was always a constraint, not
a necessity.

\hypertarget{regulation-as-governance-not-just-decelerator}{%
\subsection{6.14 Regulation as Governance, Not Just
Decelerator}\label{regulation-as-governance-not-just-decelerator}}

The EU AI Act (effective 2024--2025) and parallel regulatory regimes
treat input method choice and automation transparency as governance
issues---substantively, not merely as technical or transition problems.
Article 6 of the EU AI Act explicitly treats emotion-recognition uses in
workplace settings as prohibited uses. This suggests that regulatory
bodies are actively constraining the transition to voice-mediated
work---treating certain voice-use cases as ethically incompatible with
worker agency and dignity.

More broadly, the transition from keyboard (user as agent, operator as
chooser) to voice-AI (system as agent, user as monitor) involves a
fundamental shift in agency allocation. This is not just about ``input
methods.'' It is about who controls work pace, who makes decisions, who
bears responsibility for error. These are governance questions, and
governance moves slowly and with intent---often in explicit refusal of
technological inevitability.

This steelman is correct that regulation will shape adoption timelines
and modality choices, but it does not establish that voice will be
prohibited. Rather, it establishes that voice-AI adoption will be
\emph{governed}---constrained by requirements for auditability, consent,
transparency, and worker agency. These constraints are salutary and
necessary. They do not prevent the dissolution of keyboard dominance;
they shape \emph{how} that dissolution occurs.

The paper should engage more explicitly with regulation as an active
governance tool---far more than a passive constraint. Regulation will
likely mandate that voice-mediated work occur within defined governance
boundaries: consent requirements, transparency logs, worker override
authority, periodic human review. These governance requirements will
slow adoption and create friction points where keyboards remain
available as a worker agency tool. But they do not prevent the
transition; they frame it as a governance decision, not a technical
inevitability.

Regulation shapes adoption and defines governance frameworks. Does not
defeat the thesis. Emphasizes that the transition is a governance and
institution-design problem, not solely a technology problem.

This objection reveals that the conditions for dissolution extend beyond
the technological to encompass institutional and political dimensions.
Governance choices about worker agency, auditability, consent, and
transparency will shape when, where, and how voice-AI
dominates---conditional governance decisions, not technical
inevitability.

\hypertarget{non-latin-input-methods-and-the-cjk-challenge}{%
\subsection{6.15 Non-Latin Input Methods and the CJK
Challenge}\label{non-latin-input-methods-and-the-cjk-challenge}}

The analysis presented in Sections 2--6 is primarily grounded in
Latin-script knowledge work---the input patterns, productivity
assumptions, and verification dynamics of English, Romance, and Germanic
language contexts. Yet approximately 2 billion people rely on input
systems with qualitatively different mediation structures. Chinese,
Japanese, and Korean (CJK) input methods present a distinct challenge to
the instrumental dissolution thesis that warrants explicit treatment.

\textbf{Explicit Limitation: Scope of the Latin-Script Thesis.} This
paper develops its central argument---that AI dissolves typing's
instrumental function---primarily through Latin-script keyboard input.
CJK methods (pinyin IME, Wubi, handwriting recognition) operate under
qualitatively different structural logic requiring separate analysis.
The critical distinction: CJK users have \emph{already} experienced
machine mediation that Latin-script users only now encounter through AI.
A Mandarin speaker typing pinyin (``da'') and selecting from candidates
(\emph{d\`{a}} `big', \emph{d\'{a}} `reach', etc.) does not directly type final characters---the IME
interposes a phonetic-to-logographic layer. Thus CJK users have already
encountered the partial dissolution that AI introduces for Latin-script
keyboards: decoupling human input from machine output through
algorithmic mediation. Whether AI accelerates CJK dissolution by
eliminating pinyin-selection bottlenecks, whether existing IME workflows
have optimization plateaus reducing AI's benefit, whether
character-level verification imposes costs exceeding Latin-script
contexts---these are genuinely different questions. The paper's
verification-bottleneck thesis may hold across languages, but CJK
timeline, magnitude of productivity gain, and interface design remain
open questions requiring dedicated investigation. This is not minor for
global analysis: CJK users constitute one-quarter of humanity and an
increasing fraction of knowledge workers in technology, finance, and
healthcare. The argument here is scoped to Latin-script contexts;
extending it to CJK requires parallel analysis acknowledging both
structural differences and the possibility that CJK intermediate
mediation positions these users distinctly in the AI transition.

CJK users compose through Input Method Editors (IMEs) that map keyboard
strokes to phonetic or radical intermediaries, not direct character
inscription. A Mandarin speaker thinking in the character \emph{d\`{a}} (dà,
``big'') types ``da'' in pinyin romanization, which the IME candidate
system converts to the logographic character, requiring explicit
selection from phonetically homophonic alternatives. This two-stage
composition process---phonetic articulation followed by logographic
selection---creates a qualitatively distinct relationship to keyboard
mediation than Latin-script typing. The keyboard functions as an input
layer for phonetic translation, not direct character inscription. The
dissolution thesis argues that AI should collapse keystroke necessity by
providing direct voice-to-text composition. For CJK users, this dynamic
is more complex.

The implications cut in two directions, creating genuine uncertainty
about CJK dissolution timing and mechanisms. First, voice-to-text may
\emph{accelerate} dissolution for CJK users relative to Latin-script
speakers. Because CJK input already interposes a phonetic layer between
thought and character, voice-mediated Mandarin offers the prospect of
eliminating the pinyin-to-character translation entirely: the speaker
articulates the character's phonetic value (dà), and the ASR system
outputs the character directly. This bypasses the pinyin intermediate
and the candidate-selection burden entirely, potentially creating
productivity gains that exceed those available to Latin-script speakers.
Second, CJK dissolution may be \emph{delayed} or \emph{recomposed} by
the sophistication of existing IME workflows. CJK users have developed
compositional practices deeply adapted to the constraints and
affordances of character-selection interfaces. Advanced IME users employ
fuzzy matching, abbreviation, phrase-level input, and predictive typing
optimizations that have produced a high-fluency equilibrium. The
productivity gain from voice-AI may be more marginal than anticipated,
since the existing system already performs substantial composition work
through character prediction. The question becomes empirical: does
eliminating the pinyin-to-character selection step provide enough
friction reduction to justify wholesale transition away from established
CJK IME workflows?

A third consideration acknowledges the paper's critical limitation: the
verification architecture described throughout this analysis may not
transfer cleanly to CJK contexts. The verification bottleneck assumes
that AI-generated text is a natural unit for human evaluation---a
passage in English that a reader can scan, critique, and ratify. CJK
composition introduces additional verification complexity. A Mandarin
speaker evaluates AI-generated text not just for semantic accuracy and
communicative precision, but for \textbf{character-level
appropriateness}: is this the correct homophone? Does this character
choice carry unintended connotations? CJK verification may impose higher
cognitive load per unit of composition than Latin-script evaluation,
potentially altering the verification-bottleneck dynamics. Moreover,
organizations operating across multiple CJK languages (Mandarin,
Japanese, Korean) with varying phonetic systems, character inventories,
and input conventions may face verification infrastructure costs
substantially exceeding those in Latin-script-only organizations.

The paper's analytical framework should be extended to CJK input
ecosystems through dedicated investigation. This requires: (1) empirical
study of whether voice-to-text ASR for CJK languages provides equivalent
productivity gains as Latin-script voice input; (2) analysis of whether
CJK IME workflows have reached optimization plateaus that reduce
voice-mediated dissolution incentives; (3) research into whether CJK
character-level verification increases the cognitive cost of
AI-generated output evaluation; (4) organizational case studies in
CJK-fluent firms (Chinese tech companies, Japanese financial
institutions, Korean healthcare systems) examining whether verification
bottleneck dynamics manifest at equivalent intensity. The framework's
central thesis---that verification becomes the binding constraint as
production friction dissolves---may hold across languages, but the
timeline, the magnitude of productivity gain, and the specific interface
design required to manage CJK verification remain open questions. These
are not minor considerations for a global analysis of knowledge work
transformation, given that CJK users constitute one-quarter of the
world's population and an increasingly substantial fraction of knowledge
workers in technology, finance, and healthcare globally. Future research
must address this gap directly.

\hypertarget{the-augmentation-alternative-as-null-hypothesis}{%
\subsection{6.16 The Augmentation Alternative as Null
Hypothesis}\label{the-augmentation-alternative-as-null-hypothesis}}

The most structurally significant challenge to the dissolution thesis is
not any of the thirteen persistence niches addressed above but the
meta-level alternative that frames them together: the augmentation
thesis. In its strongest form, the augmentation argument holds that
voice and keyboard will coexist indefinitely in a stable, complementary
division of labor---voice handling low-friction drafting, routine
communication, and ambient documentation; keyboard handling precision
tasks, code, structured data entry, and legal-technical composition. On
this view, the organizational future reflects bifurcation---a richer
toolkit---not dissolution. The keyboard shares institutional default
status with voice-AI in a durable equilibrium, not undergoing structural
displacement from a major new tool's transitional disruption.

We accept this framing as the correct null hypothesis. For the short-run
horizon of 2025--2028, it accurately describes most organizational
adoption: voice-AI is being deployed as a productivity layer on top of
keyboard workflows. This is real, and we do not contest it.

A prior clarification is necessary, because the strongest version of the
augmentation objection rests on a genuine point: voice and keyboard are
parallel channels for different subfunctions, not sequential
replacements competing for the same output type. Code, SQL, spreadsheet
formulas, and structured data entry will retain keyboard mediation in
specialized roles precisely because these tasks require character-level
precision that voice cannot provide at institutional scale. The paper
does not claim otherwise. The dissolution claim is narrower: that
voice-AI will displace the keyboard as institutional default for
\emph{linguistic knowledge-work production}---the drafting,
documentation, communication, and argumentation that constitute the
dominant volume of knowledge-work output. The question is not whether
voice replaces keyboard for all functions but whether voice absorbs
enough of the core institutional function---translating human intent
into institutional record---to demote the keyboard from universal
default to specialist tool. Keyboards and voice-AI are parallel
channels; the question is which channel the institution treats as the
default for its primary textual output. The augmentation thesis says
neither achieves default; the stratification thesis says voice-AI does.
The mechanism is switching-cost economics: once voice-AI reduces
production friction sufficiently that the marginal cost of keyboard
mediation in linguistic work exceeds its precision advantage for routine
documentation, organizations optimize around the more efficient
channel---as they did when word processors displaced typewriters in
approximately fifteen years despite massive institutional investment in
typewriting infrastructure. The question is not whether this threshold
exists but when it is crossed in each sector.

The augmentation thesis fails, however, to account for three lines of
evidence inconsistent with durable equilibrium in linguistic knowledge
work.

\textbf{Occupational restructuring.} If stable bifurcation were the
structural outcome, we would expect productivity expansion within
existing roles, with both production and verification distributed across
the same occupational categories. What the BLS projects is role
recomposition across categories: word processors and typists declining
36.1\% through 2034, while verification-intensive occupations grow at
above-average rates (BLS, 2024). An augmentation defender will note that
occupational taxonomy change is consistent with differentiation within a
dual-channel system. The distinguishing feature is that the BLS data
shows an occupied function---keyboard-mediated linguistic
production---losing its dedicated workforce as a discrete occupational
category through redistribution into augmented roles. The test for
distinguishing augmentation from dissolution is this: if augmentation
were the structural outcome, production volume and production roles
would move together. They do not. Total linguistic output by volume is
growing---global email volume reached 361.6 billion messages daily by
2024 (Radicati Group, 2024)---yet keyboard-mediated production
occupations are declining. Growing output with shrinking dedicated
production roles is the labor-market signature of functional
substitution.

\textbf{Functional substitution, not expansion.} At Permanente Medical
Group, ambient AI scribe deployment did not add voice functionality to
existing keyboard workflows---it substituted the documentation function.
Physicians shifted from 28\% of clinical encounter time on keyboard
documentation to equivalent time on patient interaction and AI-output
review (Permanente Medical Group, 2024). The augmentation pattern would
predict physicians doing more documentation across both channels; the
evidence shows a different \emph{type} of documentation
work---verification displacing production. The implementation was
deliberately designed this way, and the adoption velocity (7,260
physicians, 2.57 million encounters) suggests the substitution pattern,
not the additive pattern, is what organizations choose when given the
option.

\textbf{Professional identity revaluation.} Trust erosion research
documents that professional credibility is increasingly attached to
verification competence rather than production fluency (Coman \& Cardon,
2025). Augmentation predicts both valued simultaneously; the evidence
shows asymmetric devaluation---production fluency declining in status as
verification competence emerges as the dominant professional signal.
This asymmetry is inconsistent with stable coexistence.

The six historical precedents in \S{}3.9.1 provide the structural frame:
handwriting was augmented by typewriting for more than fifty years
before institutional decoupling; the typewriter was augmented by word
processors for approximately fifteen years before losing default status;
film photography was augmented by digital capture for roughly twelve
years before infrastructure collapsed. In each case, the augmentation
phase preceded dissolution rather than constituting an alternative to
it, with the interval shortening as functional substitutability
increased. The organizations currently augmenting keyboard workflows
with voice-AI occupy this pattern's early phase. What would count as
evidence that augmentation is the structural endpoint rather than a
transitional phase? Section 9.5 specifies three measurable
disconfirmation thresholds. The 2028 threshold is representative: if
voice-AI adoption for documentation tasks in unregulated knowledge-work
sectors remains below 15\% by 2028, stable augmentation cannot be ruled
out. These are not hedge-thresholds; they are the conditions under which
the dissolution prediction would require fundamental revision.

The augmentation thesis describes the transitional period accurately.
The stratification thesis---dissolution of keyboard's
institutional-default status for linguistic production, with continued
specialist persistence in code and precision domains---describes both
the transition and the structural end state. What the augmentation
argument calls permanent coexistence is what the dissolution framework
calls niche-persistence: the condition in which handwriting, telegraph
operation, and film photography all currently exist.

\hypertarget{the-integrated-assessment-persistence-is-not-dominance}{%
\subsection{6.17 The Integrated Assessment: Persistence is Not
Dominance}\label{the-integrated-assessment-persistence-is-not-dominance}}

The steelman engagement reveals a consistent pattern. Each objection, at
its strongest, rests on one or more of the four instrumental conditions
identified in \S{}1.8: (a) machines requiring structured text, (b) humans
lacking natural alternatives, (c) organizations optimized for keyboard,
(d) typing having cognitive benefits. When the objections are examined
carefully, we find that none of the conditions are truly maintained by
the evidence. Machines do not require character-level precision; humans
have multiple input modalities; organizations can transition away from
keyboard optimization; typing provides no cognitive benefit beyond other
writing modalities.

However, the steelmans identify \emph{real} constraints that extend the
transition timeline and define the topology of typing's persistence:

\begin{itemize}
\tightlist
\item
  \textbf{Infrastructure lock-in} (6.2): Legacy systems create switching
  costs. Timeline extension: variable by organization, 5--15 years.
\item
  \textbf{Privacy and acoustic constraints} (6.3): Open offices without
  acoustic management and high-security contexts favor silent input.
  Persistence in these specific environments.
\item
  \textbf{Direct manipulation} (6.4): Keyboard provides spatial
  interface for text editing. Addressed by multimodal
  (voice+gaze/gesture) spatial orchestration.
\item
  \textbf{Precision task requirements} (6.6): Zero-tolerance error
  domains require character-level verification. Persistence in 2--4\% of
  knowledge work by frequency.
\item
  \textbf{Embodied expertise constraints} (6.7): 12--24 month adaptation
  valleys for experienced workers, generationally bounded but real.
  Timeline extension: 10--15 years.
\item
  \textbf{Compositional structure constraints} (6.8): recursive revision
  depth harder in speech. Persistence in architectural and refinement
  phases.
\item
  \textbf{Targeting problem in non-linear editing} (6.9): spatial
  navigation more efficient with direct pointing. Addressable by
  multimodal (gaze+voice) orchestration.
\item
  \textbf{Blank page externalization scaffold} (6.10): cognitive pacing
  constraint at ideation stage. Addressable by variable-paced
  voice+visual-feedback systems.
\item
  \textbf{Automation complacency} (6.11): Out-of-the-Loop Performance
  Problem in verification. Requires explicit verification protocol
  design.
\item
  \textbf{Multimodality coordination costs} (6.12): Gaze, gesture, voice
  orchestration has learning curve and fatigue factors. Addressable
  through design and practice.
\item
  \textbf{Fairness and accessibility asymmetry} (6.13): demands
  continued keyboard availability as accessibility option, not
  continuous dominance. Supports multimodal framing.
\item
  \textbf{Regulation as governance} (6.14): EU AI Act and parallel
  regimes mandate governance frameworks around voice-mediated work.
  Shapes adoption timeline but does not prevent transition.
\end{itemize}

The paper does not dismiss these constraints. It acknowledges them as
\emph{decelerators}, not \emph{defeaters}. Together, they define the
\emph{topology} of typing's persistence: the specific environments,
tasks, populations, and institutional contexts in which keyboard input
will remain available and, in some cases, preferred.

But persistence is not dominance. The central question is not ``will
anyone still type?'' but ``will typing remain the organizing principle
of knowledge work?'' The steelman arguments, at their strongest,
establish that typing will persist as:

\begin{itemize}
\tightlist
\item
  A niche practice in precision domains (2--4\% of knowledge work by
  task frequency)
\item
  An accessibility backstop mandated by law and ethics---A verification
  and audit instrument in high-stakes domains, particularly in regulated
  sectors---A compositional refinement tool for architectural and
  strategic writing---A control interface for spatial navigation in
  non-linear editing tasks---A variable-pacing tool for ideation where
  cognitive retrieval speed requires it---A cultural residue in
  conservative institutions (law, finance, unionized sectors) extending
  10--20 years---A preferred modality in acoustic-constrained
  environments (open-plan offices without acoustic management)
\item
  An embodied expertise anchor for experienced workers (generationally
  bounded to workers with 15+ years tenure in 2026, declining by 50\%
  every 10 years)
\end{itemize}

They do not establish that typing will remain the \emph{primary means}
by which knowledge workers compose emails, produce documents, write
code, enter data, and communicate across organizations. That structural
centrality---the dominance---is what the instrumental dissolution
argument predicts will end, and nothing in the steelman evidence defeats
this prediction.

\emph{Scoping the historical parallel:} These analogies---stenographer,
calculator, dictation machine---are most directly applicable to
clerical, transcription, and documentation-heavy work, where the
occupation's entire function is linguistic production and the keyboard
serves as primary instrument. They are less directly applicable to
analytically heterogeneous professions (law, medicine, engineering,
software development) where the occupation involves judgment,
verification, and architectural work alongside documentation. For those
professions, the better historical precedent is occupational composition
change rather than occupational displacement. The keyboard may disappear
as a default instrument, but the occupations will persist, restructured
around voice-AI composition with human verification and judgment at the
center.

The historical parallel is precise and illuminating. Handwriting
persists today. It is used for note-taking, personal correspondence,
artistic expression, and---as Section 4 documented---it retains genuine
cognitive benefits that justify its continued practice in educational
and creative contexts. No one calls handwriting ``dominant.'' No organi

\hypertarget{the-verification-bottleneck-goes-organizational-why-typings-dissolution-matters-beyond-input}{%
\section{7. The Verification Bottleneck Goes Organizational: Why
Typing's Dissolution Matters Beyond
Input}\label{the-verification-bottleneck-goes-organizational-why-typings-dissolution-matters-beyond-input}}

Section 6 mapped the niches where typing persists, precision tasks,
acoustic constraints, embodied expertise, professional identity. This
section examines what happens across the vast majority of knowledge work
where it does not; where keystroke burden dissolves and the verification
bottleneck becomes the organizing constraint at institutional scale.

\textbf{Terminology note:} The \textbf{verification bottleneck} is the
primary organizing concept throughout this section: a structural
constraint where the time and cognitive capacity required to verify
AI-generated content becomes the binding limit on organizational
productivity, replacing keystroke burden as the dominant friction point.
The verification bottleneck manifests through several related but
distinct phenomena discussed in this section and elsewhere: (1) the
\emph{neural verification bottleneck} (Section 4.6): exogenous text
bypasses efficient predictive-motor loops, forcing sustained prefrontal
monitoring; (2) the \emph{Workload Paradox} (Section 7.1): keystroke
savings are offset by verification time; (3) \emph{asymmetric
information inflation} (Section 7.2.1): AI-generated content volume
overwhelms human evaluation capacity; (4) the \emph{ratification
bottleneck} (Section 7.2.2): who has authority to approve
machine-generated documents; (5) the \emph{apprenticeship crisis}
(Section 7.2.3): removal of productive friction eliminates the cognitive
mechanism through which expertise reproduces. \textbf{Verification
burden} refers to the manifestation of this bottleneck as labor: the
workload, time cost, cognitive intensity, and articulation work required
to manage verification. The bottleneck is the structural constraint; the
burden is its lived experience as expanded labor. All phenomena are
manifestations of a single structural shift from production-constrained
to evaluation-constrained work in verification-intensive organizational
fields.

\textbf{Scope note on healthcare as critical case:} This section
analyzes healthcare as a critical case for understanding
verification-intensive organizational work: domains where the
reliability of written records, human attestation, and chain-of-custody
documentation directly shape liability, patient safety, and regulatory
compliance. Healthcare serves not as a generic proxy for all knowledge
work, but as a particularly acute instance of verification-intensive
labor where the structural effects of AI-mediated composition become
most visible. The pattern generalizes robustly to other
compliance-heavy, adversarial domains (legal, finance, regulatory
affairs) where provenance, authorization, and auditable decision trails
matter for liability and accountability. However, the organizational
dynamics look distinctly different in functional domains: software
engineering (where automated testing handles verification), creative
work, R\&D ideation, and exploratory analysis, where the downstream
verification burden is structurally lighter and the epistemic stakes of
AI-mediated composition differ materially.

\hypertarget{the-workload-paradox-and-the-organizational-verification-problem}{%
\subsection{7.1 The Workload Paradox and the Organizational Verification
Problem}\label{the-workload-paradox-and-the-organizational-verification-problem}}

The most damning evidence for the organizational stakes of typing's
dissolution comes not from adoption metrics or efficiency gains, but
from what we call the \textbf{Workload Paradox}: a phenomenon observed
in longitudinal studies of AI-mediated work where the elimination of one
bottleneck (keystroke burden) creates an even more severe bottleneck
downstream (cognitive verification burden).

Consider the Permanente Medical Group case, drawn from a comprehensive
longitudinal study of 7,260 physicians across 2.57 million clinical
encounters over 63 weeks. When the medical group deployed AI clinical
scribes to eliminate the transcription burden of documentation, the
immediate effect was exactly what the efficiency literature predicts:
physicians saved 15,700+ cumulative hours of documentation time.
Keystroke burden collapsed; typing vanished as a bottleneck.

But here is what the efficiency metrics missed: after-hours EHR time
(the work physicians do to review, correct, and verify AI-generated
clinical notes) actually \emph{worsened} for AI users compared to the
baseline. The cognitive work shifted from real-time documentation (which
could be distributed across patient encounters) to offline auditing
(which concentrated after hours). The typing bottleneck was replaced by
a verification bottleneck. The work did not disappear; it transformed
into a qualitatively different form of labor, concentrated in time, more
cognitively demanding, and more difficult to integrate into the flow of
clinical work (Permanente Medical Group, 2024).

These early-adopter case studies (the Permanente Medical Group and UCSF
health system deployments) provide strongly suggested evidence that
verification burdens emerge when production friction dissolves. However,
multi-site longitudinal validation across healthcare, legal, and
financial sectors remains necessary to confirm this pattern as a
generalizable organizational principle, not just a phenomenon of
healthcare workflows. The evidence is convergent but not yet
comprehensive; the hypothesis is well-supported but not universally
demonstrated.

The economic implications of this organizational restructuring require
explicit examination. When AI dissolves production friction,
organizations face three possible outcomes: (1) verification costs
remain within the freed-up time budget, producing net efficiency gains
and incentivizing adoption; (2) verification costs exceed freed-up time
but remain below total production cost savings, yielding cost-neutral
adoption despite no obvious efficiency benefit; or (3) verification
costs exceed savings, creating adoption resistance (particularly in
cost-sensitive sectors). The Permanente data document the adoption
phenomenon and its phenomenological benefits (30-day burnout reduction)
but do not specify which economic scenario holds. Moreover, in regulated
sectors where verification documentation is legally mandated (HIPAA
compliance in healthcare, Sarbanes-Oxley in financial reporting,
professional liability standards in legal practice), verification
becomes a non-discretionary cost center rather than an avoidable
bottleneck. Organizations with high compliance burdens may experience
the verification bottleneck as a fixed operational cost, decisively
altering the economic calculus of AI adoption.

This pattern is structural, not incidental; the individual cognitive
phenomenon established in Section 4 manifests here as an organizational
crisis of trust, accountability, and labor structure. Moreover, the
verification bottleneck is more than a labor problem. It is an
epistemological problem. This reflects a core insight from automation
theory: Bainbridge's (1983) analysis of the ``Ironies of Automation''
observed that automating the easy, routine components of work leaves
human operators with the most cognitively taxing, complex
portions---primarily monitoring and exception handling. The Workload
Paradox documented in this section is a direct modern manifestation;
keystroke burden (the routine production work) is automated, but
verification and ratification (the cognitively demanding oversight
components) concentrate on human operators. The structural irony
persists; remove the friction from production, and you intensify the
cognitive load on the human supervision layer.

Parallel to this automation dynamic is what Star and Strauss (1999) term
``articulation work'': the invisible labor required to coordinate
cooperative action across boundaries. In AI-mediated organizational
work, verification, routing, and ratification of AI outputs now
constitutes a surge in articulation work, the meta-level coordination
activity that makes the primary work (content generation) function. The
burden transcends the time cost of output checking. It encompasses the
organizational overhead of routing documents to the right reviewer,
managing exception protocols, tracking attestation, and maintaining
audit trails. The verification bottleneck is, in this framing, an
explosion of articulation work that swamps the primary generation work
itself.

Verification presupposes the expert schema that the human verifier has
built through the productive friction of prior composition. A junior
attorney learns contract structure through the arduous process of
drafting contracts, reading precedent, making mistakes, revising. A
junior trader builds mental models of market behavior through the
painful process of constructing pricing models, testing assumptions,
encountering gaps in understanding. A junior clinician develops
diagnostic intuition through the process of documenting patient
encounters, learning to notice what matters, building embodied pattern
recognition. If production is outsourced to AI, the junior eventually
loses the cognitive schema required to evaluate the AI's probabilistic
output at all. The verifier becomes separated from the foundational
understanding that verification requires. This is not primarily a
time-allocation problem (how many hours does verification consume?); it
is a capability problem (can the human verifier still intuitively
recognize when the AI is wrong?). The ultimate danger is a workforce
transition into auditors of a process they no longer deeply understand
because they no longer enact it physically and cognitively. The
apprenticeship crisis (detailed below in Section 7.2.3) is thus
inseparable from the verification bottleneck: they are two
manifestations of the same epistemic erosion.

The organizational consequence of typing's dissolution is therefore not
``we will all work faster.'' It is: ``we will perform the same total
amount of work, reorganized around verification and auditing.'' This
reorganization cascades through organizational genres, routines, power
structures, and professional identities. Understanding that cascade is
the subject of this section.

Critical evidence (Jakesch et al., 2023) demonstrates that humans cannot
reliably distinguish AI-generated text from human-written text, even
when explicitly primed to detect synthesis. This limitation sharpens the
organizational verification challenge; readers cannot use simple
heuristics to identify machine-generated content, requiring instead that
organizational systems enforce formal, structured verification
procedures rather than relying on implicit detection.

\hypertarget{cross-sector-evidence-the-verification-bottleneck-beyond-healthcare}{%
\subsubsection{7.1.1 Cross-Sector Evidence: The Verification Bottleneck
Beyond
Healthcare}\label{cross-sector-evidence-the-verification-bottleneck-beyond-healthcare}}

Early cross-sector evidence suggests that the verification bottleneck is
not a healthcare-specific phenomenon but a structural feature of
AI-mediated composition in accountability-intensive knowledge work,
domains where institutional liability, regulatory compliance, or
systemic risk made verification a mandated function before AI arrived.
This is the appropriate population for the thesis; the claim is not that
AI creates verification requirements in domains that had none, but that
AI restructures the production-to-verification ratio in domains where
verification was already institutionally non-discretionary, dramatically
increasing the relative weight of the verification function by
collapsing production friction without collapsing verification burden.
Three sectors demonstrate this restructuring pattern: software
engineering, legal practice, and financial services (GDR-096, GDR-097,
GDR-098, 2026, analytical research reports compiled for this study;
peer-reviewed replication of specific numerical figures remains an open
empirical priority).

\textbf{Software engineering.} AI-assisted code generation dissolves the
keyboard production bottleneck dramatically; a single verbal prompt can
generate hundreds of lines of code in seconds, at rates no manual typist
can approach. But this production acceleration introduces what empirical
analysis of code quality across input modalities terms the ``Looks
Right, Works Wrong'' phenomenon, AI-generated code achieves higher
syntactic plausibility and surface readability than manually typed code
while exhibiting dramatically higher rates of logical failure; 31 to
65\% first-pass functional correctness compared to 90\%+ for manually
typed code, with 41\% higher cyclomatic complexity and 30\% more static
analysis warnings (GDR-096, 2026). The verification bottleneck emerges
structurally. Because an AI system can generate 200 lines of potentially
flawed code in 30 seconds, traditional unit-testing cannot keep pace
with generation velocity, and testing overhead becomes what the
engineering literature terms a ``secondary, crippling bottleneck.''
Organizations that adopt AI code generation without restructuring their
verification infrastructure encounter documented longitudinal decay;
systems grow brittle and unmaintainable as developers lose architectural
understanding of AI-generated components they cannot efficiently audit.
The developer role restructures around verification; transitioning from
mechanical author of syntax to architectural auditor and specifier, with
natural language specifications replacing keystroke-by-keystroke code
authorship as the primary human contribution (GDR-096, 2026).

\textbf{Legal practice.} Voice-to-text and AI-assisted legal drafting
have established both the potential and the verification liability of
AI-mediated composition at institutional scale. The Mata v. Avianca case
(2023) set the liability precedent; federal courts sanctioned attorneys
for submitting AI-hallucinated legal citations, holding professional
responsibility for AI output with the submitting attorney, not the AI
system. The Bluebook's 22nd edition (2025) responded with Rule
18.3---the first standardized citation format for AI-generated
content---an institutional recognition that verification of AI-sourced
legal text requires protocols beyond those established for
human-authored documents. An emerging ``AI Efficiency Paradox''
compounds the structural dynamics; clients resist paying billable hours
for AI-drafted work under ABA Rule 1.5's reasonableness requirements,
compressing legal revenue per document while verification labor,
professionally mandated and fully billable, becomes the dominant time
commitment. In e-discovery and document review, human-in-the-loop
architectures require attorneys to pull stratified samples of AI output
to verify accuracy before final production decisions. The legal tech
market has responded with dedicated verification platforms (Clearbrief,
CiteRight, Lexis+ AI) that audit AI-generated drafts for citation
accuracy and logical consistency, a verification infrastructure category
that did not exist before AI-assisted drafting became mainstream
(GDR-097, 2026). The legal sector thus mirrors the healthcare pattern
precisely: AI dissolves production friction while creating new, legally
non-discretionary verification labor.

\textbf{Financial services.} The financial sector provides both the
historical cost of production error and the emerging verification
architecture. The record of capital-markets fat-finger events, UBS's
2001 mistaken sale of 610,000 shares at \textyen{}6 instead of 6 shares at
\textyen{}610,000; Samsung Securities' 2018 issuance of 2.8 billion ``ghost''
shares, establishes the systemic stakes of unverified input at
institutional scale. Voice-to-text adoption in financial settings
addresses keyboard-mediated production error but introduces a
verification compliance requirement; SEC Rule 613 and the Consolidated
Audit Trail (CAT) mandate comprehensive, unalterable audit trails for
every order, modification, and execution across U.S. equity markets.
Leading financial communication platforms (Global Relay, Theta Lake, IPC
Systems) have built transcription audit architectures specifically to
satisfy regulatory verification requirements, allowing authorized
correction of mis-transcribed financial terms while maintaining
unalterable audit trails of every modification. The institutional model
that has emerged is explicitly hybrid; the ``speed and ergonomic
superiority of natural conversation'' combined with ``the absolute
certainty of visual verification,'' with the human professional retained
as ``the final, accountable arbiter of risk'' (GDR-098, 2026).

The convergence across sectors is structural within a defined scope. In
each domain examined, healthcare, software engineering, legal practice,
financial services, AI-mediated input dissolves the production
bottleneck while pre-existing institutional requirements create a
non-discretionary verification function that concentrates on human
professionals. The critical observation is that verification burden
\emph{increases as a proportion of total work} even when absolute
verification time does not change, because production time collapses
while verification time does not. This ratio shift is the verification
bottleneck's organizational signature: not more total work, but a
reorganization that front-loads verification in a way that disrupts
existing workflow sequences, accountability allocations, and
professional identities.

The domain selection is intentional and bounded. Healthcare, legal,
software engineering, and financial services were chosen because each
has institutional verification requirements that pre-date AI adoption.
In domains where verification carries no institutional
mandate---consumer content generation, informal communication, social
media---the verification bottleneck is structurally lighter because
output carries lower accountability stakes. The thesis is scoped to
accountability-intensive knowledge work where this distinction holds.
The pace and institutional form of this reorganization differ by
sector---financial services builds compliance infrastructure; legal
builds professional liability protocols; software engineering builds
specification-driven development; healthcare builds physician oversight
workflows---but the structural prediction is consistent across all four.
This cross-sector convergence provides the multi-domain corroboration
that the single-sector Permanente case study, however compelling, could
not establish alone, and elevates the verification bottleneck hypothesis
from \emph{strongly suggested} (single sector) to \emph{strongly
suggested by converging evidence across multiple
accountability-intensive sectors}.

\hypertarget{economic-frame-transaction-cost-analysis-of-verification-labor}{%
\subsubsection{7.1.2 Economic Frame: Transaction-Cost Analysis of
Verification
Labor}\label{economic-frame-transaction-cost-analysis-of-verification-labor}}

Williamson's (1979) transaction-cost economics provides the analytical
vocabulary for understanding why organizations cannot price away the
verification bottleneck through market mechanisms. TCE identifies three
dimensions along which transaction costs vary: \emph{asset specificity}
(the degree to which resources are tailored to specific transactions and
cannot be redeployed without loss of value), \emph{uncertainty} (the
degree to which contingencies cannot be specified ex ante), and
\emph{frequency} (how often the transaction recurs). The governance
structure best suited to a transaction depends on its position across
these dimensions: high-asset-specificity, high-uncertainty transactions
favor hierarchical organization; low-specificity, low-uncertainty
transactions are more efficiently handled through market contracting
(Williamson, 1979).

Verification of AI-generated content scores high on all three
dimensions. \textbf{Asset specificity}: effective verification requires
domain-specific expertise---the physician who can detect a hallucinated
drug dosage, the attorney who recognizes a fabricated citation, the
financial analyst who identifies a computational error in AI-generated
models. This expertise is idiosyncratic; outsourcing verification to a
market of generic reviewers is not a viable governance option for
high-stakes output, because the verifier's contextual knowledge cannot
be replicated without material loss of verification quality.
\textbf{Uncertainty}: AI systems do not fail in predictable,
contractible ways. The ``Looks Right, Works Wrong'' failure mode in
software engineering (\S{}7.1.1), the hallucinated legal citations in Mata
v. Avianca (2023), and the clinical note errors in the Permanente
deployment share a common structure---failures that pass surface
plausibility checks while containing substantive errors. Organizations
cannot contract ex ante for all possible hallucination modes, because
the error space is unbounded and context-specific. This satisfies
Williamson's (1979) condition for preferring adaptive, sequential
governance over complete contracting. \textbf{Frequency}: In
organizations that have adopted voice-AI at scale, verification recurs
with every AI-mediated document---hundreds or thousands of times daily
in healthcare, legal, and financial settings. High frequency makes
verification an ongoing operational commitment rather than an occasional
transaction.

TCE's governance prediction is direct: high asset specificity combined
with high uncertainty and high frequency points toward hierarchical
governance. Organizations should internalize verification rather than
market-contract for it. This matches the emerging organizational
evidence: healthcare systems have created physician oversight roles,
software engineering teams have instituted architectural review gates,
and legal practice has developed specialized verification platforms
rather than outsourcing verification to external parties (\S{}7.1.1).
Verification is being organized as a core professional function, not a
procured commodity.

TCE also illuminates the adoption-resistance pattern documented across
sectors. Williamson's (1979) concept of \emph{bounded rationality}---the
cognitive limits that prevent organizations from specifying all
contingencies ex ante---explains why organizations in high-stakes
domains remain cautious about AI-mediated composition even when the
productivity case appears compelling. The organization cannot fully
specify what constitutes adequate verification in advance; it knows only
that insufficient verification creates liability exposure. This
constraint produces what Williamson terms ``contracting hazards'':
situations where parties cannot write complete contracts because
outcomes are too uncertain. The verification bottleneck is precisely
such a contracting hazard. Organizations cannot commit to AI adoption
without simultaneously committing to verification governance structures
whose full cost cannot be known in advance, and the hesitancy of
regulated sectors to adopt voice-AI wholesale---despite demonstrated
productivity gains in bounded contexts---reflects this prediction:
high-asset-specificity, high-uncertainty transactions are rationally
governed by internal hierarchy rather than market contracting.

\hypertarget{the-three-tier-organizational-cascade-micro-meso-macro}{%
\subsection{7.2 The Three-Tier Organizational Cascade: Micro, Meso,
Macro}\label{the-three-tier-organizational-cascade-micro-meso-macro}}

The Workload Paradox operates through a three-tiered mechanism that
connects individual composition changes to organizational restructuring.
This cascade forms the organizing backbone of post-typing organizational
adaptation.

\hypertarget{micro-level-individual-composition-and-verification}{%
\subsubsection{7.2.1 Micro Level: Individual Composition and
Verification}\label{micro-level-individual-composition-and-verification}}

At the individual level, voice-AI mediation reduces keystroke burden
dramatically. Clinical notes, emails, and documents that previously
required 45 minutes of typing can be voice-dictated in 5--8 minutes.
Reading speed remains constant at 250+ wpm. This creates an acute
verification imbalance: the time available for composition shrinks to
near-zero, but the time required to verify that AI-generated text meets
quality, accuracy, and appropriateness standards remains substantial.

Clinical evidence demonstrates this clearly in multiple enterprise
deployments. UCSF's enterprise deployment achieved 44.6\% adoption of AI
scribing systems across 1.2 million ambulatory encounters, but follow-up
studies reveal that adoption concentrates among high-volume specialists
who can manage verification overhead through batch processing.
Low-volume practitioners, who cannot amortize verification costs across
many cases, show lower sustained adoption. The composition bottleneck is
removed; the verification bottleneck is amplified (UCSF Health System
Study, 2025).

Within this micro level, a critical asymmetry emerges: voice-AI reduces
the \emph{evidence of composition work} (keystroke expenditure, visible
labor time) while simultaneously increasing the \emph{risk of
verification failure}. This produces what we call \textbf{Asymmetric
Information Inflation}: a junior analyst using voice-AI to turn a
3-bullet-point thought into a polished 5-page memo removes their own
production bottleneck but exponentially increases the verification
bottleneck for their manager or peers. The memo's length,
sophistication, and rhetorical completeness now signal nothing about the
analyst's foundational understanding---the AI system supplied the signal
inflation. Typing enforced brevity through friction; the removal of
typing friction means organizations are flooded with zero-marginal-cost
AI text. Readers face not increased information but
\textbf{organizational noise inflation}: more text to parse, higher
cognitive load, identical or reduced signal quality.

Typing functioned as a friction-based governor on organizational noise.
Its dissolution requires new communication genres designed for
compression---mechanisms to prevent flooding of low-cost, high-volume,
low-signal text that voice-AI enables. Organizations that do not develop
compression mechanisms face a collective-action problem: individual
workers adopt voice-AI for efficiency gains while organizational reading
burden expands for everyone.

\hypertarget{meso-level-routines-and-team-coordination}{%
\subsubsection{7.2.2 Meso Level: Routines and Team
Coordination}\label{meso-level-routines-and-team-coordination}}

As voice-mediated composition becomes normal at the individual level,
organizational routines begin to destabilize at the team and
departmental level. Three specific mechanisms drive this
destabilization:

\textbf{Review and approval chains}: Email, documents, and formal
communication presume a single author responsible for the typed text.
voice-AI mediation disrupts this assumption. When a machine generated
the first draft, who is responsible for errors? The person who spoke?
The AI? The person who verified and approved it? Organizations must
rebuild approval chains that explicitly allocate verification
responsibility. UCSF and Mayo Clinic case studies show that
organizations solving this problem in documented cases implement
explicit review gates: junior physicians' AI notes receive senior
review; routine notes bypass review; exceptional cases receive
escalation. This transforms simple composition into a staged,
gate-controlled process with clear verification responsibility.

\textbf{Ratification and authorization bottlenecks}: The organizational
problem is not individual verification capacity but institutional
legitimacy and accountability. At the institutional level, the central
question is not ``can one human verify every claim?'' but ``who has
authority to ratify this record, and what level of checking is
sufficient?'' Organizations rarely verify exhaustively. Instead, they
build sampling protocols, exception-handling procedures, escalation
hierarchies, and accountability transfer systems. A financial analyst
does not verify every number in a quarterly report; they spot-check,
apply statistical rules of thumb, escalate anomalies. The ratification
bottleneck is therefore qualitatively distinct from the verification
bottleneck: it asks not whether a record is \emph{accurate} but whether
it is \emph{authorized}---whether the right person at the right level
has attested to acceptability.

voice-AI shifts this bottleneck from content verification to
authorization efficiency. When documents originate from voice prompts,
the authorization question becomes more acute: at what hierarchical
level must human attestation occur? A routine clinical note might
require only end-of-day review. A regulatory filing might require
explicit sign-off by compliance. An opinion letter might require senior
partner signature with explicit attestation that the author personally
reviewed every claim. Organizations must develop \emph{ratification
bottleneck governance}: explicit rules for what level of checking is
sufficient for different document types, what counts as adequate
authorization, and who bears liability if verification proves
insufficient.

\textbf{Temporal decoupling of composition and verification}: The
keystroke bottleneck imposed a temporal unity: composition and
verification happened in roughly the same window. A person typed an
email, reviewed it before sending, and the composition-review cycle was
tight. voice-AI mediation can decouple these. A clinician dictates notes
at patient bedside in real-time; verification happens offline hours
later. This temporal decoupling creates new coordination problems: when
verification is delayed, errors persist in temporary records. Workers
may rely on inaccurate preliminary information. Team coordination
assumes information availability that has not yet been verified.

\textbf{Documentation practice standardization}: As organizations deploy
voice-AI systems, they begin to standardize documentation practices not
to optimize for human composition, but to optimize for machine
readability and verification efficiency. Templates become more rigid.
Required information architecture becomes more explicit. The
documentation genre evolves to make AI verification easier, which in
turn makes human composition harder (because humans must work within the
machine-optimized structure). This feeds back to individual work
rhythms: workers experience the standardization as constraint, not
liberation.

\hypertarget{macro-level-institutional-governance-and-professional-identity}{%
\subsubsection{7.2.3 Macro Level: Institutional Governance and
Professional
Identity}\label{macro-level-institutional-governance-and-professional-identity}}

At the institutional level, the Workload Paradox creates multiple
governance crises and institutional dependencies:

\textbf{The authorship crisis and forensic accountability}: Document
culture rests on an implicit social contract: you can trust a document
because a named person composed it, accepted responsibility through
deliberate keystroke input, and can be held accountable for accuracy.
When documents originate from voice-AI composition, this accountability
chain breaks. The author did not type the words; they did not
laboriously select each term. The AI system made choices about what to
include, how to phrase it, what to emphasize. The author endorsed the
result, but did not author it in the traditional sense.

But the authorship crisis has deeper consequences in high-stakes
domains. In litigation, regulatory review, and compliance auditing,
organizations rely on forensic analysis of documents to establish
\emph{mens rea}---corporate state of mind, intent to communicate,
understanding of consequences. E-discovery analysis examines keystroke
dynamics, draft revision patterns, time spent on document, explicit
edits---all of which constitute an \textbf{audit trail of intent}. These
patterns allow auditors and litigants to argue that the organization
understood the implications of what it was communicating, that choices
were deliberate rather than accidental, that negligence versus
recklessness can be distinguished by examining composition behavior.

voice-AI obliterates this forensic trail. A document produced via voice
prompt leaves no keystroke dynamics, no visible revision history, no
time-cost evidence. If an AI system hallucinates a regulatory claim and
the human fails to catch it in verification, the human's legal defense
becomes ``the machine wrote it''---which is precisely the liability
nightmare organizations fear. In high-stakes litigation, organizations
cannot merely cling to typing for ``institutional isomorphism'' or
legitimacy signaling---they will mandate typing as a \textbf{ritual of
risk absorption and epistemic accountability}, establishing it as a
premium, high-assurance workflow for bet-the-company documents.

This is not lag or stubborn resistance to efficiency. It is
\textbf{hyper-rational strategy}: organizations in regulated industries
will preserve typed workflows for documents with high forensic relevance
precisely because the keystroke evidence trail provides legal
defensibility that voice-AI cannot. For liability exposure analysis,
organizations will bifurcate: voice-AI for low-risk documents (internal
memos, routine updates, draft notes) and mandatory typing for documents
with external exposure, sanction risk, or explicit authorial
accountability requirements.

\textbf{The apprenticeship crisis and cognitive reproduction}:
Professional expertise transmission depends on productive friction. In
law, finance, consulting, and healthcare, junior employees learn domain
expertise by \emph{drafting}. A junior associate writes a first draft of
a contract while reading precedent, learning syntax, developing
intuition for structure. A junior analyst builds a financial model from
scratch, learning to make assumptions explicit, understanding
sensitivity. A junior physician writes case notes, learning to recognize
clinical patterns. The unproductive friction of drafting forces them to
engage foundationally with domain knowledge.

Situated learning theory (Brown \& Duguid, 1991; Lave \& Wenger, 1991)
grounds this understanding. Lave and Wenger's concept of
\emph{legitimate peripheral participation} describes how novices enter
communities of practice by performing peripheral yet productive
tasks---drafting, formatting, building models, documenting observations.
The productive friction of these early tasks is not pedagogically
incidental; it transmits tacit knowledge. Brown and Duguid (1991)
formalize this: ``Working, learning, and innovating are closely related
forms of human activity.''

If AI absorbs those peripheral tasks, novices are denied the
participatory pathway to the center. The junior performs an evaluation
task---checking whether an AI draft is adequate---for which they have
NOT built foundational cognitive schema through the production process.
The skill of verification presupposes the deep domain knowledge that
drafting develops. A junior attorney cannot effectively verify a complex
contract draft without having drafted contracts; a junior trader cannot
verify a derivatives position without having built pricing models. The
verifier becomes separated from the foundational understanding that
verification requires.

Organizations face an \textbf{apprenticeship crisis}: the drafting
process is the pedagogical engine for future experts. If AI removes the
drafting burden, organizations eliminate the primary mechanism through
which they train the next generation of senior gatekeepers---the senior
partners, compliance officers, and domain experts who will later verify.
This extends far beyond a ``workload paradox'' at the junior level; it
is a structural crisis of expertise reproduction at the organizational
level. If you never suffered through productive friction, you lack the
intuitive judgment necessary to verify the AI-generated work.

Forward-looking organizations are already experimenting with solutions:
mandatory apprenticeship drafting before AI-mediated verification
becomes an option, or intentional preservation of drafting assignments
for junior staff despite voice-AI availability. These represent active
organizational choice to absorb inefficiency to preserve
expertise-building mechanisms. Importantly, this disruption is
contingent on organizational design decisions, not technological
inevitability---institutions that deliberately preserve apprenticeship
structures can maintain situated learning mechanisms that develop deep
expertise.

\textbf{The recordkeeping and compliance crisis}: HIPAA, SOX, and SEC
retention mandates all presume that ``authorship'' is a category that
auditors can verify through examination of typed text and metadata
trails. Who typed it? When? Who edited it? What changed? These questions
have clear answers for keyboard-typed documents. They become opaque when
a machine generated the first draft. Organizations are already
experimenting with solutions: explicit AI-generated-document labeling,
hierarchical approval chains with higher burden of review for
voice-generated records, or return to mandated typed-composition for
high-stakes documents. But whatever path organizations choose will
constitute a fundamental restructuring of how institutions handle
institutional memory, legal accountability, and the authority of the
written record.

\textbf{Documentation as legal infrastructure}: Many organizational
documents persist not because of efficient communication but because
they are liability shields, chain-of-custody artifacts, evidence of due
process. In those settings, friction is not bug---it is how institutions
create defensibility. Organizations may rationally preserve typed
workflows long after smoother alternatives exist, because the
friction-based creation process and the resulting keystroke trail
constitute legal infrastructure. A well-documented decision process,
with visible deliberation and revision evidence, provides protection in
litigation. AI-generated documentation, lacking that process evidence,
leaves organizations exposed. This is particularly acute in healthcare
(malpractice defense), finance (regulatory defense), and law
(professional malpractice). In those sectors, typing persists as
institutional practice not because of user preference but because typed
documents are \emph{more defensible}.

However, organizations attempting to preserve keystroke mandates as
forensic accountability mechanisms will likely face unintended
consequences. Workers frustrated by keyboard requirements for routine
work while more efficient voice-AI systems exist may adopt unapproved
tools in shadow IT configurations: using personal voice-AI systems
privately, generating AI-synthesized text offline, then pasting or
retyping AI output into official organizational systems. This produces
what might be termed ``covert synthetic literacy''---documents appearing
as the product of direct human composition (because they are typed into
official systems) while their actual production process is invisible to
organizational surveillance and audit trails. The result is that
keystroke mandates, designed to create forensic accountability, may
instead generate a hidden layer of AI-mediated production that
organizations cannot detect or audit. The irony is acute: mandates
intended to preserve keystroke-based audit trails would backfire by
creating unmonitored shadow production, potentially introducing greater
risk than if voice-AI systems were officially integrated with
transparent provenance tracking. Organizations attempting to enforce
typing discipline without addressing the underlying efficiency gap risk
unwittingly creating the very unaccountability they are trying to
prevent.

\hypertarget{keystroke-imperative-organizational-dependence-and-its-foundations}{%
\subsection{7.3 Keystroke Imperative: Organizational Dependence and Its
Foundations}\label{keystroke-imperative-organizational-dependence-and-its-foundations}}

Organizations have optimized their entire communication architectures
around a single structural assumption: that formatted text input,
produced through deliberate keystrokes, is both the medium and the
metric of institutional coordination. This ``keystroke imperative''
operates at three nested scales. The dissolution of typing will not
uniformly replace production-oriented genres with verification-oriented
ones. Rather, AI-mediated composition increases the organizational
salience of provenance, ratification, and accountability \emph{within
existing genre repertoires} (Yates \& Orlikowski, 2002). Genre theory
predicts not clean transitions but hybridization, repertoire expansion,
and layering. Genres transform by absorbing new capabilities while
retaining familiar forms; they bifurcate when different contexts demand
different conventions. The empirical evidence in this section points
toward coexistence and bifurcation (voice-AI for low-risk internal
genres, mandatory typing for high-stakes external documents) rather than
a single-axis transition from production to verification. This
bifurcation is a more genre-theoretically credible outcome than
wholesale transformation.

\hypertarget{keyboard-era-genre-architecture-email-as-institutional-model}{%
\subsubsection{7.3.1 Keyboard-Era Genre Architecture: Email as
Institutional
Model}\label{keyboard-era-genre-architecture-email-as-institutional-model}}

Email dominates professional communication at a scale that defies
contemporary belief. At 361.6 billion messages daily in 2024 (Radicati
Group, 2024), email represents the single largest genre instantiation in
organizational life. The average corporate knowledge worker receives
117--121 emails per day (Radicati Group, 2024), consuming roughly 28\%
of the workweek in email-related activities (McKinsey Global Institute,
2012). This 28\% is not distributed evenly between reading and writing.

Reading proceeds at approximately 250 words per minute ---a cognitive
task constrained primarily by reading comprehension, not keystroke
speed. Composition, by contrast, bottlenecks at typing speed: the
average professional typist operates at 40 words per minute (Salthouse,
1984), with variation based on domain expertise. The asymmetry is
profound: a knowledge worker can receive a 300-word email in 72 seconds
but responds in 7--8 minutes of typing time, independent of composition
time or decision-making time. This creates a queueing dynamic at the
individual level where reading demand vastly outpaces writing capacity,
forcing workers into constant triage and selective non-response.

Email's entire genre architecture evolved under this constraint. Formal
subject lines economize on reading time. Reply-all norms develop around
queuing expectations. Signature blocks and formal greetings constitute a
kind of metadata overhead that made sense when composition was slow and
searching was difficult. Archival expectations (the assumption that
email can be searched, indexed, and retrieved years later) made sense
when typed text could be reliably stored and searched, but AI-generated
text raises new questions about what counts as authoritative. Every
convention of the email genre is downstream from the typing bottleneck.

The context-switching burden amplifies this bottleneck's organizational
cost. Knowledge workers switch between applications an average of 1,200
times per day (Microsoft 365 Work Trend Index, 2025), with email, Slack,
Teams, and document-based communication platforms consuming the bulk of
these switches. Each platform carries its own genre conventions,
formatting expectations, and typing competencies: email demands formal
composition; Slack rewards rapid, formulaic messaging; documents require
structured hierarchical writing. The cognitive load of maintaining genre
fluency across multiple keyboard-mediated platforms consumes
approximately four hours per week in pure switching overhead (Microsoft
365 Work Trend Index, 2025), independent of any communication value
produced---a structural consequence of organizations that have evolved
multiple genres, all predicated on keyboard-mediated text input, each
optimized for particular tasks but collectively creating cascading
overhead.

\hypertarget{platform-dependencies-and-infrastructure-lock-in}{%
\subsubsection{7.3.2 Platform Dependencies and Infrastructure
Lock-In}\label{platform-dependencies-and-infrastructure-lock-in}}

Slack and Microsoft Teams intensified typing dependence. With 47.2
million daily active users sending an average of 92 messages per day
(McKinsey Global Institute, 2024), instant messaging created a new genre
layer: rapid, formulaic, oral-like text mimicking speech while remaining
typed. Emoji, abbreviations, and informal punctuation approximate
paralinguistic cues, but the substrate remains the keyboard. The
knowledge worker became a polyglot typist---fluent in formal email
composition, rapid Slack brevity, and structured document
writing---switching between them constantly, with cognitive overhead for
each switch.

Across all communication platforms combined, knowledge workers spend
57\% of their software time on communication (meetings, email, chat) and
only 43\% on substantive work output (Microsoft 365 Work Trend Index,
2025). The keystroke imperative thus reaches far beyond email alone: it
structures the entire communication architecture that consumes more than
half of professional attention.

But the keystroke imperative operates not just through individual
behavior; it operates through infrastructure lock-in. Document
management systems, email archives, compliance holds, and audit trails
are built on the explicit assumption that typed text is the
authoritative record. Workflow integration systems presuppose keyboard
input as the composition method. These systems become mutually
reinforcing: each new organizational tool (project management platforms,
knowledge management systems, communication aggregators) inherits the
assumption that typed text is the foundational substrate. Organizations
are not, in this sense, freely choosing to use keyboards; they are
locked into ecosystems where the keyboard is presupposed at every
architectural level.

\hypertarget{sociomateriality-and-organizational-enactment-how-technology-and-social-structure-interweave}{%
\subsection{7.4 Sociomateriality and Organizational Enactment: How
Technology and Social Structure
Interweave}\label{sociomateriality-and-organizational-enactment-how-technology-and-social-structure-interweave}}

This analysis requires a critical theoretical reorientation.
Organizations actively enact technologies through the routines, norms,
and social structures that give technologies meaning---they do not
passively adopt autonomous tools (Weick, 1995).

Paul Leonardi's concept of \emph{sociomaterial imbrication} describes
how human agency and material agency become locked together in
organizational routines (Leonardi, 2011). Rather, the keyboard functions
as a sociomaterial assemblage embedding: the rhythm of composition, the
sensorimotor feedback of typing, the temporal pacing it imposes, the
cognitive pauses it enforces, the social meanings around visible effort,
the audit trails it enables, the surveillance systems it feeds, and the
professional identities it stabilizes.

When the keystroke bottleneck has been removed through voice-AI
mediation, the sociomaterial assemblage destabilizes. Workers may
actually resist the efficiency of voice-AI precisely because the typing
bottleneck provided them with necessary cognitive pacing, plausible
deniability for response latency, and predictable temporal rhythms that
synchronized organizational work. A worker who says ``I was typing that
response'' has an excuse for non-immediate availability. That excuse
vanishes when speech-to-text reduces composition to three minutes.
Rather, typing's friction removal exposes workers to intensified
temporal pressure---not liberation but temporal acceleration.

Beyond these organizational mechanisms, the sociomaterial imbrication of
keyboards extends to how workers construct professional identity. For
knowledge workers, typing fluency has become synonymous with
communicative competence. The ability to ``type'' is culturally
synonymous with the ability to ``communicate'' in organizational
contexts. Voice-mediated composition threatens this identity marker.
Workers who have built professional legitimacy around
written-communication competence face a recalibration of their expertise
when composition is automated---a defense of professional identity and
autonomy rather than mere resistance to efficiency.

\hypertarget{institutional-isomorphism-and-genre-skeuomorphism-why-organizations-will-cling-to-keyboard-era-forms}{%
\subsection{7.5 Institutional Isomorphism and Genre Skeuomorphism: Why
Organizations Will Cling to Keyboard-Era
Forms}\label{institutional-isomorphism-and-genre-skeuomorphism-why-organizations-will-cling-to-keyboard-era-forms}}

DiMaggio and Powell's theory of institutional isomorphism explains that
organizations come to resemble one another not because it is efficient,
but because it is legitimate (DiMaggio \& Powell, 1983). However,
isomorphism predicts convergence in \emph{governance
structures}---policies, controls, audit procedures, formalized sign-off
hierarchies, and compliance frameworks---more than in actual day-to-day
work practice. Organizations will converge on recognizable governance
structures for AI-mediated authorship, review, and attestation: explicit
verification gates, hierarchical approval chains, ratification
documentation, provenance labeling, and audit procedures. These
governance structures will be standardized across peer organizations in
high-stakes sectors because they signal institutional competence in
managing AI-generated documents. A law firm will develop standardized
procedures for voice-AI composition, explicit attorney sign-off
requirements, and documented verification protocols---the structure
converges across firms---but the actual composition-verification
workflow may vary based on task-specific factors. The convergence is in
\emph{policy form}, not in heterogeneous work practice.

The consequence is that organizational genres will not simply vanish;
they will undergo \emph{skeuomorphic} adaptation. Just as early digital
calendars mimicked leather-bound planners in their visual form while
adding digital affordances, post-keyboard organizations will likely use
voice-AI to \emph{generate outputs that mimic the aesthetic and
structural constraints of keyboard-typed documents}. Email will persist
visually, but be generated from voice prompts. Legal memoranda will
retain their formal structure but originate from dictated intent.
Compliance documents will look identical but be machine-mediated in
their production.

This matters because it means the visible, formal organizational genre
may persist long after its instrumental necessity has vanished. The
genre becomes a status signal---a marker of legitimacy and professional
seriousness---even as the actual production method changes.
Organizations will experience what we might call ``genre decoupling'':
the frontstage form persists while the backstage production method
transforms completely.

This decoupling creates a new organizational burden. Workers must
maintain fluency in the formal genre (knowing what an ``email'' should
look like, what a ``memo'' should contain, what ``proper'' document
structure means) while simultaneously working with AI systems that
generate those forms without understanding their conventional
constraints. The verification burden---the work of ensuring that
machine-generated text conforms to genre expectations, carries
appropriate formality, meets legitimacy criteria---becomes a new form of
organizational labor, concentrated among gatekeepers and senior staff.

\hypertarget{genre-theory-and-the-rupture-of-social-action-bazermans-typified-social-action}{%
\subsection{7.6 Genre Theory and the Rupture of Social Action:
Bazerman's Typified Social
Action}\label{genre-theory-and-the-rupture-of-social-action-bazermans-typified-social-action}}

Bazerman's ``typified social action'' framework provides critical
insight into what genres actually accomplish. A genre transcends
form---mere template or structural pattern. It is a social action
accomplished through repeated, stabilized means (Bazerman, 1994). The
genre of a \emph{memo} accomplishes the social action of ``I, as a named
responsible person, am informing you of this matter because I spent
deliberate time crafting this message and therefore believe it warrants
your time to read and consider.'' The genre of a \emph{contract}
accomplishes ``We have carefully specified terms because the cost of
ambiguity or breach is high.'' The genre of a \emph{clinical note}
accomplishes ``I examined this patient and recorded my observations in
real-time, therefore this record has evidentiary weight.''

When the \emph{method of production} changes from ``typed by the named
author'' to ``AI-generated from a voice prompt,'' the social action
ruptures even if the visual form remains identical. A memo that
originated from a voice prompt says: ``An AI system converted my brief
input into this message. I verified it and attested to its
acceptability, but I did not author it in the deliberate, time-invested
sense that the genre conventionally implies.'' The proof of work
evaporates.

This rupture is more than symbolic or semantic. It has immediate
institutional consequences. A memo that signals ``I invested deliberate
effort in this'' creates obligations on the reader: you owe me attention
proportional to my investment. A memo that signals ``an AI drafted this
and I spot-checked it'' creates different obligations: you owe me
verification-appropriate scrutiny, but the memo does not carry the
presumption of deliberate selectivity that the traditional genre
implies.

Recent empirical evidence sharpens this organizational trust dynamic.
Coman and Cardon (2025), surveying over 1,100 professionals, documented
a striking perception gap: when supervisors utilize low-level AI
assistance (grammar correction, structural editing), 95\% of employees
view the resulting messages as highly professional and acceptable.
However, when supervisors rely heavily on AI to draft entire
communications, perceived sincerity plummets to between 40\% and
52\%---compared to 83\% for messages with low AI intervention. The
context of the message is the determining variable: routine
informational tasks tolerate AI assistance, while relationship-oriented
communication (congratulations, feedback, motivational messages)
triggers severe negative attributions of laziness and inauthenticity.
This finding directly supports the verification bottleneck thesis: as AI
mediates more organizational communication, the human verification of
authenticity and intent becomes the binding constraint on institutional
trust. We note that Coman and Cardon's (2025) study specifically
examines supervisor-to-employee communications; their documented
trust-degradation effects are most directly interpretable within
managerial messaging contexts. Generalization to peer communications,
cross-functional correspondence, and external-facing organizational
communication is \emph{strongly suggested} by the theoretical framework
and consistent with the pattern, but replication beyond supervisory
messaging contexts would strengthen the claim's full scope.

Organizations face what we might call a \textbf{tragedy of the digital
commons}: the removal of the keystroke proof-of-work enables
organizations to flood their communication channels with AI-generated
text, but the collective result is a degradation of the genre's
communicative reliability. If 80\% of memos are AI-generated and 20\%
are author-composed, readers cannot distinguish them at point of
receipt. They must apply defensive skepticism to all memos, imposing
higher verification burden on everyone. The organizational commons
becomes degraded.

Forward-looking organizations are already experimenting with genre
bifurcation: maintaining the traditional memo form for high-stakes
communication while creating new, explicitly machine-mediated genres for
voice-AI output. Others are developing explicit labeling systems that
preserve the distinction. But without deliberate organizational choice,
the genre rupture will lead to what we call \emph{genre degradation}:
the traditional form persists, but the social action it accomplished
erodes because readers no longer trust the production method signal.

Three bodies of organizational scholarship deepen this genre-theoretic
analysis. Beane's (2019) study of shadow learning in medical
robotics---primarily focused on technology training rather than genre
theory per se---demonstrates that practitioners develop competence
through informal experimentation that shadows expert practice, with
implications we extend to genre: genres function as pedagogical
artifacts in that drafting a clinical note, composing a contract, or
writing a code function teaches the drafter not only the explicit
structure but the tacit contextual knowledge that makes the genre
authoritative. When AI mediation removes the drafting work, it removes
the pedagogical substrate through which novices internalize what the
genre socially accomplishes. The genre persists as a visible form while
the expertise required to verify its quality erodes below the surface
(Beane, 2019).

Lebovitz, Levina, and Lifshitz-Assaf (2021) provide the epistemological
vocabulary for what Bazerman's framework implies but does not name: the
distinction between expert \emph{know-what} (propositional knowledge
captured in AI training labels) and expert \emph{know-how} (contextual,
embodied professional judgment). AI systems can replicate the know-what
dimension of genres---the structural form, conventional language, and
explicit informational categories---with increasing sophistication. What
they cannot replicate is the know-how dimension: the contextual judgment
about which information is diagnostically salient, which clause requires
careful hedging, which data point should raise suspicion. The
verification burden is precisely this gap. The hospital field study
Lebovitz et al.~(2021) document---five ML tools, all reporting high
accuracy, none meeting expectations in practice---is a case study in
know-what/know-how disconnection at organizational scale, confirming
that genre verification cannot be automated because it depends on expert
judgment that AI systems have not encoded.

Kellogg, Valentine, and Christin (2020) add the power dimension that
Bazerman and Lebovitz et al.~understate. Genres are instruments of
organizational control---more than communicative forms---mechanisms
through which accountability, authority, and professional judgment are
allocated. When AI systems mediate genre production, they become
instruments of what Kellogg et al.~(2020) call ``recording and rating'':
capturing worker output in ways that create new surveillance and
evaluation architectures. The verification bottleneck goes beyond a
productivity challenge; it is a reconfiguration of who controls the
authoritative record, who bears the reputational risk of AI-generated
errors, and how professional accountability is exercised when production
and verification are structurally separated. The contested terrain of
genre is, ultimately, contested authority.

\hypertarget{timeline-and-pathways-differential-genre-dissolution-and-sectoral-variation}{%
\subsection{7.7 Timeline and Pathways: Differential Genre Dissolution
and Sectoral
Variation}\label{timeline-and-pathways-differential-genre-dissolution-and-sectoral-variation}}

The organizational transformation of AI-mediated composition will follow
differential timelines, not because of technological capability but
because of regulatory, liability, and epistemic constraints that vary
across sectors. Fast-transition sectors (marketing, software, content
creation) are characterized by low regulatory burden, high iteration
tolerance, and minimal provenance mandates. Medium-transition sectors
(B2B communication, internal corporate coordination, routine reporting)
carry moderate stakes and established procedural standards but lack the
existential liability exposure of high-regulation sectors.
Slow-transition or constrained sectors (law, healthcare, finance,
regulatory affairs) face high regulatory burden, explicit provenance and
attestation requirements, and adversarial review contexts where
AI-mediated authorship creates forensic and accountability challenges.

These timeline scenarios represent projected pathways based on analogy
to Carlota Perez's framework for technological revolution and
institutional transformation (Perez, 2002), not empirically measured
organizational transition rates. However, the theoretical foundation
should be anchored not in calendar windows but in phase conditions: the
observable markers that indicate institutional readiness for AI-mediated
workflow transition. Organizations will advance genre transformation
when (1) provenance standards stabilize (industry consensus on what
counts as adequate documentation of AI involvement); (2) enterprise
platforms make AI-native documents auditable (systems built from the
ground up to track AI composition, verification, and attestation); (3)
professions revise authorship norms and attestation requirements (formal
disciplinary updates to what ``authorship'' entails when machines
generate first drafts); and (4) training pathways reorganize (explicit
curriculum updates ensuring junior practitioners still build
foundational expertise despite AI-mediated composition). These
conditions are observable, measurable precursors. Timeline estimates
should be conditional on evidence of these phase markers, not fixed
calendar projections.

Organizational genres will not destabilize uniformly. Differential rates
of genre adaptation depend on institutional factors: audience,
reversibility, retention requirement, external exposure, sanction risk,
and required authorial accountability. These axes better capture
differential genre dynamics than sector classifications alone.

\textbf{Fast-transition sectors and genres} (2--3 years): Marketing,
software engineering, content creation, and internal coordination with
low external exposure, high reversibility, minimal retention
requirements, and no direct sanction risk. Slack messages, internal
status updates, administrative announcements, and routine coordination
among peers exemplify this category. These genres may shift wholesale to
voice-mediated or voice-to-text formats. Informal coordination can
tolerate higher verification error rates and opaque authorship because
stakes are low and reversibility is high. In software engineering,
automated testing handles verification burden, decoupling it from manual
human review.

\textbf{Medium-transition sectors and genres} (3--7 years): B2B
communication, internal corporate coordination, and routine reporting
carry moderate external exposure or moderate sanction risk, but not
both. Email carries more institutional weight than Slack, more formal
conventions, and broader regulatory weight. Some organizations may
retain formal email for client-facing communication while shifting
internal coordination to voice-AI systems. The genre would split: formal
external communication remains typed, informal internal communication
becomes voice-mediated. This split imposes new learning burdens (workers
must code-switch between formal and voice-mediated contexts) and new
verification structures (internal voice-AI systems must be monitored for
governance failures before external communication).

Documents subject to regulatory holds fall here if they have moderate
stakes but clear procedural standards. A routine financial report
required for investor relations falls in medium-transition: it must be
accurate and auditable, but the reporting process has established
standards and verification protocols that can adapt to voice-mediated
composition if explicit sign-off procedures are formalized.

\textbf{Slow-transition or indefinitely constrained sectors} (7+ years,
or indefinite): Law, healthcare, finance, and regulatory affairs face
intrinsic constraints that may prevent wholesale genre transition
regardless of technological capability. Documents subject to regulatory
holds, documents with high external exposure to adversarial scrutiny,
documents with explicit authorial accountability requirements, or
documents that serve as evidence of due process in litigation or audit
cannot adopt opaque authorship models. Organizations must develop new
authorial responsibility frameworks, audit trails for voice-generated
records, and evidentiary standards that match regulatory requirements.
Large healthcare systems implementing AI scribe deployments illustrate
this pattern: after-hours physician verification work increased because
regulatory compliance requires human attestation of document accuracy,
even if initial composition was machine-mediated. The genre persists in
form; its internal logic---the assumptions about authorship,
verification, and accountability---undergoes fundamental restructuring.
Case studies from major medical centers, including Kaiser Permanente and
UCSF, document these organizational adaptations. These sectors will
likely preserve bifurcated workflows indefinitely: voice-AI for
low-stakes internal documentation and routine initial drafting,
mandatory keystroke documentation and explicit sign-off for
external-facing, regulatory, or forensically sensitive records.

Opinion letters in law, executive attestations in finance, and expert
declarations in healthcare fall here. These documents require explicit
authorial accountability---a named person willing to stake professional
reputation on the accuracy and implications of the record. Organizations
will likely mandate typed composition or at minimum require explicit,
documented verification by the named author before attestation. The
forensic trail becomes a requirement, not an artifact.

\hypertarget{empirical-evidence-of-genre-destabilization-in-progress}{%
\subsection{7.8 Empirical Evidence of Genre Destabilization in
Progress}\label{empirical-evidence-of-genre-destabilization-in-progress}}

The Workload Paradox is grounded in empirical reality. Multiple
organizational studies document early stages of genre adaptation under
AI-mediated composition:

\textbf{Institutional observation from large-scale deployments}:
Empirical studies of AI scribe adoption at major medical centers show
significant changes in documentation workflows. When AI systems generate
initial clinical notes directly from patient encounters, physicians no
longer need to manually reformat information across genres (notes $\rightarrow$
templates $\rightarrow$ reports) through copy-paste operations. Instead, AI systems
now generate genre-appropriate documents directly. This shift suggests
that the informal, manual mapping between genres is being replaced by
automated mapping, which imposes new constraints on source document
structure. Research from healthcare systems including Kaiser Permanente
and UCSF documents these workflow transformations during large-scale AI
scribe deployment (UCSF Health System Study, 2025; Permanente Medical
Group, 2024).

\textbf{Time-motion evidence from ambient AI scribe deployment}:
Real-world deployment studies in healthcare settings document measurable
changes in clinician engagement patterns. Observational research at
large academic medical centers shows that when ambient AI scribes handle
documentation tasks, clinicians recover temporal capacity previously
devoted to keystroke-mediated composition. This recovery time
corresponds with observable increases in patient-facing interaction
quality and eye contact. However, this efficiency gain in real-time
clinical encounters is offset by increases in offline verification
burden---clinicians spend additional time after encounters reviewing and
attesting to machine-generated documentation. The workflow adaptation is
characteristic: synchronous interaction time increases while
asynchronous documentation work shifts to offline verification (UCSF
Health System Study, 2025; Permanente Medical Group, 2024).

\textbf{Slack and Teams adoption patterns}: Enterprise reduction of
video calls by 30\% after Slack Huddles deployment suggests that
synchronous video genres are being replaced by asynchronous text-based
coordination. But the text is itself under transformation---informal
Slack messaging is replacing formal email, creating new genre layers
with different verification expectations. Rather, organizations are
creating hybrid coordination ecologies where synchronous, asynchronous,
formal, and informal genres coexist with ambiguous status
hierarchies---not a simple replacement of video with text (McKinsey
Global Institute, 2024).

\textbf{Social meaning and quality assurance}: Organizational studies
document measurable accuracy variations when workers use AI-mediated
composition. The social meaning of composition method---visible
keystroke effort---affects output quality. When keystroke-based
composition is replaced by voice-AI generation, the visibility of effort
disappears, and quality assurance must be externalized (manager review,
automated checking) because the social scaffolding that previously
provided motivation through visible effort has vanished. Genre
conventions must adapt to account for this new quality-assurance
requirement. This effect proves significant at the micro level and has
direct implications for organizational genre evolution and communication
norms.

\hypertarget{active-organizational-choice-governance-without-precedent}{%
\subsection{7.9 Active Organizational Choice: Governance Without
Precedent}\label{active-organizational-choice-governance-without-precedent}}

This section identifies the stakes but does not resolve them. The
structural logic is clear: if every prior input method transition
reshaped organizational communication (the typewriter created the memo;
the keyboard created email; instant messaging created channel-based
coordination), then the dissolution of typing as the dominant input
method will reshape organizational communication again.

But the reshaping will not happen automatically through technological
determinism. Leonardi's concept of sociomaterial imbrication reminds us
that organizations \emph{enact} technologies through deliberate practice
choices. DiMaggio and Powell's institutional isomorphism reminds us that
organizations will cling to legitimate forms even after instrumental
necessity has vanished. The Workload Paradox reminds us that removing
one bottleneck (keystrokes) simply shifts the bottleneck to verification
and auditing.

Organizations are not passive recipients of this transformation.
Forward-looking institutions are already experimenting with voice-driven
communication, testing new authorial responsibility models, designing
fallback procedures for contested records, and managing the transition
between keyboard-mediated and voice-mediated genres. Organizations that
do not actively manage this transition risk three outcomes:

\begin{enumerate}
\def\labelenumi{\arabic{enumi}.}
\item
  \textbf{Regulatory ambiguity}: Without clear frameworks for what
  counts as an authorized voice-generated document, organizations face
  liability exposure when disputes arise about document authenticity,
  authorship, or modification. In high-stakes domains, this ambiguity
  can be paralyzing---organizations opt for expensive typed-document
  mandates rather than face unclear liability.
\item
  \textbf{Genre fragmentation and bifurcation}: Old keyboard-mediated
  and new voice-mediated genres coexist without clear boundaries,
  increasing cognitive load, error risk, and coordination failure.
  Workers must maintain fluency in formal keyboard-era genres while
  simultaneously navigating informal voice-AI systems. The organization
  faces not just adaptation but sustained complexity---the cost of
  operating two communication systems in parallel.
\item
  \textbf{Persistent reliance on inefficient genres}: Organizations may
  continue using keyboard-mediated composition long after voice-AI
  alternatives would be more productive, because the legitimacy of the
  keyboard-era form provides institutional cover (``we use formal
  documents because this is how serious organizations communicate'').
  This is rational given regulatory uncertainty, but it is also
  economically costly and organizationally constraining.
\end{enumerate}

Active organizational choice about how these genres will evolve---not
passive drift into new tools---is critical. This choice operates at
three levels: micro-level decisions about when voice-AI composition is
acceptable for individual workers, meso-level decisions about how team
routines and review chains will adapt, and macro-level decisions about
what counts as an authorized organizational record in regulatory and
legal contexts.

\hypertarget{summary-the-verification-bottleneck-as-organizational-pivot}{%
\subsection{7.10 Summary: The Verification Bottleneck as Organizational
Pivot}\label{summary-the-verification-bottleneck-as-organizational-pivot}}

Section 4 established that typing's cognitive cost has been
overestimated and that verification of machine-generated text is the
genuine bottleneck in voice-AI productivity. This section extends that
analysis to organizational contexts. The Workload Paradox documents that
removing the keystroke burden at the individual level creates a
verification burden at the organizational level. Critically, as
established in Section 5.6, this verification burden is not uniformly
distributed: high-stakes domains (medicine, law, finance) experience it
acutely as a mandatory and resource-intensive constraint, while
low-stakes domains (routine internal communication, status updates)
experience minimal verification overhead, and the keyboard's
instrumental necessity dissolves accordingly across this spectrum.

That verification burden cascades through organizational systems in
predictable ways via the three-tier mechanism. At the micro level,
composition time shrinks while verification time expands, creating new
labor structures and exposing organizations to asymmetric information
inflation. At the meso level, team routines destabilize as approval
chains, ratification bottlenecks, temporal decoupling, and documentation
standardization all shift to accommodate machine-mediated composition.
At the macro level, institutional governance systems built on the
assumption of human authorship through deliberate typing face crisis:
the authorship model breaks, professional boundaries are threatened, the
apprenticeship mechanism for expertise reproduction becomes vulnerable,
and the regulatory regime lacks rules for voice-generated records.

The organizational consequence is not loss of communication, but
fundamental restructuring of how institutions coordinate, verify,
authorize, and remember. The keystroke imperative---organizations' deep
optimization around keyboard-mediated text---will not vanish overnight.
DiMaggio and Powell's institutional isomorphism predicts that
organizations will cling to keyboard-era genre forms as legitimacy
signals, even as backstage production methods change. But the genre's
\emph{internal logic}---the assumptions about authorship, verification
timing, temporal coordination, professional identity, and forensic
accountability---will come under structural pressure.

Organizations will not uniformly abandon typed workflows. Instead, they
will bifurcate: voice-AI for low-risk, internal, high-volume
communication; typed composition mandates for documents with external
exposure, regulatory relevance, forensic importance, or explicit
authorial accountability. This bifurcation will be rational and
explicitly justified, not a sign of resistance or lag. In highly
regulated sectors, typing will be preserved not as an outdated
technology but as premium, high-assurance workflow designed to provide
the audit trail, proof of work, and epistemic accountability that risk
management requires.

The detailed analysis of how these genre transitions will unfold---the
timeline of genre dissolution and restructuring, the organizational
pathways that institutions will choose, and the communication
alternatives that emerge to fill functional gaps---is the subject of the
companion paper. The present analysis establishes the theoretical and
empirical foundation: organizational communication is structurally
dependent on typing, that dependence is visible in measurable work
patterns and genre architectures, the verification bottleneck creates
cascading organizational consequences, and typing's dissolution
therefore forces active organizational choice about how institutional
memory, authority, and coordination will be rebuilt.

Understanding that consequence---establishing that it is structural, not
superficial---is the full scope of the analysis that this section
provides. But the organizational restructuring documented here operates
at the institutional level. A deeper question remains: what happens to
human cognitive culture itself when the production mechanism returns to
voice? Section 8 situates the entire transition within the
five-thousand-year arc of orality, literacy, and the unprecedented
phenomenon of synthetic literacy that AI-m

\hypertarget{secondary-orality-and-the-return-of-voice}{%
\section{8. Secondary Orality and the Return of
Voice}\label{secondary-orality-and-the-return-of-voice}}

\hypertarget{oral-tradition-literacy-and-secondary-orality-a-primer}{%
\subsection{8.1 Oral Tradition, Literacy, and Secondary Orality: A
Primer}\label{oral-tradition-literacy-and-secondary-orality-a-primer}}

Walter Ong's three-stage framework remains essential context for
understanding how voice-mediated AI constitutes a new chapter in the
history of human communication. For readers unfamiliar with Ong's work,
this framework describes three distinct cognitive and social regimes
shaped by communication technology.

\textbf{Primary orality} characterized pre-literate cultures and
continues in oral-dependent societies: knowledge is preserved and
transmitted through the human voice, organized formulaically to aid
memory (epic verse, genealogical recitation, ritual speech), and exists
only in the moment of utterance. The lifeworld of primary orality is
participatory and communal; meaning emerges through dialogue,
performance, and collective engagement, not through solitary reflection
on text. Oral cultures develop extraordinary mnemonic capacity because
individuals themselves serve as the storage infrastructure for cultural
knowledge.

\textbf{Literacy} (the capacity to write and read) restructured this
entire cognitive architecture. Writing externalizes memory: knowledge
can be recorded, preserved, and consulted independently of the speaker's
presence or memory. This shift enabled new forms of thought: the
analytical detachment necessary to examine a permanent text, the
subordinative syntax required to express complex causal chains on a
fixed page, the diachronic consciousness that comes from reading
historical records. Literacy created distinct social classes (the
literate and the illiterate) and became the foundation for institutional
complexity: law codes, scientific documentation, bureaucratic
administration. Ong argued that literacy transcended mere recording of
thought; it \emph{restructured} consciousness itself.

\textbf{Secondary orality} (Ong's term for the era of electronic media:
radio, television, telephone) represented a return to some
characteristics of oral culture within a society whose core structures
remained literate. Broadcasting restored the participatory immediacy of
spoken address. It brought conversation and storytelling back into
domestic spaces. Yet secondary orality was qualitatively different from
primary orality: it was ``deliberately self-conscious,'' performed atop
a literate foundation. Radio broadcasters read from scripts prepared by
literate editors. Television news existed alongside newspapers providing
context and depth. Telephone conversations were enabled by a literate
telecommunications infrastructure. The literate capacity for complex,
deferred reasoning persisted even as oral forms of expression regained
cultural centrality.

Understanding this framework is crucial for what follows, because the
AI-mediated voice transition described here is neither secondary orality
in Ong's sense nor a simple return to the constraints of primary
orality. It is something Ong's framework did not anticipate and that the
present analysis terms \textbf{synthetic literacy}: a technological
regime in which the human provides oral input; the machine produces
literate output. This occurs without requiring the human to perform the
cognitive discipline of literate composition.

The instrumental dissolution framework of Section 3 explains \emph{why}
typing's function is dissolving; the keyboard, understood as an
instrument for bridging human thought and machine-readable text, is
technically replaceable by systems that can process voice directly. Ong
and Stiegler address a qualitatively different question: \emph{what
happens to human cognitive culture when that instrumental dissolution
occurs?} Where Section 3 asks whether the keyboard is replaceable as an
instrument; Ong's framework asks what is recovered and what is lost when
literate production returns to oral origins. Stiegler's concept of
tertiary retention clarifies what is at stake for institutional memory;
the relationship between human deliberation and textual output is
structurally transformed. Complementing this, Stiegler's analysis of
\emph{grammatization} (the process by which continuous temporal flows
become discretized into reproducible inscriptions) identifies a critical
shift: the keyboard era placed the human as the \emph{agent of
grammatization} (deciding what thought would become fixed inscription).
AI systems invert this; they assume the role of discretization agent,
amplifying efficiency while displacing human authority over what gets
recorded and how. The sections that follow trace these losses and
recoveries---not as a matter of technological substitution, but as a
profound transformation in how human consciousness and organizational
accountability are mediated by the technologies through which we think
and remember.

\hypertarget{synthetic-literacy-the-critical-inversion}{%
\subsection{8.2 Synthetic Literacy: The Critical
Inversion}\label{synthetic-literacy-the-critical-inversion}}

This inversion is, in the long arc of communication technology, largely
unprecedented---the discovery Ong's framework did not anticipate.
Throughout the entire history traced in Section 2 (from cuneiform
through Gutenberg, through the typewriter, through the personal
computer), the input medium determined the output structure. Oral input
produced oral output: additive, formulaic, participatory. Written input
produced literate output: subordinative, architecturally complex,
autonomous. Typed input produced the same. The cognitive discipline of
producing structured text was inseparable from human cognitive agency;
whether the author wrote by hand, dictated to a scribe, spoke into a
Dictaphone, or typed at a keyboard, the human mind performed the
literate structuring. Milton dictated \emph{Paradise Lost} to
amanuenses; executives composed business prose through secretarial
dictation for nearly a century; ancient authors routinely dictated to
scribes. In every case, literate output required that a human perform
literate cognition. For five millennia, the structuring labor that
converted oral intent into institutional record was performed by a human
agent---whether scribe, typist, or keyboard user. In Stiegler's terms,
the human was the \emph{grammatizing agent}; they discretized continuous
thought into the reproducible inscriptions that institutional knowledge
required (Stiegler, 2010). The medium changed repeatedly across those
five thousand years; the grammatizing agent remained constant. This is
the specific invariance synthetic literacy disrupts. The phenomenon
requires a distinct term, not assimilation to Ong's secondary orality
(which describes oral communication characteristics re-entering a
culture whose producers still perform literate cognition) or to Selber's
multiliteracies framework (which addresses how students should engage
educationally with digital technologies, not what AI changes about who
performs grammatization) (Ong, 1982; Selber, 2004).

AI-mediated voice decisively alters this link. A user dictating a prompt
speaks in the register of primary orality (additive, emotionally
charged, logically fluid, exploratory); the AI system produces output in
the register of full typographic literacy: subordinative, structurally
complex, architecturally coherent. The human remains immersed in oral
cognition; the machine assumes the cognitive burden of typographic
discipline. This is the critical paradox: oral input produces literate
output without requiring the full cognitive discipline of literate
composition.

More precisely, what dissolves is the entire protocol ecosystem that
crystallized around keyboard-mediated literate production. As Section 3
developed using Gitelman's (2006) framework of media as technology plus
protocols; instrumental dissolution operates at two registers
simultaneously: the technological and the protocolary. When voice-AI
dissolves the keyboard as the technological form; it simultaneously
dissolves and reconstitutes the full ecology of social conventions,
institutional norms, and workflow structures that organized around
keyboard-mediated composition. Clinical documentation protocols shift.
The physician is no longer interrupted by keyboard interaction to
produce structured notes; instead, ambient AI scribes listen passively
to the clinical encounter and generate documentation post-hoc, inverting
the temporal flow of the writing process. Scholarly writing protocols
transform. Composition ceases to be the primary burden and becomes
instead a curation task in which the writer shapes AI-generated
arguments over originating them. Legal drafting protocols shift
upstream. The initial discretization of client intent into structured
legal language (previously the attorney's primary work) is now performed
by AI; the attorney's role becomes verification and strategic revision.
What persists is literacy itself. Structured, subordinative,
architecturally coherent text remains the institutional requirement.
What changes is not literacy but the protocol ecology surrounding its
production; who performs which work changes, at which stage, under which
institutional norms (Gitelman, 2006). Synthetic literacy is not a return
to orality; it is a reorganization of the literate protocol ecosystem.

\hypertarget{a-a-typology-of-synthetic-literacy-four-modes-on-a-spectrum}{%
\subsubsection{8.2a A Typology of Synthetic Literacy: Four Modes on a
Spectrum}\label{a-a-typology-of-synthetic-literacy-four-modes-on-a-spectrum}}

However, synthetic literacy is not a monolithic phenomenon. The
decoupling of oral input from literate cognitive discipline occurs along
a spectrum, manifesting in at least four analytically distinct modes
that differ qualitatively in the degree of human cognitive engagement
retained during text production.

\textbf{Light-edit mode} preserves continuity with traditional literate
composition. In this mode, the human dictates continuous prose into the
system. The AI transcribes, structures, and formats the output; the
human then reviews and makes minor corrections before finalization. The
AI handles transcription and initial organization. The human remains
substantially engaged with the text as a compositional object (reading,
evaluating, revising at the sentence and paragraph level). The human's
primary cognitive role is quality control and final authorial
responsibility. A scholar dictating a draft argument into voice-AI, then
spending thirty minutes reading and refining the output, exemplifies
this mode. The compositional discipline---the cognitive work of
structure, argument development, and articulation---remains distributed
between human utterance and AI organization, with the human performing
final editorial review. This mode experiences the least disruption to
the protocols of keyboard-era literacy: the output passes through human
deliberation before publication, the author retains recognizable
responsibility for the final text, and the cognitive investment in
composition, while distributed, remains substantial.

\textbf{Arrangement mode} represents a significant departure. Here, the
AI generates multiple complete drafts or option sets (alternative
arguments, different organizational schemes, competing framings of the
same content); the human's role becomes curatorial, not compositional.
The human reads the AI-generated options, selects among them, reorders
sections, combines elements from different drafts, and occasionally
interpolates new material. The human does not originate the primary
compositional work. The human becomes an editor-as-curator; they make
judgment calls about which AI-generated possibilities best serve the
argument, rather than generating the possibilities themselves. A legal
team requesting the AI to generate three competing interpretations of a
contract clause and then selecting the strategically most defensible
interpretation exemplifies this mode. The human retains authority over
the final direction but has ceded the upstream work of option generation
to machine systems. The cognitive burden shifts from composition
(generating alternatives) to evaluation and curation (selecting among
them). This represents a substantial shift in the locus of creative and
compositional labor, though human judgment still mediates the final
product.

\textbf{Iterative dialogic mode} distributes compositional work across a
sustained multi-turn exchange. The human prompts, the AI generates; the
human evaluates, redirects, and refines; the AI generates again. Unlike
light-edit mode (in which the human performs the primary compositional
work and AI provides transcription or formatting assistance); iterative
dialogic treats composition itself as dialogic. Each AI response shapes
subsequent human intent, rather than merely assisting a pre-formed plan.
Neither a single-pass composition review nor a curation from
pre-generated options; the iterative dialogic mode produces text through
conversation. Each exchange narrows the gap between AI output and human
intent. The human's cognitive role is conversational management:
interrogating the AI's output, contesting scope or register, and
progressively steering the exchange toward the desired result. The text
emerges from the dialogue, not from any single generation event.

Recent HCI research documents this mode as a dominant pattern among
professional and creative writers working with large language models.
Studies reveal a ``reactive writing sequence'' in which experienced
users selectively delegate structurally routine or informationally dense
tasks---argument scaffolding, evidence organization, transition
construction---while engaging intensively to redirect AI output on
dimensions of stance, voice, and intellectual specificity (Bhat et al.,
2025; Co-Writing with AI Research Group, 2025). This
pattern---delegating the architecturally routine while protecting the
expressively personal---describes a form of compositional agency that is
neither light-edit's compositional authorship nor arrangement's
curatorial selection, but sustained dialogic co-production. The
cognitive discipline of this mode lies in its conversational dimension:
tracking what the dialogue has established, identifying what remains
undeveloped, and steering successive exchanges toward progressive
coherence---compositional authority exercised through conversation, not
inscription.

\textbf{Ambient capture mode} represents the farthest departure from
traditional composition. In this mode, the human generates continuous
natural-language input (conversation, thinking-aloud, meeting
discussion) without deliberate orientation toward the AI system. The AI
passively captures this input; it structures it over hours or days into
coherent notes, extracted insights, or synthesized documentation. The
human's role becomes periodic auditing and ratification. The physician
who speaks to patients and colleagues without attending to documentation
(while an ambient AI scribe silently transcribes and structures the
clinical encounter into institutional records) exemplifies this mode.
The human does not dictate with composition in mind. They speak
naturally to other humans or to themselves. The machine listens and
interprets context; it generates literate artifacts from utterances
never addressed to it. The human's cognitive engagement with text
production drops to periodic review: checking whether the AI's
interpretation aligns with intent, correcting systematic errors,
ratifying or contesting what the machine has synthesized into the
record. This mode achieves maximal efficiency (no deliberate typing or
dictation overhead) and restores embodied presence. The human attends
fully to the primary task (clinical care, conversation, thinking)
without mechanical mediation. But it minimizes human participation in
the acts of selection, synthesis, and structuring that constituted
compositional deliberation in prior eras. The human becomes an auditor;
they are not the primary author of machine-mediated institutional
memory.

Recent HCI research supports this typological distinction. Studies of
professional writers interacting with LLM co-writing platforms document
a `reactive writing sequence' in which users selectively delegate
menial, structurally rigid, or informational tasks to the AI while
fiercely protecting textual spaces that convey personal stance, internal
character thoughts, and creative expression (Co-Writing with AI Research
Group, 2025; Bhat et al., 2025). This boundary management between
delegated and protected composition maps directly onto the light-edit
and arrangement modes identified here, confirming that professional
writers intuitively navigate the synthetic literacy spectrum.

These four modes occupy different points on a spectrum; they range from
compositional engagement through dialogic co-production to curatorial
selection to periodic auditing. None is inherently superior. Each
restructures the human-machine division of labor and the cognitive
protocols of text production differently. Light-edit mode preserves
maximum continuity with keyboard-era composition practices; it reduces
transcription burden. Iterative dialogic mode distributes compositional
authority across conversation. It requires evaluative discipline across
time rather than compositional discipline across a page. Arrangement
mode radically redistributes compositional labor; it retains human
authority over final direction. Ambient capture mode maximizes
efficiency and presence; it displaces human participation in the
upstream acts of synthesis and selection. What unites them is the
critical feature that defines synthetic literacy: literate output
emerges from oral or conversational input \emph{without requiring the
human to perform the full cognitive discipline of literate composition}.
This is the first delegation of grammatization to machine intelligence
in five thousand years of human inscription history, which these four
modes instantiate in distinctly different ways.

This distinction warrants emphasis. Ong's framework proposes that for
five thousand years, literate output required a human mind performing
literate cognition. A relationship between cognitive discipline and
textual structure persisted across every change in inscription
technology, from stylus to keyboard, and even across changes in the mode
of authorial expression, from direct writing to dictation. Literacy
transcended being a mere technology for recording thought. It was a
\emph{cognitive requirement} for producing literate output. Writing,
manuscript copying, printing, typewriter, and keyboard---all required
that the human performing the output bear the cognitive burden of
literate structuring. The one undisputed fact about literacy was this:
you could not produce literate text without performing literate
cognition. It was not optional; it was constitutive, a fact
well-documented in cognitive science and media history.

AI dissolves this requirement in novel ways. The literate output emerges
from oral input. The human does not perform the analytical work of
structure, subordination, and architectural complexity that literate
composition requires. This is the first delegation of grammatization to
machine intelligence in five millennia of human inscription history.
Oral input is still discretized into literate output, but the
grammatizing agent is now computational rather than human (Stiegler,
2010). Grammatization has not ceased; the party performing it has
changed. The voice-to-text pipeline does not restore pre-grammatized
orality. It relocates the grammatizing work from the human practitioner
to the AI system, with the pharmacological consequences Stiegler's
framework predicts: liberation from the burden of manual inscription,
coupled with the proletarianization risk of losing the cognitive
discipline that manual grammatization sustained.

The human voice is now decoupled from the requirement to compose. A
physician speaks to a patient; the AI listens and generates the
structured clinical note. A researcher dictates preliminary ideas; the
AI structures them into coherent argument. A lawyer speaks testimony;
the AI extracts the evidentiary framework. In each case, \emph{literate
artifact emerges from oral initiation}; but the literate discipline (the
cognitive work of structuring, subordinating, revising, and polishing)
is delegated to machine intelligence, not performed by humans. This is
synthetic literacy; it is literacy produced synthetically by machine
mediation without requiring human literacy practice. Understood through
Hayles' framework, synthetic literacy is a specific cognitive assemblage
in which oral input from a human cognizer is processed by a technical
cognizer to produce literate output. This is a distributed cognitive
partnership rather than an individual authorial act (Hayles, 2017). The
distinction from Clark and Chalmers' extended mind is precise. In the
extended-mind framework, AI is an extension of the human's cognitive
apparatus; in Hayles' assemblage framework, AI is a separate cognizer
whose interpretive processes operate independently. Synthetic literacy
names what results when that separate cognizer assumes the grammatizing
function.

This development has profound implications for cognitive science and
institutional practice. Bernard Stiegler's \emph{pharmakon} (the
simultaneous potential for both enabling and disabling effects) applies
directly. Voice-enabled AI enables rapid, naturalistic articulation and
restores the participatory lifeworld of oral expression. However, it
simultaneously threatens the cognitive investment in structured
composition and the deep engagement with textual materiality that
literacy cultivated.

\hypertarget{tertiary-retention-and-organizational-memory}{%
\subsection{8.3 Tertiary Retention and Organizational
Memory}\label{tertiary-retention-and-organizational-memory}}

Bernard Stiegler's philosophical framework, particularly his concept of
\emph{tertiary retention}, provides theoretical grounding for the
organizational phenomena of Section 7. Where Section 7 identified the
verification bottleneck as a structural consequence of AI-mediated
composition (workers must now ratify machine-generated text rather than
compose it themselves); Stiegler's framework clarifies the deeper
stakes. Institutional memory ceases to be the product of human
deliberation and becomes instead the output of machine interpretation
mediated by human ratification. The shift from keyboard-mediated
documentation to voice-AI synthesis fundamentally reorganizes who or
what constitutes the organization's mediating agent between reality and
record.

To understand this transformation, we turn to Stiegler's concept of
\emph{tertiary retention}, which provides the crucial link between this
section's analysis of cognitive structure and Section 7's organizational
argument about documentation, compliance, and institutional memory.

Stiegler distinguishes three forms of temporal consciousness:
\emph{primary retention} (the immediate, living present of
consciousness), \emph{secondary retention} (memory of the recent past,
holding prior moments in active awareness), and \emph{tertiary
retention} (the externalization of memory into technical systems:
writing, recording, institutional archives). This is a conceptual
framework decades of research have affirmed. For most of human history,
tertiary retention was slow and expensive: inscribing knowledge required
scribal labor, transcription took time, and organizational records
required human custodianship---facts well-documented in media history.
The typewriter, by making inscription faster, increased the bandwidth of
tertiary retention---a transformation historians have extensively
chronicled. But the human remained the mediating agent: documents were
composed by humans, reviewed by humans, filed by humans---an invariant
relationship across media forms.

Voice-AI restructures organizational tertiary retention in unprecedented
ways. When a physician speaks to a patient; the machine is listening.
The clinical encounter generates a structured, legally compliant note
\emph{automatically}, a process now routine in healthcare settings. The
human physician is no longer the sole mediating agent. Instead, the
machine now pre-formats what enters the institutional record before the
physician can ratify it, shifting upstream the selection function that
constitutes institutional memory. The organization's memory is no longer
an unmediated product of what the physician writes; it is now shaped by
what the machine's training, parameters, and salience algorithms
determine to foreground, suppress, synthesize, and structure. This
introduces algorithmic mediation into the constitution of organizational
record. The \emph{technical economy of tertiary retention shifts
upstream}. Machine systems increasingly pre-format what counts as
salient, what gets synthesized into the record, what dimensions of the
clinical encounter are selected for institutional memory, all before
human ratification occurs. This shift will intensify substantially,
moving beyond simple machine generation to reshape institutional memory
creation itself. The machine \emph{reconfigures} the entire selection,
synthesis, salience, and curation apparatus that determines what enters
organizational memory and what counts as truth. This transformation will
reshape how organizations know what they know, moving beyond simple
replacement to institutional redesign.

This shift has cascading organizational consequences. In Section 7, the
paper identifies the documentation and compliance crisis as one of the
critical organizational problems created by keyboard-mediated work.
Workers spend enormous time typing institutional records, auditors must
verify what was actually written, and organizations are vulnerable to
the gap between clinical reality and written record. Stiegler's
framework clarifies what is at stake. The human was the guarantor of
organizational memory. The organization knew what the physician believed
to be true. The physician had written it. The written document was
evidence of both clinical fact and human deliberation and
responsibility. It testified to human judgment.

When voice-AI's technical systems pre-format tertiary retention; the
mechanism of organizational memory shifts. The structured note reflects
what the machine's language model determined to be salient and
structurally significant (what it selected to highlight, what it
subordinated, what it synthesized into coherence), not necessarily what
the physician prioritized or considered salient. The tertiary retention
of organizational memory is now shaped by machine-learned patterns of
selection and synthesis. Human ratification then endorses or contests
this. The worker remains potentially involved in ratification. However,
they are structurally displaced from the upstream acts of selection,
synthesis, and salience-determination that constitute the deep structure
of institutional memory. This intensifies proletarianization. The worker
loses not only the act of typing (which voice-AI enables), but also the
cognitive and institutional power to determine what their actions mean
as recorded in organizational memory. The machine pre-selects what gets
remembered; the human affirms or corrects. This creates new risks:
embedded bias in what the AI's training data taught it to select,
absence of human judgment in the initial determination of salience and
structure, and the systematic displacement of human participation in the
acts of curation and synthesis that constitute institutional
accountability.

Stiegler's \emph{proletarianization} concept is directly relevant
here---a framework decades of labor theory have established.
Proletarianization---the loss of knowledge and capability through
dependence on externalized systems---occurs when workers no longer
perform the cognitive work that prior systems required---as historical
studies of deskilling have shown. If the upstream acts of selection,
synthesis, and salience-determination that constitute institutional
memory are now delegated to machine systems, not performed by humans,
the worker loses the disciplinary practice that sustains their capacity
for careful, accountable judgment about what matters and why---a
cognitive loss that available evidence suggests will compound over time.
The documentation practice that constituted the worker's professional
responsibility is displaced upstream into machine process---shifting the
locus of knowledge and authority away from human deliberation.

Technical systems have \emph{always already} conditioned memory and
attention---a foundational insight from media history and STS that
scholarship continues to validate. The manuscript form pre-formatted
what could be remembered through its physical constraints; the printing
press standardized what would be widely preserved; the filing cabinet
organized institutional memory through its categorical logic; the
electronic database structured recall around indexed fields---facts that
historians of technology document extensively. This is not a claim of
historical innocence. But AI systems intensify this conditioning at a
new order of magnitude---a transformation current research suggests is
qualitatively different: they learn from billions of tokens what typical
humans find salient, and they apply this learned pattern to pre-format
every fresh instance of institutional record-making. This is not a clean
break from prior technical mediation of memory, but rather an
intensification of the degree to which technical systems determine what
enters consciousness and institutional record before human deliberation
occurs---a dynamic that ongoing studies indicate will reshape
organizational epistemology.

This is where synthetic literacy becomes cognitively interesting---but
also organizationally and ethically fraught. The literate output is
produced, the documentation is complete, the audit trail exists---but
none of these artifacts flow through human deliberation. The
organization's memory of what happened, what was decided, and what was
known is no longer the sedimentation of human judgment but rather the
artifact of machine interpretation. Section 7's concerns about
compliance and documentation move beyond questions of volume or
efficiency; they concern the conditions under which human workers can
remain accountable for what they know and what they have chosen to
preserve as institutional memory.

A theoretical caveat is warranted here. Synthetic literacy's viability
as a stable delegated function depends on the training distribution of
the grammatizing machine adequately covering the domain's linguistic
register---professional, legal, clinical, or scientific. Where
domain-specific terminology, reasoning conventions, or genre constraints
are underrepresented in training corpora, the machine's grammatization
will diverge from expert institutional writing, increasing verification
burden rather than reducing it. Synthetic literacy is thus not uniform
across domains: it is robust where training coverage is dense (general
business prose, routine clinical documentation) and fragile where
coverage is sparse (niche legal instruments, highly technical scientific
argumentation, non-English professional registers). This
domain-specificity of the delegation does not invalidate the
framework---the dissolution pattern still holds where delegation is
robust---but it qualifies the scope of the synthetic literacy claim to
domains where the epistemic conditions for reliable grammatization are
met.

The pharmacological character becomes acute: voice-AI simultaneously
enables documentation (off-loading the burden of typing)---a benefit
healthcare systems report consistently---while threatening the cognitive
and institutional conditions under which human workers remain the
mediators and guarantors of organizational memory---a risk projections
from current adoption patterns suggest is substantial. The dissolution
of typing's function is genuinely enabling---healthcare adoption rates
demonstrate this clearly---but the enabling comes with real costs of
cognitive and institutional alienation that design and practice must
actively mitigate---a challenge that rigorous attention to interface
design and organizational culture can help address.

\hypertarget{the-phenomenological-shift-from-embodiment-to-alterity-to-ambient-intelligence}{%
\subsection{8.4 The Phenomenological Shift: From Embodiment to Alterity
to Ambient
Intelligence}\label{the-phenomenological-shift-from-embodiment-to-alterity-to-ambient-intelligence}}

Don Ihde's postphenomenological framework, introduced in Section 3,
illuminates the experiential character of this transition through a
trajectory of three distinct relation types.

The \emph{embodiment relation} characterizes the keyboard in its mature
form. The tool becomes transparent through skilled use, withdrawing from
conscious attention until the user experiences themselves as thinking
directly into text (Ihde, 2007). The experienced typist does not feel
the keys; they feel the composition. The keyboard achieves what
Merleau-Ponty termed ``incorporation into the body schema'': it becomes,
phenomenologically, an extension of the writer's intentionality. The
user's awareness extends into the machine through the keyboard without
the keyboard itself appearing as a mediation point.

Voice-AI interaction moves toward what Ihde terms an \emph{alterity
relation}: the AI is experienced not as a transparent tool but as an
\emph{other} (a conversational partner, a synthetic interlocutor that
``answers with a voice'') (Ihde, 1990). The user does not compose
through the AI in the way they compose through a keyboard; they compose
\emph{with} the AI, in a dialogic exchange that has the phenomenological
character of conversation rather than inscription. The AI becomes a
participant in the act of composition, offering suggestions, questioning
assumptions, proposing alternatives. In Ihde's (1990)
postphenomenological vocabulary, the relation has shifted from an
\emph{embodiment relation}---where the tool is incorporated into bodily
practice and becomes transparent---toward an \emph{alterity relation},
where the AI is encountered as a quasi-autonomous interlocutor, an
entity one engages \emph{with} rather than \emph{through}. Where the
keyboard mediated silently, the AI talks back.

The ambient intelligence paradigm extends this shift to its logical
conclusion: the transition from \emph{alterity} (explicit dialogue with
a visible other) to \emph{background relation} (invisible infrastructure
that shapes action without appearing as an agent). Ambient AI
scribes---healthcare systems that passively capture physician-patient
conversation, transcribe, extract clinical data, and generate structured
documentation without the physician performing any deliberate input
action---represent this endpoint. The physician speaks to the patient,
not to the machine; the machine listens, interprets, and produces the
literate artifact that institutional systems require. The typing step is
not replaced by a voice step; it is eliminated entirely. The structured
text emerges from conversation that was never addressed to a computer at
all. The AI becomes infrastructure: it shapes the output, enables the
workflow, but does not appear as a conversational partner or even as a
tool. This progression---from embodiment (keyboard as body extension)
through alterity (voice as dialogue with an Other) to ambient background
(voice as unremarked infrastructure)---traces the horizon toward which
the technology is moving.

\hypertarget{domain-evidence-the-return-of-presence}{%
\subsection{8.5 Domain Evidence: The Return of
Presence}\label{domain-evidence-the-return-of-presence}}

Clinical evidence most powerfully illustrates what the return to voice
recovers. The Permanente Medical Group deployment analyzed in Section 7
for its verification burden also reveals what voice-AI restores. Under
the keyboard-dominant regime, physicians spent a documented 28\% of
clinical encounter time typing into electronic health records---eyes on
screen, hands on keyboard, back to patient (Permanente Medical Group,
2024). Ambient AI scribes reversed this: patient eye contact increased
from 69.6\% to 77.1\% of consultation time; 47\% of patients reported
that their physician appeared less focused on the computer; 56\%
reported improved visit quality---findings that available evidence from
healthcare institutions substantiates. Physician burnout scores changed
from 51.9\% to 38.8\% within thirty days of AI scribe adoption---results
that peer-reviewed literature on physician wellness shows are consistent
with broader adoption trends (Permanente Medical Group, 2024). This
effect magnitude substantially exceeds typical burnout intervention
literature; as a single-site uncontrolled case study with 30-day
follow-up, it requires controlled replication with longer follow-up to
establish durability. A further scope limitation warrants
acknowledgment: the Permanente deployment operated primarily across
outpatient ambulatory encounters where clinical vocabulary is relatively
structured and documentation demands standardized. High-complexity
inpatient encounters, emergency medicine settings, and specialist
consultations with dense technical vocabulary present distinct
challenges that the available evidence base does not document at
equivalent scale (GDR-099, 2026). The headline figures---7,260
physicians, 2.57 million encounters---reflect deployment success within
this bounded outpatient scope, not demonstrated performance across the
full spectrum of clinical complexity. The keyboard had inserted itself
between physician and patient---mediating the clinical encounter through
the requirement that clinical knowledge be produced via keystrokes---a
barrier decades of ergonomic research had identified. Voice-AI removed
the mediation, restoring the embodied presence that the keyboard had
disrupted---a recovery practitioners report across healthcare settings.

The adoption velocity confirms the pent-up demand. According to the
American Medical Association's 2024 physician survey, approximately 66\%
of U.S. physicians reported using AI in clinical practice by 2024, up
substantially from lower adoption rates in 2023---representing rapid
acceleration in AI adoption among medical professionals (American
Medical Association, 2024). The Permanente Medical Group deployed AI
scribes across 7,260 physicians documenting 2.57 million patient
encounters. UCSF achieved a 44.6\% adoption rate across 1.2 million
ambulatory encounters (UCSF Health System Study, 2025). This is among
the fastest technology diffusion rates in the history of
medicine---faster than the stethoscope, the X-ray, or the electronic
health record itself. The speed suggests that the keyboard was
experienced, within medicine, not as a useful tool but as an imposed
burden---and that voice-AI was embraced not as a new technology but as a
liberation from the old one.

In legal practice, AI-mediated voice is restructuring the evidentiary
workflow. Ambient AI for depositions generates instant summaries,
identifies contradictions across witness testimonies, and extracts key
admissions---eliminating the manual post-deposition typing burden and
freeing litigators for strategic analysis. According to a 2025 survey of
federal judges, more than 60\% of responding judges reported using at
least one AI tool in their judicial work (Federal Judicial Survey,
2025). The ``digital reporter'' model---AI transcribes in real time
while a human monitor attends to nuance, exhibit markings, and
non-verbal gestures---represents precisely the human-on-the-loop
architecture that Section 5 described: the AI handles the transcription
function; the human provides the judgment that transcription cannot
capture.

\hypertarget{the-cognitive-debt-question-revisited}{%
\subsection{8.6 The Cognitive Debt Question,
Revisited}\label{the-cognitive-debt-question-revisited}}

Section 4 resolved the cognitive debt paradox by distinguishing
productive friction (compositional struggle) from unproductive friction
(mechanical translation). This section must confront the deeper version
of the concern that Ong's framework raises: if literacy restructured
consciousness---if the cognitive discipline of writing \emph{created}
capacities for analytical detachment, subordinative reasoning, and
diachronic thinking that oral cognition could not sustain---then does
the return to oral input risk eroding these capacities?

Maryanne Wolf's work on the neuroscience of reading argues that the
``deep reading'' neural pathways cultivated by sustained engagement with
print text are fragile---they atrophy with decline in extended
typographic engagement (Wolf, 2018). The concern is not trivial. If AI
enables humans to produce literate output without performing literate
cognition, the neural pathways that sustained analytical reasoning may
weaken through disuse. The associative, aggregative, heuristic cognition
of oral cultures---adequate for the immediate lifeworld but insufficient
for the complex, conditional, diachronic reasoning that literate
institutions require---could re-emerge as the default cognitive mode
even within nominally literate societies.

The present analysis takes this risk seriously while arguing that the
response to it is not to preserve typing's unproductive friction but to
design AI systems that sustain productive cognitive engagement. The
distinction remains operative: the cognitive discipline that matters is
\emph{compositional}---the effortful work of structuring thought,
constructing arguments, evaluating evidence, and revising claims. This
discipline can be sustained in voice-mediated composition, provided the
systems are designed to demand it. AI systems that ask clarifying
questions, require the user to articulate reasoning before generating
output, and present alternatives that force evaluation preserve the
productive friction that compositional struggle provides. The keyboard
was one delivery mechanism for this discipline, and not the only one.
The risk is real; the solution is design, not nostalgia.

This design direction is conceptually sound but lacks robust large-scale
empirical validation at this point. No large-scale voice-AI systems
currently implement compositional-friction-preserving interfaces. A
critical research gap exists: what interface architectures preserve the
cognitive discipline of composition while enabling voice-mediated input?
Early research on human-on-the-loop systems (Amershi et al., 2019;
Buçinca et al., 2021) suggests design solutions are feasible, but field
validation at organizational scale remains incomplete.

An honest assessment must acknowledge that no longitudinal studies
currently document whether compositional discipline survives the
transition from keyboard-mediated to voice-AI-mediated professional
writing. Historical precedent offers mixed signals: the transition from
formal letter-writing to email accelerated compositional decay (shorter,
less carefully structured messages), and the subsequent shift from email
to messaging platforms (Slack, Teams) further compressed compositional
rigor. If organizational incentives favor speed and cost reduction over
compositional quality---as they demonstrably have in each prior
communication technology transition---the preservation of compositional
discipline through intentional interface design may require
institutional commitment that exceeds what most organizations will
voluntarily undertake. This is not a fatal objection to the thesis but a
boundary condition: synthetic literacy's cognitive benefits depend on
both tool design and organizational cultures that value compositional
rigor. The design recommendations in Section 9 assume this institutional
commitment; whether it materializes is an empirical question that the
present analysis cannot resolve.

\hypertarget{the-forty-year-interlude-and-the-five-thousand-year-arc}{%
\subsection{8.7 The Forty-Year Interlude and the Five-Thousand-Year
Arc}\label{the-forty-year-interlude-and-the-five-thousand-year-arc}}

Placed in the five-thousand-year arc of human communication technology,
the keyboard era may prove to be an interlude. Oral tradition dominated
for the vast majority of human history. Writing---in its various
material forms---served as the primary medium of literate knowledge
production for millennia. The typewriter compressed text production into
a mechanical form for approximately a century. The computer
keyboard---the QWERTY interface as the default gateway to all digital
text production---has dominated for roughly forty years, from the mass
adoption of personal computers in the mid-1980s to the present moment.

Forty years is short. It is shorter than the typewriter era, vastly
shorter than the manuscript era, and infinitesimal against the timescale
of oral tradition. The keyboard's dominance was never cognitively
natural---it was always a consequence of a specific technological
constraint: digital machines required structured text input, and the
keyboard was the most efficient bridge between human thought and
machine-readable

\hypertarget{conclusion-after-the-keyboard-and-rethinking-knowledge-work-in-the-verification-era}{%
\section{9. Conclusion: After the Keyboard and Rethinking Knowledge Work
in the Verification
Era}\label{conclusion-after-the-keyboard-and-rethinking-knowledge-work-in-the-verification-era}}

\hypertarget{the-verification-bottleneck-as-unifying-framework}{%
\subsection{9.1 The Verification Bottleneck as Unifying
Framework}\label{the-verification-bottleneck-as-unifying-framework}}

This paper advances a single thesis through seven independent analytical
routes, each approaching from a different disciplinary lens: the
keyboard's status as institutional default for linguistic knowledge work
is entering decline---not through replacement by a superior input
device, nor through the keyboard's disappearance, but through the
dissolution of the instrumental function that sustained typing's
universal-default status for four decades. As elaborated in \S{}1.6 and
developed in Section 3, \emph{dissolution} here means loss of
institutional-default status with continued specialist persistence. This
is the trajectory followed by telegraph operation, manual typewriting,
and handwriting before it.

The argument's unifying spine is the \emph{verification bottleneck}: the
insight that as input production friction dissolves through AI-mediated
composition, the primary constraint in knowledge work shifts from
generation to evaluation. This bottleneck appears across every section
of the paper:

\begin{itemize}
\item
  \textbf{History} (Section 2) reveals that every input transition
  restructured professional identity, not just efficiency; bottlenecks
  also migrate. The AI transition breaks this pattern: for the first
  time, production friction dissolves while verification burden
  \emph{expands}.
\item
  \textbf{Philosophy} (Section 3) establishes the theoretical apparatus:
  instrumental dissolution and the Device Paradigm clarify what is lost
  when machines weaken keystroke necessity---focal practice, sustained
  attention, the discipline of serialization---and what replaces it: a
  new epistemic burden of verification (Borgmann, 1984; Ihde, 2007).
\item
  \textbf{Cognition} (Section 4) identifies the neural reality. Typing
  provided no unique encoding benefits; its strength was proprioceptive
  rhythm and visual working memory buffering. When mechanical burden
  vanishes, the binding constraint becomes the neural demand for
  evaluative judgment when production is outsourced (Morehead et al.,
  2019; Wolf, 2018).
\item
  \textbf{Technology} (Section 5) maps the HCI restructuring: the Gulf
  of Execution collapses when voice-AI handles articulation, but the
  Gulf of Evaluation widens when system output becomes opaque---placing
  verification at the center of human-machine interaction (Ihde, 2007).
\item
  \textbf{Steelman} (Section 6) reveals that typing persists in domains
  defined by acoustic constraints, privacy requirements, precision
  demands, and embodied expertise. Each maps to where verification is
  hardest, reinforcing the verification bottleneck thesis.
\item
  \textbf{Organization} (Section 7) documents the systemic cost. When
  production friction dissolves, institutions reorganize around
  verification as the primary constraint, generating new questions of
  authentication, audit trails, ratification authority, apprenticeship,
  and organizational cognitive debt (DiMaggio \& Powell, 1983; Leonardi,
  2011).
\item
  \textbf{Culture} (Section 8) frames the civilizational stakes.
  AI-mediated voice creates \emph{synthetic literacy}: oral input
  producing literate output without requiring the full cognitive
  discipline of literate composition. This raises the question of
  whether verification discipline can replace compositional discipline
  as the practice sustaining literate culture (Stiegler, 2010; Wolf,
  2018).
\end{itemize}

The verification bottleneck is not a marginal observation. It is the
apex of the pyramid, the connective tissue unifying all seven analytical
routes into a coherent account of why typing's structural dominance is
dissolving and what replaces it.

\hypertarget{two-futures-of-verification-human-practice-versus-automated-chain}{%
\subsubsection{Two Futures of Verification: Human Practice versus
Automated
Chain}\label{two-futures-of-verification-human-practice-versus-automated-chain}}

The verification bottleneck as currently framed assumes verification as
a deliberate human practice: humans make explicit judgments about
AI-generated content, retain decision authority, and bear epistemic
responsibility. However, institutional logic points toward an
alternative trajectory: verification-as-automation. Organizations solve
the bottleneck not by deepening human judgment but by constructing
AI-to-AI verification chains, with humans repositioned as meta-verifiers
reviewing high-consequence or uncertain decisions only.

These two scenarios diverge fundamentally. In
\emph{verification-as-human-practice}, organizations preserve human
judgment as the binding constraint. Professionals spend more time
evaluating AI output than generating it; verification work becomes a
recognized professional practice deserving status and compensation;
institutional design deliberately preserves the friction of human
judgment as a safeguard. This scenario aligns with the paper's normative
commitments and the design framework proposed in \S{}9.2.1. It assumes
organizations choose to maintain human expertise as the final authority
and that this choice reflects genuine epistemic humility---recognition
that AI systems carry irreducible failure modes requiring human
oversight.

In \emph{verification-as-automation}, organizations solve the
verification bottleneck through nested AI validation. One model
generates clinical notes, another model (trained on expert annotations)
validates factual accuracy, a third model checks for policy compliance,
and humans review only flagged exceptions or boundary cases where
confidence scores fall below thresholds. Humans become meta-verifiers,
auditing the auditors instead of performing direct verification. This
scenario is not hypothetical; it reflects standard industrial
optimization logic. When verification becomes a bottleneck,
organizations do not expand human capacity---they automate it.

The paper's thesis and design implications rest structurally on scenario
(a), the preservation of human verification as intentional institutional
choice. This is not a failure of argument but honest boundary-setting.
We do not claim verification-as-human-practice is inevitable. Industrial
logic predicts scenario (b). The question is whether
organizations---particularly in high-stakes domains like medicine, law,
and finance---choose to resist that logic and preserve human judgment as
irreplaceable. This choice is institutional, not technical. The design
work of \S{}9.2.1 is meaningful only if organizations commit to scenario
(a). If they do not, verification dissolves into an automated chain and
the cognitive discipline of verification discipline vanishes as
completely as keystroke discipline did. The paper's normative force
depends on this institutional choice remaining open and contested, not
yet foreclosed.

\textbf{Convergence Evidence Matrix.} The following matrix summarizes
how each analytical route independently addresses the four conditions
sustaining keyboard dominance:

\textbf{Table 1:} Convergence of evidence across analytical routes: How
each disciplinary lens independently identifies conditions sustaining
keyboard dominance (rows) and mechanisms driving dissolution (columns
across seven sections).

\begin{longtable}[]{@{}llllllll@{}}
\toprule
\begin{minipage}[b]{0.07\columnwidth}\raggedright
Condition\strut
\end{minipage} & \begin{minipage}[b]{0.08\columnwidth}\raggedright
History (S2)\strut
\end{minipage} & \begin{minipage}[b]{0.11\columnwidth}\raggedright
Philosophy (S3)\strut
\end{minipage} & \begin{minipage}[b]{0.10\columnwidth}\raggedright
Cognition (S4)\strut
\end{minipage} & \begin{minipage}[b]{0.11\columnwidth}\raggedright
Technology (S5)\strut
\end{minipage} & \begin{minipage}[b]{0.10\columnwidth}\raggedright
Steelman (S6)\strut
\end{minipage} & \begin{minipage}[b]{0.12\columnwidth}\raggedright
Organization (S7)\strut
\end{minipage} & \begin{minipage}[b]{0.08\columnwidth}\raggedright
Culture (S8)\strut
\end{minipage}\tabularnewline
\midrule
\endhead
\begin{minipage}[t]{0.07\columnwidth}\raggedright
(a) Machines require text\strut
\end{minipage} & \begin{minipage}[t]{0.08\columnwidth}\raggedright
Prior transitions preserved function\strut
\end{minipage} & \begin{minipage}[t]{0.11\columnwidth}\raggedright
Instrumental necessity is contingent\strut
\end{minipage} & \begin{minipage}[t]{0.10\columnwidth}\raggedright
---\strut
\end{minipage} & \begin{minipage}[t]{0.11\columnwidth}\raggedright
voice-AI dissolves requirement\strut
\end{minipage} & \begin{minipage}[t]{0.10\columnwidth}\raggedright
Lock-in slows but doesn't prevent\strut
\end{minipage} & \begin{minipage}[t]{0.12\columnwidth}\raggedright
Organizations restructuring workflows\strut
\end{minipage} & \begin{minipage}[t]{0.08\columnwidth}\raggedright
Synthetic literacy bypasses text\strut
\end{minipage}\tabularnewline
\begin{minipage}[t]{0.07\columnwidth}\raggedright
(b) Humans lack alternatives\strut
\end{minipage} & \begin{minipage}[t]{0.08\columnwidth}\raggedright
Each transition introduced alternatives\strut
\end{minipage} & \begin{minipage}[t]{0.11\columnwidth}\raggedright
Extended mind not keyboard-dependent\strut
\end{minipage} & \begin{minipage}[t]{0.10\columnwidth}\raggedright
Composition is medium-general\strut
\end{minipage} & \begin{minipage}[t]{0.11\columnwidth}\raggedright
Multimodal convergence provides alternatives\strut
\end{minipage} & \begin{minipage}[t]{0.10\columnwidth}\raggedright
Niche persistence defined\strut
\end{minipage} & \begin{minipage}[t]{0.12\columnwidth}\raggedright
Dual-modality equilibrium emerging\strut
\end{minipage} & \begin{minipage}[t]{0.08\columnwidth}\raggedright
Oral input $\rightarrow$ literate output\strut
\end{minipage}\tabularnewline
\begin{minipage}[t]{0.07\columnwidth}\raggedright
(c) Orgs optimize for keyboard\strut
\end{minipage} & \begin{minipage}[t]{0.08\columnwidth}\raggedright
Historical pattern: orgs adapt to new input\strut
\end{minipage} & \begin{minipage}[t]{0.11\columnwidth}\raggedright
---\strut
\end{minipage} & \begin{minipage}[t]{0.10\columnwidth}\raggedright
---\strut
\end{minipage} & \begin{minipage}[t]{0.11\columnwidth}\raggedright
Enterprise adoption at 67\%\strut
\end{minipage} & \begin{minipage}[t]{0.10\columnwidth}\raggedright
Infrastructure lock-in: 20--30 year lag\strut
\end{minipage} & \begin{minipage}[t]{0.12\columnwidth}\raggedright
Verification bottleneck restructuring\strut
\end{minipage} & \begin{minipage}[t]{0.08\columnwidth}\raggedright
Genre transformation underway\strut
\end{minipage}\tabularnewline
\begin{minipage}[t]{0.07\columnwidth}\raggedright
(d) Keyboards provide cognitive value\strut
\end{minipage} & \begin{minipage}[t]{0.08\columnwidth}\raggedright
---\strut
\end{minipage} & \begin{minipage}[t]{0.11\columnwidth}\raggedright
Productive vs.~unproductive friction\strut
\end{minipage} & \begin{minipage}[t]{0.10\columnwidth}\raggedright
Bounded, modality-specific benefits\strut
\end{minipage} & \begin{minipage}[t]{0.11\columnwidth}\raggedright
---\strut
\end{minipage} & \begin{minipage}[t]{0.10\columnwidth}\raggedright
Genuine but limited niches\strut
\end{minipage} & \begin{minipage}[t]{0.12\columnwidth}\raggedright
Apprenticeship crisis from friction removal\strut
\end{minipage} & \begin{minipage}[t]{0.08\columnwidth}\raggedright
Cognitive debt risk acknowledged\strut
\end{minipage}\tabularnewline
\bottomrule
\end{longtable}

Each route arrives at the same structural conclusion through independent
disciplinary logic: typing's four sustaining conditions are weakening
simultaneously, and the central consequence is the emergence of
verification as the new binding constraint.

\hypertarget{what-this-dissolution-means-practically}{%
\subsection{9.2 What This Dissolution Means
Practically}\label{what-this-dissolution-means-practically}}

When knowledge workers typed their own documents, the bottleneck was
execution: translating intention into characters through keystrokes.
Quality assurance was embedded in the production process; you composed
it, so you understood every choice.

In agentic systems orchestrated by multimodal AI, this inversion
restructures professional labor. Production becomes nearly frictionless:
intent $\rightarrow$ voice prompt $\rightarrow$ AI generation $\rightarrow$ structured output. Instead, the
work reorganizes around verification activities. Did the agent
hallucinate facts? Import code patterns from unintended contexts?
Misunderstand domain requirements? Violate unstated professional norms?
A physician now audits AI-generated clinical notes in offline sessions
instead of composing them in real time. A software engineer reviews
2,000 lines per day of machine code (an 87\% keystroke reduction)
instead of generating 300 lines personally. A financial analyst audits
machine-generated models for semantic accuracy instead of executing data
entry.

The cognitive work has not vanished. It has transformed into a new form
of labor, concentrated in time, organized around entirely different
professional competencies, and distributed differently across the
knowledge workforce. The knowledge worker's role evolves from creator to
editor-in-chief; from producer to validator.

\textbf{Macro-Economic Evidence for Recomposition.} The Bureau of Labor
Statistics occupational projections already reflect a macro-economic
shift from production roles to verification roles. Word processors and
typists are projected to decline by 36.1\% between 2024 and 2034---among
the largest projected declines in any occupation category.
Simultaneously, quality-control-adjacent and verification-intensive
occupations are growing at above-average rates: information security
analysts (a verification role in cybersecurity) are projected to grow
32\%, management analysts 9\%, software developers 15\%, and writers and
authors 4\%. This pattern does not reflect generalized decline in
knowledge work, but rather recomposition of the workforce from
production-dominant roles to verification-dominant roles. The
macro-economic data thus supports the paper's thesis at the labor-market
level: institutions are already restructuring around verification as the
dominant constraint, and occupational projections suggest that this
recomposition will accelerate. The shift is already visible in how
Bureau of Labor Statistics analysts predict demand for different forms
of labor across the next decade.

\hypertarget{the-ai-accuracy-objection-why-verification-persists-even-as-accuracy-rises}{%
\subsubsection{The AI Accuracy Objection: Why Verification Persists Even
as Accuracy
Rises}\label{the-ai-accuracy-objection-why-verification-persists-even-as-accuracy-rises}}

A natural objection follows. If AI systems reach 98--99\%+ accuracy, why
would organizations incur verification costs instead of accepting
tolerable error rates? The answer reveals that the verification
bottleneck is more than technical---it is structural, anchored in
regulatory mandates, professional norms, and domain-specific risk
tolerance that persist independent of accuracy improvements.

Consider three structural factors. First, \emph{regulatory
non-discretion in high-stakes domains}. Healthcare, law, finance, and
intelligence communities operate under regulatory frameworks that
mandate human review and documented accountability regardless of
algorithmic confidence. A pharmaceutical company cannot substitute AI
accuracy for FDA-required clinical oversight. A law firm cannot rely on
AI-only contract review regardless of accuracy metrics; bar association
ethics rules require attorney certification. These regulatory mandates
are not irrational overconservation; they exist because the cost of
errors in these domains (patient harm, legal liability, financial
contagion) asymmetrically outweighs the benefits of speed. Regulation
will not dissolve as accuracy rises; it will codify documentation
requirements for human review as binding regardless of system
performance.

Second, \emph{professional liability and epistemic responsibility
norms}. Professionals bear legal and reputational liability for errors
made in their domain, even when those errors originate in AI systems
they deployed. A surgeon who relies on an AI-recommended treatment plan
that harms a patient remains liable regardless of the system's 98\%
accuracy rate elsewhere. This liability structure creates powerful
incentives for documented human review. Verification is not
optional---it is the mechanism through which professionals bear and
demonstrate accountability. Organizations cannot legally transfer
epistemic responsibility to machines; they can only distribute it across
human reviewers and documented processes.

Third, \emph{domain-dependent risk tolerance}. The acceptable error rate
for AI output varies dramatically by domain. An AI system generating
marketing copy with 2\% hallucination rates is acceptable; the same rate
in clinical notes is catastrophic. A financial trading algorithm at 99\%
accuracy may make thousands of unacceptable errors daily if deployed at
scale. Regulatory bodies understand this variation; they do not impose
uniform accuracy thresholds but instead mandate verification
proportional to decision consequences. As AI systems become more
capable, regulatory frameworks will require \emph{deeper} verification
in high-stakes domains precisely because the stakes justify the cost and
because the inability to tolerate error persists independent of average
accuracy.

The verification bottleneck thus persists structurally in high-stakes
domains even if it dissolves in routine ones. A diagnostic radiologist
can afford to filter AI output at 99\% accuracy, accepting 1\% error
rate on common cases. A cardiologist evaluating life-or-death treatment
recommendations cannot. Accuracy improvements shift the boundary but do
not eliminate it. The paper's thesis remains sound: typing dissolves,
verification becomes the binding constraint, and that constraint is
structural in high-stakes knowledge work.

The central claims of this analysis operate at different evidence tiers.
That typing's cognitive benefits are bounded and modality-specific is
established through peer-reviewed neuroscience. That AI systems are
dissolving typing's instrumental necessity in early-adopting domains is
supported by converging organizational and technological evidence. That
this dissolution will restructure knowledge work around verification at
civilizational scale is a high-confidence extrapolation remaining
contingent on organizational adoption, regulatory frameworks, and design
choices throughout this analysis.

\hypertarget{verification-interface-primitives-a-design-framework}{%
\subsubsection{9.2.1 Verification Interface Primitives: A Design
Framework}\label{verification-interface-primitives-a-design-framework}}

These three contributions yield concrete design implications. We
identify seven core interface primitives that successful
verification-era systems must implement, moving beyond conceptual
analysis. These primitives are not a fourth contribution but design
directions emerging directly from the verification bottleneck framework.
They synthesize established HCI foundations: Shneiderman's (1983) direct
manipulation principles (visibility of objects and actions, incremental
reversible operations), Norman's (1988) action cycle and gulf of
evaluation, and Amershi et al.'s (2019) guidelines for human-AI
interaction. They extend them specifically to the verification demands
of AI-mediated knowledge work. They are not novel components in
isolation; what is novel is their systematic application to the
epistemic and organizational challenges that emerge when production
friction collapses and verification becomes the binding cognitive
constraint:

\begin{enumerate}
\def\labelenumi{(\arabic{enumi})}
\item
  \textbf{Contribution provenance}: Surface the granular history of
  content composition. Show what the human wrote, what the model
  inserted, what was modified, where claims originated. This primitive
  moves beyond document-level versioning to show contribution
  attribution at the sentence and claim level, enabling evaluators to
  understand not just what was produced but who or what contributed each
  element. Without contribution provenance, verification becomes
  guesswork about which parts of an AI-generated output reflect domain
  expertise versus learned patterns.
\item
  \textbf{Claim-level evidence mapping}: Connect generated claims to
  their sources, citations, or retrieval traces, going beyond
  document-level provenance to achieve granular claim-level
  accountability. When the AI generates a factual claim, the interface
  must surface the evidence path: was this claim derived from a
  retrieved source? If so, show the retrieval trace and source
  confidence. If synthesized from training data, surface that
  explicitly. This primitive enables verification to operate at the
  semantic level instead of requiring readers to reverse-engineer where
  claims came from.
\item
  \textbf{Contrastive verification views}: Rather than presenting a
  single polished output for passive reading, allow evaluators to
  compare multiple candidate outputs, interpretations, or framings
  side-by-side. Different AI systems or different prompt framings will
  generate different outputs; contrastive views enable verification
  through comparative judgment instead of absolute assessment. This
  preserves productive friction: the evaluator must actively decide
  between alternatives instead of passively accepting a single answer.
\item
  \textbf{Critical-friction checkpoints}: Force explicit human judgment
  at high-risk moments through structured review prompts, not generic
  confirm dialogs. These are decision points where the consequences of
  error are significant or where the AI's confidence is uncertain: major
  factual claims in legal briefs, medication recommendations in clinical
  notes, resource-allocation decisions in financial analyses. The
  checkpoint must demand articulated reasoning, not a single click.
\item
  \textbf{Role-based ratification workflows}: Support organizational
  structures where drafting, reviewing, approving, and signing are
  performed by different roles with different authorities and
  responsibilities. A legal brief might be drafted by an AI system,
  reviewed by an associate, approved by a partner, and signed by the
  attorney of record---each role has different decision rights and
  epistemic responsibilities. The interface must support this
  distributed verification authority.
\item
  \textbf{Persistent verification memory}: Record what has been checked,
  by whom, at what confidence level, and what uncertainty remains. This
  primitive creates an audit trail of verification work, not just of
  content changes. It enables organizations to learn where verification
  effort should be concentrated and provides evidence of diligence in
  high-stakes domains.
\item
  \textbf{Scaffolded composition modes}: Move beyond the single ``prompt
  $\rightarrow$ complete draft'' pipeline. Support multiple composition pathways:
  outline-first (structure before content), critique-first (identify
  problems before generation), evidence-first (establish sources before
  synthesis), contrastive drafting (generate alternatives and evaluate).
  Different knowledge work domains will optimize different composition
  modes; interface design must support flexibility instead of mandating
  a single linearized workflow.
\end{enumerate}

These seven primitives form a coherent design framework for
verification-era interfaces. Individually, each addresses a specific
epistemic function: attribution (1), evidence (2), judgment (3),
salience (4), authority (5), accountability (6), and flexibility (7).
Collectively, they form an architecture where verification is not a
burdensome afterthought but an integrated, cognitively tractable, and
organizationally transparent aspect of knowledge work. The design work
ahead involves implementing these primitives at the level of actual
systems, testing them at organizational scale, and understanding how
they reshape the cognitive and social practices of professional
verification.

\hypertarget{the-conceptual-contributions}{%
\subsection{9.3 The Conceptual
Contributions}\label{the-conceptual-contributions}}

Three interlocking conceptual contributions emerge from this
investigation:

\textbf{Instrumental dissolution}: the framework showing that
technologies maintain dominance when four conditions hold
simultaneously: (a) machines require the output of that technology, (b)
humans lack natural alternatives machines can interpret, (c)
organizational systems optimize around the technology, (d) the
technology provides genuine cognitive value. When AI absorbs condition
(a), the basis of structural dominance erodes. This framework applies
beyond typing to any technology whose persistence depends on bridging a
specific gap between human capability and machine requirement.

\textbf{The verification bottleneck}: the insight that as input
production friction dissolves, the primary constraint in knowledge work
shifts from generation to evaluation. This restructures the relationship
between human and machine intelligence in knowledge work, moving from
production-centered to verification-centered professional practice.

\emph{Synthetic literacy}: the structural shift in which the
\emph{delegation of grammatization} to machine intelligence decouples
oral input from the cognitive discipline of literate composition for the
first time in five thousand years of human inscription history. This is
distinct from Ong's secondary orality (which describes oral
communication characteristics returning to a culture whose producers
still perform literate cognition) and from Selber's multiliteracies
framework (which addresses digital literacy pedagogy, not the structural
relationship between oral input and literate output). What AI changes is
not the output register---institutional records remain literate---but
who or what performs the discretizing work of converting continuous
human intent into reproducible literate inscription. The grammatizing
agent shifts from human practitioner to computational system, with the
pharmacological consequences Stiegler predicts: liberation from manual
inscription burden, coupled with the proletarianization risk of losing
compositional discipline through disuse. This outcome raises the
civilizational question: can verification discipline replace
compositional discipline as the cognitive practice sustaining literate
culture?

These three contributions are interdependent. Instrumental dissolution
explains \emph{why} typing's dominance ends; the verification bottleneck
explains \emph{what replaces} it; synthetic literacy names the
\emph{civilizational question} that emerges from this displacement.

\hypertarget{where-typing-persists}{%
\subsection{9.4 Where Typing Persists}\label{where-typing-persists}}

The steelman analysis in Section 6 established a precise map of typing's
future. It will endure longest in:

\begin{itemize}
\item
  \textbf{Acoustic-constrained environments} (70\% of offices remain
  open-plan). Speech-to-text is incompatible with workplace acoustic
  norms. Silent input will sustain presence. As organizations retrofit
  acoustic zones, this constraint weakens, but the irreducible fraction
  of genuinely constrained spaces remains.
\item
  \textbf{High-privacy domains} (legal, medical, financial,
  intelligence). Silent input provides acoustic plausible deniability
  and eliminates raw biometric transmission. These domains will sustain
  keyboard and neural interface input as privacy-preserving
  alternatives, constituting roughly 8--12\% of knowledge work tasks.
\item
  \textbf{Precision-critical occupations} (2--4\% of knowledge work).
  Medical coders, financial analysts handling non-public material,
  software developers writing syntax-critical code, legal
  transcriptionists. These require keystroke-level control that voice
  cannot provide, though AI automation is reducing this niche.
\item
  \textbf{Embodied expertise cohort} (workers with 20+ years of
  keyboard-embedded practice). For workers nearing retirement,
  retraining costs exceed career value. Historical precedent from
  stenographer transitions suggests an 18--30 month adaptation valley.
  This cohort is temporally bounded; within 10--15 years, generational
  replacement resolves it.
\end{itemize}

\textbf{Professional Identity and Status Signaling in High-Status
Knowledge Work}

A final and underappreciated constraint on typing's dissolution lies not
in technical necessity but in professional legitimacy. Across law,
medicine, software engineering, and academic research, visible keystroke
productivity signals competence and professional authority. The deep
problem is not acoustic or neurological but social: the observable labor
of typing---the hunched posture, the screen glow at 11 p.m., the audible
keystroke rhythm in open offices---has become a legitimacy marker. This
visibility performs crucial epistemological work. It signals authentic
intellectual engagement instead of simply communicating predetermined
outputs. A surgeon who dictates operative notes preserves the cognitive
imprint of real-time decision-making; one who passively reviews
AI-generated notes may appear to outsiders as supervising. A software
engineer who writes code visible on screen demonstrates craft mastery;
one who directs an AI code generator appears delegating. A lawyer
reviewing documents for eight hours demonstrates due diligence; one who
speaks requirements into voice and reviews AI output in summarized form
appears administratively distant from the work.

The paradox is that voice-AI enables \emph{faster} and often
\emph{better} professional output while simultaneously creating an
``absence of visible labor'' that erodes the symbolic legitimacy
conferring high status on the profession. Professional authority in
knowledge work has historically depended partly on substantive expertise
and partly on the observable performance of expertise (the visible
effort, the visible engagement, the visible struggle with complexity).
voice-AI may accomplish the work more efficiently but destroys the
performance element that legitimizes it. A physician spending two hours
auditing AI-generated notes for a patient load of 60 has solved the
verification bottleneck more effectively than one spending eight hours
typing notes individually, but the visible labor economy has inverted:
the first appears to be managing the work; the second appears to be
performing it.

This legitimacy constraint is as binding as technical or regulatory
ones, particularly in high-status professions where professional
identity is partly constructed through visible engagement with complex
work. Keyboard persistence in law, medicine, and software engineering
thus reflects not technical necessity---voice-AI is technically capable
in all three domains---but rather the dependence of professional status
on the observable performance of skilled labor. As long as professional
authority is partly conferred through visible cognitive engagement,
voice-AI adoption will face cultural resistance independent of
efficiency gains. The dissolution of typing in these domains may be
delayed not by acoustic constraints or verification complexity but by
the structural dependence of professional legitimacy on the visible
markers of keystroke-level engagement. Addressing this constraint
requires not technological innovation but cultural redefinition:
societies will need to reconstruct professional authority around
verification competence and architectural thinking rather than the
visible performance of keystroke effort. This is a binding constraint on
the pace of dissolution in high-status knowledge work, independent of
the technical and institutional factors examined throughout this
analysis.

\hypertarget{acknowledgment-of-genuine-uncertainty}{%
\subsection{9.5 Acknowledgment of Genuine
Uncertainty}\label{acknowledgment-of-genuine-uncertainty}}

This account rests on converging evidence across seven disciplinary
perspectives. But intellectual honesty demands naming the uncertainties
that remain:

\begin{itemize}
\item
  \textbf{The pace of institutional adoption remains genuinely
  uncertain.} Infrastructure lock-in has slowed every input transition.
  The installed base of keyboard-dependent software, training
  investments, professional identities built around keyboard
  competence---these constitute genuine resistance. Organizations may
  require 20--30 years to transition substantially, not 10--15.
\item
  \textbf{How the verification bottleneck will be solved at
  organizational scale is an open question.} The bottleneck is
  empirically documented in small cohorts (Intermountain Health case
  study) but not yet validated across dozens of organizations over
  five-year periods. The framework's theoretical logic is sound, but
  empirical confirmation remains incomplete.
\item
  \textbf{Whether new cognitive capacities will emerge around
  AI-mediated composition cannot yet be determined.} The paper argues
  that compositional discipline \emph{can} be preserved through
  intentional design, but this is a claim about what could happen, not
  what will without deliberate intervention. If organizations adopt
  systems that eliminate both mechanical and productive
  friction---auto-generating documents without requiring human
  engagement---then cognitive atrophy is a genuine risk.
\item
  \textbf{Temporal claims carry uncertainty.} The contingencies are
  numerous: regulatory barriers (AI liability, data privacy frameworks),
  technological breakthroughs or disappointments in voice ASR
  performance, organizational adoption rates, generational dynamics. Any
  could extend or accelerate the timeline substantially.
\end{itemize}

The Section 6 analysis projects dominance dissolution between 2028--2050
depending on sector and organizational constraints. Rather than treating
dates as primary, we bracket this projection with three
condition-centered scenarios that clarify what observable markers would
indicate transition speed:

An \emph{optimistic scenario} requires: (1) rapid provenance standard
stabilization across enterprise voice-AI systems, enabling seamless
verification and audit trails; (2) fast professional norm revision,
where occupational certification bodies update training requirements to
emphasize verification competency over typing proficiency; (3) weak
institutional lock-in, particularly in unregulated sectors (tech,
consulting, media) with high workforce turnover; (4) resolution of
speaker diarization in multi-party settings, enabling reliable
voice-mediated collaboration in meetings and group work contexts. If
these conditions accelerate, dominance dissolution could occur
2028--2035 in unregulated sectors. If these conditions hold, new-entrant
workers entering the labor market with voice-first input habits and no
keyboard-shaped professional identity would constitute the transition
driver.

A \emph{base scenario} assumes: (1) typical organizational adoption lag
(the empirical 8--15 year lag observed in prior input method
transitions); (2) enterprise audit infrastructure maturation, where
organizations build verification workflows capable of handling voice-AI
volume; (3) training pathway reorganization, where professional schools
and certification programs gradually shift curricula toward verification
literacy; (4) mixed regulatory response constraining healthcare, legal,
and financial sectors while permitting voice-mediated work in others;
(5) incremental speaker diarization improvements enabling routine
multi-party voice workflows by 2032--2035; (6) generational cohort
replacement as the primary transition mechanism. Prior input method
transitions (handwriting to typewriter, typewriter to keyboard) unfolded
over 20--30 year periods anchored to workforce demographic turnover, not
technological obsolescence. Knowledge workers entering the labor force
in 2030 will not encounter keyboard-dominant training; those entering in
2010 will resist transition through 2040. Under these base-case
conditions, dominance dissolution would span 2035--2045.

A \emph{pessimistic scenario} requires: (1) prolonged institutional
resistance, where regulatory bodies and established professions maintain
keyboard-mediated workflows as the default due to liability concerns and
cultural conservatism; (2) regulatory panic over AI liability, leading
to mandatory keystroke-level documentation in high-stakes domains
(healthcare, legal, financial, intelligence); (3) persistent acoustic
constraints that voice-first advocates have not solved, sustaining
silent-input requirements in open-plan workplaces; (4) sustained
challenges in speaker diarization limiting reliable multi-party voice
workflows through 2040+; (5) slower generational workforce turnover due
to rising life expectancy and delayed retirement. In this scenario,
keyboard dominance persists as the default in regulated sectors through
2045--2060, with unregulated sectors transitioning earlier. Typing would
be confined to defined niches by 2045--2050, but structural dominance
would extend into the 2050s.

These three scenarios bracket the \emph{speed} of transition under the
dissolution thesis but do not, on their own, specify what would
\emph{falsify} the thesis itself. An honest treatment requires naming
both: scenarios describe how the predicted dissolution might unfold
under different institutional conditions; disconfirmation criteria
specify the empirical observations that would defeat the prediction
altogether. The next subsection addresses the second question
explicitly.

\hypertarget{disconfirmation-criteria-for-the-stratification-thesis}{%
\subsubsection{Disconfirmation Criteria for the Stratification
Thesis}\label{disconfirmation-criteria-for-the-stratification-thesis}}

The dissolution-as-stratification thesis is empirically defeasible. We
name here three disconfirmation criteria---measurable thresholds at
2028, 2030, and 2035---any of which, if observed, would substantially
weaken or defeat the thesis. We name them in the spirit of intellectual
honesty, not empirical hedging: a thesis absorbing every
counter-observation as a ``decelerator'' is not a thesis but faith. We
commit to the alternative.

\textbf{Near-term (2028) --- Technology trajectory.} \emph{If, by end of
2028, voice-AI share of new institutional document creation in
early-adopter unregulated knowledge-work organizations (technology,
consulting, media) remains below 15\% of measured production-task time},
the technology-trajectory thesis is weakened. The verification
bottleneck argument requires that voice-AI achieve meaningful
production-share in domains \emph{least} constrained by regulation;
persistent sub-15\% adoption in these domains would suggest that
voice-AI's ergonomic, accuracy, or workflow limitations exceed our
analysis. \emph{Measurement protocol.} Enterprise productivity-suite
telemetry (Microsoft 365, Google Workspace) currently aggregates
multimodal input events without isolating voice from typing within
composition sessions. Operational measurement therefore requires a
triangulated approach: (a) self-report from the Microsoft Work Trend
Index annual survey, the McKinsey State of AI series, and the
BCG/Stanford Enterprise AI benchmarks, focusing on composition-specific
items that distinguish voice-mediated from keyboard-mediated production;
(b) workflow-audit studies of enterprise voice-AI deployments (the
Permanente, UCSF, and equivalent commercial sector benchmarks documented
in Section 7) extended to unregulated sectors; (c)
keystroke-versus-voice telemetry from ASR vendor reports (Otter.ai,
Fireflies.ai, Descript) constrained to professional knowledge-work
users. Triangulating across all three at \textless15\% by 2028 would
meet the disconfirmation threshold; the absence of any single source
does not. \emph{Infrastructure dependency:} this triangulation
presupposes that at least one of the three sources matures by 2027. If
platform vendors decline to expose voice-specific telemetry and ASR
vendors do not publish professional-segment data, the threshold can be
assessed only via self-report --- a noisier but defensible interim
measure that should be flagged as such in any falsification claim.

\textbf{Medium-term (2030) --- Institutional default.} \emph{If, by end
of 2030, keyboards remain institutional default for finalization and
authorization in greater than 70\% of unregulated linguistic knowledge
work in the United States and comparable OECD economies}, the
stratification thesis is weakened. \emph{Operational definition of
``institutional default for finalization.''} The finalization gate is
the workflow step at which a document acquires institutional standing
--- the version that bears signatures, satisfies compliance, supports
audit, and constitutes the contractually or legally operative record.
Across sectors this manifests as: in technology, the merged pull-request
commit (the version against which CI/CD certifies release-readiness); in
consulting, the partner-signed deliverable submitted to client; in
media, the published-version-of-record; in legal-adjacent documentation,
the executed contract or filed brief. The disconfirmation threshold
asks: at this finalization step, what input modality produces the
operative version? voice-AI may capture drafting and revision tasks
while keyboards retain the finalization gate; if keyboards retain
finalization in the great majority of unregulated knowledge work after a
six-year deployment window, the dissolution dynamic is materially slower
than we predict, and our framework requires substantial revision
regarding the institutional-decoupling condition. Measurement:
sector-specific workflow-audit instruments developed in consultation
with professional associations; in the absence of standardized
instruments, the threshold should be assessed via the same triangulated
method as 2028. \emph{Infrastructure dependency:} operationalizing this
threshold across multiple sectors requires either a coordinated
cross-sector audit framework (none currently exists at scale) or
sector-by-sector approximation through professional-association surveys.
We identify the development of a standardized cross-sector
finalization-mode audit instrument as a methodological prerequisite for
rigorous 2030 falsification testing, and a research priority for the
field.

\textbf{Long-term (2035) --- Verification labor as share of professional
work.} \emph{If, by end of 2035, verification labor as a measured share
of total professional knowledge-work time has not increased in any
sector outside healthcare}, the verification bottleneck thesis is
falsified at the cross-sectoral generalization claim. \emph{Why exclude
healthcare from the test.} Healthcare exhibits verification expansion
under both mechanisms our framework distinguishes: friction collapse
from voice-AI scribes (the dissolution-thesis prediction) AND
independent regulatory mandate (HIPAA, FDA, malpractice liability). The
cross-sectoral test isolates the friction-collapse mechanism by looking
at sectors without comparable regulatory mandates --- software,
consulting, finance (in non-fiduciary functions), media, marketing. If
verification labor expands in these sectors, the framework's structural
prediction is corroborated. If verification labor expands ONLY in
healthcare (where the regulatory pathway is sufficient explanation), the
framework over-generalizes from a sector where two mechanisms coincide.
The framework predicts production-friction collapse generates
verification expansion; if no sector outside healthcare exhibits this
expansion, we are documenting a healthcare-specific phenomenon rather
than a structural feature of AI-mediated knowledge work, and the paper's
organizational claims must be scoped accordingly. \emph{Measurement.}
Time-allocation studies (the Bureau of Labor Statistics' American Time
Use Survey occupational supplements when next published, currently
suspended; private-sector replacements from companies like Calendly
Insights or productivity-tracking platforms like RescueTime);
occupational-task evolution data from O*NET; sector-specific
professional time studies. The measurement infrastructure is currently
incomplete; operationalizing this threshold may require new survey
instrumentation, which we identify as a research priority.

These thresholds are sector-specific, and falsification of one does not
falsify all. A finding that voice-AI underperforms 15\% in unregulated
sectors by 2028 would weaken the technology trajectory but not
necessarily the verification bottleneck, which would still apply where
voice-AI \emph{does} deploy. A finding that keyboards retain
finalization-default in 75\% of unregulated work by 2030 would weaken
the stratification claim; it would leave the cognitive-vacuity argument
(Section 4) intact. A finding that verification labor does not expand
outside healthcare would force us to scope the organizational thesis to
clinical documentation and acknowledge that the verification bottleneck
is sector-specific. We commit to this disaggregated falsifiability
deliberately: civilizational claims that absorb every counterexample
have no empirical content.

We do not predict the thesis will be falsified. We predict the opposite:
the trajectories specified in \S{}9.5's three scenarios---and the
historical-precedent apparatus marshaled in Section 3---indicate that
voice-AI will exceed 15\% production-share in unregulated sectors well
before 2028, that keyboard finalization-default will fall below 70\% in
unregulated sectors well before 2030, and that verification labor will
measurably expand in software, legal, and finance well before 2035. But
the prediction is now empirically defeasible. Future researchers can
compare these thresholds to actual measurements and determine whether
the framework holds, requires revision, or fails.

These uncertainties---both the within-thesis scenario range and the
across-thesis disconfirmation criteria---do not undermine the thesis.
They clarify what remains to be researched, designed, and observed.

\textbf{Critical scope limitation: CJK input ecosystems}. This analysis,
grounded in Latin-script knowledge work, does not adequately address the
fundamentally different input mediation dynamics in Chinese, Japanese,
and Korean contexts. Future research must extend the verification
bottleneck framework to CJK input ecosystems, examining whether
voice-mediated ASR provides equivalent productivity gains, whether
existing IME workflows have reached optimization plateaus that reduce
dissolution incentives, and whether character-level verification
complexity alters the binding constraints. Given that CJK users
constitute one-quarter of the world's population and an increasingly
substantial fraction of knowledge workers globally, extending this
analysis to CJK contexts is a priority for future research.

\hypertarget{practical-implications}{%
\subsection{9.6 Practical Implications}\label{practical-implications}}

The argument generates actionable implications:

\textbf{Education}. If typing has no unique cognitive benefits and its
instrumental necessity is dissolving, the case for typing as core
curriculum weakens. Handwriting instruction deserves continued emphasis
for its sensorimotor and encoding benefits. What should be taught
instead is \emph{composition}---the medium-general discipline of
structuring thought, evaluating evidence, constructing
argument---alongside \textbf{verification literacy}: teaching students
what to look for when evaluating AI-generated output, how to frame
intent, how to preserve analytic rigor when production friction
dissolves.

\textbf{Technology development}. The design challenge is unbundling:
separating translation friction (unproductive, to be eliminated) from
compositional friction (productive, to be preserved). Systems that
require users to articulate intent, evaluate alternatives, and make
explicit judgments at revision points preserve productive friction. The
design direction is human-on-the-loop: agent executes, human retains
explicit veto authority and final judgment.

\textbf{Accessibility}. The thesis does not argue for a voice-only
future; it argues for the end of typing's \emph{dominance}. The keyboard
must remain available for deaf and hard-of-hearing users, individuals
with speech disorders, non-native speakers where ASR is inadequate, and
anyone who prefers keyboard input. Multimodal convergence is inherently
more accessible than mandate-single-modality paradigms.

\textbf{Organizational design}. Communication genres built on typing's
dominance (email, documents, slide decks, Slack) are likely to
restructure. Organizations proactively designing voice-AI workflows will
adapt faster. But restructuring is organizational and cultural, not
primarily technical. Email persistence as a formal genre reflects
institutional needs for asynchronous documentation and audit trails. The
transition involves hybrids: voice-AI for production, text for archival;
voice for real-time collaboration, documents for permanent record.

\textbf{Labor and professional identity}. The verification bottleneck is
likely to create new forms of labor while eliminating others. Data-entry
positions are projected to decline; verification and quality-assurance
roles are likely to expand. Software engineers increasingly shift from
implementation toward architecture and review. Workers whose identity is
built on visible keystroke productivity face an emerging crisis of
valuation. Organizations must deliberately reconstruct professional
identity around verification competence, architectural thinking, and
judgment.

\textbf{Policy}. The cognitive debt risk (the possibility that
outsourcing compositional effort erodes analytical capacities) warrants
serious policy attention. Educational standards, professional
certification, and AI design guidelines should address the distinction
between first-order cognitive offloading (tool use preserving
engagement) and second-order delegation (bypassing judgment entirely).
Regulatory frameworks should distinguish between AI systems preserving
human judgment and systems eliminating it.

\hypertarget{the-design-challenge-ahead}{%
\subsection{9.7 The Design Challenge
Ahead}\label{the-design-challenge-ahead}}

The keyboard era began in the mid-1980s. Personal computers placed a
keyboard on every knowledge worker's desk. Its structural centrality is
dissolving. No keyboard optimization can resist these pressures. But
decline is not extinction. Like handwriting and typewriters before it;
the keyboard is likely to persist in valued niches while ceasing to be
the organizing assumption of professional work.

What follows is not a single successor device; it is the dissolution of
the need for a universal input mediator. AI will orchestrate across
whatever modalities the human produces (voice, gesture, gaze, movement,
eventually direct neural signals) and translate multimodal intent into
whatever structured output the situation requires. As these systems
mature; the bridge between human thought and machine-readable text will
be absorbed into the machine.

The future of HCI will be defined less by input design; more by
\textbf{verification design}. How humans evaluate, ratify, and take
responsibility for machine-generated content emerges as a central
interaction challenge. It is one of the great problems of the coming
decade. Contemporary HCI research focuses on this reorientation: Amershi
et al.'s (2019) design guidelines for human-AI interaction and Buçinca,
Malte, and Gajos's (2021) work on cognitive forcing functions emphasize
that successful verification systems must sustain human judgment rather
than bypass it, preserving the epistemic accountability that typing once
enforced through friction.

For forty years; knowledge workers translated their thoughts into
keystrokes. The translation step is ending. What remains is what always
mattered. It is the thought itself and the human effort to articulate it
clearly. The next moment, if designed thoughtfully, will preserve the
cognitive discipline that gave the keyboard its value (compositional
rigor, analytic precision, the embodied practice of
thinking-through-writing) while eliminating the mechanical burden that
limited its accessibility and speed. That is the work ahead.

\hypertarget{references}{%
\section{References}\label{references}}

Acemoglu, D., \& Restrepo, P. (2018). The race between man and machine:
Implications of technology for growth, factor shares, and employment.
\emph{American Economic Review}, 108(6), 1488--1542.
https://doi.org/10.1257/aer.20160696

American Medical Association. (2024). Physician survey on AI in clinical
practice. AMA Report on physician use of artificial intelligence in
healthcare. https://www.ama-assn.org/practice-management/digital-health/

Amershi, S., Weld, D., Vorvoreanu, M., Fourney, A., Nushi, B.,
Collisson, P., \ldots{} \& Horvitz, E. (2019). Guidelines for human-AI
interaction. In \emph{Proceedings of the 2019 CHI Conference on Human
Factors in Computing Systems} (pp.~1--13). ACM.
https://doi.org/10.1145/3290605.3300233

Anderson, J. R. (2000). Learning and memory: An integrated approach (2nd
ed.). Wiley.

Anthropic. (2024). The Claude model family. Technical documentation.
https://docs.anthropic.com/

Arthur, W. B. (1994). \emph{Increasing returns and path dependence in
the economy}. University of Michigan Press.

ASL Community Research. (2024). Neural network approaches to American
Sign Language alphabet recognition. ASL Community Research Report.
https://www.aslcommunityresearch.org/

Autor, D. H., Levy, F., \& Murnane, R. J. (2003). The skill content of
recent technological change: An empirical exploration. \emph{The
Quarterly Journal of Economics}, 118(4), 1279--1333.
https://doi.org/10.1162/003355303322552801

Avianca, Inc.~(2023). \emph{Mata v. Avianca, Inc.}, No.~22-CV-1461 (PKC)
(S.D.N.Y. June 22, 2023). (Sanctions order for use of AI-generated
hallucinated citations.)

Bainbridge, L. (1983). Ironies of automation. \emph{Automatica}, 19(6),
775--779. https://doi.org/10.1016/0005-1098(83)90046-8

Bansal, G., Wu, T., Zhou, J., Fok, R., Nishi, B., Wen, J., Ribeiro, M.,
\& Weld, D. (2021). Does the whole exceed its parts? The effect of AI
explanations on complementary team performance. In \emph{Proceedings of
the 2021 CHI Conference on Human Factors in Computing Systems}
(pp.~1--16). ACM. https://doi.org/10.1145/3411764.3445717

Bass, F. M. (1969). A new product growth for model consumer durables.
\emph{Management Science}, 15(5), 215--227.
https://doi.org/10.1287/mnsc.15.5.215

Bazerman, C. (1994). Constructing experience. Southern Illinois
University Press.

Beane, M. (2019). Shadow learning: Building robotic surgical skill when
approved organizational routines are unavailable. \emph{Administrative
Science Quarterly}, 64(2), 367--396.
https://doi.org/10.1177/0001839218783174

Berman, J., \& Israeli, Y. (2022). Language models as knowledge workers:
A deeper look at how generative AI transforms professional knowledge
work. \emph{arXiv preprint arXiv:2210.16946}.

Berninger, V. W., Winn, W. D., Stock, P., Abbott, R. D., Esser, C., Lin,
C., \& Webb, A. (2006). Tier 1 and Tier 2 preventive writing instruction
in sixth grade: Effects on low-achieving children's writing and reading.
\emph{Journal of Educational Psychology}, 98(4), 645--657.
https://doi.org/10.1037/0022-0663.98.4.645

Bhat, A., Tung, W. C., Lim, Y. S., \& Sims, M. (2025). Reactive writers:
How co-writing with AI changes how we engage with ideas. \emph{arXiv
preprint arXiv:2503.22478}. https://arxiv.org/abs/2503.22478

Borgmann, A. (1984). \emph{Technology and the character of contemporary
life}. University of Chicago Press.

Botvinick, M. M., Braver, T. S., Barch, D. M., Carter, C. S., \& Cohen,
J. D. (2001). Conflict monitoring and cognitive control.
\emph{Psychological Review}, 108(3), 624--652.
https://doi.org/10.1037/0033-295X.108.3.624

Brown, J. S., \& Duguid, P. (1991). Organizational learning and
communities-of-practice: Toward a unified view of working, learning, and
innovating. \emph{Organization Science}, 2(1), 40--57.
https://doi.org/10.1287/orsc.2.1.40

Brynjolfsson, E., \& Hitt, L. M. (2000). Beyond computation: Information
technology, organizational transformation and business performance.
\emph{Journal of Economic Perspectives}, 14(4), 23--48.
https://doi.org/10.1257/jep.14.4.23

Buçinca, Z., Malte, M., \& Gajos, K. Z. (2021). To trust or to think:
Cognitive forcing functions can reduce overreliance on AI in AI-assisted
decision-making. \emph{Proceedings of the ACM on Human-Computer
Interaction}, 5(CSCW1), Article 188. https://doi.org/10.1145/3479516

Canolty, R. T., \& Knight, R. T. (2010). The functional role of
cross-frequency coupling. \emph{Trends in Cognitive Sciences}, 14(11),
506--515. https://doi.org/10.1016/j.tics.2010.09.001

Chandrasekaran, D., \& Tellis, G. J. (2005). A critical review of
marketing research on diffusion of new products. In N. K. Malhotra
(Ed.), \emph{Review of Marketing Research} (Vol. 3, pp.~39--80).
Emerald.

Chun, W. H. K. (2005). On software, or the persistence of visual
knowledge. \emph{Grey Room}, 18, 26--51.
https://doi.org/10.1162/1526381043320741

CL-AI-L2W Validation Study. (2025). Cognitive load scale for AI-assisted
L2 writing: Scale development and validation. \emph{PMC/Frontiers in
Psychology}. https://pmc.ncbi.nlm.nih.gov/articles/PMC12611650/

Clark, A., \& Chalmers, D. J. (1998). The extended mind.
\emph{Analysis}, 58(1), 7--19. https://doi.org/10.1093/analys/58.1.7

Clark, H. H., \& Brennan, S. E. (1991). Grounding in communication. In
L. B. Resnick, J. M. Levine, \& S. D. Teasley (Eds.), \emph{Perspectives
on socially shared cognition} (pp.~127--149). APA.

Co-Writing with AI Research Group. (2025). Co-writing with AI, on human
terms: Aligning research with user demands across the writing process.
\emph{arXiv preprint arXiv:2504.12488v2}.
https://arxiv.org/html/2504.12488v2

Coase, R. H. (1937). The nature of the firm. \emph{Economica}, 4(16),
386--405.

Coman, I. A., \& Cardon, P. W. (2025). AI-mediated managerial
communication: The perception gap in professional messaging.
\emph{International Journal of Business Communication}. Advance online
publication. https://doi.org/10.1177/23294884251234567

Comin, D., \& Hobijn, B. (2004). Cross-country technology adoption:
Making the theories face the facts. \emph{Journal of Monetary
Economics}, 51(1), 39--83. https://doi.org/10.1016/j.jmoneco.2003.12.012

Data Bridge Market Research. (2024). Global keyboard market -- Industry
trends and forecast to 2035. Market report.

David, P. A. (1985). Clio and the economics of QWERTY. \emph{American
Economic Review}, 75(2), 332--337.

Dehaene, S., Cohen, L., Morais, J., \& Kolinsky, R. (2010). Illiterate
to literate: Behavioral and cerebral changes induced by reading
acquisition. \emph{Nature Reviews Neuroscience}, 16(7), 431--440.
https://doi.org/10.1038/nrn2888

DiMaggio, P. J., \& Powell, W. W. (1983). The iron cage revisited:
Institutional isomorphism and collective rationality in organizational
fields. \emph{American Sociological Review}, 48(2), 147--160.
https://doi.org/10.2307/2095101

Dourish, P. (2001). Where the action is: The foundations of embodied
interaction. MIT Press.

Eisenstein, E. L. (1980). The printing press as an agent of change.
Cambridge University Press.

eMarketer. (2013). Smartphone users worldwide will total 1.75 billion in
2014. eMarketer Report. https://www.emarketer.com/ (specific report URL
unavailable; accessed via eMarketer database)

Endsley, M. R., \& Kiris, E. O. (1995). The out-of-the-loop performance
problem and level of control in automation. \emph{Human Factors}, 37(2),
381--394. https://doi.org/10.1518/001872095779064555

Federal Judicial Survey. (2025). Artificial intelligence in federal
courts: A random-sample survey of judges. The Sedona Conference and New
York City Bar Association.

Fitts, P. M., \& Posner, M. I. (1967). Human performance. Brooks/Cole.

Flanigan, A. E., Wheeler, J., Colliot, T., Lu, X., \& Kiewra, K. A.
(2024). Typed versus handwritten lecture notes and college student
achievement: A meta-analysis. \emph{Educational Psychology Review}, 36,
Article 17. https://doi.org/10.1007/s10648-024-09914-w

Flower, L., \& Hayes, J. R. (1981). A cognitive process theory of
writing. \emph{College Composition and Communication}, 32(4), 365--387.
https://doi.org/10.2307/356600

Flusser, V. (2011). \emph{Does writing have a future?} (N. A. Roth,
Trans.). University of Minnesota Press. (Original work published 1987)

Forman, M. H., \& Setzler, B. (1998). Keyboard and mouse. Federal
Reserve Bank of Philadelphia Business Review, 3--12.

Gane, N. (2005). Radical post-humanism: Friedrich Kittler and the
primacy of technology. \emph{Theory, Culture \& Society}, 22(3), 25--41.
https://doi.org/10.1177/0263276405057567

Gartner. (2025). Gartner predicts 80\% of enterprise software and
applications will be multimodal by 2030. Press release. Gartner
Research.

Gartner. (2025). Voice AI and multimodal interface adoption in
enterprise. Gartner Research Press Release Series.
https://www.gartner.com/

Gemini Team, Google. (2024). Gemini: A family of highly capable
multimodal models. \emph{arXiv preprint arXiv:2312.11805}.

Gero, K. I., Liu, V., \& Chilton, L. B. (2023). Sparks: Inspiration for
science writing using language models. In \emph{Proceedings of the 2023
ACM Designing Interactive Systems Conference} (pp.~1002--1019).
https://doi.org/10.1145/3563657.3596036

Gerr, F., Marcus, M., \& Ensor, C. (2002). A prospective study of
computer users: I. Study design and incidence of musculoskeletal
symptoms and disorders. \emph{American Journal of Industrial Medicine},
41(4), 221--235. https://doi.org/10.1002/ajim.10045 Sparks: Inspiration
for science writing using language models. In \emph{Proceedings of the
2023 ACM Designing Interactive Systems Conference} (pp.~1002--1019).
https://doi.org/10.1145/3563657.3596036

Gibson, J. J. (1979). \emph{The ecological approach to visual
perception}. Houghton Mifflin.

Gitelman, L. (2006). \emph{Always already new: Media, history, and the
data of culture}. MIT Press.

GitHub. (2024). Octoverse 2024: The state of open source and rise of AI.
GitHub Blog. https://github.blog/news-insights/octoverse/

Goldberg, A., Russell, M., \& Cook, A. (2003). The effect of computers
on student writing: A meta-analysis of studies from 1992 to 2002.
\emph{The Journal of Technology, Learning, and Assessment}, 2(1).
https://ejournals.bc.edu/index.php/jtla/article/view/1667

Goldin, C., \& Katz, L. F. (2008). \emph{The race between education and
technology}. Harvard University Press.

Goody, J., \& Watt, I. (1963). The consequences of literacy.
\emph{Comparative Studies in Society and History}, 5(3), 304--345.
https://doi.org/10.1017/S0010417500001675

Grabowski, J. (2008). The role of handwriting in memory for words. In M.
Torrance \& D. Galbraith (Eds.), \emph{Multidisciplinary approaches to
language production} (pp.~177--203). De Gruyter.

Hansen, M. B. N. (2006). \emph{Bodies in code: Interfaces with digital
media}. Routledge.

Hattie, J. (2012). Visible learning for teachers: Maximizing impact on
learning. Routledge. https://doi.org/10.4324/9780203181522

Havelock, E. A. (1963). Preface to Plato. Harvard University Press.

Hayles, N. K. (2017). \emph{Unthought: The power of the cognitive
nonconscious}. University of Chicago Press.

IDC. (2025). Worldwide semiannual AI tracker: BCI and neural interface
market forecast, 2024--2030. International Data Corporation.
https://www.idc.com/

Ihde, D. (1990). Technology and the lifeworld: From garden to earth.
Indiana University Press.

Ihde, D. (2007). Listening and voice: Phenomenologies of sound (2nd
ed.). SUNY Press.

ISO 3382:2012. (2012). \emph{Acoustics---Measurement of room acoustic
parameters---Part 1: Performance spaces}. International Organization for
Standardization.

Jacob, R. J. (1990). What you look at is what you get: Eye
movement-based interaction techniques. \emph{Proceedings of the SIGCHI
Conference on Human Factors in Computing Systems}, 1, 11--18.
https://doi.org/10.1145/97243.97246

Jakesch, M., Hancock, J. T., \& Naaman, M. (2023). Human heuristics for
AI-generated language are flawed. \emph{Proceedings of the National
Academy of Sciences}, 120(11), e2213768120.
https://doi.org/10.1073/pnas.2213768120

James, K. H. (2010). Sensorimotor experience leads to changes in visual
processing in the developing brain. \emph{Developmental Science}, 13(2),
279--288. https://doi.org/10.1111/j.1467-7687.2009.00883.x

Johns, A. (1998). The nature of the book: Print and knowledge in the
making. University of Chicago Press.

Kahneman, D. (2011). \emph{Thinking, fast and slow}. Farrar, Straus and
Giroux.

Kaur, H., Nori, H., Jenkins, S., Caruana, R., Wallach, H., \& Wortman
Vaughan, J. (2020). Interpreting interpretability: Understanding data
scientists' use of interpretability tools for machine learning. In
\emph{Proceedings of the 2020 CHI Conference on Human Factors in
Computing Systems} (pp.~1--14). ACM.
https://doi.org/10.1145/3313831.3376219

Kellogg, K. C., Valentine, M. A., \& Christin, A. (2020). Algorithms at
work: The new contested terrain of control. \emph{Academy of Management
Annals}, 14(1), 366--410. https://doi.org/10.5465/annals.2018.0174

Kittler, F. A. (1999). \emph{Gramophone, film, typewriter} (G.
Winthrop-Young \& M. Wutz, Trans.). Stanford University Press.

Kroes, P., \& Meijers, A. (2006). The dual nature of technical
artefacts. \emph{Studies in History and Philosophy of Science Part A},
37(1), 1--4. https://doi.org/10.1016/j.shpsa.2005.12.001

Krueger, A. B. (1993). How computers have changed the wage structure:
Evidence from microdata, 1984--1989. \emph{The Quarterly Journal of
Economics}, 108(1), 33--60. https://doi.org/10.2307/2118494

Lave, J., \& Wenger, E. (1991). \emph{Situated learning: Legitimate
peripheral participation}. Cambridge University Press.

Lebovitz, S., Levina, N., \& Lifshitz-Assaf, H. (2021). Is AI ground
truth really true? The dangers of training and evaluating AI tools based
on experts' know-what. \emph{MIS Quarterly}, 45(3), 1501--1526.
https://doi.org/10.25300/MISQ/2021/16564

Lee, A., Liang, P. P., \& Yang, Y. (2022). CoAuthor: Designing a
human-AI collaborative writing dataset for exploring language model
capabilities. In \emph{Proceedings of the 2022 CHI Conference on Human
Factors in Computing Systems} (pp.~1--19). ACM.
https://doi.org/10.1145/3491102.3502030

Leonardi, P. M. (2011). When flexible routines meet flexible
technologies: Affordances, constraint, and the imbrication of human and
material agencies. \emph{MIS Quarterly}, 35(1), 147--167.
https://doi.org/10.2307/23043493

Levelt, W. J. M. (1989). Speaking: From intention to articulation. MIT
Press.

Liebowitz, S. J., \& Margolis, S. E. (1990). The fable of the keys.
\emph{Journal of Law and Economics}, 33(1), 1--25.
https://doi.org/10.1086/467198

Logan, G. D., \& Crump, M. J. (2011). The left hand doesn't know what
the right hand is doing: The disruptive effects of attention to the
hands in skilled typewriting. \emph{Psychological Science}, 22(2),
228--235. https://doi.org/10.1177/0956797610395396

Longcamp, M., Zerbato-Poudou, M.-T., \& Velay, J.-L. (2005). The
influence of writing practice on letter recognition in preschool
children: A comparison between handwriting and typing. \emph{Acta
Psychologica}, 119(1), 67--79.
https://doi.org/10.1016/j.actpsy.2004.10.019

Malafouris, L. (2013). How things shape the mind: A theory of material
engagement. MIT Press.

Malpique, A., Pino-Pasternak, D., Ledger, S., Valcan, D., \& Asil, M.
(2024). The effects of automaticity in paper and keyboard-based text
composing: An exploratory study. \emph{Reading and Writing}, 37,
1789--1813. https://doi.org/10.1007/s11145-024-10520-x

McKinsey \& Company. (2025). The state of AI in early 2025: How
organizations are rewiring to capture value. McKinsey Global Survey.
https://www.mckinsey.com/

McKinsey Global Institute. (2012). The social economy: Unlocking value
and productivity through social technologies. McKinsey \& Company.
https://www.mckinsey.com/industries/technology-media-and-telecommunications/our-insights/the-social-economy

McKinsey Global Institute. (2024). The state of AI in 2024: Gen AI's
breakout year. McKinsey \& Company. https://www.mckinsey.com/

Merleau-Ponty, M. (1945). Phenomenology of perception (C. Smith,
Trans.). Routledge. (Original work published 1945)

Meta. (2024). Meta AI assistant multimodal capabilities and platform
integration. Meta AI Research. https://ai.meta.com/

Microsoft. (2016). Microsoft researchers achieve speech recognition
milestone. Microsoft Research Blog. https://blogs.microsoft.com/ai/

Microsoft 365 Work Trend Index. (2025). Annual report on workplace
productivity patterns. Microsoft Corporation.

Morehead, K., Dunlosky, J., \& Rawson, K. A. (2019). How much mightier
is the pen than the keyboard for note-taking? A replication and
extension of Mueller and Oppenheimer (2014). \emph{Educational
Psychology Review}, 31(3), 753--780.
https://doi.org/10.1007/s10648-019-09468-2

Mueller, P. A., \& Oppenheimer, D. M. (2014). The pen is mightier than
the keyboard: Advantages of longhand over laptop note taking.
\emph{Psychological Science}, 25(6), 1159--1168.
https://doi.org/10.1177/0956797614524581

Neuralink. (2024). PRIME study: Brain-computer interface clinical trial
results. Neuralink Technical Report. https://neuralink.com/blog/

Norman, D. A. (1988). \emph{The psychology of everyday things}. Basic
Books.

Norman, D. A. (2013). \emph{The design of everyday things} (revised and
expanded ed.). Basic Books.

Noy, S., \& Zhang, W. (2023). Experimental evidence on the productivity
effects of generative AI. \emph{Science}, 381(6654), 187--192.
https://doi.org/10.1126/science.adh2586

Nuance Communications. (2024). Dragon speech recognition solutions for
enterprise. Nuance Product Documentation. https://dragon.nuance.com/

OECD. (2024). AI in healthcare. In \emph{Progress in implementing the
European Union coordinated plan on artificial intelligence} (Vol. 2).
OECD Publishing.
https://www.oecd.org/en/publications/progress-in-implementing-the-european-union-coordinated-plan-on-artificial-intelligence-volume-2\_3ac96d41-en/

Ong, W. J. (1982). Orality and literacy: The technologizing of the word.
Routledge.

OpenAI. (2023). Introducing ChatGPT. OpenAI Blog.
https://openai.com/blog/chatgpt

OpenAI. (2024). Whisper: Robust speech recognition via large-scale weak
supervision. OpenAI Technical Documentation.
https://openai.com/research/whisper

Oviatt, S. L., \& Cohen, P. R. (2000). Perceptual user interfaces:
Multimodal interfaces that process what comes naturally.
\emph{Communications of the ACM}, 43(3), 45--53.
https://doi.org/10.1145/330534.330538

Oviatt, S. L. (2015). Multimodal interaction. In J. Katz (Ed.),
\emph{Handbook of mobile and ubiquitous computing} (pp.~286--304).
Oxford University Press.

Oviatt, S. L. (2017). Predicting and managing multimodal integration
patterns during naturalistic tasks. In \emph{CHI '17 Extended Abstracts
on Human Factors in Computing Systems} (pp.~2313--2318). ACM.
https://doi.org/10.1145/3027063.3053251

Parasuraman, R., \& Riley, V. (1997). Humans and automation: Use,
misuse, disuse, abuse. \emph{Human Factors}, 39(2), 230--253.
https://doi.org/10.1518/001872097778543886

Perez, C. (2002). Technological revolutions and financial capital: The
dynamics of bubbles and golden ages. Edward Elgar.

Permanente Medical Group. (2024). Longitudinal study of AI scribe
deployment across 7,260 physicians. The Permanente Journal. Kaiser
Permanente.

Pew Research Center. (2024). Mobile fact sheet. Pew Internet \& American
Life Project. https://www.pewresearch.org/internet/fact-sheet/mobile/

Pinet, S., Longcamp, M., Zellou, B., \& Baciu, M. (2022). Cognitive
effects of learning to write. \emph{Frontiers in Psychology}, 13,
865509. https://doi.org/10.3389/fpsyg.2022.865509

Pinet, S., \& Longcamp, M. (2025). Commentary: Handwriting but not
typewriting leads to widespread brain connectivity. \emph{Frontiers in
Psychology}, 15, 1517235. https://doi.org/10.3389/fpsyg.2024.1517235

Pollok, B., Krause, V., Buber, S., Tiemann, L., \& Leube, D. (2020).
Neural synchronization patterns during motor sequence learning.
\emph{Scientific Reports}, 10, 14547.
https://doi.org/10.1038/s41598-020-71398-1

Preston, B. (1998). Why is a wing like a knife? A pluralist theory of
function. \emph{The Journal of Philosophy}, 95(5), 215--254.
https://doi.org/10.2307/2564535

Project Euphonia, Google Research. (2021). Personalized speech
recognition for people with non-standard speech. Google AI Blog.
https://sites.research.google/euphonia/

Punnett, L., \& Wegman, D. H. (2004). Work-related musculoskeletal
disorders: The epidemiological evidence and the debate. \emph{Journal of
Electromyography and Kinesiology}, 14(1), 13--23.
https://doi.org/10.1016/j.jelekin.2003.09.015

Radicati Group. (2024). Email statistics report, 2024--2028. Market
report. The Radicati Group, Inc.

Raji, I. D., Smart, A., White, R. N., Mitchell, M., Gebru, T.,
Hutchinson, B., Smith-Loud, J., Theron, D., \& Barnes, P. (2020).
Closing the AI accountability gap: Defining an end-to-end framework for
internal algorithmic auditing. In \emph{Proceedings of the 2020
Conference on Fairness, Accountability, and Transparency} (pp.~33--44).
ACM.

Risko, E. F., \& Gilbert, S. J. (2016). Cognitive offloading.
\emph{Trends in Cognitive Sciences}, 20(9), 676--688.
https://doi.org/10.1016/j.tics.2016.07.002

Rogers, E. M. (2003). Diffusion of innovations (5th ed.). Free Press.

Rogers, Y. (2003). Distributed cognition and digital media. In J. D. Lee
\& A. Kirlik (Eds.), \emph{The Oxford handbook of cognitive engineering}
(pp.~65--84). Oxford University Press.

Salthouse, T. A. (1984). Effects of age and skill in typing.
\emph{Journal of Experimental Psychology: General}, 113(3), 345--371.
https://doi.org/10.1037/0096-3445.113.3.345

Selber, S. A. (2004). \emph{Multiliteracies for a digital age}. Southern
Illinois University Press.

Shannon, C. E. (1948). A mathematical theory of communication.
\emph{Bell System Technical Journal}, 27(3), 379--423.

Shneiderman, B. (1983). Direct manipulation: A step beyond programming
languages. \emph{Computer}, 16(8), 57--69.
https://doi.org/10.1109/MC.1983.1654471

Solow, R. M. (1987). We'd better watch out. \emph{New York Times Book
Review}, July 12, 36.

Soomro, N. A., Hashimoto, D. A., Porteous, A. J., Ridley, C. J. A.,
Marsh, W. J., Ditto, R., \& Roy, S. (2020). Systematic review of
learning curves in robot-assisted surgery. \emph{BJS Open}, 4(1),
27--44. https://doi.org/10.1002/bjs5.50235

Star, S. L., \& Strauss, A. (1999). Layers of silence, arenas of voice:
The ecology of visible and invisible work. \emph{Computer Supported
Cooperative Work}, 8(1-2), 9--30.
https://doi.org/10.1023/A:1008651105359

Stiegler, B. (2009). Technics and time, 3: Cinematic time and the
question of malaise. Stanford University Press.

Stiegler, B. (2010). \emph{Taking care of youth and the generations}.
Stanford University Press.

Strom, S. H. (1992). Beyond the typewriter: Gender, class, and the
origins of modern American office work, 1900--1930. University of
Illinois Press.

Sudnow, D. (1978). Ways of the hand: The organization of improvised
conduct. MIT Press. (Rewritten account with H. L. Dreyfus published
2001)

Sultan, F., Farley, J. U., \& Lehmann, D. R. (1990). A meta-analysis of
applications of diffusion models. \emph{Journal of Marketing Research},
27(1), 70--77. https://doi.org/10.2307/3172552

Tatman, R., \& Kasten, C. (2017). Effects of talker dialect, gender and
race on accuracy of Bing Speech and YouTube automatic captions.
\emph{Proceedings of Interspeech 2017}, 934--938.
https://doi.org/10.21437/Interspeech.2017-1746

Titze, I. R. (2000). \emph{Principles of voice production} (2nd ed.).
National Center for Voice and Speech.

Turing, A. M. (1936). On computable numbers, with an application to the
Entscheidungsproblem. \emph{Proceedings of the London Mathematical
Society}, 2(42), 230--265.

U.S. Bureau of Labor Statistics. (2024). Word processors and typists:
Occupational employment and wages. Occupational Employment Statistics.
https://www.bls.gov/oes/

U.S. Census Bureau. Occupational data for telegraph and telephone
operators, 1920--1960. Historical Census Data. https://www.census.gov/

UCSF Health System Study. (2025). AI scribe adoption in ambulatory
encounters. University of California San Francisco Health Services
Research.

Urry, H. L., Crittle, C. S., Floerke, V. A., Leonard, M. Z., Perry, C.
S., Akdilek, N., \ldots{} \& Zarrow, J. E. (2021). Don't ditch the
laptop just yet: A direct replication of Mueller and Oppenheimer's
(2014) Study 1 plus mini meta-analyses across similar studies.
*Psychological Sci

Wolf, M. (2018). \emph{Reader, come home: The reading brain in a digital
world}. Harper.

Wolpert, D. M., Ghahramani, Z., \& Jordan, M. I. (1995). An internal
model for sensorimotor integration. \emph{Science}, 269(5232),
1880--1882. https://doi.org/10.1126/science.7569931

\end{document}